\documentclass[a4paper,10pt]{article}
\usepackage[a4paper,includeheadfoot,margin=2.54cm]{geometry}
\usepackage[utf8]{inputenc}
\usepackage[affil-it]{authblk}

\usepackage{graphicx}
\usepackage{placeins}
\usepackage{epstopdf}
\usepackage{todonotes}
\usepackage[hyphens]{url}
\usepackage{hyperref}
\usepackage{lineno}
\usepackage{upgreek}
\usepackage{units}
\usepackage{multirow}
\usepackage{array}
\usepackage{amssymb}
\usepackage{amsmath}
\usepackage{textcomp}
\usepackage{pdfpages}
\usepackage{booktabs}
\usepackage{lscape}
\usepackage{rotating}
\usepackage{xspace}
\usepackage{tablefootnote}
\usepackage{mathtools}
\usepackage{enumitem}
\usepackage[export]{adjustbox}
\usepackage{float}

\usepackage{caption}
\usepackage{subcaption}
\usepackage{grffile}

\hypersetup{
    colorlinks=true,
    linkcolor=blue,
    filecolor=blue,
    urlcolor=blue,
    citecolor=blue
}

\usepackage[notlof,notlot,nottoc]{tocbibind}

\newcommand{\wrap}[1]{\ensuremath{#1}\xspace}
\newcommand{\ifb}{\wrap{\textrm{fb}^{-1}}}

\newcommand{\pt}{\wrap{p_\textrm{T}}}
\newcommand{\lumi}{\wrap{\mathcal{L}}}
\newcommand{\Whizard}[1][]{\textsc{Whizard}\ifthenelse{\not\equal{#1}{}}{\,#1}{}\xspace}
\newcommand{\Pythia}[1][]{\textsc{Pythia}\ifthenelse{\not\equal{#1}{}}{\,#1}{}\xspace}
\newcommand{\Tauola}{\wrap{\textsc{Tauola}}}
\newcommand{\CirceTwo}{\wrap{\textsc{Circe}2}}
\newcommand{\GuineaPig}{\wrap{\textsc{Guinea-Pig}}}
\newcommand{\MarlinReco}{\wrap{\textsc{MarlinReco}}}
\newcommand{\GeantFour}{\wrap{\textsc{Geant}4}}
\newcommand{\DDFourhep}{\wrap{\textsc{DD4hep}}}
\newcommand{\sqrts}{\wrap{\sqrt{s}}}
\newcommand{\kfold}{\wrap{k\textrm{-fold}}}
\newcommand{\BR}{\wrap{\textrm{BR}}}
\newcommand{\CLs}{\wrap{\textrm{CL}_s}}

\newcommand{\Hss}{\wrap{h\rightarrow s\bar{s}}}
\newcommand{\Hgg}{\wrap{h\rightarrow gg}}
\newcommand{\Zqq}{\wrap{Z\rightarrow q\bar{q}}}
\newcommand{\Zll}{\wrap{Z\rightarrow\ell\bar{\ell}}}
\newcommand{\Zinv}{\wrap{Z\rightarrow\nu\bar{\nu}}}

\newcommand{\oC}{\wrap{^{\circ}\textrm{C}}}

\newcommand\snowmass{
    \begin{center}\rule[-0.2in]{\hsize}{0.01in} \\
        \rule{\hsize}{0.01in} \\
        \vskip 0.1in Submitted to the Proceedings of the US Community Study \\
        on the Future of Particle Physics (Snowmass 2021) \\
        \rule{\hsize}{0.01in} \\
        \rule[+0.2in]{\hsize}{0.01in}
    \end{center}
}

\title{\textbf{Strange quark as a probe for new physics in the Higgs sector}}

\author[a]{Alexander~Albert}
\author[b]{Matthew~J.~Basso}
\author[a]{Samuel~K.~Bright-Thonney}
\author[c,d]{Valentina~M.~M.~Cairo}
\author[e]{Chris~Damerell}
\author[f]{Daniel~Ega{\~n}a-Ugrinovic}
\author[g]{Ulrich~Einhaus}
\author[h]{Ulrich~Heintz}
\author[i]{Samuel~Homiller}
\author[g]{Shin-ichi~Kawada}
\author[h]{Jingyu~Luo}
\author[j]{Chester~Mantel}
\author[k]{Patrick~Meade}
\author[a]{Jose~Monroy}
\author[h]{Meenakshi~Narain}
\author[b]{Robert~S.~Orr}
\author[a]{Joseph~Reichert}
\author[a]{Anders~Ryd}
\author[j]{Jan~Strube}
\author[c]{Dong~Su}
\author[c]{Ariel~G.~Schwartzman}
\author[l]{Tomohiko~Tanabe}
\author[m]{Junping~Tian}
\author[h]{Emanuele~Usai}
\author[c]{Jerry~Va'vra}
\author[c]{Caterina~Vernieri}
\author[c]{Charles~C.~Young}
\author[a]{Rui~Zou}

\affil[a]{Department of Physics, Cornell University, Ithaca, New York 14850, USA}
\affil[b]{Department of Physics, University of Toronto, 60 Saint George Street, Toronto, Ontario, Canada}
\affil[c]{SLAC National Accelerator Laboratory, 2575 Sand Hill Road, Menlo Park, California 94025-7015, USA}
\affil[d]{Experimental Physics Department, CERN, Geneva, Switzerland}
\affil[e]{Particle Physics Department, STFC Rutherford Appleton Laboratory, Harwell Science and Innovation Campus, Didcot, United Kingdom}
\affil[f]{Perimeter Institute for Theoretical Physics, Waterloo, Ontario N2L 2Y5, Canada}
\affil[g]{Deutsches Elektronen-Synchrotron DESY, Notkestr. 85, 22607 Hamburg, Germany}
\affil[h]{Department of Physics, Brown University, 182 Hope Street, Providence, Rhode Island 02912, USA}
\affil[i]{Department of Physics, Harvard University, Cambridge, Massachusetts 02138, USA}
\affil[j]{Department of Physics, University of Oregon, 1371 E 13th Avenue, Eugene, Oregon, USA}
\affil[k]{C. N. Yang Institute for Theoretical Physics, Stony Brook University, Stony Brook, New York 11794, USA}
\affil[l]{Energy Accelerator Research Organisation (KEK), 1-1 Oho, Tsukuba, Ibaraki, 305-0801, Japan}
\affil[m]{International Center for Elementary Particle Physics (ICEPP), University of Tokyo, Hongo 7-3-1, Bunkyo-ku, Tokyo, 113-0033, Japan}

\date{\today}

\begin{document}

\snowmass

{\let\newpage\relax\maketitle}

\noindent\textbf{Contact Information:} \\
Matthew~J.~Basso (\href{mailto:mbasso@physics.utoronto.ca}{mbasso@physics.utoronto.ca}) \\
Valentina~M.~M.~Cairo (\href{mailto:valentina.maria.cairo@cern.ch}{valentina.maria.cairo@cern.ch}) \\
Jerry~Va'vra (\href{mailto:jjv@slac.stanford.edu}{jjv@slac.stanford.edu}) \\

\centerline{\emph{This is a preliminary study performed in the framework of the ILD concept group.}}

\newpage

\begin{abstract}

This paper describes a novel algorithm for tagging jets originating from the hadronisation of strange quarks (strange-tagging) with the future International Large Detector (ILD) at the International Linear Collider (ILC). It also presents the first application of such a strange-tagger to a Higgs to strange (\Hss) analysis with the $P(e^-,e^+) = (-80\%,+30\%)$ polarisation scenario, corresponding to \unit[900]{\ifb} of the initial proposed \unit[2000]{\ifb} of data which will be collected by ILD during its first 10~years of data taking at \sqrts = \unit[250]{GeV}. Upper limits on the Standard Model Higgs-strange coupling strength modifier, $\kappa_s$, are derived at the 95\% confidence level to be 7.14. The paper includes as well a preliminary study of a Ring Imaging Cherenkov (RICH) system capable of discriminating between kaons and pions at high momenta (up to \unit[25]{GeV}), and thus enhancing strange-tagging performance at future Higgs factory detectors.

\end{abstract}

\section{Introduction}
\label{sec:intro}

The experimental program at the Large Hadron Collider (LHC)~\cite{LHC} has strongly established Yukawa couplings of the \unit[125]{GeV} Higgs ($h$) to the third-generation of fermions~\cite{ATLASHbb, CMSHbb, ATLASHtt, CMSHtt}. The ATLAS and CMS experiments~\cite{ATLAS, CMS} have recently reported evidence that the Higgs boson decays into two muons~\cite{ATLASHmm, CMSHmm}, which indicates for the first time that the Higgs boson interacts with second-generation leptons. At the same time, this is just a first step and not yet a complete exploration of the second-generation Yukawa couplings, because these rare Higgs decay modes (i.e., to charm or strange quarks) are very challenging or nearly impossible to detect with the current detector capabilities. Furthermore, the overwhelming multi-jet production rate at the LHC inhibits the study of strange, up, and down quark couplings with inclusive $h\rightarrow q\bar{q}$ decays, in addition to the dominant $h\rightarrow b\bar{b}$ decay mode.

At the LHC, new algorithms for the identification of jets originating from the hadronisation of $c$-quarks ($c$-tagging) are gradually becoming available and enabling new searches for the decay of the Higgs boson to charm quarks~\cite{ATLASHcc:2018, CMSHcc:2019, ATLASHcc:2022, CMSHcc:2022}. Less literature, however, is available about searches of Higgs boson decays to strange, up, and down quarks~\cite{Nakai2020, Erdmann2019, Duarte-Campderros2018, CMSHcs, Djouadi2009, ATLASZandHToPhiOrRho, CMSHToZAndPhiOrRho, Bedeschi:2022rnj}.

Searches for exclusive Higgs boson decays to a $\phi$ or $\rho(770)$ meson and a $Z/\gamma$ have been suggested and experimentally tested~\cite{ATLASZandHToPhiOrRho, CMSHToZAndPhiOrRho} as a probe of the Higgs boson couplings to strange and up/down quarks, respectively. For these Higgs couplings there are no projections available and it will most likely remain out of direct experimental reach unless they are enhanced compared to Standard Model (SM) expectations. 

In fact, when considering Beyond the Standard Model (BSM) scenarios that allow for extended Higgs sectors, the possibilities open up dramatically. Models with additional Higgs doublets have new Yukawa matrices which need not be directly proportional to the SM fermion masses, provided that they do not lead to large flavour-changing neutral currents. One such class of models are those exhibiting spontaneous flavour violation (SFV)~\cite{Egana-Ugrinovic:2018znw}, which allows for new Yukawa couplings either to the up or the down quarks with no relation to the quark masses. A two Higgs doublet model with up-type SFV, for example, could thus have large couplings to the $d$ and $s$ quarks, and the new Higgs states would be produced in quark fusion, with decays to gauge and Higgs bosons and quarks~\cite{Egana-Ugrinovic:2019, Egana-Ugrinovic:2021}. If the observed \unit[125]{GeV} Higgs boson is an admixture of a SM-like Higgs and one of the new Higgs states, its couplings to the first or second generation quarks can be significantly larger than predicted in the SM, leading to large deviations in the Higgs boson branching ratios. A second class of BSM models exists where the first- and second-generation fermions get their mass from a second Higgs doublet while the observed \unit[125]{GeV} Higgs couples predominantly to the third-generation~\cite{2HDM}. This results in very different decay branching ratios of the additional heavy Higgs bosons ($H$). The largest production mode of the neutral Higgs bosons would be from a $c\bar{c}$ initial state, while the charged Higgs bosons would be predominantly produced from a $c\bar{s}$ initial state. The most interesting decay modes include $H/A \rightarrow c\bar{c}$, $t\bar{c}$, $\mu\bar{\mu}$~\cite{ATLASHAmm, CMSAmm}, and $\tau\mu$~\cite{ATLASHmutauetau, CMSHmutauetau} and $H^\pm \rightarrow c\bar{b}$, $c\bar{s}$~\cite{CMSHcs2020}, and $\mu\nu$.

\begin{figure}
    \centering
    \begin{subfigure}{0.70\textwidth}
        \centering
        \includegraphics[width=1.\textwidth]{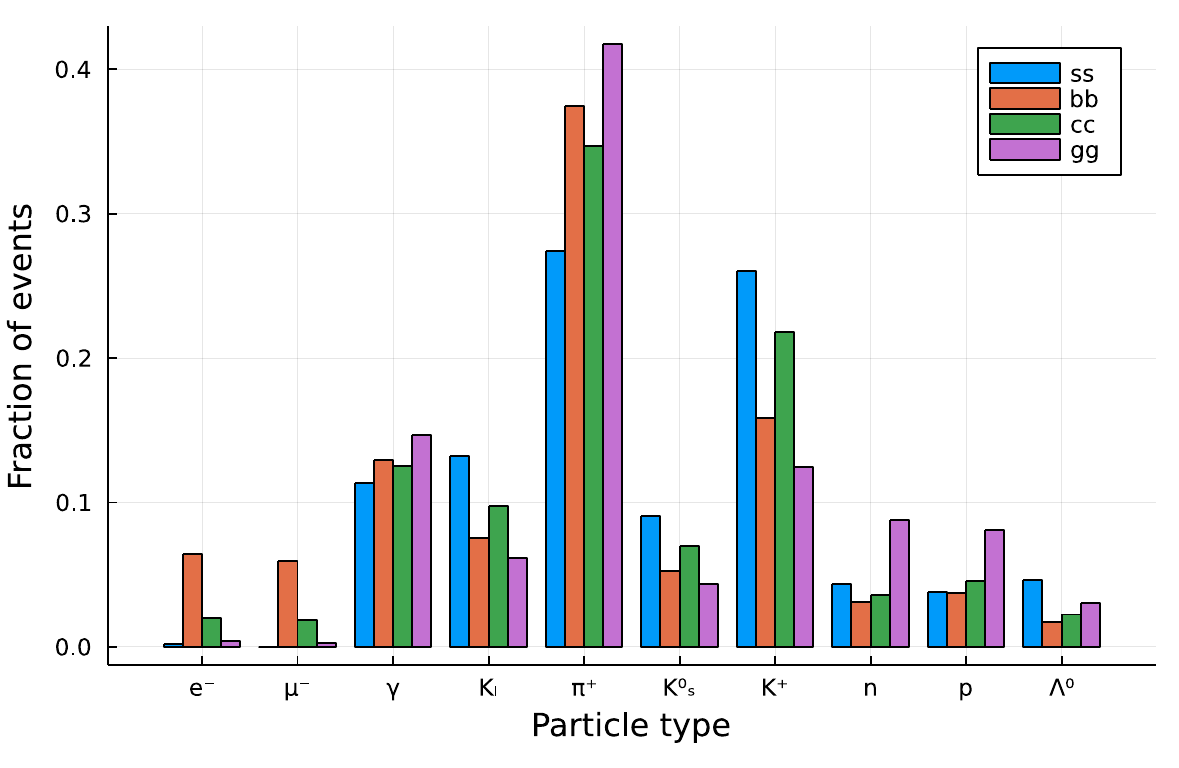}
        \caption{Leading particle fractions}
        \label{fig:leading_particle_MCParticle}
    \end{subfigure} \\
    \begin{subfigure}{0.70\textwidth}
        \centering
        \includegraphics[width=1.\textwidth]{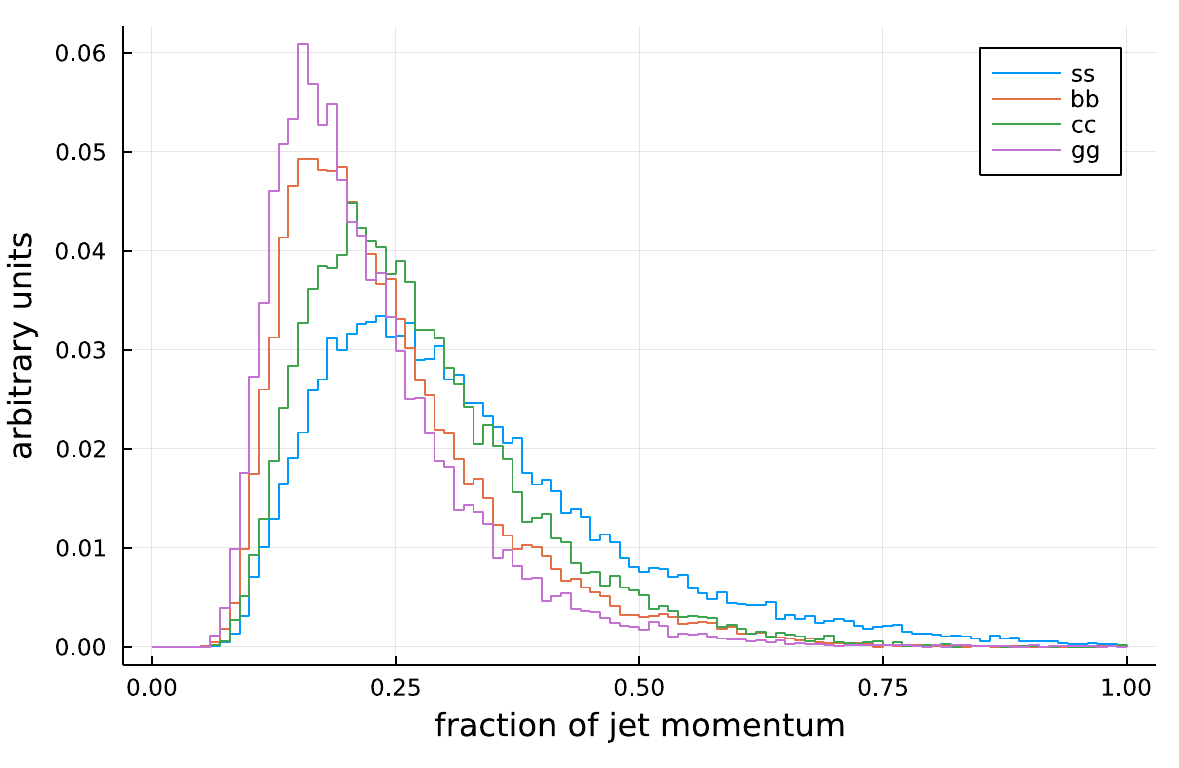}
        \caption{Fraction of jet momentum carried by leading particle}
        \label{fig:leading_particle_p_frac}
    \end{subfigure} \\
    \caption{(a) Leading particle fractions and (b) the fraction of the jet's momentum carried by the leading particle for reconstructed jets from $h\rightarrow s\bar{s}/b\bar{b}/c\bar{c}/gg$ events. The events were generated using \Whizard~\cite{Whizard, Whizardv2.8.4} and reconstructed with iLCSoft~\cite{iLCSoft} using a Silicon Detector~\cite{SiDLOI} simulation. In (a), all of the bars of a particular colour sum to 1 by definition. Neutrinos and very soft ($E < \unit[0.05]{GeV}$) particles are excluded.}
    \label{fig:leading_particle}
\end{figure}

Tagging strange jets comes with some difficulty, however. As shown in Table~\ref{tab:jet_flavours}, bottom and charm jets can be differentiated based on the presence of 2 or 1 secondary vertices. Strange jets, which, excluding $V^0$s, have 0 secondary vertices, are only differentiated from light (i.e., up or down) jets based on the ability to reliably tag the presence of a strange hadron within the jet. Strange hadrons are also most often the leading particle in strange jets, as evident from Fig.~\ref{fig:leading_particle_MCParticle}. And from Fig.~\ref{fig:leading_particle_p_frac}, we see that the leading particle more often carries a larger fraction of the strange jet's momentum as compared to other jet flavours. Separation of strange jets from other flavours only begins when the leading particle carries a jet momentum fraction $\gtrsim0.2$, which translates to leading particle momenta $\gtrsim 0.2 \times 0.5 \times \unit[125]{GeV} = \unit[12.5]{GeV}$. Accordingly, technology enabling kaon-pion discrimination, and specifically at moderate to high particle momentum (i.e., \unit[$>$10]{GeV}), is highly relevant at future detectors for measurements of decays to strange jets.

\begin{table}[htbp]
    \centering
    \caption{Defining features for the different categories of quark jets. N.B. the number of strange hadrons is defined as the number originating from the initial strange quark, and ``Light'' jets refer only to those originating from up or down quarks.}
    \label{tab:jet_flavours}
    \begin{tabular}{ccc}
        \toprule
        Jet flavour & Number of secondary vertices & Number of strange hadrons \\
        & (excluding $V^0$s) & (e.g., $K^\pm$, $K^0_{L/S}$, and $\Lambda^0$) \\
        \midrule
        Bottom  & 2 & $\geq$1 \\
        Charm   & 1 & $\geq$1 \\
        Strange & 0 & $\geq$1 \\
        Light   & 0 & 0 \\
        \bottomrule
    \end{tabular}
\end{table}

The work presented in this paper describes the novel development of a flavour tagging algorithm capable of tagging jets that originate from the hadronisation of strange quarks (strange-tagging). This allows us to tag for the first time exclusive Higgs to strange decays and opens new opportunities in direct \Hss searches. If used in conjunction with $c$-tagging, it also allows to probe new physics models. The tagger is then applied to a SM (\unit[125]{GeV}) \Hss analysis using $e^+e^-$ collisions at \sqrts = \unit[250]{GeV}, estimating the prospects for Higgs-strange coupling strength measurements.

Strange tagging itself is of interest in the context of the ILC study questions~\cite{ILCStudyQs} proposed for Snowmass~2021~\cite{Snowmass}. The study presented here is conducted in the context of the International Linear Collider (ILC)~\cite{ILCTDR}, an electron-positron collider proposed to be built in Japan. Nevertheless, the results are easily applicable to future experiments at other electron-positron machines being considered by the high-energy physics community.

The paper is organised as follows:

\begin{itemize}
    \item Section~\ref{sec:ild} describes the International Large Detector (ILD), a proposed detector at the ILC and the detector used for the contained studies;
    \item Section~\ref{sec:mc} describes the Monte Carlo samples included in the study;
    \item Section~\ref{sec:tagger} describes the development and validation of a jet flavour tagger using a neural network;
    \item Section~\ref{sec:analysis} describes the application of the jet flavour tagger to a SM \Hss analysis with ILD at the \sqrts = \unit[250]{GeV} ILC run;
    \item Section~\ref{sec:alt_detector} describes a detector proposal utilising a RICH system which would maximise particle identification (PID) at high momenta and thus boost strange tagging performance;
    \item Section~\ref{sec:conclusion} describes the conclusions and next steps.
\end{itemize}

\noindent N.B. throughout the paper, ``light'' quarks refer exclusively to up and down quarks -- strange quarks are \emph{excluded} from this classification. This is similarly true for ``light'' jets, which refer exclusively to jets originating from up and down quarks.

\section{The ILD detector}
\label{sec:ild}

The International Large Detector (ILD) is one of two detector concepts proposed at the ILC~\cite{ILDLOI, ILCDetTDR, ILD, ILDConceptGroup:2020}, the other being the Silicon Detector (SiD)~\cite{SiDLOI, ILCDetTDR}.

Closest to the interaction point, ILD has 3~double-layer pixel detectors for vertexing followed by 2~double-layer pixel detectors, a Time Projection Chamber (TPC), and 1~double-layer strip detector for tracking. The TPC additionally provides PID via measurements of energy loss from charged particles due to ionisation ($dE/dx$) which is envisioned to be completed by time-of-flight (TOF) measurements in the TPC's silicon envelope or in the electromagnetic calorimeter. The low material budget of the TPC is highly desirable for low momentum tracking of particles and its PID capabilities make ILD the most promising detector layout for strange tagging at the ILC. A forward tracking detector comprised of silicon pixel and strip discs provides tracking acceptance starting at a polar angle of $4.8^\circ$.

Immediately beyond the tracking system, ILD has high granularity sampling calorimeters for particle flow reconstruction~\cite{PFCalo}. The precise design of the electromagnetic and hadronic calorimeters is still under study.

The tracking and calorimetry systems are contained within in a solenoid providing a \unit[3.5]{T} (or \unit[4]{T}, depending on the detector model) magnetic field. A surrounding iron yoke instruments muon detection.

\section{Monte Carlo simulation}
\label{sec:mc}

\begin{figure}
    \centering
    \includegraphics[width=0.6\textwidth]{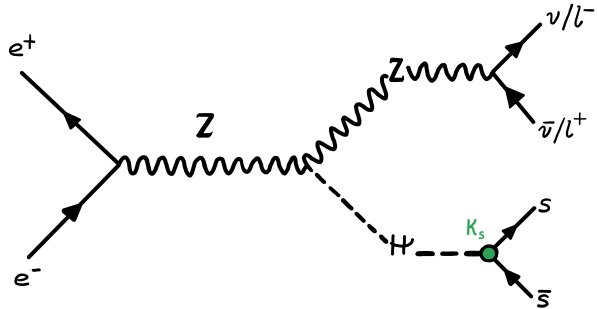}
    \caption{Tree-level Feynman diagram for production of a Higgs boson in association with a $Z$ boson. The Higgs decays hadronically to strange quarks (with Yukawa coupling strength modifier $\kappa_s$) and the $Z$ decays leptonically to charged leptons or neutrinos. Drawing by F.~Cairo.}
    \label{fig:feynman}
\end{figure}

The main Higgs boson production mechanism at \sqrts = \unit[250]{GeV} at the ILC is production in association with a $Z$ boson (``associated production''), $Zh$ (or $hZ$ -- each is used interchangeably throughout this paper). Accordingly, associated production is considered in this paper for generating the signal and some of the background events in both the \Zinv and \Zll decay channels. The tree-level Feynman diagram for $Zh$ is shown in Fig.~\ref{fig:feynman}. To a smaller degree, interference from $ZZ$- and $WW$-fusion contribute signal to the $Z\rightarrow e^+e^-$ and $Z\rightarrow\nu_e\bar{\nu}_e$ decay channels, respectively; however, only interference from $ZZ$-fusion is included in this paper.\footnote{Interference from $WW$-fusion is excluded because our \Zinv signal samples only include $\nu_{\mu}\bar{\nu}_{\mu}$ and $\nu_{\tau}\bar{\nu}_{\tau}$ final states -- if the $\nu_{e}\bar{\nu}_{e}$ final state was included, the interference would be built into the \Whizard cross section. Accordingly, the cross section from \Whizard is scaled by a factor of 1.5 to extrapolate to flavour-inclusive final states. However, this assumes only $s$-channel production of the neutrinos.}

The Monte Carlo (MC) events used in this study were generated at centre-of-mass energy \sqrts = \unit[250]{GeV} using \Whizard[2.8.5]~\cite{Whizard, Whizardv2.8.5} interfaced with \Pythia[6.4]~\cite{Pythia6.4} for showering/hadronisation, \Tauola for polarised $\tau$ lepton decays~\cite{Tauola1990, Tauola2006, Tauola2012}, and \GuineaPig~\cite{GuineaPig} and \CirceTwo~\cite{Circe, Circe2} for beam spectrum effects.

The generated events were reconstructed with \MarlinReco~\cite{MarlinReco}, using a full simulation of the ILD detector based on \GeantFour~\cite{Geant4} in the \DDFourhep framework~\cite{DD4hepEPJ, DD4hep}. The reconstructed events were persistified as DSTs, an LCIO\footnote{LCIO version~02-15-04~\cite{LCIOv02-15-04} was used for processing all of the input miniDSTs analysed in this paper.}~\cite{LCIO} event data model, and further refined as miniDSTs~\cite{miniDSTs}, a slimmed version of DST which also includes the results of the LCFIPlus\footnote{LCFIPlus is a software framework for vertex and jet finding as well as for jet flavour tagging at future $e^+e^-$ linear colliders.}~\cite{LCFIPlus} $b$-, $c$-, and $o$- (i.e., ``other'' -- strange, light, or gluon) jet tagger scores.

Low-\pt pileup from $\gamma\gamma\rightarrow\textrm{hadrons}$ events was simulated using the cross section model of Chen-Barklow-Peskin~\cite{pileupxs}, reconstructed using the full ILD simulation, and overlaid onto all hard scatter events.

All MC samples are generated using 100\%~left-handed-(LH-) polarised electron beams and 100\%~right-handed-(RH-) polarised positron beams. We consider the ILC running scenario at \sqrts = \unit[250]{GeV} using 80\%~LH-polarised electron beams (i.e., $P_{e^-} = -80\%$) and 30\%~RH-polarised positron beams (i.e., $P_{e^+} = +30\%$), abbreviated as $P(e^-,e^+) = (-80\%,+30\%)$. The total cross section $\sigma_{P(e^-,e^+)}$ for an arbitrary polarisation scenario is given by:

\begin{equation}
    \begin{split}
        \sigma_{P(e^-,e^+)} = \,\,\,\,\,\,\,\, & \frac{1 - P_{e^-}}{2} \frac{1 + P_{e^+}}{2}\sigma_{LR} + \frac{1 + P_{e^-}}{2} \frac{1 - P_{e^+}}{2}\sigma_{RL} \\ + & \frac{1 - P_{e^-}}{2} \frac{1 - P_{e^+}}{2}\sigma_{LL} + \frac{1 + P_{e^-}}{2} \frac{1 + P_{e^+}}{2}\sigma_{RR} \,,
    \end{split}
\end{equation}

\noindent where the first subscript on the $\sigma_{XY}$ cross sections indicates the handedness of the electron beam and the second subscript indicates the handedness of the positron beam. As we only have samples available for 100\% LH-polarised electron and 100\%-polarised positron beams (i.e, $\sigma_{LR}$), we are \emph{only} able to estimate the contribution from $\sigma_{LR}$ to the total cross section: $0.585\sigma_{LR}$ for $P(e^-,e^+) = (-80\%,+30\%)$. This is an nonphysical selection, and additional signal and background are missing as a result. However, $\sigma_{LR}$ is the most important contribution to the total cross section and the upper limits on the Higgs-strange coupling strength modifier, estimated later in Section~\ref{sec:analysis}, are still expected to be valid and to only improve with the inclusion of additional polarisation states and running scenarios.\footnote{In other words, ``upper limits'' on the upper limits.}

For \sqrts = \unit[250]{GeV} running scenario, an integrated luminosity \lumi of \unit[2000]{\ifb} is expected, as per the ILC physics programme~\cite{ILCPhysProg} and ILC Snowmass white paper~\cite{ILCSMWhitePaper}. According to the white paper, which is the most up-to-date source on the proposed physics programme, only 45\% of the \unit[2000]{\ifb} is expected to be operated in $P(e^-,e^+) = (-80\%,+30\%)$ polarisation scenario. Our expected luminosity $\lumi$ is therefore \unit[900]{\ifb}.

Using the LR cross sections and the expected luminosity, each sample is normalised prior to applying any analysis cuts, where the event weights are modified as:

\begin{equation}
    w^\prime_i = 0.585 \times \frac{\lumi\sigma_{LR}}{\sum_j w_j}w_i \,\forall\, i \,,
\end{equation}

\noindent where $w_i$ is the weight for event $i$ and $\lumi = \unit[900]{\ifb}$. N.B. there is no estimate available for the SM \Hss branching ratio (BR), $\BR[\Hss]_\textrm{SM}$ -- instead, it is estimated by scaling the SM $h\rightarrow c\bar{c}$ BR, $\BR[h\rightarrow c\bar{c}]_\textrm{SM}$, by the square of the ratio of the strange quark mass over the charm quark mass, $M_s/M_c$:

\begin{equation}
    \begin{split}
    \BR[\Hss]_\textrm{SM} & \approx \left(\frac{M_s}{M_c}\right)^2 \times \BR[h\rightarrow c\bar{c}]_\textrm{SM} \\
    & = 11.72^{-2} \times 0.0291 \\
    & = 2 \times 10^{-4} \,.
    \end{split}
\end{equation}

\noindent The ratio, $M_s/M_c = 11.72^{-1}$, is taken from the Particle Data Group (PDG)~\cite{PDG}. A similar procedure yields $\BR[h\rightarrow d\bar{d}]_\textrm{SM} \approx 5 \times 10^{-7}$ and $\BR[h\rightarrow u\bar{u}]_\textrm{SM} \approx 1 \times 10^{-7}$ using $M_s/M_d \sim 20$ and $M_u/M_d \approx 0.47$, also taken from the PDG~\cite{PDG}.

The signal and background MC samples used in this study are shown in Table~\ref{tab:samples}. Also shown are the raw numbers of events as well as the LR cross sections, per sample.

\begin{table}[H]
    \centering
    \caption{MC processes considered in the \Hss analysis, including raw statistics and cross sections. N.B. the samples were generated at \sqrts = \unit[250]{GeV} and the cross sections assume 100\% LH-polarised electron beams and 100\% RH-polarised positron beams. The cross sections include the corresponding BRs for the indicated decays.{\protect\footnotemark} In the non-Higgs processes, ``$n\!f$'' denotes the number ($n$) of fermions ($f$) in the final state. In $Z(\rightarrow\ell\bar{\ell})h(\rightarrow{\textrm{other}})$, ``other'' denotes any non-hadronic SM decay. The $ZZ/WW$ process covers the interference of $ZZ$ and $WW$ final states, e.g., $u\bar{u}d\bar{d}$ or $c\bar{c}s\bar{s}$.}
    \label{tab:samples}
    \begin{tabular}{lrr}
        \toprule
        Process name & Raw events [a.u.] & LR cross section [fb] \\
        \midrule
        $Z(\rightarrow\nu\bar{\nu})h(\rightarrow{s\bar{s}})$         &    500,000 & 0.021 \\
        \midrule
        $Z(\rightarrow\nu\bar{\nu})h(\rightarrow{b\bar{b}})$         &    500,000 & 58.1 \\
        $Z(\rightarrow\nu\bar{\nu})h(\rightarrow{c\bar{c}})$         &    499,800 &  2.9 \\
        $Z(\rightarrow\nu\bar{\nu})h(\rightarrow{u\bar{u}})$         &    499,800 & $1 \times 10^{-5}$ \\
        $Z(\rightarrow\nu\bar{\nu})h(\rightarrow{d\bar{d}})$         &    500,000 & $5 \times 10^{-5}$ \\
        $Z(\rightarrow\nu\bar{\nu})h(\rightarrow{gg})$               &    499,800 &  8.6 \\
        \midrule
        $Z(\rightarrow\ell\bar{\ell})h(\rightarrow{s\bar{s}})$       &    224,000 & 0.011 \\
        \midrule
        $Z(\rightarrow\ell\bar{\ell})h(\rightarrow{b\bar{b}})$       &    872,380 & 29.8 \\
        $Z(\rightarrow\ell\bar{\ell})h(\rightarrow{c\bar{c}})$       &     43,334 &  1.5 \\
        $Z(\rightarrow\ell\bar{\ell})h(\rightarrow{u\bar{u}})$       &    218,600 & $6 \times 10^{-6}$ \\
        $Z(\rightarrow\ell\bar{\ell})h(\rightarrow{d\bar{d}})$       &    209,400 & $3 \times 10^{-5}$ \\
        $Z(\rightarrow\ell\bar{\ell})h(\rightarrow{gg})$             &    123,225 &  4.4 \\
        $Z(\rightarrow\ell\bar{\ell})h(\rightarrow{\textrm{other}})$ &    460,688 & 15.9 \\
        \midrule
        $2f$ $Z$ hadronic                                            & 25,354,400 & 127,965 \\
        $4f$ $ZZ$ hadronic                                           &  7,099,000 &   1,405 \\
        $4f$ $WW$ hadronic                                           & 14,790,600 &  14,866 \\
        $4f$ $ZZ/WW$ hadronic                                        & 18,494,200 &  12,389 \\
        $2f$ $Z$ leptonic                                            & 24,500,000 &  21,214 \\
        $4f$ $ZZ$ semileptonic                                       &  4,199,600 &     838 \\
        $4f$ single $Z$ semileptonic                                 &  6,999,600 &   1,423 \\
        \bottomrule
    \end{tabular}
\end{table}
\footnotetext{The cross sections are consistent with other sources, in particular Table~2 of Tomohisa Ogawa's thesis~\cite{Ogawa}: e.g., $s$-channel ${\nu\bar{\nu}}h$ production has $\sigma_{P(e^-,e^+)=(-80\%,+30\%)} = \unit[61.6]{fb} = 0.585\sigma_{LR} + 0.035\sigma_{RL}$ and $\sigma_{P(e^-,e^+)=(+80\%,-30\%)} = \unit[41.6]{fb} = 0.035\sigma_{LR} + 0.585\sigma_{RL}$. Solving yields $\sigma_{LR} = \unit[101.4]{fb}$, which is consistent (up to the BR) with the $Z(\rightarrow\nu\bar{\nu})h$ cross section in Table~\ref{tab:samples} above, and $\sigma_{RL} = \unit[65.0]{fb}$.}

As a back-of-the-envelope calculation, assuming \unit[2000]{\ifb} of data collected at the ILC after 10~years of data-taking and a Higgs boson production cross section of about \unit[200]{fb}, $\sim$400,000~Higgs bosons would be produced where only 80 of those feature a \Hss event. As a point of comparison, $\sim$200,000 $h\rightarrow b\bar{b}$ and $\sim$12,000 $h\rightarrow c\bar{c}$ events are expected.

\begin{figure}[H]
    \centering
    \begin{subfigure}{0.7\textwidth}
        \centering
        \includegraphics[width=1.\textwidth]{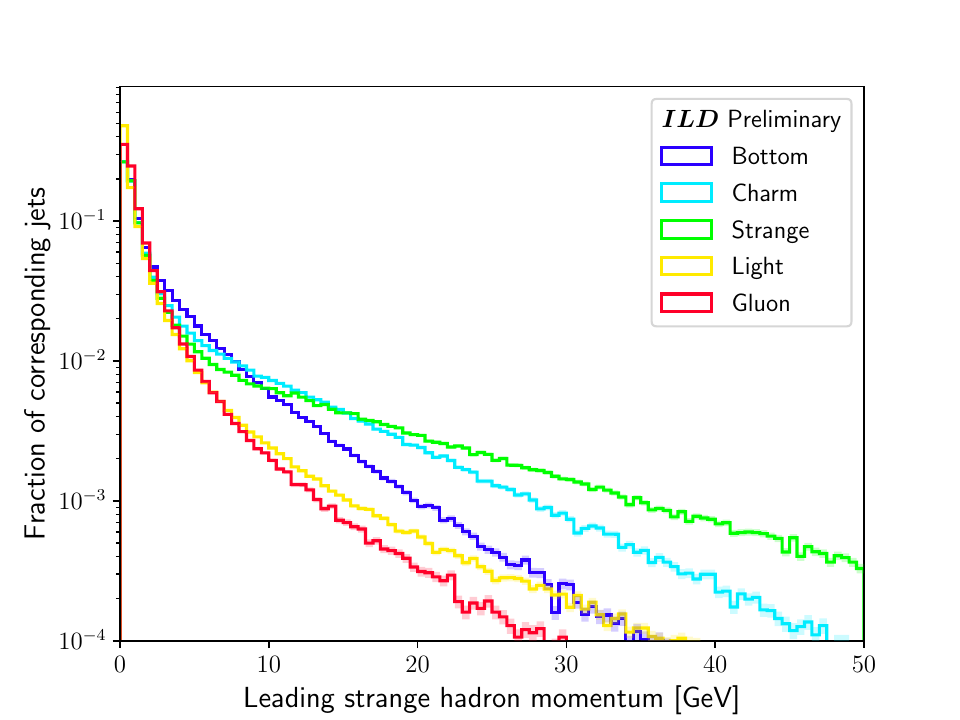}
        \caption{Differential}
        \label{fig:leading_strange_p_hist}
    \end{subfigure} \\
    \begin{subfigure}{0.7\textwidth}
        \centering
        \includegraphics[width=1.\textwidth]{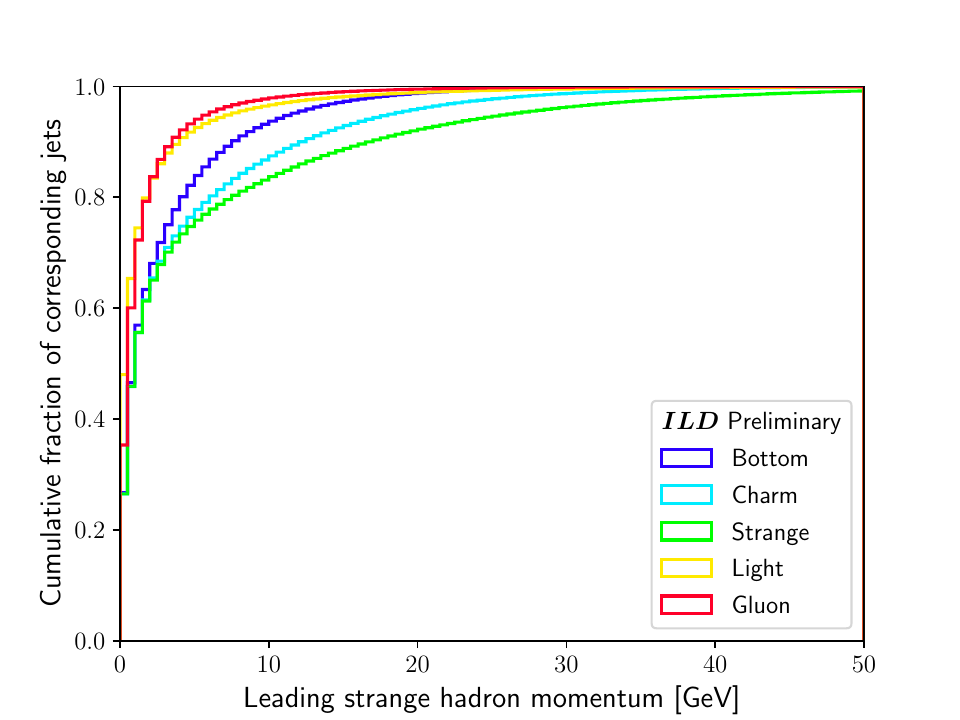}
        \caption{Cumulative}
        \label{fig:leading_strange_p_cumul}
    \end{subfigure} \\
    \caption{Differential and cumulative distributions of the momentum of the leading strange particle in the leading or subleading momentum jet of the $h(\rightarrow q\bar{q}/gg)Z(\rightarrow\nu\bar{\nu})$ events described in Table~\ref{tab:samples}. The choice of leading or subleading jet is random. The leading strange particle is identified by iterating over the momentum-ordered PFOs in the jet and selecting the first PFO which is truth-matched to a strange hadron. If no strange particle is found, a momentum of \unit[0]{GeV} is assigned. The sum-of-weights for each class is normalised to 1.}
    \label{fig:leading_strange_p}
\end{figure}

Fig.~\ref{fig:leading_strange_p} shows the momentum of the leading strange particle in jets from $h(\rightarrow q\bar{q}/gg)Z(\rightarrow\nu\bar{\nu})$ events. From Fig.~\ref{fig:leading_strange_p_hist}, we see that separation of strange jets from other flavours becomes possible above approximately \unit[15]{GeV}, which is consistent with our conclusion from SiD simulated events. And from Fig.~\ref{fig:leading_strange_p_cumul}, above \unit[10]{GeV}, we are targeting approximately 25\% of all strange jets.

\FloatBarrier

\section{Jet flavour tagger}
\label{sec:tagger}

In order to better tag strange-, light-, and gluon-initiated jets, an artificial neural network (ANN) was developed in Keras~\cite{Keras} using the TensorFlow backend~\cite{Tensorflow}. The goal of tagging each jet by their flavour of progenitor particle (i.e., $b$, $c$, $s$, $u/d$, or $g$) inspires the use of a multiclassifier. The multiclassifier assigns a probability of a jet belonging to each possible output class (i.e., outputs a vector of size 5), and these probabilities logically sum up to 1 per jet.

\subsection{Inputs}

The training is performed on the $Z(\rightarrow\nu\bar{\nu})h(\rightarrow q\bar{q}/gg)$ samples from Table~\ref{tab:samples}. All events are required to have $N_\textrm{jets} \ge 2$ and $N_\textrm{leptons} = 0$. The training is performed using only one jet per event, where the leading or subleading momentum jet is randomly chosen. Per process, 250,000 raw MC events are used -- additionally, the $h\rightarrow u\bar{u}$ and $h\rightarrow d\bar{d}$ processes are combined into a single class, $h\rightarrow\textrm{light}$.

As input to the ANN, several jet-level variables are chosen:

\begin{itemize}
    \item kinematics: momentum $p$, pseudorapidity $\eta$, azimuthal angle $\phi$, and mass $m$;
    \item LCFIPlus tagger results: $b$- (``BTag''), $c$- (``CTag''), and $o$-tag (``OTag'') scores as well as jet category;
    \item number of Particle Flow Objects (PFOs -- these are the particles which are grouped into the jet).
\end{itemize}

\noindent In addition to jet-level variables, it is prudent to include variables at the level of the PFOs contained within the jet. The 10 leading momentum particles contained within the jet have their kinematics redefined relative to the jet's axis and their momentum and mass scaled by the momentum of the jet. Per-particle, the following variables are also chosen as inputs:

\begin{itemize}
    \item kinematics: $p$, $\eta$, $\phi$, and $m$;
    \item charge $q$;
    \item truth likelihoods: $L(e^\pm)$, $L(\mu^\pm)$, $L(\pi^\pm)$, $L(K^{0/\pm})$, $L(p^\pm)$.
\end{itemize}

\noindent The ILD detector will provide PID information per PFO, including electron ($e^\pm$), muon ($\mu^\pm$), pion ($\pi^\pm$), kaon/strange hadron ($K^{0/\pm}$), and proton ($p^\pm$) likelihoods, $L$. However, the reconstructed likelihoods utilising the $dE/dx$ and TOF information were not available in the inputs at the time of the study.~\emph{Truth} likelihoods are assigned instead, representing a best-case scenario in terms of PID. The 5 truth likelihoods are assigned a binary number by comparing the absolute value of the (particle- or truth-level) PDG ID~\cite{PDG} of the PFO to the PDG ID(s) of each particle class:

\begin{itemize}
    \item electrons: 11;
    \item muons: 13;
    \item pions: 211;
    \item kaons and strange hadrons: 310, 321, and 3122 (includes $V^0$s: $K^0_s$ and $\Lambda^0$);
    \item protons: 2212;
\end{itemize}

\noindent where 1 is assigned if one of the PDG IDs match and 0 is assigned otherwise. Distributions of the inputs for each class are provided in Figs.~\ref{fig:inputs_jet_1} through \ref{fig:inputs_PFO_2} in Appendix~\ref{app:training_plots}.

\subsection{Architecture}

PFO-level inputs motivate the use of a recurrent neural network (RNN), which can handle input events where the jet has fewer than 10 constituent particles (in these rare cases, the input vectors of particles are padded to size 10 with zero-initialised variables). A similar architecture (using a different flavour of RNN) has been used for studying the maximum strange tagging performance at hadron machines~\cite{StrangeTagHadCol}. The RNN consist of 3 layers using gated recurrent units (GRUs)~\cite{GRU}. The output from the RNN is concatenated with the jet-level inputs and serve as inputs to a multilayer perceptron (MLP) with 3 layers. Each layer of the MLP uses a scaled exponential linear unit (SELU)~\cite{SELU} activation, which has the beneficial property of self-normalising inputs. As the network is a multiclassifier, the sensible choice of output activation is the softmax function:

\begin{equation}
    [\vec{f}(\vec{x})]_i = \frac{\exp([\vec{x}]_i)}{\sum_{i=1}^5 \exp([\vec{x}]_i)} \,\forall\, i=1,\ldots,5 \,,
\end{equation}

\noindent where $\vec{f}$ is the softmax activation function, $\vec{x}$ is the input vector, and $[\ldots]_i$ denotes the $i$-th value of a vector. The output vector is of size 5, as there are 5 jet flavour classes, and sums to 1, by definition.

\begin{figure}[htbp]
    \centering
    \includegraphics[width=0.7\textwidth]{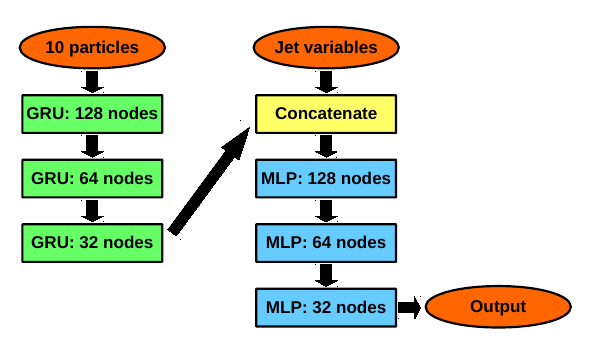}
    \caption{A cartoon of the network architecture used for the jet flavour tagger ANN. The arrows denote the flow of vectors through the network.}
    \label{fig:network}
\end{figure}

A pictorial representation of the network's architecture, including the number of nodes per layer, is shown in Fig.~\ref{fig:network}.

\subsection{Training and validation}

To train the network, input events are first split, where 90\% of all events per class are reserved for training and 10\% are reserved for testing. Within the training dataset, events are split according to even and odd event numbers. A two-way {\kfold}ing procedure is used, where the network is trained using only odd events and simultaneously validated\footnote{The validation consists of evaluating performance metrics (e.g., loss, accuracy, mean squared error, etc.) using the current ANN weights for both the training and validation events, per-epoch. The metrics for both the training and validation events should roughly follow one another, assuming no overtraining is present.} using only even events (``\kfold~0''), and then the network is trained using only even events and simultaneously validated using only odd events (``\kfold~1''). In this way, the entire training dataset may be used. If the input vector for a given event is $\vec{x}$ and that same event has an event number $n$, then the output of the  (\emph{post}-training) tagger, $\vec{F}$, is:

\begin{equation}
    \vec{F}(\vec{x}) = \left\{
        \begin{array}{ll}
            \vec{F}_{\kfold\,0}(\vec{x}) \,, & n~\textrm{mod}~2 = 0 \\
            \vec{F}_{\kfold\,1}(\vec{x}) \,, & n~\textrm{mod}~2 = 1 \\
        \end{array}
    \right. \,,
    \label{eqn:tagger}
\end{equation}

\noindent where $\vec{F}_{\kfold\,0}$ is the output of network trained on \kfold~0
and $\vec{F}_{\kfold\,1}$ is the output of network trained on \kfold~1. In this way, we avoid bias by ensuring the tagger is \emph{never} applied to the same events it was trained on. N.B. we have written the output as a vector-valued function to emphasise that the tagger is a multiclassifier.

A categorical cross-entropy loss function is chosen, and the network is trained using the Adam~\cite{Adam} optimiser with a learning rate of 0.0005 and a batch size of 1024. Each class is re-normalised to have the same sum-of-weights. Early stopping is applied to prevent overtraining.

\begin{figure}[htbp]
    \centering
    \begin{subfigure}{0.49\textwidth}
        \centering
        \includegraphics[width=1.\textwidth]{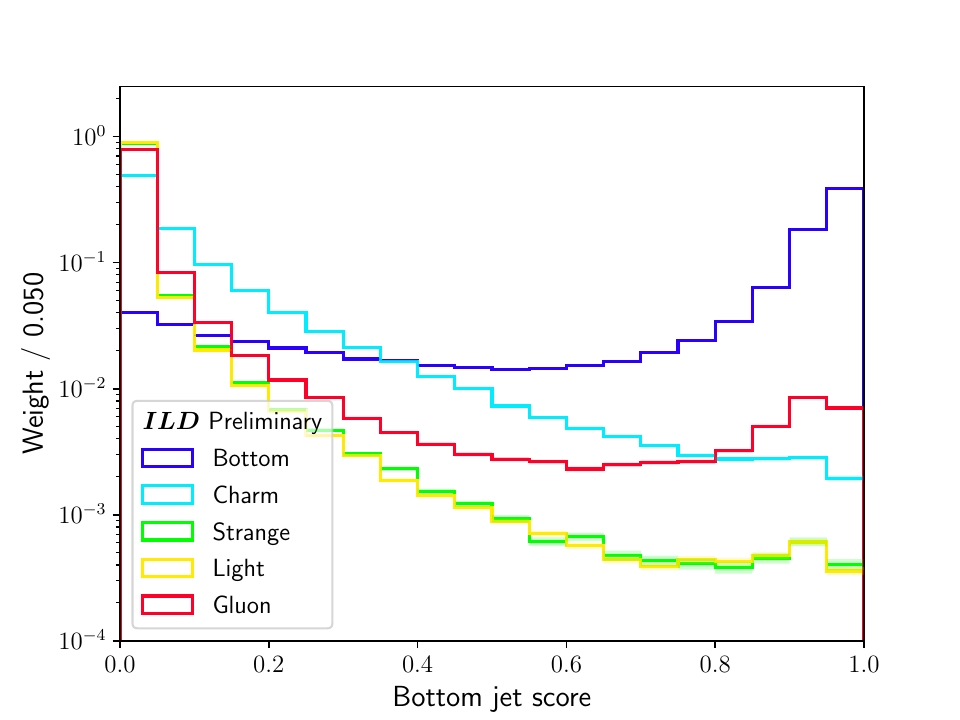}
        \caption{$b$-jet score}
    \end{subfigure}
    \hfill
    \begin{subfigure}{0.49\textwidth}
        \centering
        \includegraphics[width=1.\textwidth]{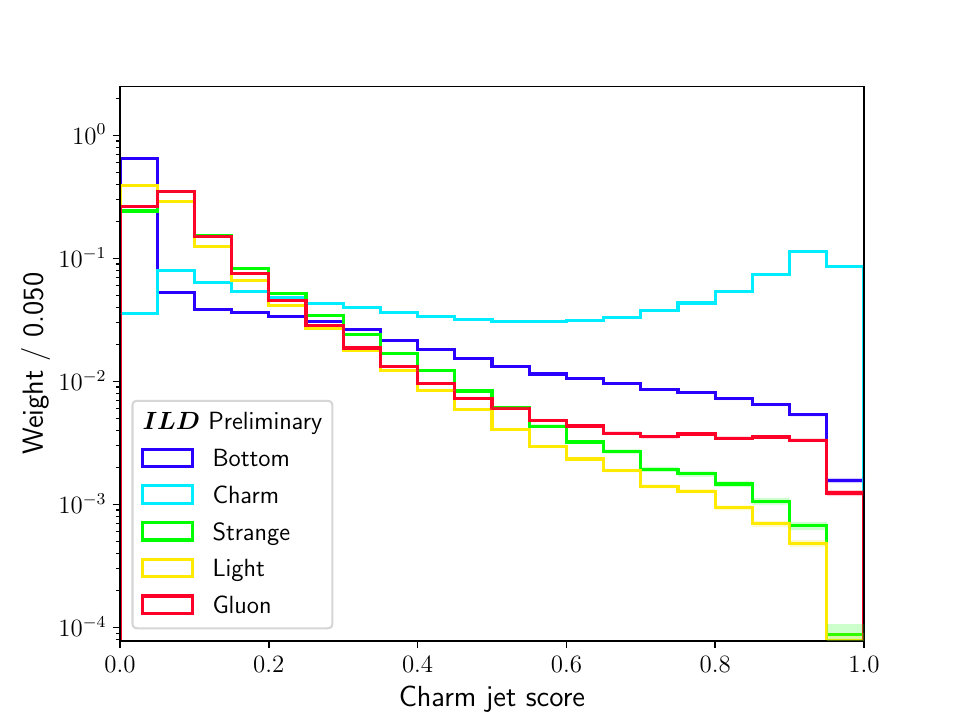}
        \caption{$c$-jet score}
    \end{subfigure} \\
    \begin{subfigure}{0.49\textwidth}
        \centering
        \includegraphics[width=1.\textwidth]{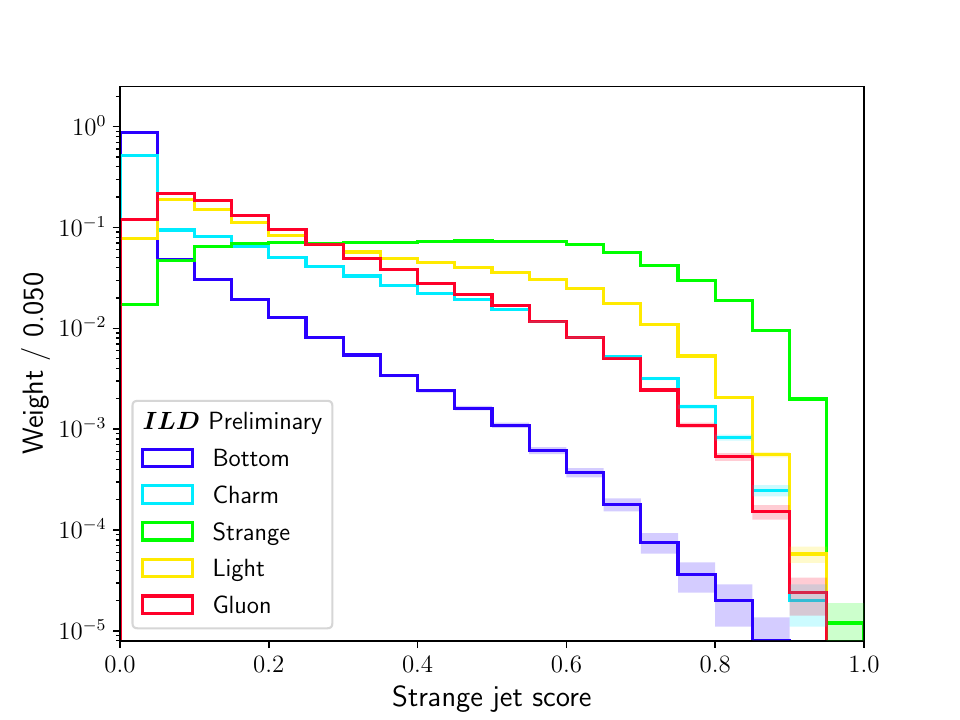}
        \caption{$s$-jet score}
    \end{subfigure}
    \hfill
    \begin{subfigure}{0.49\textwidth}
        \centering
        \includegraphics[width=1.\textwidth]{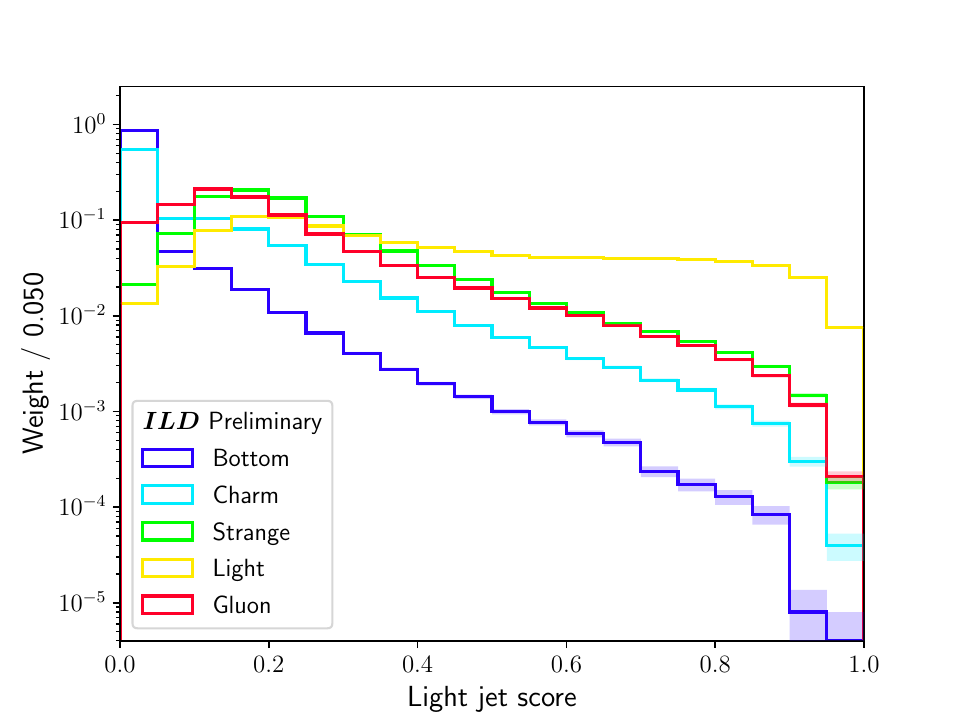}
        \caption{Light-jet score}
    \end{subfigure} \\
    \begin{subfigure}{0.49\textwidth}
        \centering
        \includegraphics[width=1.\textwidth]{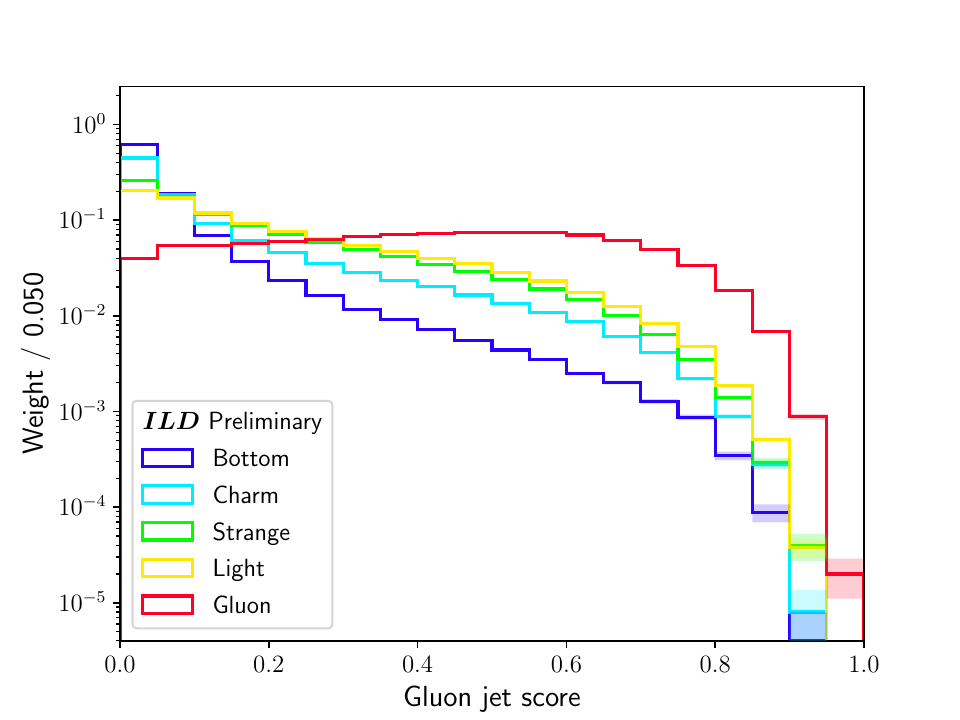}
        \caption{$g$-jet score}
    \end{subfigure} \\
    \caption{Distributions for each output node of the described jet flavour tagger, Eq.~\ref{eqn:tagger}. The sum-of-weights for each class is normalised to 1 and logarithmic $y$-axis scales are used. The error bars correspond to MC statistical uncertainties.}
    \label{fig:output_nodes}
\end{figure}

Eq.~\ref{eqn:tagger}, which combines the networks trained using {\kfold}s~0 and 1, is plotted for each output node in Fig.~\ref{fig:output_nodes}. Additionally, the output nodes for the networks trained using {\kfold}s~0 and 1 are independently plotted in Figs.~\ref{fig:traintest0} and \ref{fig:traintest1}, respectively, in Appendix~\ref{app:training_plots}. Each network is applied to both the ``training'' events (90\% -- includes both the actual events used in training as well as those used in validation) and to the testing events (10\%). In all distributions, the training and testing curves are in good agreement with one another, indicating that no overfitting occurred.

\begin{figure}
    \centering
    \includegraphics[width=0.9\textwidth]{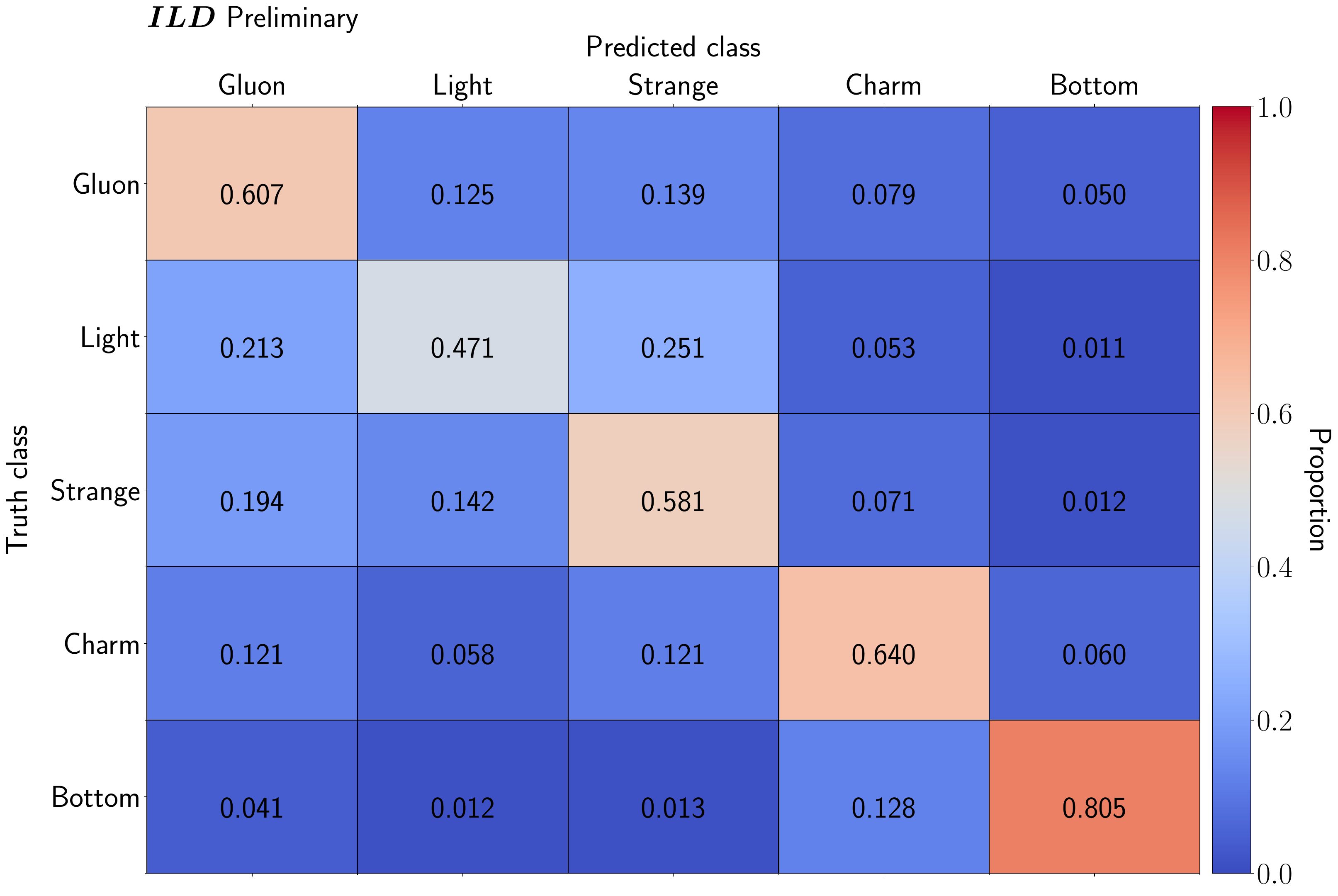}
    \caption{Confusion matrix for the output of the described jet flavour tagger, Eq.~\ref{eqn:tagger}. Each truth class (i.e., row) is normalised to 1.}
    \label{fig:confusion_matrix}
\end{figure}

There is clear discrimination of $b$- and $c$-jets. Additionally, there is a capacity for \emph{independently} tagging light-, $s$- and $g$-jets, but the separation power is somewhat reduced in comparison to $b$- and $c$-jets as these classes are more often confused with one another. This is demonstrated by the confusion matrix shown in Fig.~\ref{fig:confusion_matrix}, where the off-diagonal terms are of order 10--25\% in the upper 3$\times$3 (i.e., gluon, light, and strange) matrix compared to off-diagonal terms of order 5--10\% in the lower 2$\times$2 (i.e., charm and bottom) matrix. The cells connecting the upper 3$\times$3 and lower 2$\times$2 matrices are also of order 5--10\%, indicating little confusion between gluon/light/strange jets and charm/bottom jets.

\begin{figure}[htbp]
    \centering
    \begin{subfigure}{0.49\textwidth}
        \centering
        \includegraphics[width=1.\textwidth]{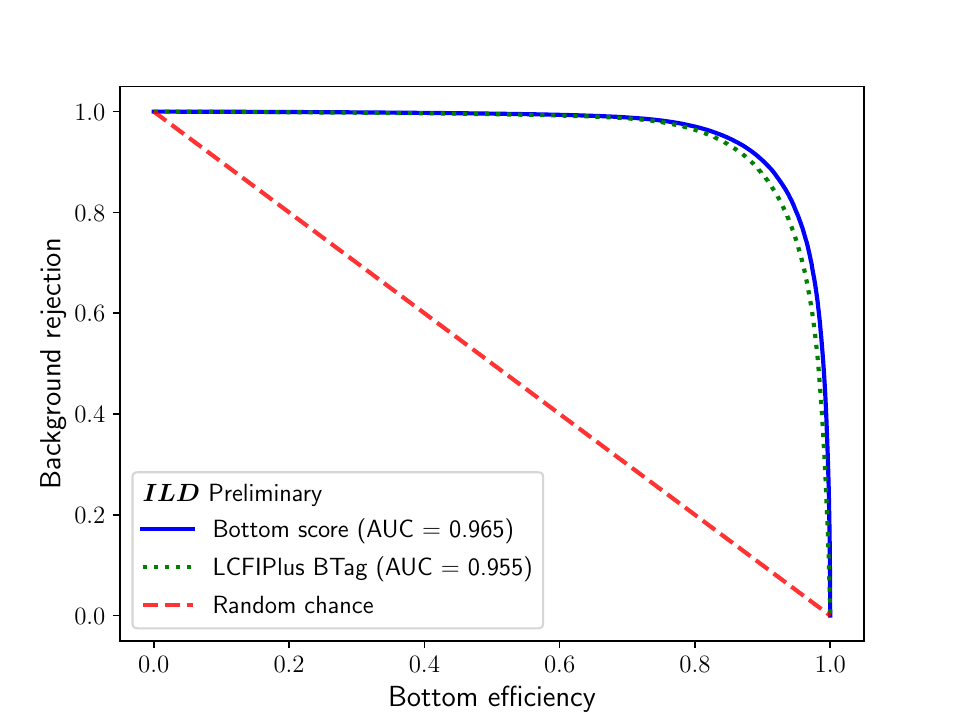}
        \caption{$b$-jet score}
    \end{subfigure}
    \hfill
    \begin{subfigure}{0.49\textwidth}
        \centering
        \includegraphics[width=1.\textwidth]{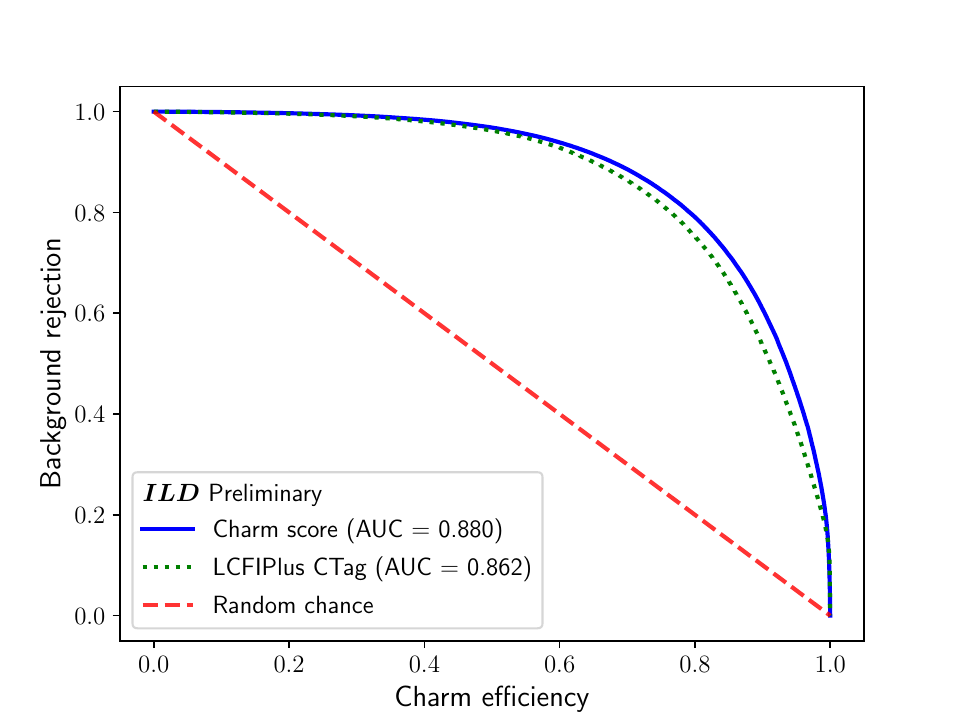}
        \caption{$c$-jet score}
    \end{subfigure} \\
    \begin{subfigure}{0.49\textwidth}
        \centering
        \includegraphics[width=1.\textwidth]{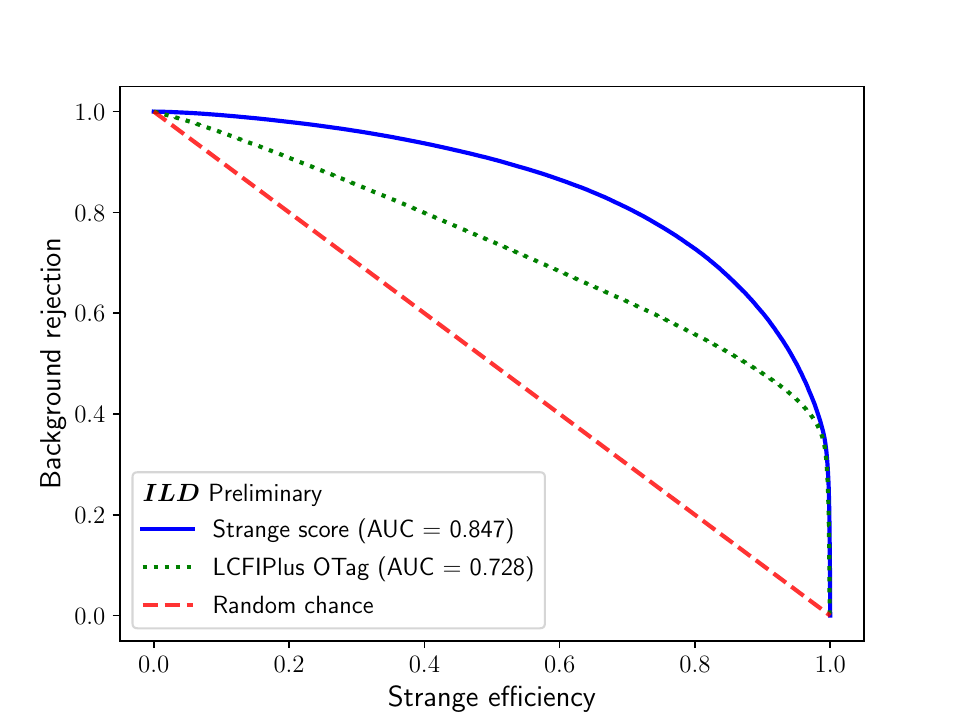}
        \caption{$s$-jet score}
    \end{subfigure}
    \hfill
    \begin{subfigure}{0.49\textwidth}
        \centering
        \includegraphics[width=1.\textwidth]{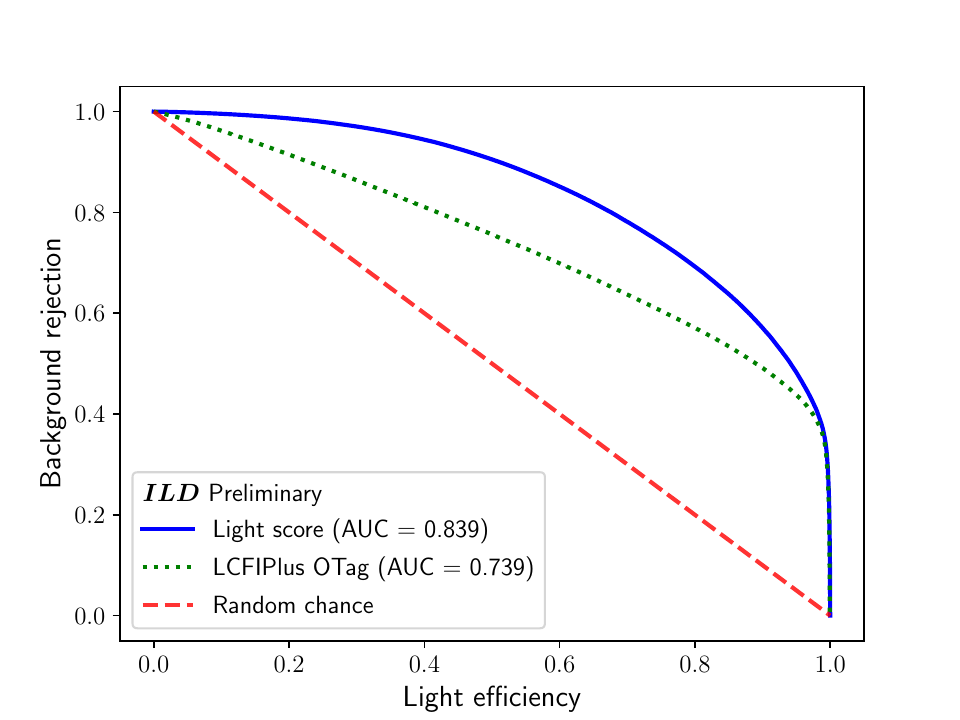}
        \caption{Light-jet score}
    \end{subfigure} \\
    \begin{subfigure}{0.49\textwidth}
        \centering
        \includegraphics[width=1.\textwidth]{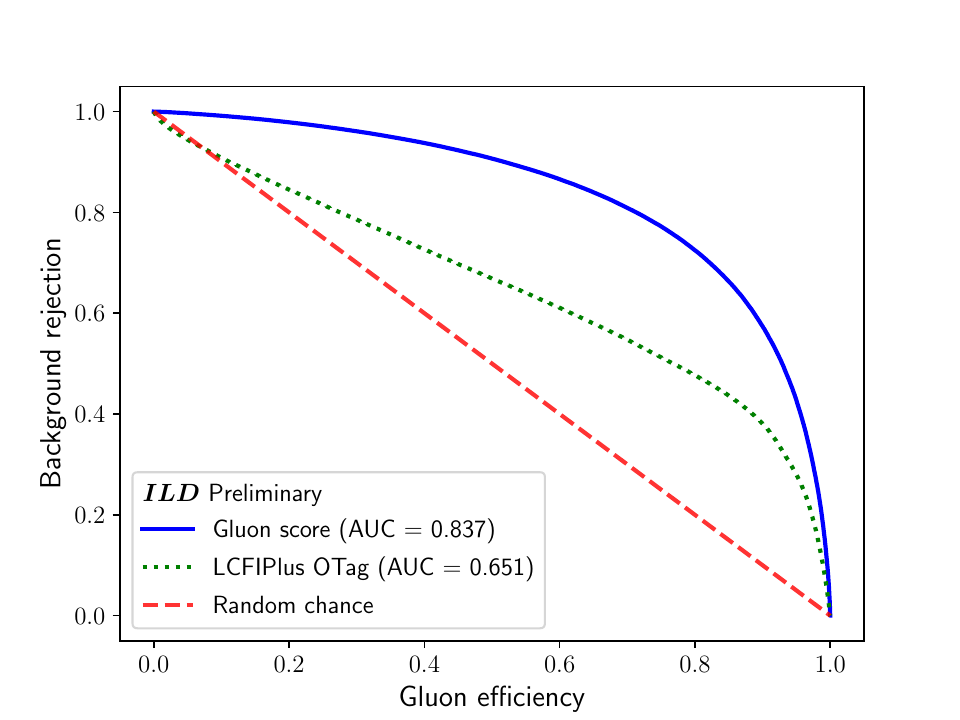}
        \caption{$g$-jet score}
    \end{subfigure} \\
    \caption{ROC curves for each output node of the described jet flavour tagger, Eq.~\ref{eqn:tagger}. Also shown on each graph is the ROC curve for the appropriate LCFIPlus tagger: ``BTag'' for the $b$-jet node, ``CTag'' for the $c$-jet node, and ``OTag'' for the light-, $s$-, and $g$-jet nodes. The area under the curve (AUC) is given for each tagger -- ideally, AUC = 1 (i.e., 100\% background rejection with 100\% signal efficiency). The sum-of-weights for each class is normalised to 1. The ``Background'' in a given plot corresponds to all classes not targeted by that node of the tagger.}
    \label{fig:roc}
\end{figure}

In order to quantify the performance of each network, the receiver-operator characteristic (ROC) curves (i.e., background rejection as a function of signal efficiency) are also calculated using Eq.~\ref{eqn:tagger} and shown in Fig.~\ref{fig:roc}. Alongside the tagger's ROC curves, the corresponding LCFIPlus results are also shown. Small improvements are seen for the $b$- and $c$-jet output nodes -- likely, the tagger is simply returning the input LCFIPlus tagger scores with small enhancements due to the truth PID on the jet's constituent PFOs. However, large improvements are observed for light-, $s$-, and $g$-jet tagging when using the multiclassifier over the LCFIPlus OTag.

\begin{figure}[htbp]
    \centering
    \begin{subfigure}{0.49\textwidth}
        \centering
        \includegraphics[width=1.\textwidth]{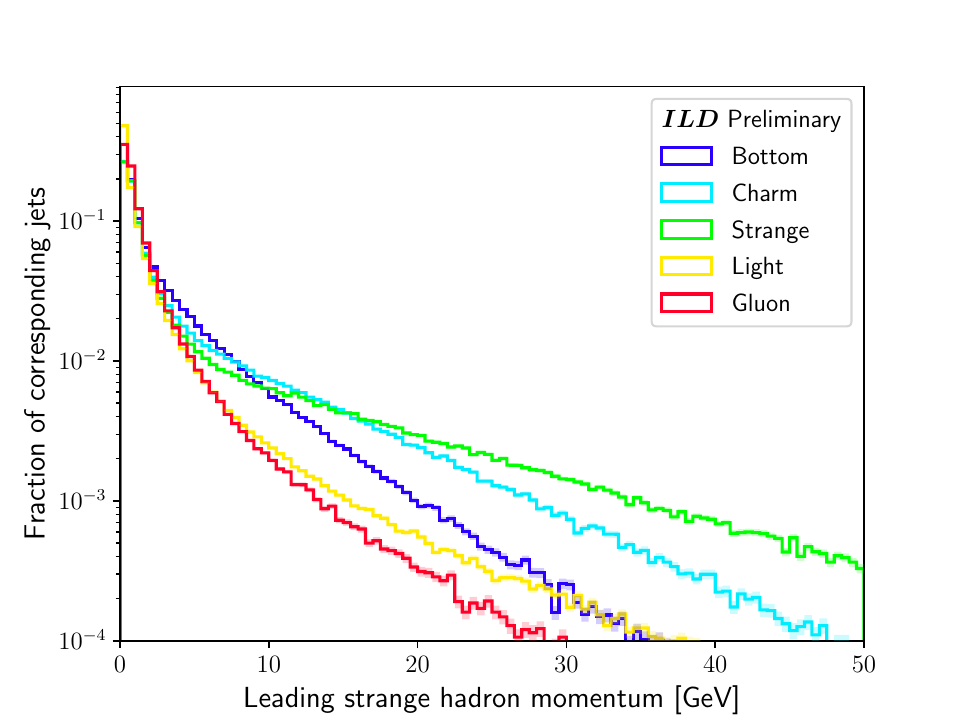}
        \caption{$s$-jet score $> 0.0$}
    \end{subfigure}
    \hfill
    \begin{subfigure}{0.49\textwidth}
        \centering
        \includegraphics[width=1.\textwidth]{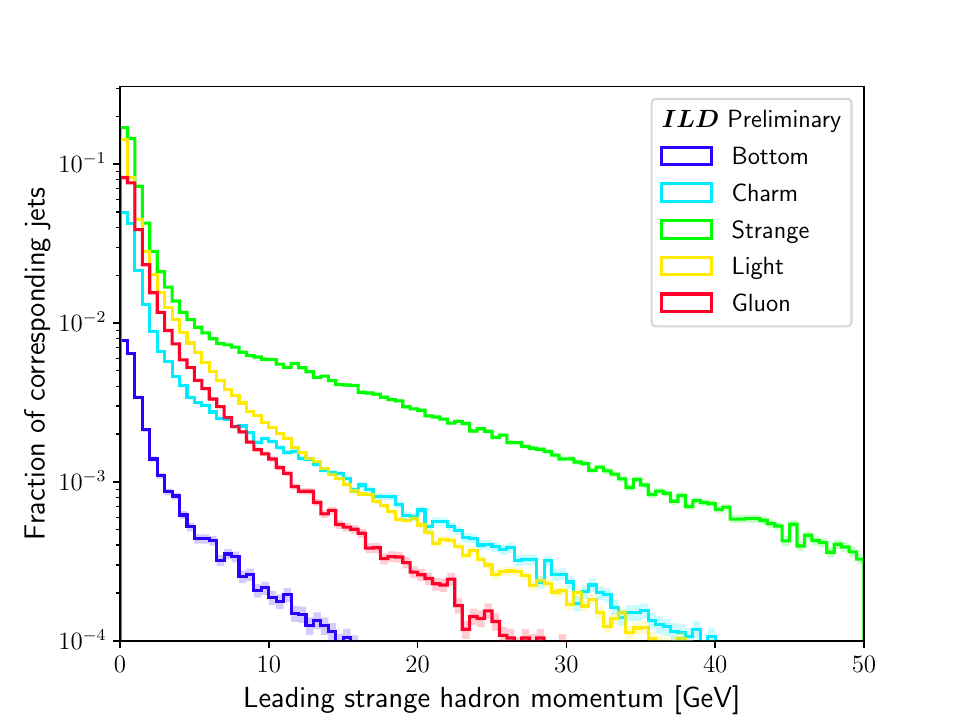}
        \caption{$s$-jet score $> 0.2$}
    \end{subfigure} \\
    \begin{subfigure}{0.49\textwidth}
        \centering
        \includegraphics[width=1.\textwidth]{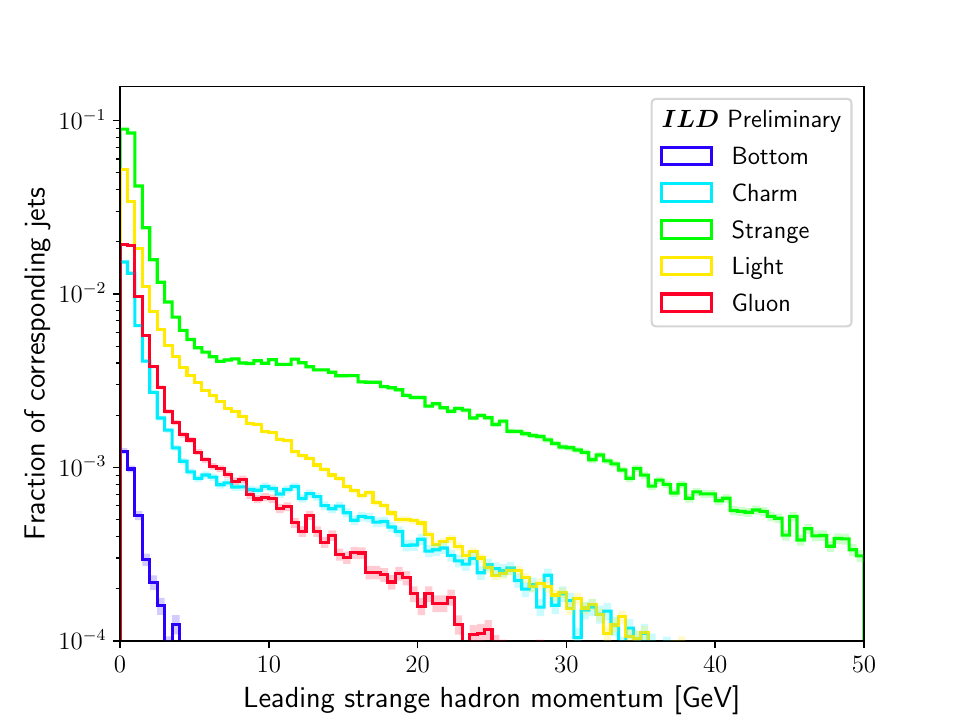}
        \caption{$s$-jet score $> 0.4$}
    \end{subfigure}
    \hfill
    \begin{subfigure}{0.49\textwidth}
        \centering
        \includegraphics[width=1.\textwidth]{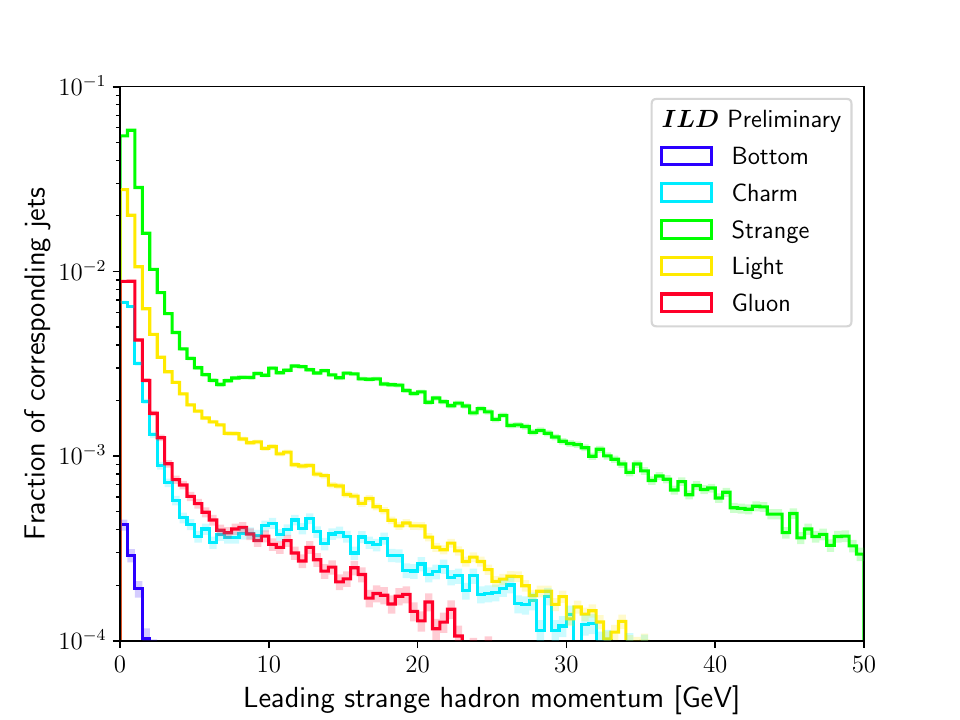}
        \caption{$s$-jet score $> 0.5$}
    \end{subfigure} \\
    \begin{subfigure}{0.49\textwidth}
        \centering
        \includegraphics[width=1.\textwidth]{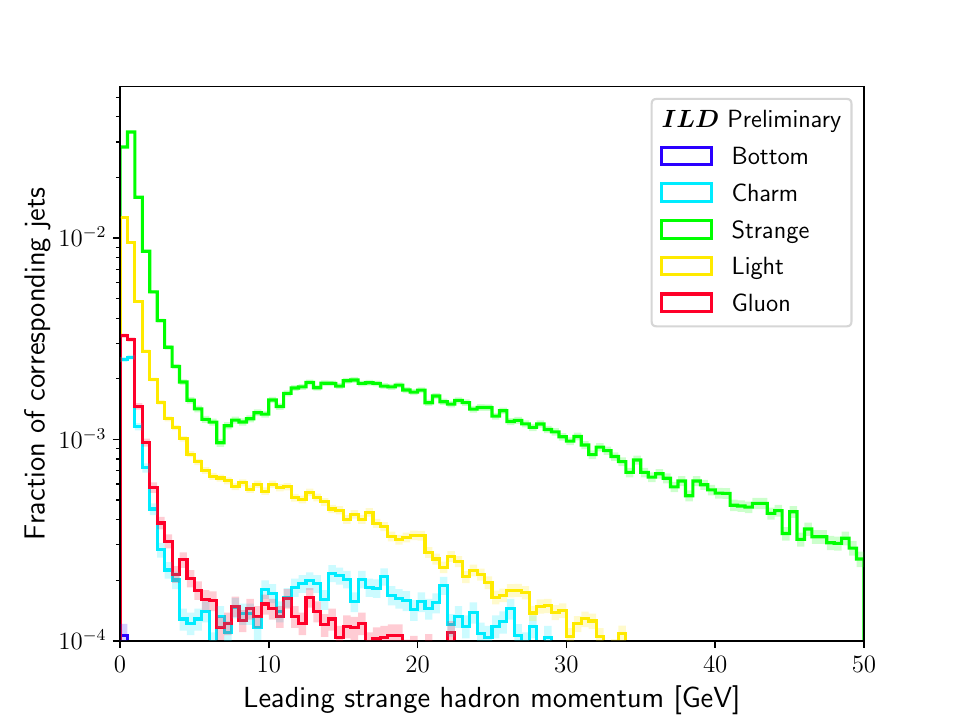}
        \caption{$s$-jet score $> 0.6$}
    \end{subfigure}
    \hfill
    \begin{subfigure}{0.49\textwidth}
        \centering
        \includegraphics[width=1.\textwidth]{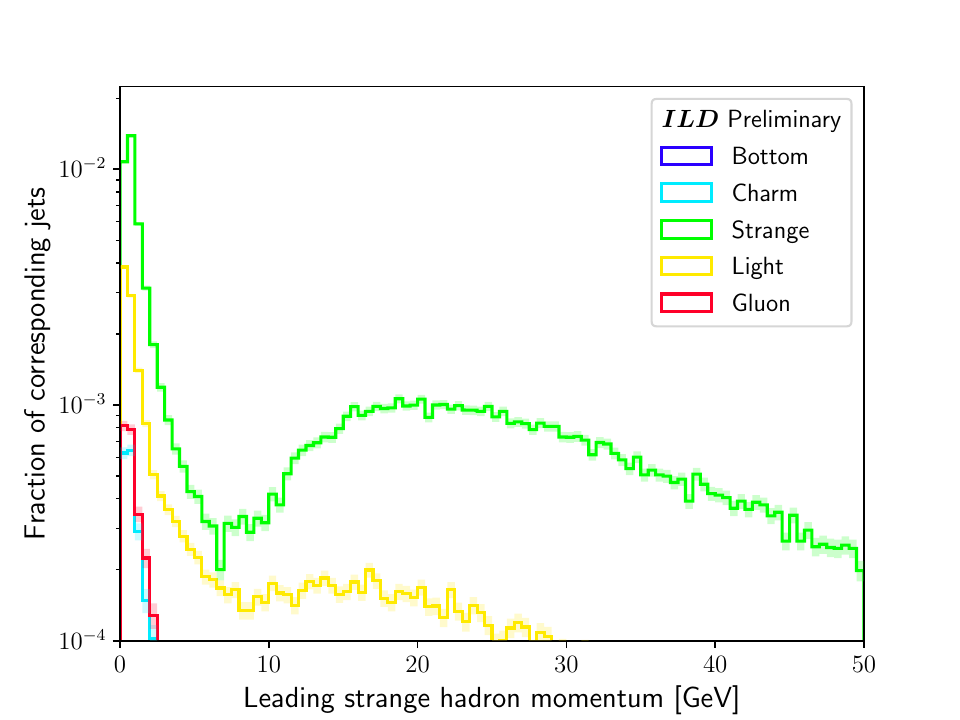}
        \caption{$s$-jet score $> 0.7$}
        \label{fig:leading_strange_p_RNN_cuts_0p7}
    \end{subfigure} \\
    \caption{Distributions of the momentum of the leading strange particle in jets from $h(\rightarrow q\bar{q}/gg)Z(\rightarrow\nu\bar{\nu})$ events. The distributions are shown for different choices of cut on $s$-jet score of the described jet flavour tagger, Eq.~\ref{eqn:tagger}. The momentum of the leading strange particle is determined by following the same procedure as for Fig.~\ref{fig:leading_strange_p}. The sum-of-weights for each class is normalised to 1 in (a) but is \emph{not} renormalised following the application of cuts in (b) through (e).}
    \label{fig:leading_strange_p_RNN_cuts}
\end{figure}

We have included plots of the leading strange hadron momentum, following the same procedure as for Fig~\ref{fig:leading_strange_p}, for different choices of cut on the $s$-jet score. These are shown in Fig.~\ref{fig:leading_strange_p_RNN_cuts}. As we cut tighter on the $s$-jet score (i.e., generate a region purer in strange jets), the fraction of jets with leading strange hadrons with momentum above \unit[10]{GeV} increases. Looking specifically at Fig.~\ref{fig:leading_strange_p_RNN_cuts_0p7}, the weights in the bins with momentum below \unit[10]{GeV} are an order-of-magnitude larger than the weights in the bins with momentum above \unit[10]{GeV}; however, there are approximately an order-of-magnitude more bins above \unit[10]{GeV} than below. These two effects roughly cancel, leading to a 50/50\% split of strange jets with leading strange hadron momentum above/below \unit[10]{GeV}. What this indicates is that having PID for particles with momentum greater than \unit[10]{GeV}, as already predicted in Section~\ref{sec:intro}, is \emph{paramount} for tagging strange jets. Appendix~\ref{app:no_PID_tagger_ana} describes a more detailed study of the effect of including PID information for specific momentum ranges, and a proposal on how this may be achieved is described in Section~\ref{sec:alt_detector}.

\FloatBarrier

\section{Higgs to strange analysis}
\label{sec:analysis}

The jet flavour tagger described in Section~\ref{sec:tagger} is applied to a search for SM Higgs decaying to strange quarks (\Hss), using all of the MC samples described in Table~\ref{tab:samples}. The parameter of interest (POI) for the analysis is the Higgs-strange quark coupling strength modifier, $\kappa_s$, which tunes the SM \Hss BR, $\BR[\Hss]_\textrm{SM}$, as:

\begin{equation}
    \BR[\Hss] = \mu(\kappa_s) \times \BR[\Hss]_\textrm{SM} \,,
    \label{eqn:branching_ratio}
\end{equation}

\noindent where $\BR[\Hss]$ is the modified BR and $\mu(\kappa_s)$ is our signal strength modifier as a function of $\kappa_s$, given by\footnote{The signal strength modifier has the same form as that used by ATLAS for measuring the Higgs-charm quark coupling -- for instance, see Eq.~1 of Ref.~\cite{VHbbcc}.}:

\begin{equation}
    \mu(\kappa_s) = \frac{\kappa_s^2}{\kappa_s^2 \times \BR[\Hss]_\textrm{SM} + (1 - \BR[\Hss]_\textrm{SM})} \,.
    \label{eqn:mu}
\end{equation}

\noindent The denominator is to account for the modification of the total decay width of the Higgs given the modified \Hss decay width. The coupling strength modifier is understood within the context of the kappa framework, the experimental tool for exploring the properties of the Higgs~\cite{LHCHXSWGRec, LHCHXSWGProp}. When $\kappa_s = 1$, the SM BR is recovered. N.B. in the limit $\BR[\Hss]_\textrm{SM} \ll 1$ in Eq.~\ref{eqn:mu} (which is a valid assumption), Eq.~\ref{eqn:branching_ratio} reduces to the intuitive result: $\BR[\Hss] \approx \kappa_s^2 \times \BR[\Hss]_\textrm{SM}$.

\subsection{Kinematic selections}
\label{sec:selections}

The measurement of \Hss is performed using the associated production mode in two channels based on the decay of the $Z$: \Zinv and \Zll. The kinematic selections for each channel, detailed in Table~\ref{tab:selections}, are designed to be orthogonal and to reduce the dominant $Z$, $VV$, and $h\rightarrow b\bar{b}/c\bar{c}/gg$ backgrounds. The cuts on the number of PFOs per event and per jet reduce the $\Hgg$ backgrounds -- in general, gluon jets have a higher track multiplicity than quark jets. N.B. the $h(\rightarrow s\bar{s})Z(\rightarrow\nu\bar{\nu})$ and $h(\rightarrow s\bar{s})Z(\rightarrow\ell\bar{\ell})$ processes are combined to define the signal template for both channels (with orthogonality applied via the object counting cuts).

\begin{table}[H]
    \centering
    \caption{Kinematic selections for \Zinv and \Zll channels of the \Hss analysis. The selections are grouped into categories serving specific purposes.}
    \label{tab:selections}
    \resizebox{\textwidth}{!}{
        \begin{tabular}{c||l|r|r}
            \toprule
            Category & Selection & \Zinv & \Zll \\
            \midrule
            \multirow{3}{*}{Object counting} & Number of leptons, $N_\textrm{leptons}$ & 0 & $\geq 2$ \\
            & Number of jets, $N_\textrm{jets}$ & $\geq 2$ & $\geq 2$ \\
            & Leading 2 leptons are SFOS\tablefootnote{``SFOS'' $\coloneqq$ ``same-flavour, opposite-sign''.} & -- & True \\
            \midrule
            \multirow{11}{*}{$2f$ $Z$ rejection} & Leading jet momentum, $p_{j_0}$ & $\in \unit[[40, 110]]{\,\textrm{GeV}}$ & $\in \unit[[60, 105]]{\,\textrm{GeV}}$ \\
            & Subleading jet momentum, $p_{j_1}$ & $\in \unit[[30, 80]]{\,\textrm{GeV}}$ & $\in \unit[[35, 75]]{\,\textrm{GeV}}$ \\
            & Dijet mass, $M_{jj}$ & $\in \unit[[120, 140]]{\,\textrm{GeV}}$ & $\in \unit[[115, 145]]{\,\textrm{GeV}}$ \\
            & Dijet energy, $E_{jj}$ & $\in \unit[[125, 155]]{\,\textrm{GeV}}$ & $\in \unit[[130, 156]]{\,\textrm{GeV}}$ \\
            & Missing mass, $M_\textrm{miss}$ & $\in \unit[[75, 120]]{\,\textrm{GeV}}$ & -- \\
            & Dijet/missing-$p^\mu$ angular separation, ${\Delta}R_{jj,\textrm{miss}}$\tablefootnote{${\Delta}R \equiv \sqrt{\Delta\eta^2 + \Delta\phi^2}$, where $\Delta\eta$ is the rapidity separation and $\Delta\phi$ is the azimuthal separation.} & $\in [3.1, 4.0]$\tablefootnote{If the 4-vectors of jets summed to ($\sqrt{s}$, 0, 0, 0), then ${\Delta}R_{jj\textrm{,miss}} = \pi$ -- the fact that this isn't true implies there are additional PFOs or tracks in the event not grouped into either jet.} & -- \\
            & Dijet azimuthal separation, $\Delta\phi_{jj}$ & $> \unit[1.25]{rad}$ & $> \unit[1.75]{rad}$ \\
            & Leading lepton momentum, $p_{\ell_0}$ & -- & $\in \unit[[40, 90]]{\,\textrm{GeV}}$ \\
            & Subleading lepton momentum, $p_{\ell_1}$ & -- & $\in \unit[[20, 60]]{\,\textrm{GeV}}$ \\
            & Dilepton mass, $M_{\ell\bar{\ell}}$ & -- & $\in \unit[[80, 100]]{\,\textrm{GeV}}$ \\
            & Dilepton energy, $E_{\ell\bar{\ell}}$ & -- & $\in \unit[[85, 115]]{\,\textrm{GeV}}$ \\
            & Recoil mass, $M_\textrm{recoil}$\tablefootnote{$M_\textrm{recoil} \equiv (p^\mu_\textrm{COM} - p^\mu_Z)^2 = (p^\mu_\textrm{COM} - p^\mu_{\ell\bar{\ell}})^2$, where $p^\mu_\textrm{COM}$ is the center-of-mass (COM) 4-momentum and $p^\mu_Z$ is the 4-momentum of the $Z$ boson.} & -- & $\in \unit[[122, 155]]{\,\textrm{GeV}}$ \\
            \midrule
            \multirow{4}{*}{$h\rightarrow b\bar{b}/c\bar{c}$ rejection} & Leading jet LCFIPlus BTag score, $\textrm{score}^{j_0}_b$ & $< 0.20$ & $< 0.1$ \\
            & Subleading jet LCFIPlus BTag score, $\textrm{score}^{j_1}_b$ & $< 0.20$ & $< 0.1$ \\
            & Leading jet LCFIPlus CTag score, $\textrm{score}^{j_0}_c$ & $< 0.35$ & $< 0.3$ \\
            & Subleading jet LCFIPlus CTag score, $\textrm{score}^{j_1}_c$ & $< 0.35$ & $< 0.3$ \\
            \midrule
            \multirow{2}{*}{$4f$ $VV$ rejection} & $2\rightarrow3$ jet transition variable, $y_{23}$\tablefootnote{In jet clustering, the jet transition variable, $y_{N\rightarrow M}$, is the value of the upper cut on the distance parameter between final state particles, $y_{ij}$, at which the event goes from a $N$-jet event to a $M$-jet event where $M = N + 1$.} & $< 0.010$ & $< 0.050$ \\
            & $3\rightarrow4$ jet transition variable, $y_{34}$ & $< 0.002$ & $< 0.005$ \\
            \midrule
            \multirow{3}{*}{$h\rightarrow gg$ rejection} & Number of PFOs in event, $N_\textrm{PFOs}^\textrm{event}$ & $\in [30, 60]$ & $\in [30, 70]$\\
            & Number of PFOs in leading jet, $N_\textrm{PFOs}^{j_0}$ & $\in [10, 40]$ & $\in [10, 40]$ \\
            & Number of PFOs in subleading jet, $N_\textrm{PFOs}^{j_1}$ & $\in [9, 37]$ & $\in [10, 40]$ \\
            \bottomrule
        \end{tabular}
    }
\end{table}

\begin{figure}[htbp]
    {\centering
    \begin{subfigure}{0.49\textwidth}
        \centering
        \includegraphics[width=1.\textwidth]{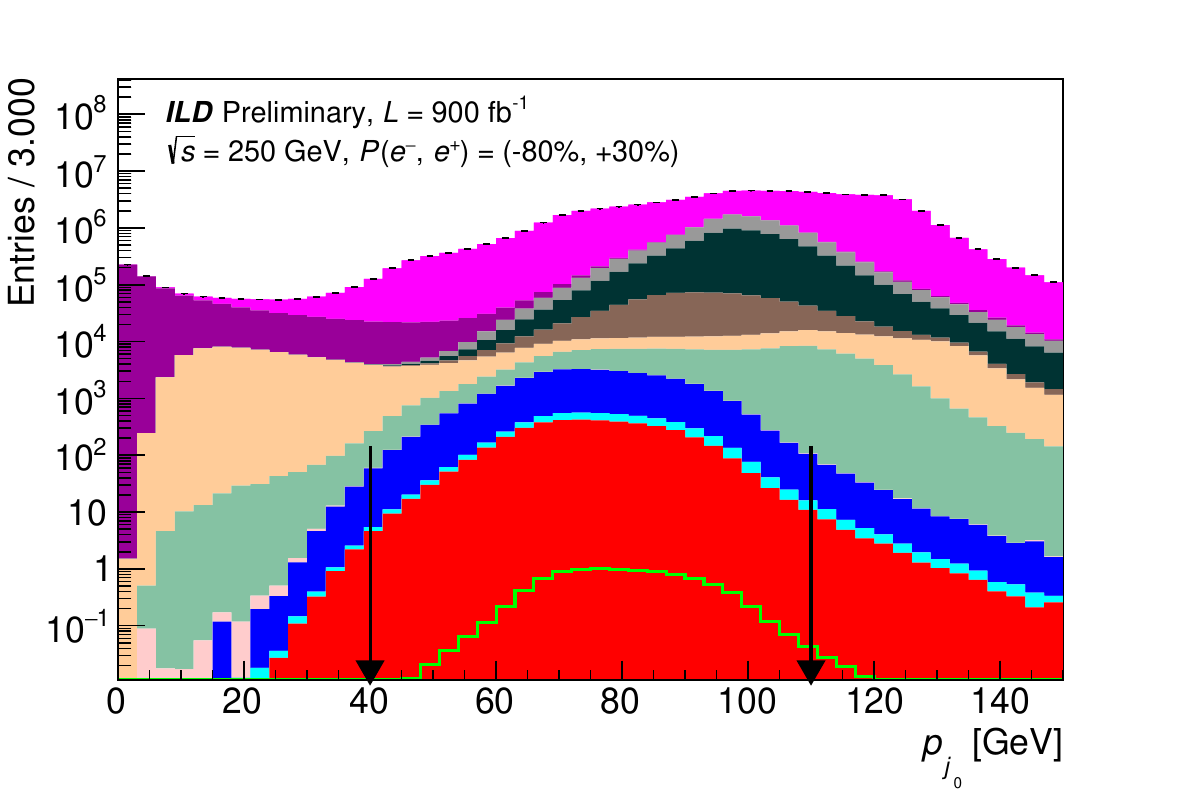}
        \caption{Leading jet momentum $p_{j_0}$}
    \end{subfigure}
    \hfill
    \begin{subfigure}{0.49\textwidth}
        \centering
        \includegraphics[width=1.\textwidth]{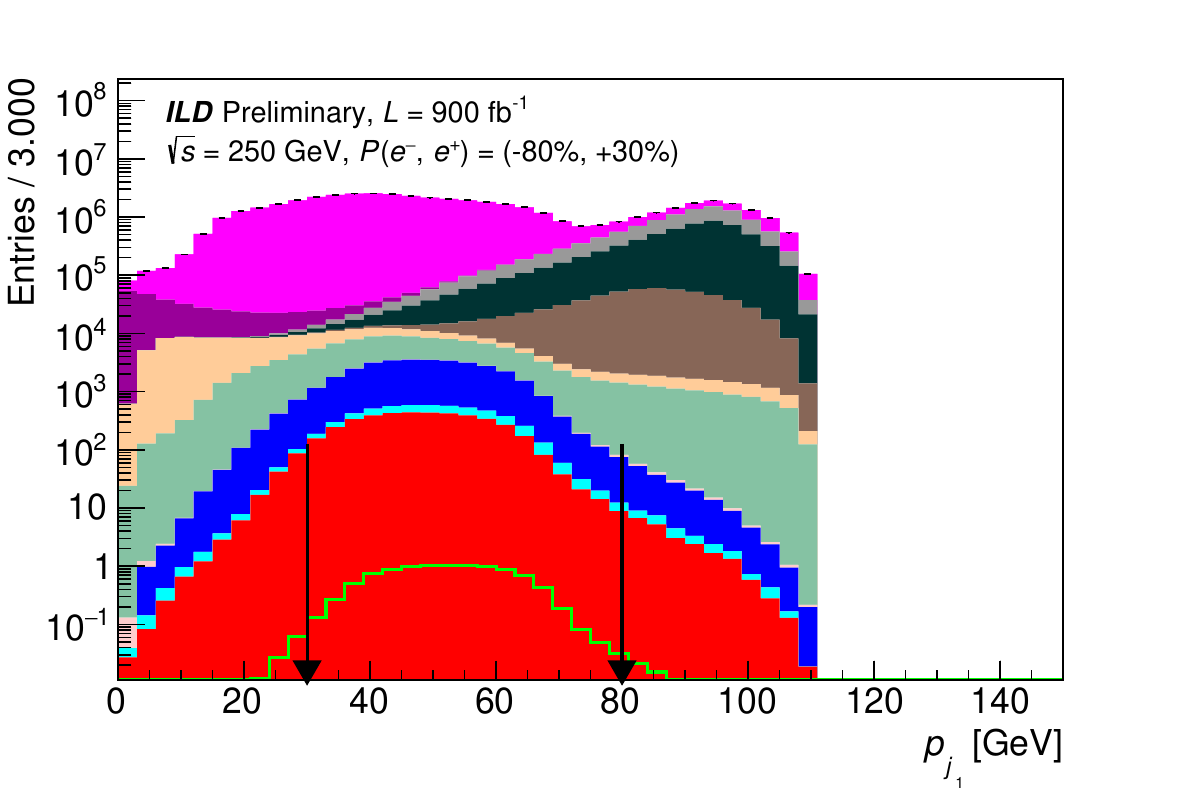}
        \caption{Subleading jet momentum $p_{j_1}$}
    \end{subfigure} \\
    \begin{subfigure}{0.49\textwidth}
        \centering
        \includegraphics[width=1.\textwidth]{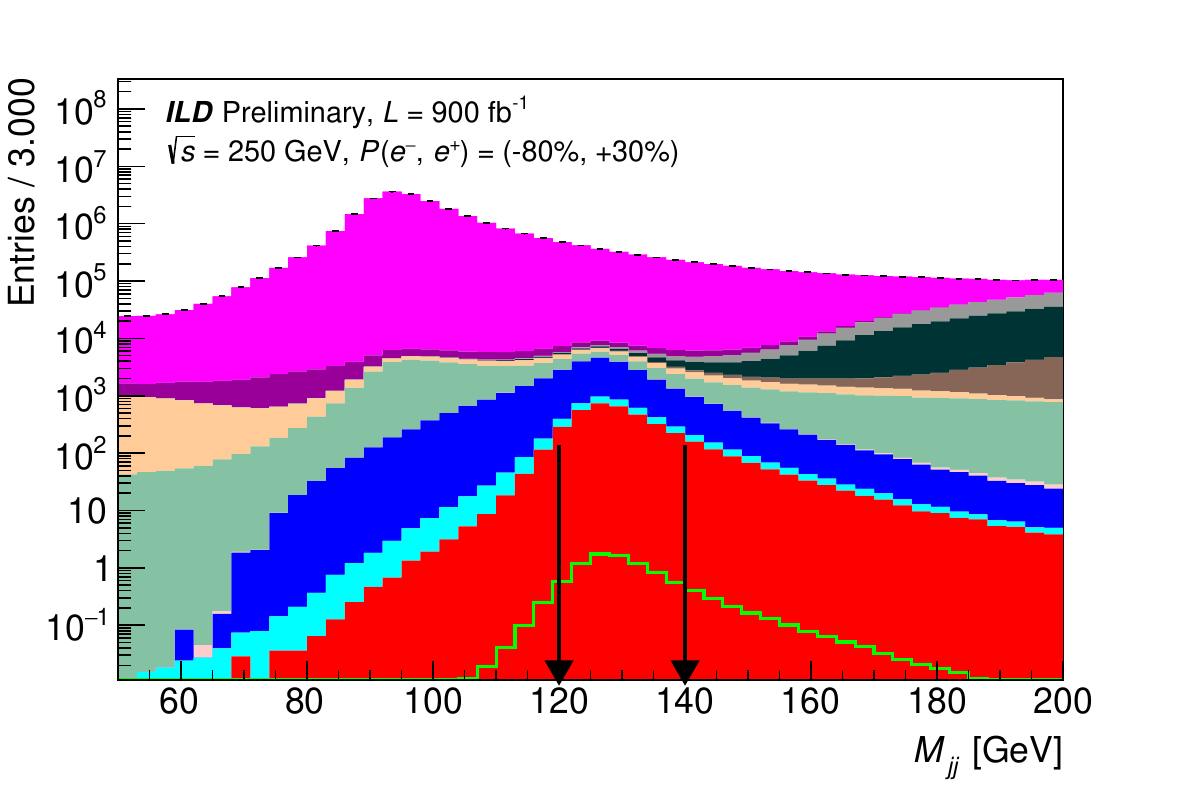}
        \caption{Dijet mass $M_{jj}$}
    \end{subfigure}
    \hfill
    \begin{subfigure}{0.49\textwidth}
        \centering
        \includegraphics[width=1.\textwidth]{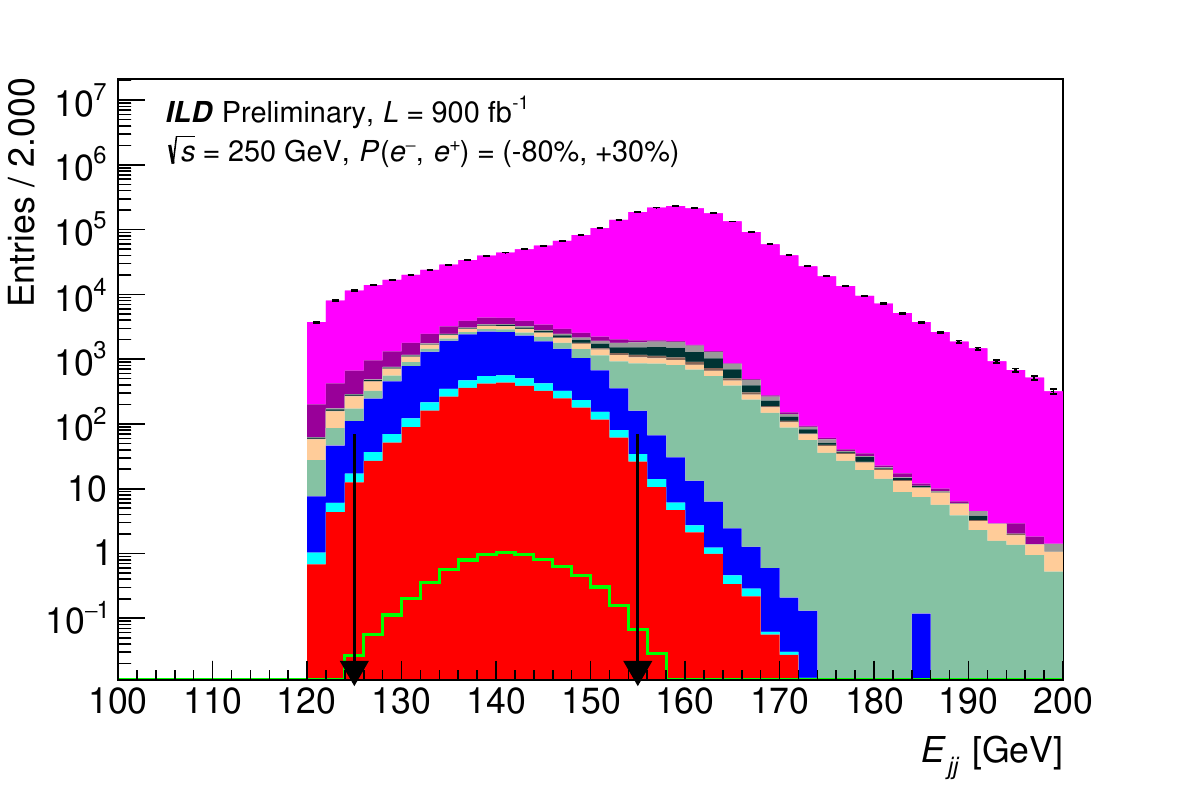}
        \caption{Dijet energy $E_{jj}$}
    \end{subfigure} \\
    }
    \begin{subfigure}{0.49\textwidth}
        \centering
        \includegraphics[width=1.\textwidth]{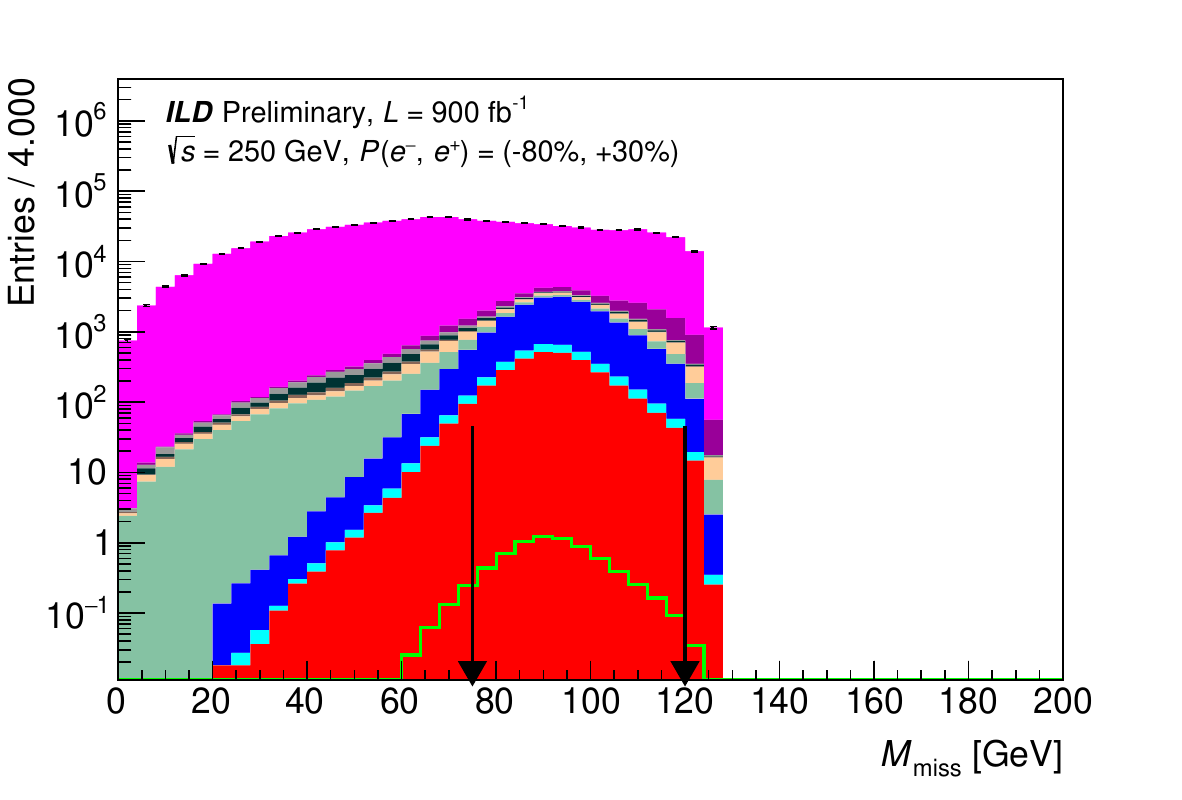}
        \caption{Missing mass $M_\textrm{miss}$}
    \end{subfigure}
    \begin{subfigure}{0.23\textwidth}
        \centering
        \includegraphics[width=1.\textwidth]{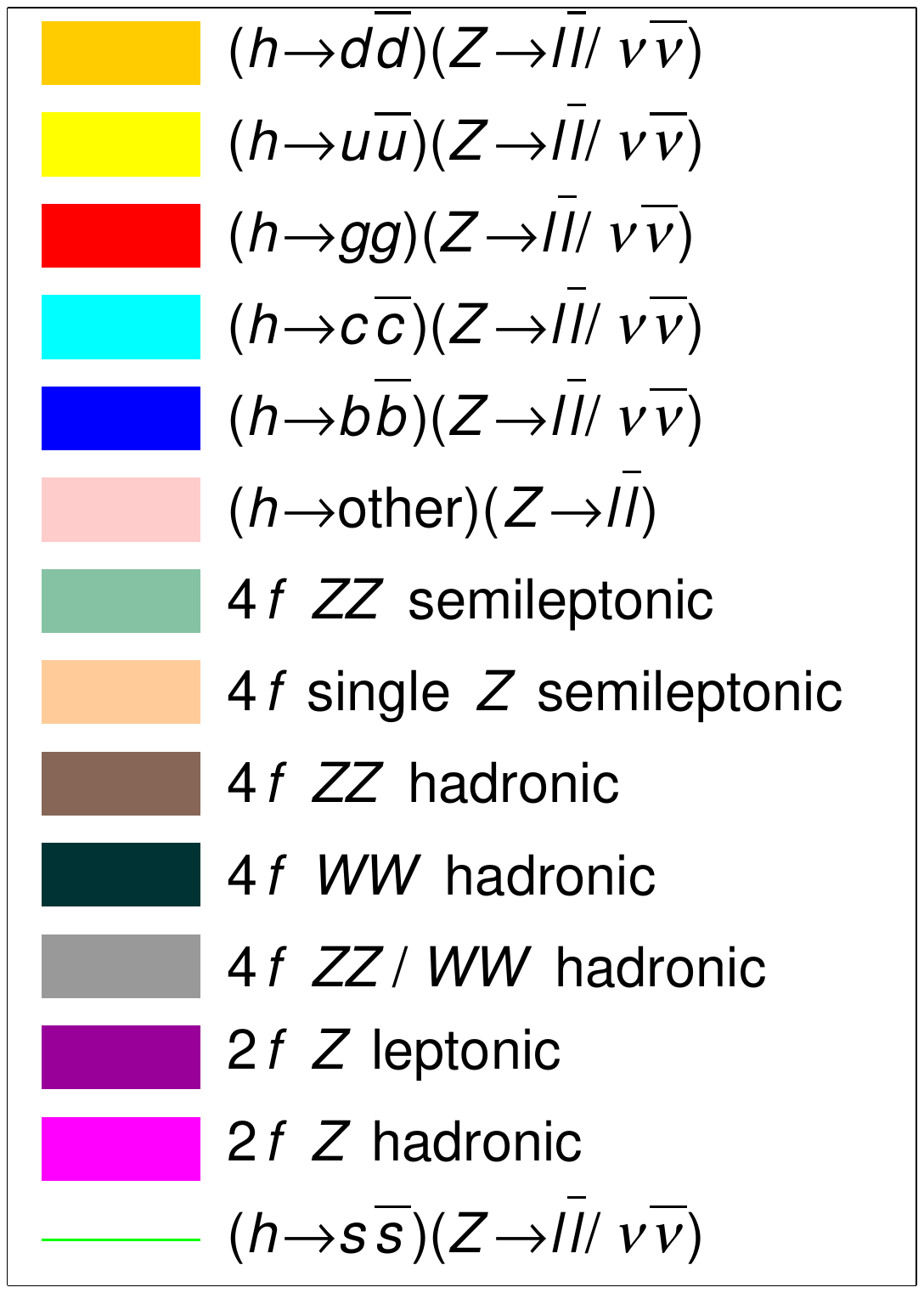}
    \end{subfigure} \\
    \caption{Histograms of the variables used in the kinematic selections of the \Zinv channel, as described in Table~\ref{tab:selections}. Each histogram is given at the level of its corresponding selection but \emph{before} that selection is applied. The arrows represent the placement of the selection cuts, and the error bars represent the MC statistical uncertainties. The sum-of-weights per process is normalised to the SM cross section. N.B. the $h(\rightarrow s\bar{s})Z(\rightarrow\ell\bar{\ell}/\nu\bar{\nu})$ signal is unstacked.}
    \label{fig:histograms_Zinv_1}
\end{figure}

\begin{figure}[htbp]
    {\centering
    \begin{subfigure}{0.49\textwidth}
        \centering
        \includegraphics[width=1.\textwidth]{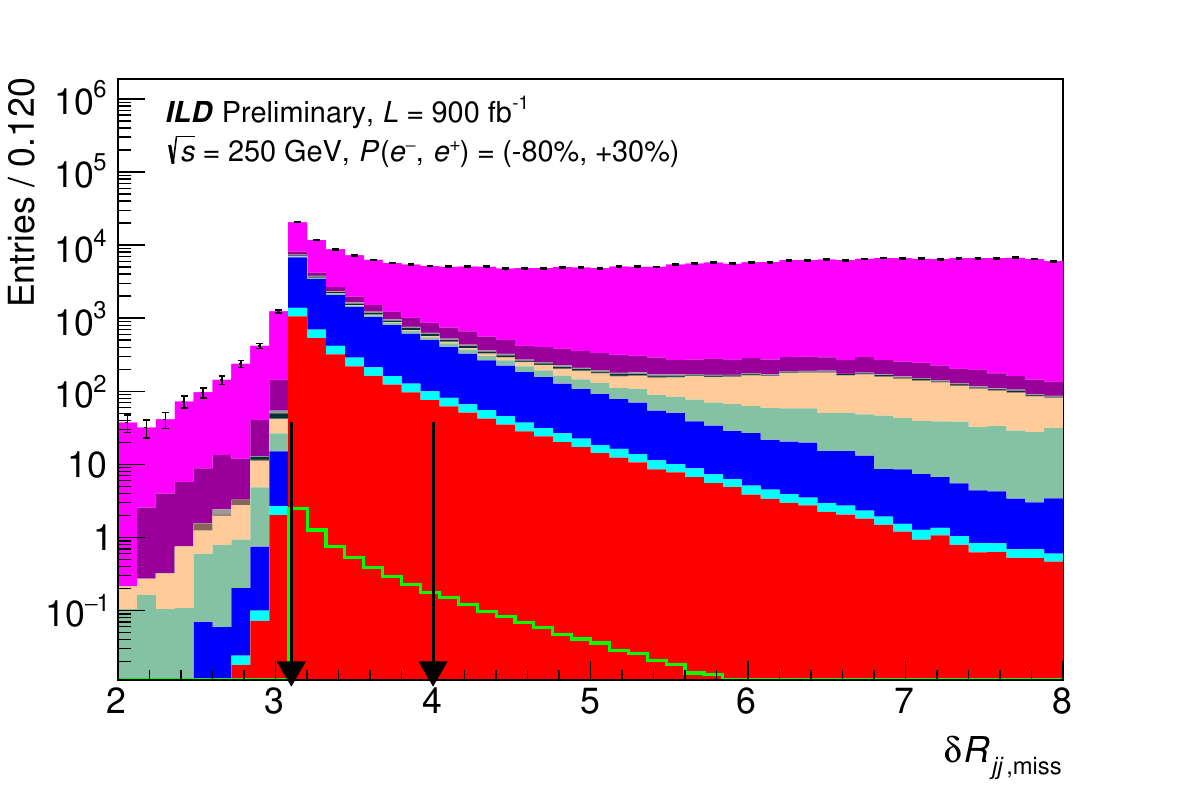}
        \caption{Angular separation ${\Delta}R_{jj\textrm{,miss}}$}
    \end{subfigure}
    \hfill
    \begin{subfigure}{0.49\textwidth}
        \centering
        \includegraphics[width=1.\textwidth]{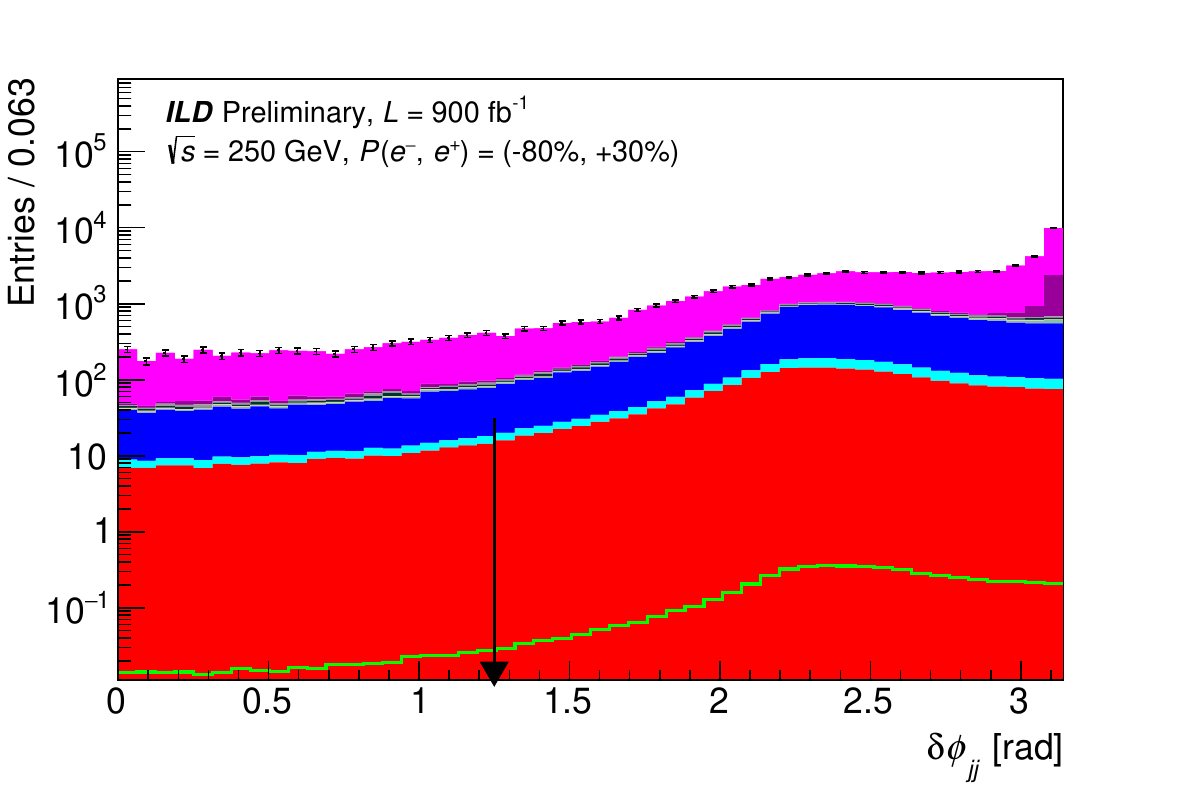}
        \caption{Dijet azimuthal separation $\Delta\phi_{jj}$}
    \end{subfigure} \\
    \begin{subfigure}{0.49\textwidth}
        \centering
        \includegraphics[width=1.\textwidth]{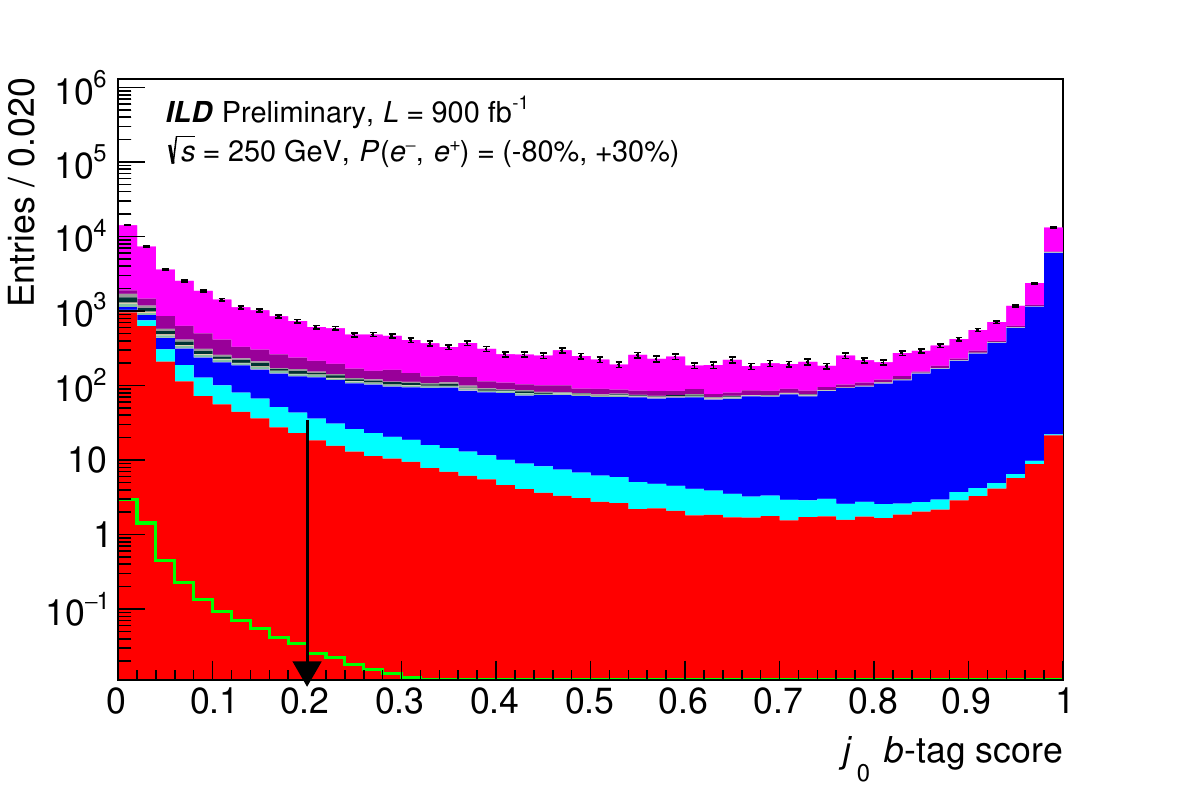}
        \caption{Leading jet BTag score}
    \end{subfigure}
    \hfill
    \begin{subfigure}{0.49\textwidth}
        \centering
        \includegraphics[width=1.\textwidth]{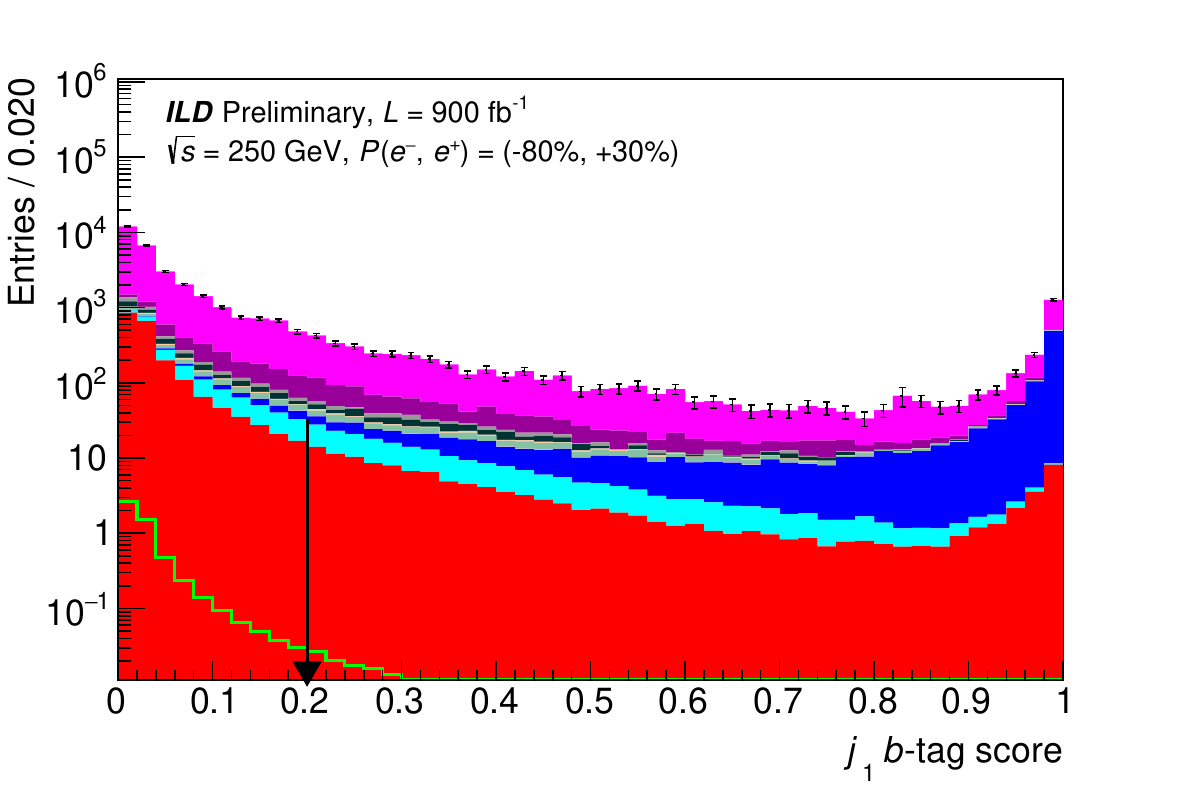}
        \caption{Subleading jet BTag score}
    \end{subfigure} \\
    }
    \begin{subfigure}{0.49\textwidth}
        \centering
        \includegraphics[width=1.\textwidth]{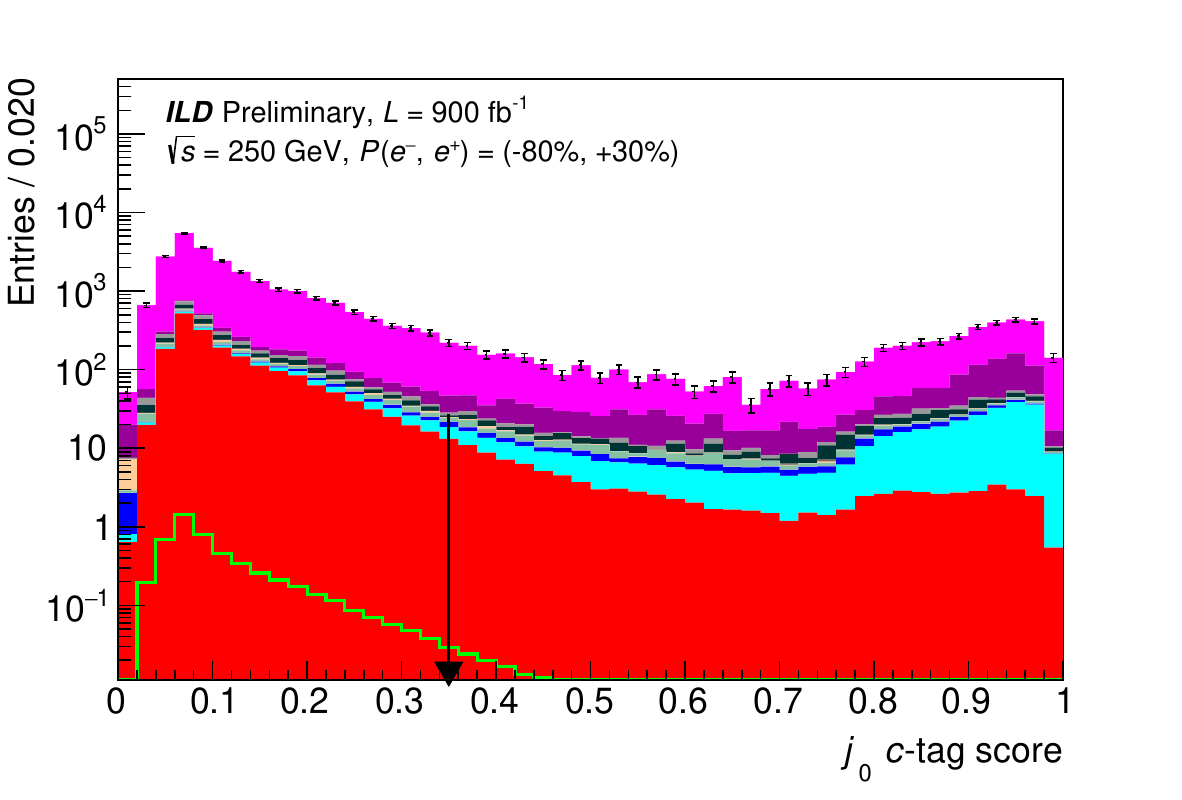}
        \caption{Leading jet CTag score}
    \end{subfigure}
    \begin{subfigure}{0.23\textwidth}
        \centering
        \includegraphics[width=1.\textwidth]{figures/analysis_Zinv/legend.pdf}
    \end{subfigure} \\
    \caption{Histograms of the variables used in the kinematic selections of the \Zinv channel, as described in Table~\ref{tab:selections}. Each histogram is given at the level of its corresponding selection but \emph{before} that selection is applied. The arrows represent the placement of the selection cuts, and the error bars represent the MC statistical uncertainties. The sum-of-weights per process is normalised to the SM cross section. N.B. the $h(\rightarrow s\bar{s})Z(\rightarrow\ell\bar{\ell}/\nu\bar{\nu})$ signal is unstacked. A continuation of Fig.~\ref{fig:histograms_Zinv_1}.}
    \label{fig:histograms_Zinv_2}
\end{figure}

\begin{figure}[htbp]
    {\centering
    \begin{subfigure}{0.49\textwidth}
        \centering
        \includegraphics[width=1.\textwidth]{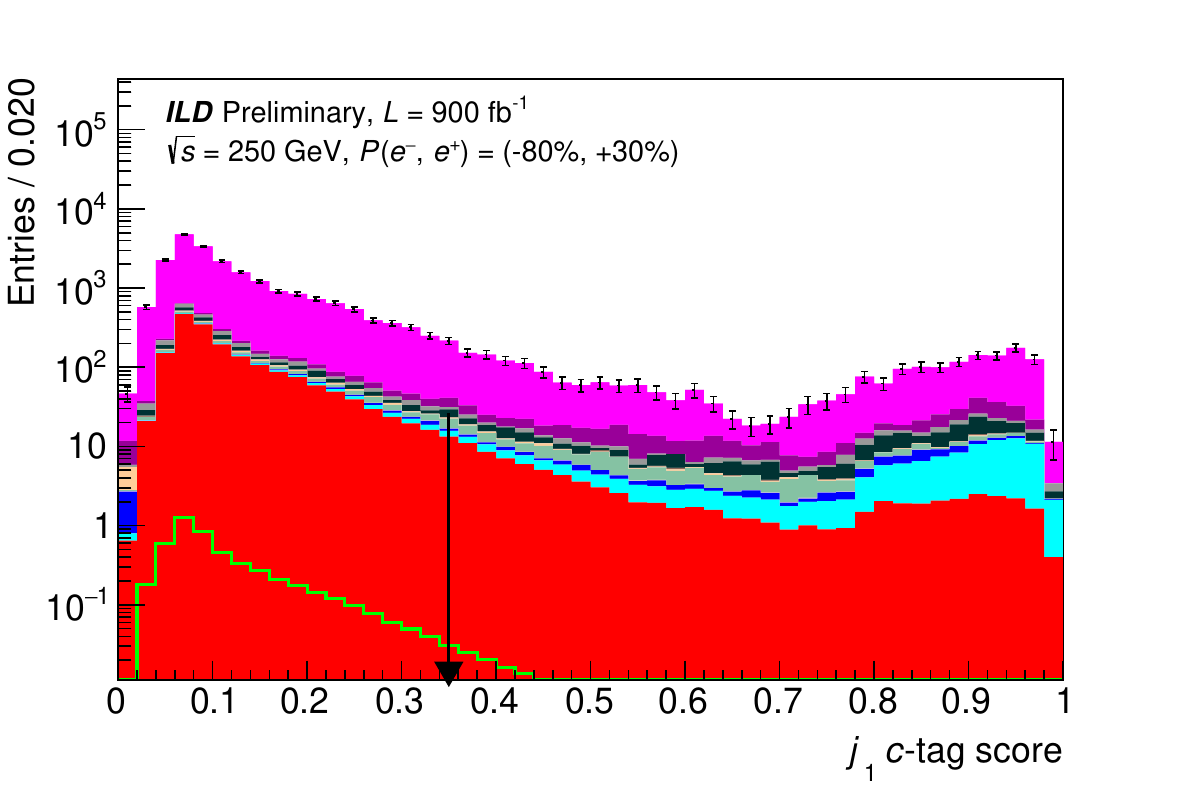}
        \caption{Subleading jet CTag score}
    \end{subfigure}
    \hfill
    \begin{subfigure}{0.49\textwidth}
        \centering
        \includegraphics[width=1.\textwidth]{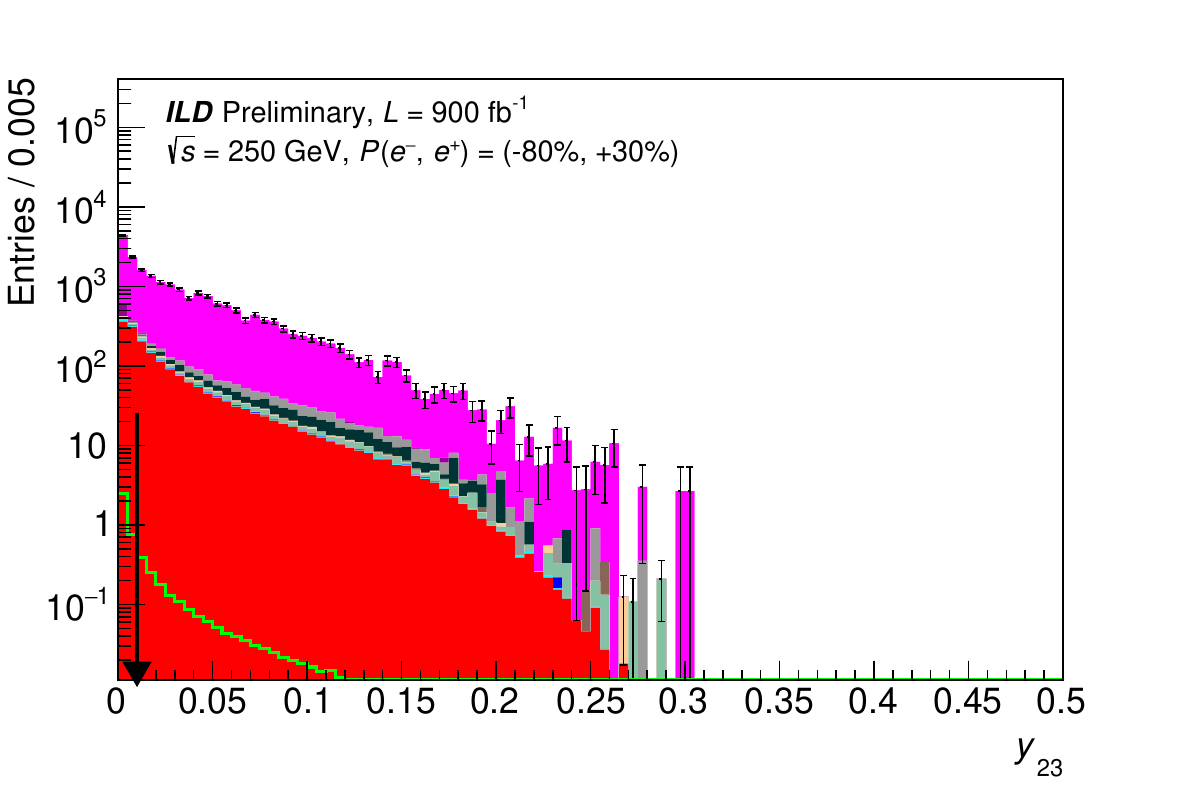}
        \caption{$2\rightarrow3$ jet transition variable $y_{23}$}
    \end{subfigure} \\
    \begin{subfigure}{0.49\textwidth}
        \centering
        \includegraphics[width=1.\textwidth]{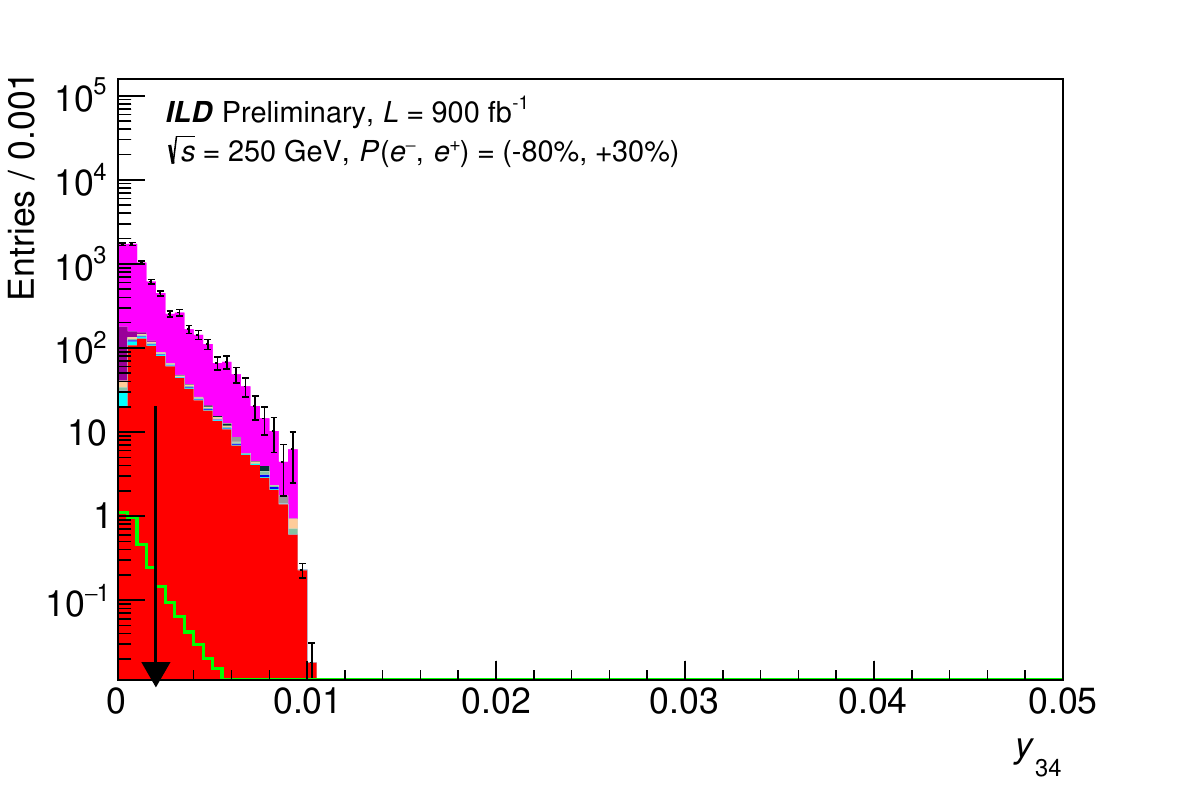}
        \caption{$3\rightarrow4$ jet transition variable $y_{34}$}
    \end{subfigure}
    \hfill
    \begin{subfigure}{0.49\textwidth}
        \centering
        \includegraphics[width=1.\textwidth]{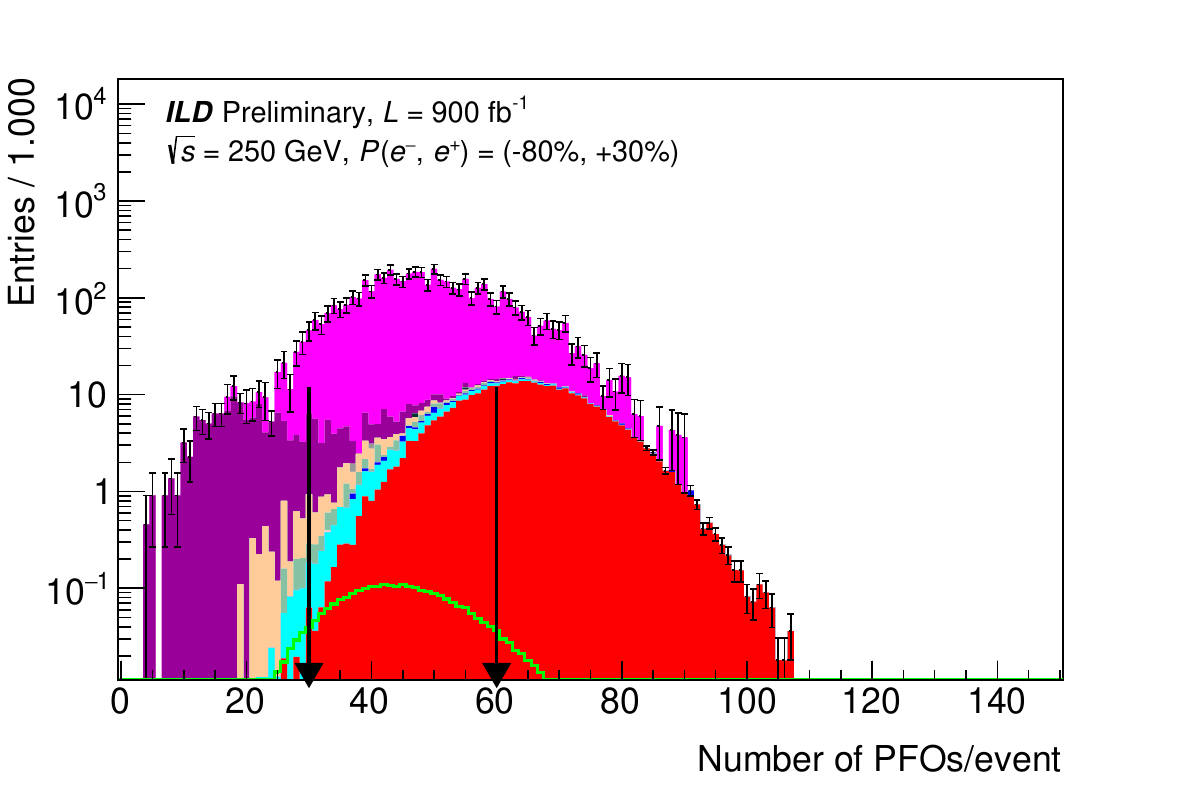}
        \caption{Number of PFOs in event}
    \end{subfigure} \\
    }
    \begin{subfigure}{0.49\textwidth}
        \centering
        \includegraphics[width=1.\textwidth]{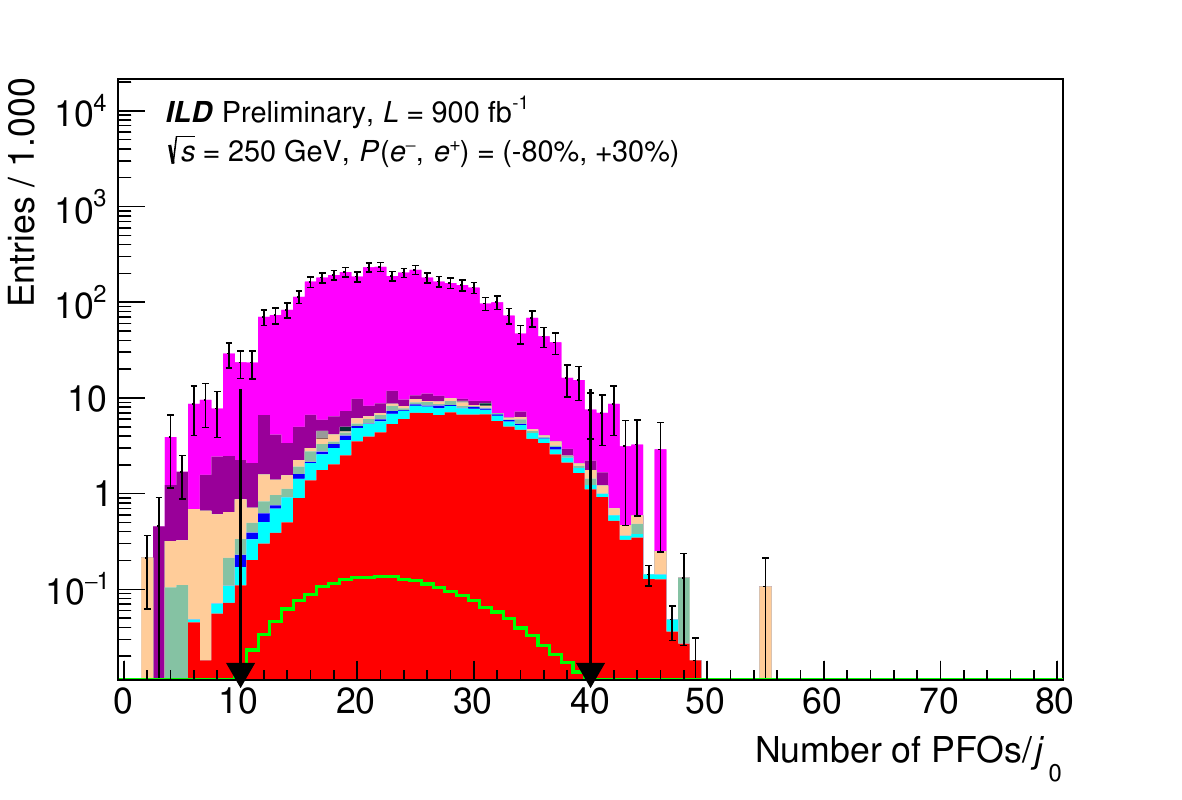}
        \caption{Number of PFOs in leading jet}
    \end{subfigure}
    \begin{subfigure}{0.23\textwidth}
        \includegraphics[width=1.\textwidth]{figures/analysis_Zinv/legend.pdf}
    \end{subfigure}
    \caption{Histograms of the variables used in the kinematic selections of the \Zinv channel, as described in Table~\ref{tab:selections}. Each histogram is given at the level of its corresponding selection but \emph{before} that selection is applied. The arrows represent the placement of the selection cuts, and the error bars represent the MC statistical uncertainties. The sum-of-weights per process is normalised to the SM cross section. N.B. the $h(\rightarrow s\bar{s})Z(\rightarrow\ell\bar{\ell}/\nu\bar{\nu})$ signal is unstacked. A continuation of Fig.~\ref{fig:histograms_Zinv_2}.}
    \label{fig:histograms_Zinv_3}
\end{figure}

\begin{figure}[htbp]
    \begin{subfigure}{0.49\textwidth}
        \centering
        \includegraphics[width=1.\textwidth]{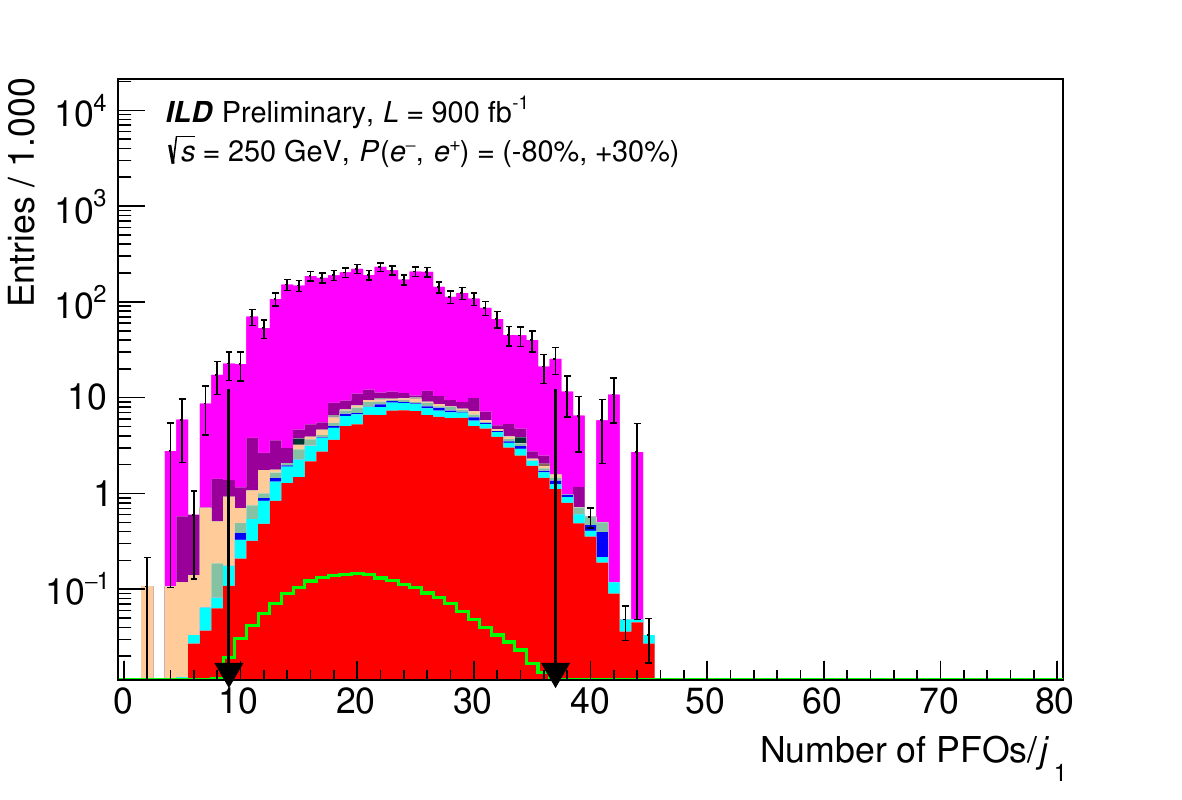}
        \caption{Number of PFOs in subleading jet}
    \end{subfigure}
    \begin{subfigure}{0.23\textwidth}
        \includegraphics[width=1.\textwidth]{figures/analysis_Zinv/legend.pdf}
    \end{subfigure}
    \caption{Histograms of the variables used in the kinematic selections of the \Zinv channel, as described in Table~\ref{tab:selections}. Each histogram is given at the level of its corresponding selection but \emph{before} that selection is applied. The arrows represent the placement of the selection cuts, and the error bars represent the MC statistical uncertainties. The sum-of-weights per process is normalised to the SM cross section. N.B. the $h(\rightarrow s\bar{s})Z(\rightarrow\ell\bar{\ell}/\nu\bar{\nu})$ signal is unstacked. A continuation of Fig.~\ref{fig:histograms_Zinv_3}.}
    \label{fig:histograms_Zinv_4}
\end{figure}

The cutflow for the \Zinv channel is plotted in Fig.~\ref{fig:cutflow_Zinv} -- the full cutflow table is given in Table~\ref{tab:cutflow_Zinv} in Appendix~\ref{app:cutflows}. Histograms of the variables included as part of this channel's selections (showing the evolution of the yields as each selection is applied) are shown in Figs.~\ref{fig:histograms_Zinv_1} through \ref{fig:histograms_Zinv_4}. From Table~\ref{tab:cutflow_Zinv}, we see the signal efficiency for our selections is 13\% while our background efficiency is 0.004\%. Even with the high background rejection, $\Zqq$ is still highly dominant with $\sim$3,300~events compared to the $\sim$2~events expected for \Hss. Therefore, improvements to the sensitivity of the analysis are expected to be accompanied by improved rejection of $\Zqq$. The $\Hgg$ process is the dominant Higgs background with $\sim$100~events.

\begin{figure}[htbp]
    {\centering
    \begin{subfigure}{0.49\textwidth}
        \centering
        \includegraphics[width=1.\textwidth]{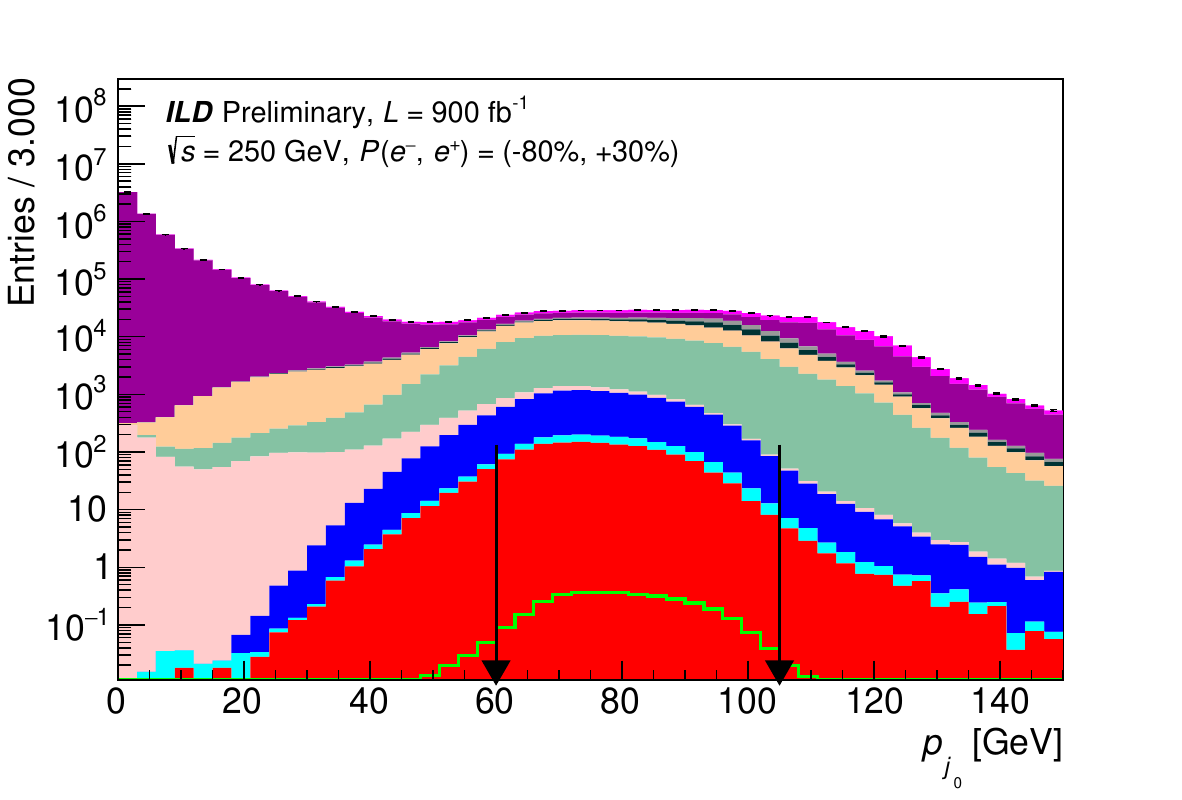}
        \caption{Leading jet momentum $p_{j_0}$}
    \end{subfigure}
    \hfill
    \begin{subfigure}{0.49\textwidth}
        \centering
        \includegraphics[width=1.\textwidth]{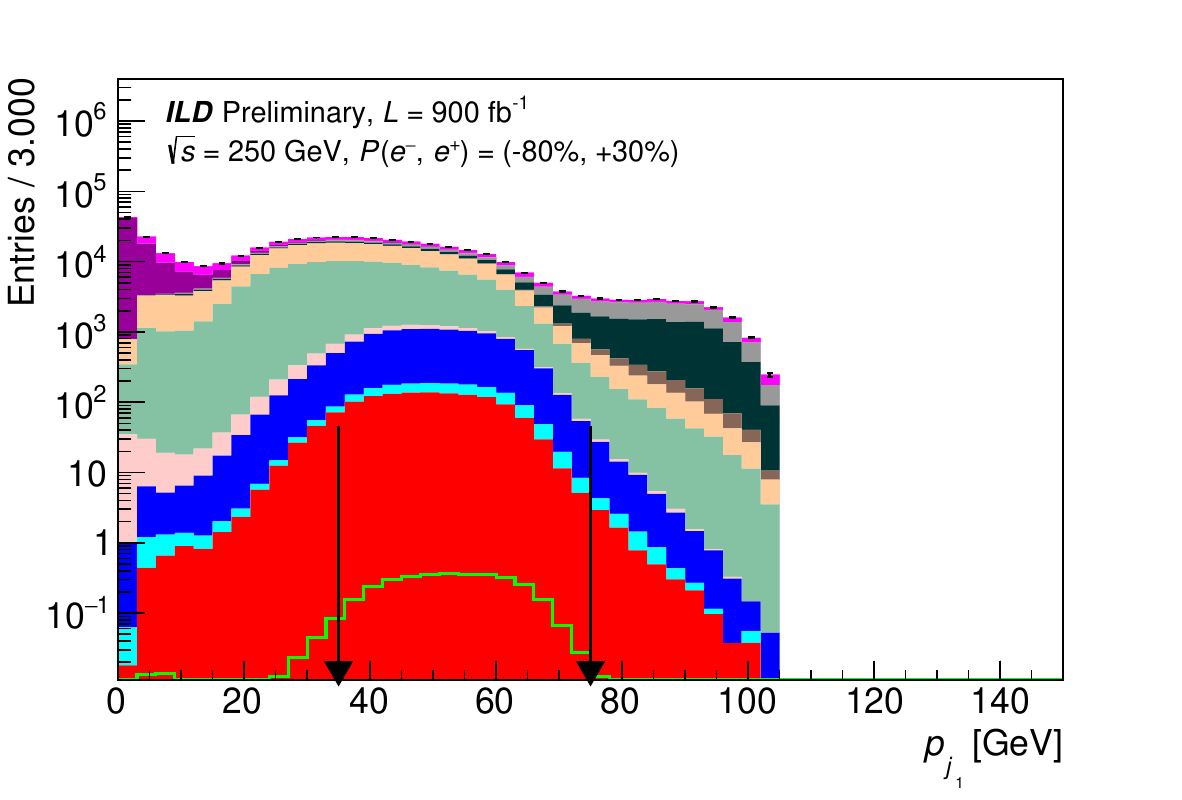}
        \caption{Subleading jet momentum $p_{j_1}$}
    \end{subfigure} \\
    \begin{subfigure}{0.49\textwidth}
        \centering
        \includegraphics[width=1.\textwidth]{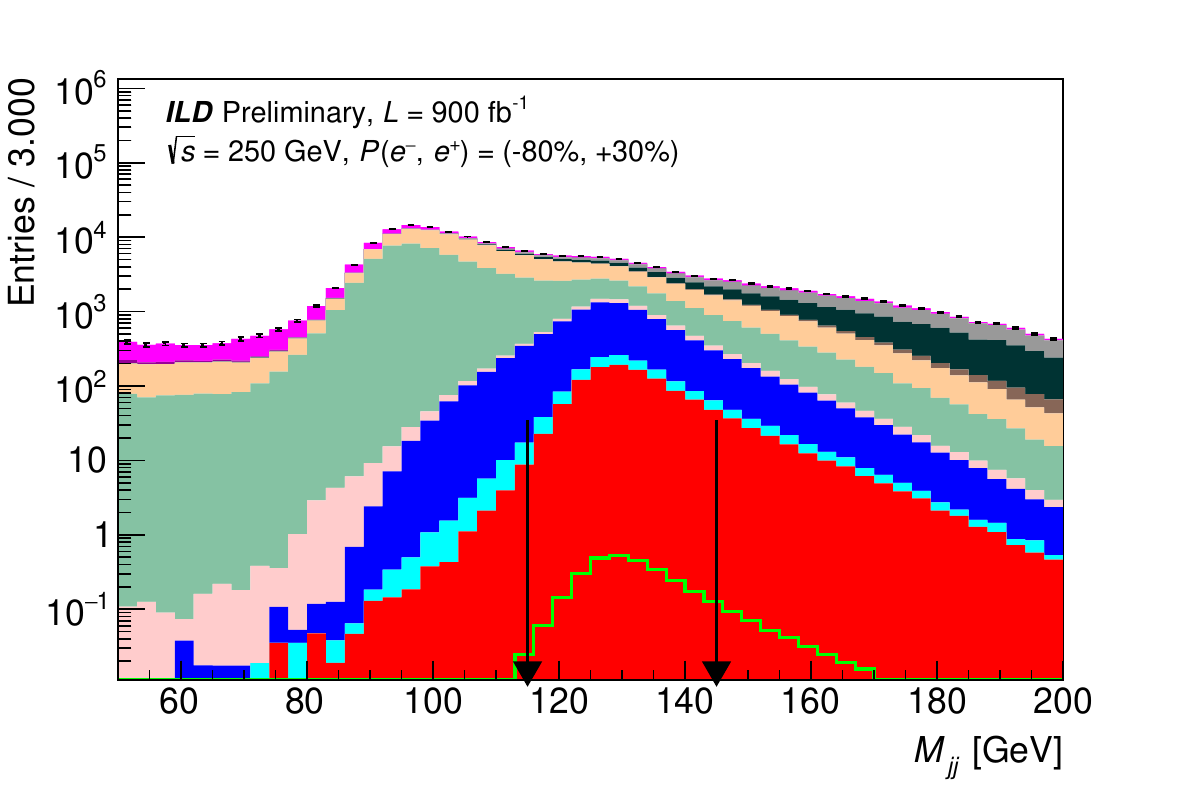}
        \caption{Dijet mass $M_{jj}$}
    \end{subfigure}
    \hfill
    \begin{subfigure}{0.49\textwidth}
        \centering
        \includegraphics[width=1.\textwidth]{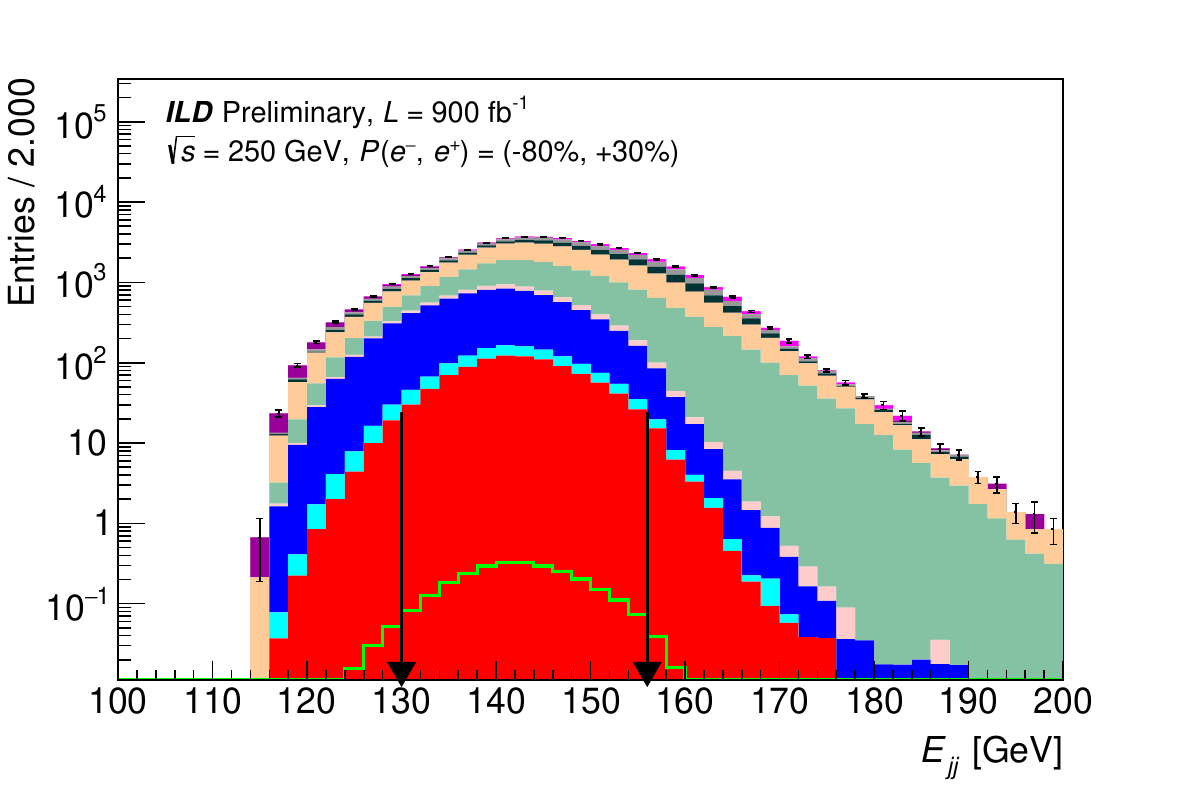}
        \caption{Dijet energy $E_{jj}$}
    \end{subfigure} \\
    }
    \begin{subfigure}{0.49\textwidth}
        \centering
        \includegraphics[width=1.\textwidth]{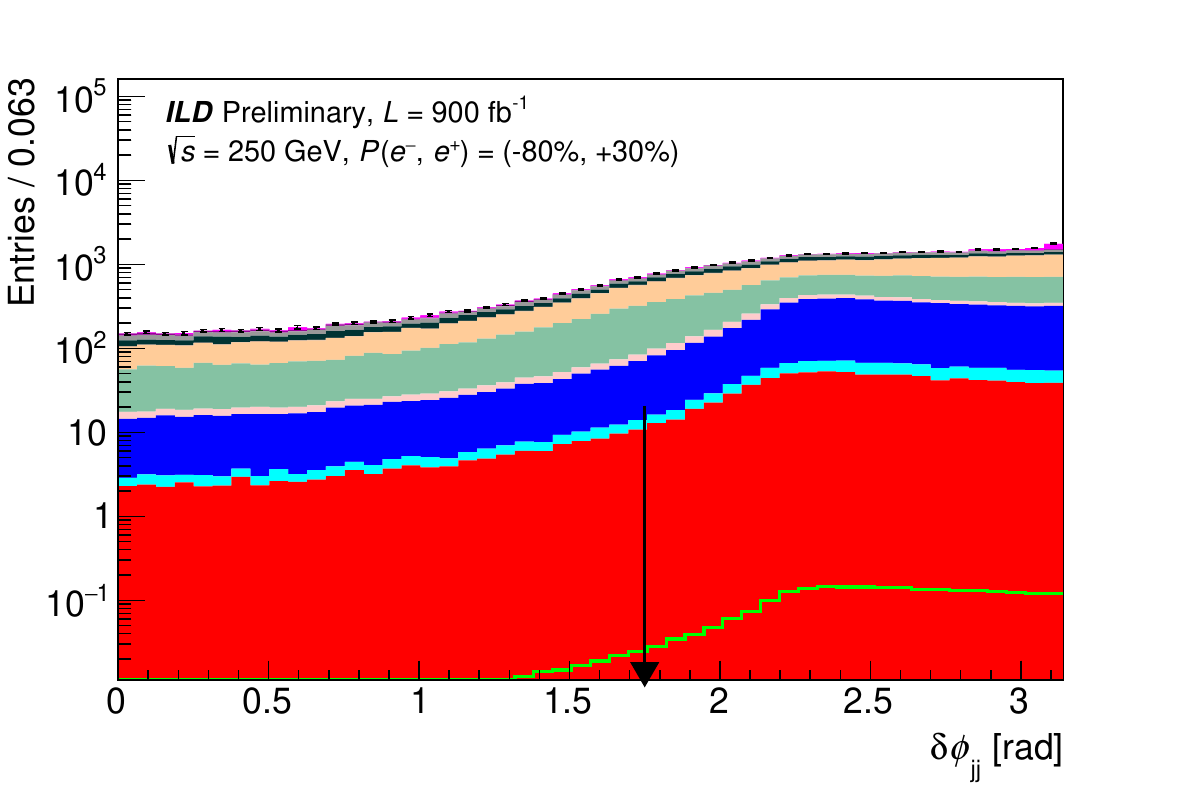}
        \caption{Dijet azimuthal separation $\Delta\phi_{jj}$}
    \end{subfigure}
    \begin{subfigure}{0.23\textwidth}
        \centering
        \includegraphics[width=1.\textwidth]{figures/analysis_Zinv/legend.pdf}
    \end{subfigure} \\
    \caption{Histograms of the variables used in the kinematic selections of the \Zll channel, as described in Table~\ref{tab:selections}. Each histogram is given at the level of its corresponding selection but \emph{before} that selection is applied. The arrows represent the placement of the selection cuts, and the error bars represent the MC statistical uncertainties. The sum-of-weights per process is normalised to the SM cross section. N.B. the $h(\rightarrow s\bar{s})Z(\rightarrow\ell\bar{\ell}/\nu\bar{\nu})$ signal is unstacked.}
    \label{fig:histograms_Zll_1}
\end{figure}

\begin{figure}[htbp]
    {\centering
    \begin{subfigure}{0.49\textwidth}
        \centering
        \includegraphics[width=1.\textwidth]{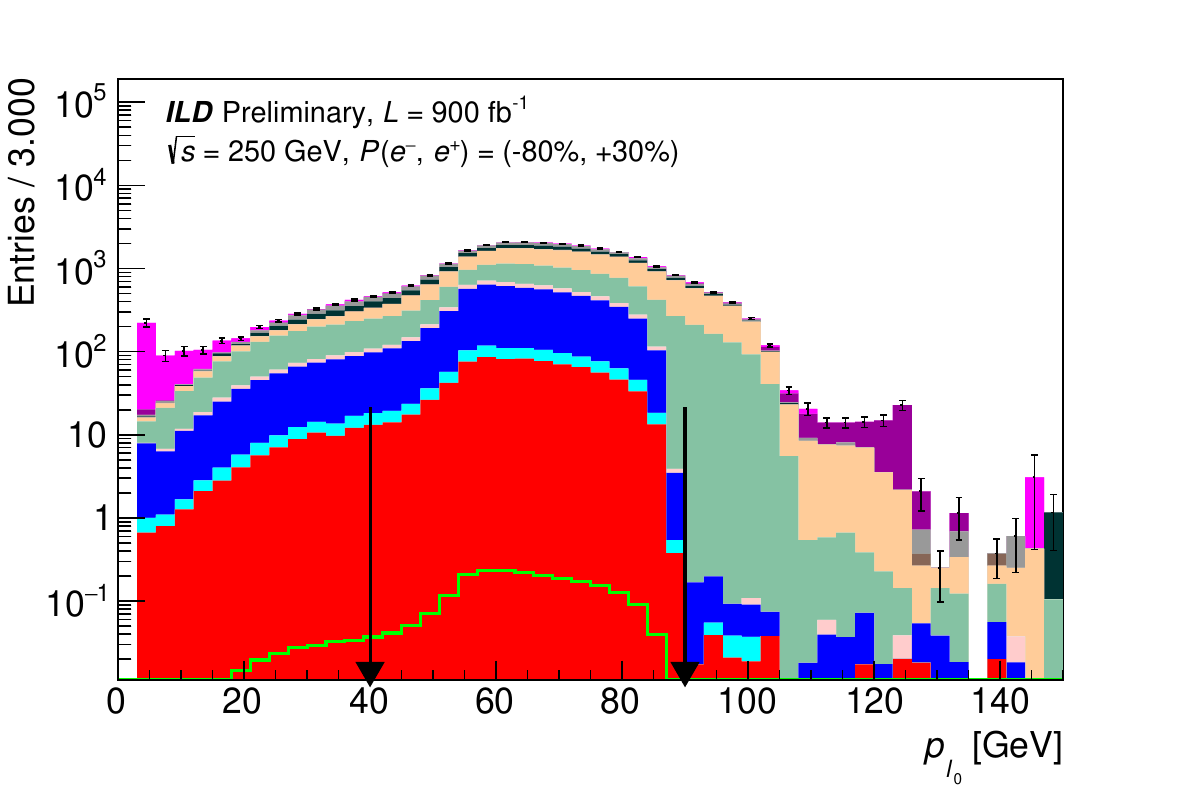}
        \caption{Leading lepton momentum $p_{\ell_0}$}
    \end{subfigure}
    \hfill
    \begin{subfigure}{0.49\textwidth}
        \centering
        \includegraphics[width=1.\textwidth]{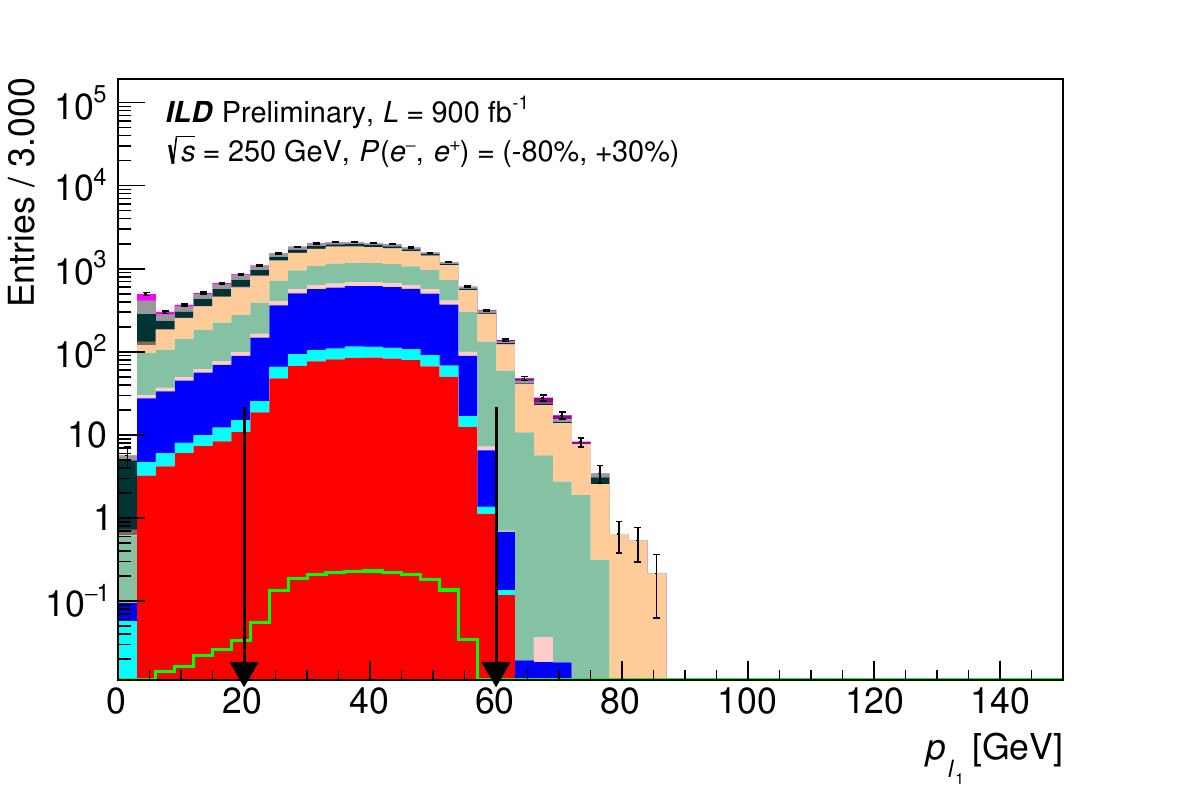}
        \caption{Leading lepton momentum $p_{\ell_1}$}
    \end{subfigure} \\
    \begin{subfigure}{0.49\textwidth}
        \centering
        \includegraphics[width=1.\textwidth]{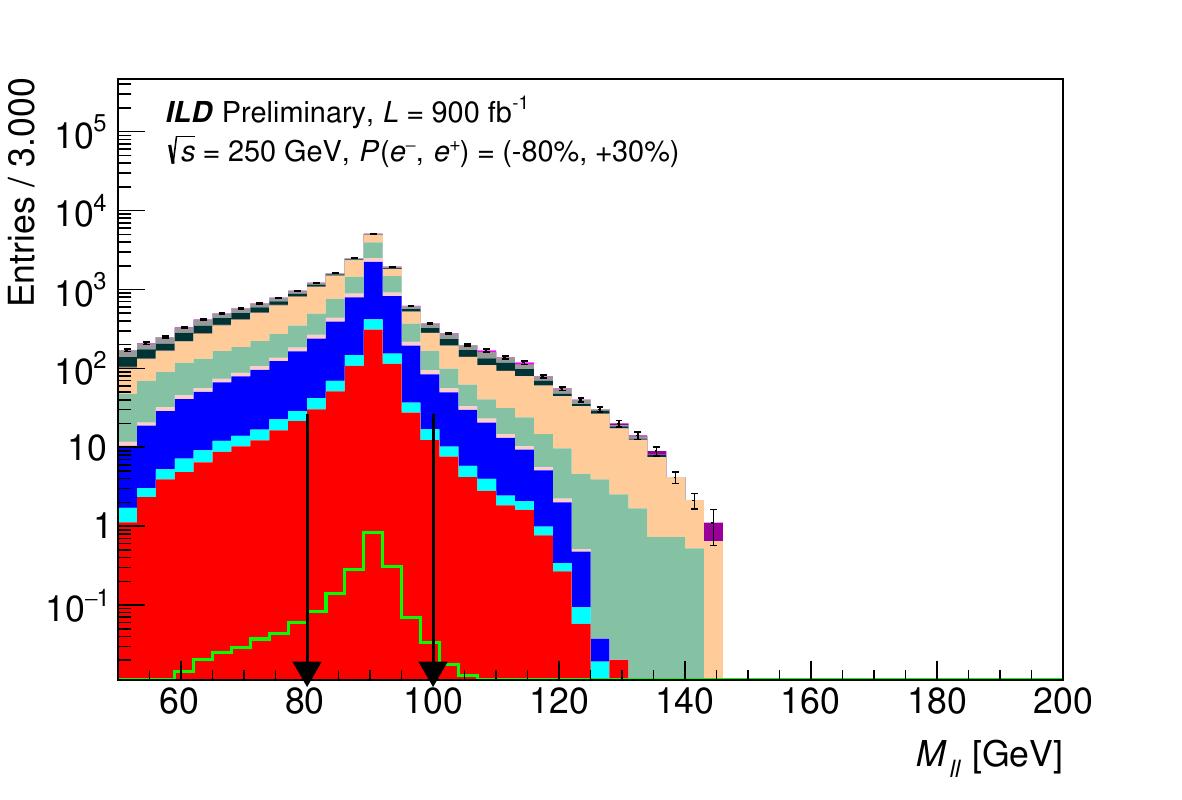}
        \caption{Dilepton mass $M_{\ell\bar{\ell}}$}
    \end{subfigure}
    \hfill
    \begin{subfigure}{0.49\textwidth}
        \centering
        \includegraphics[width=1.\textwidth]{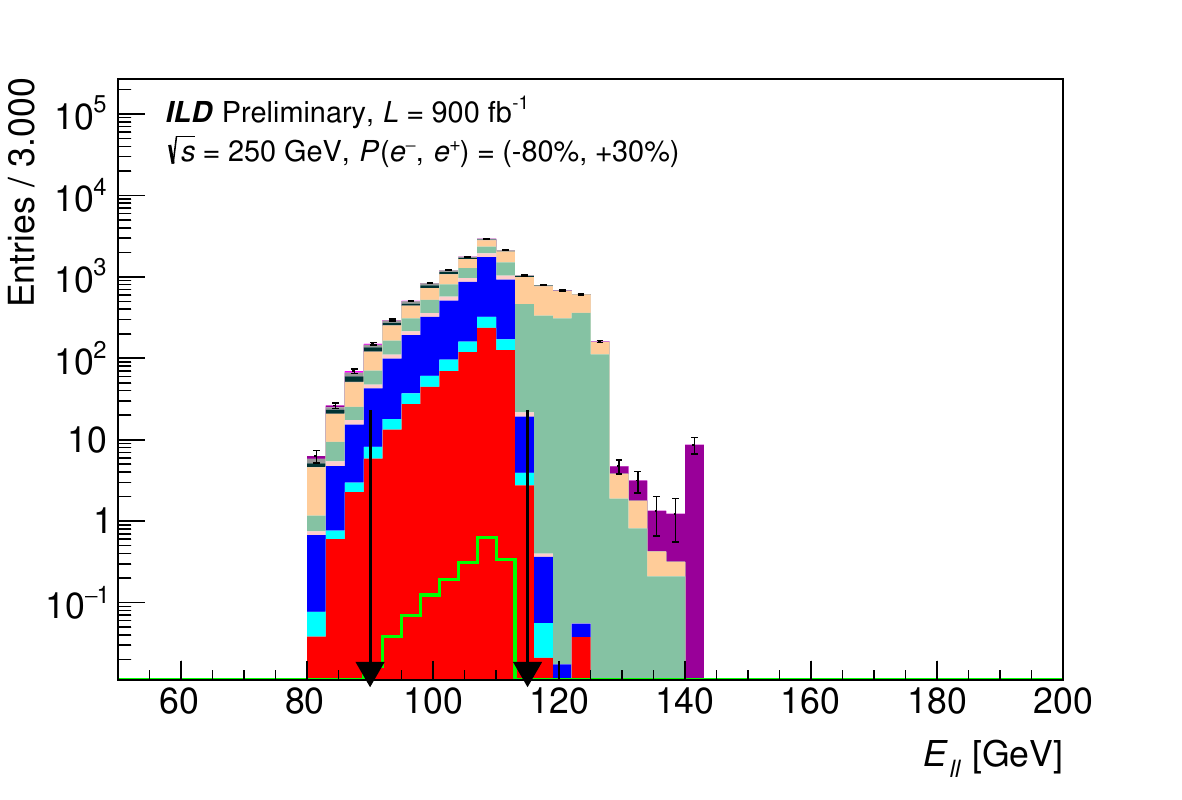}
        \caption{Dilepton energy $E_{\ell\bar{\ell}}$}
    \end{subfigure} \\
    }
    \begin{subfigure}{0.49\textwidth}
        \centering
        \includegraphics[width=1.\textwidth]{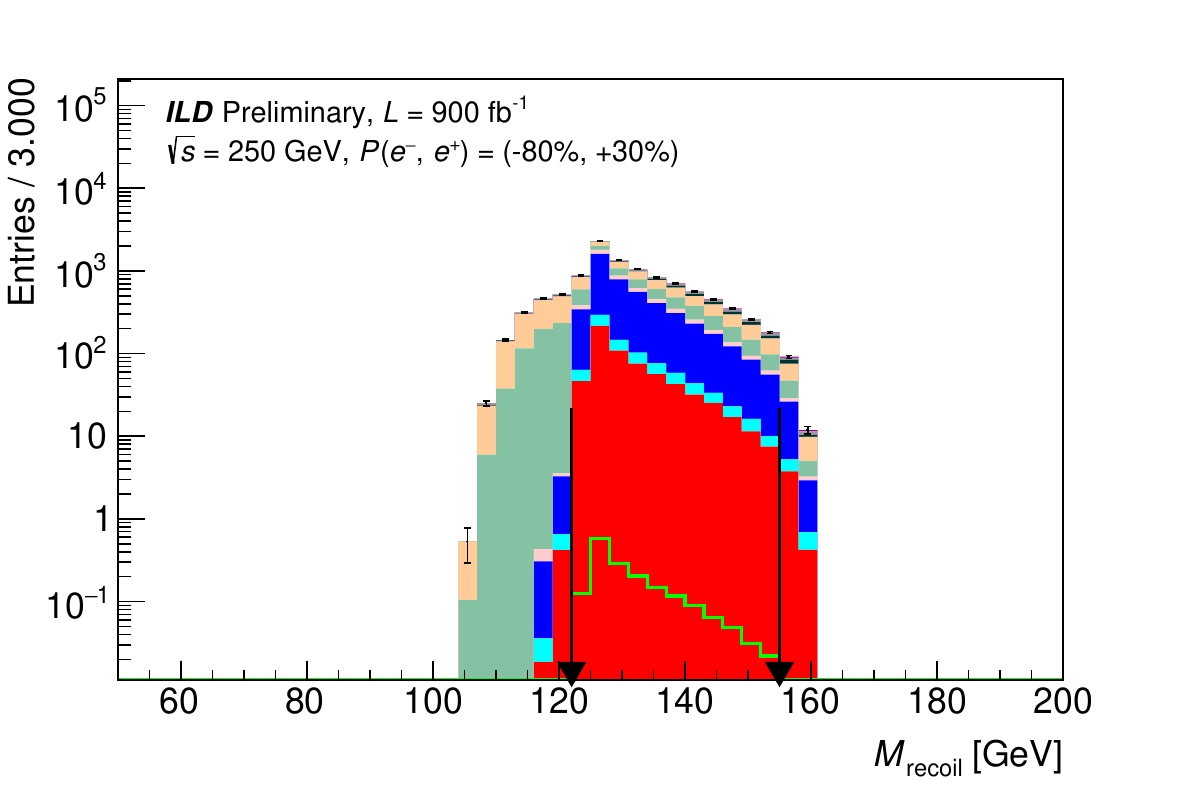}
        \caption{Recoil mass $M_\textrm{recoil}$}
    \end{subfigure}
    \begin{subfigure}{0.23\textwidth}
        \centering
        \includegraphics[width=1.\textwidth]{figures/analysis_Zinv/legend.pdf}
    \end{subfigure} \\
    \caption{Histograms of the variables used in the kinematic selections of the \Zll channel, as described in Table~\ref{tab:selections}. Each histogram is given at the level of its corresponding selection but \emph{before} that selection is applied. The arrows represent the placement of the selection cuts, and the error bars represent the MC statistical uncertainties. The sum-of-weights per process is normalised to the SM cross section. N.B. the $h(\rightarrow s\bar{s})Z(\rightarrow\ell\bar{\ell}/\nu\bar{\nu})$ signal is unstacked. A continuation of Fig.~\ref{fig:histograms_Zll_1}.}
    \label{fig:histograms_Zll_2}
\end{figure}

\begin{figure}[htbp]
    {\centering
    \begin{subfigure}{0.49\textwidth}
        \centering
        \includegraphics[width=1.\textwidth]{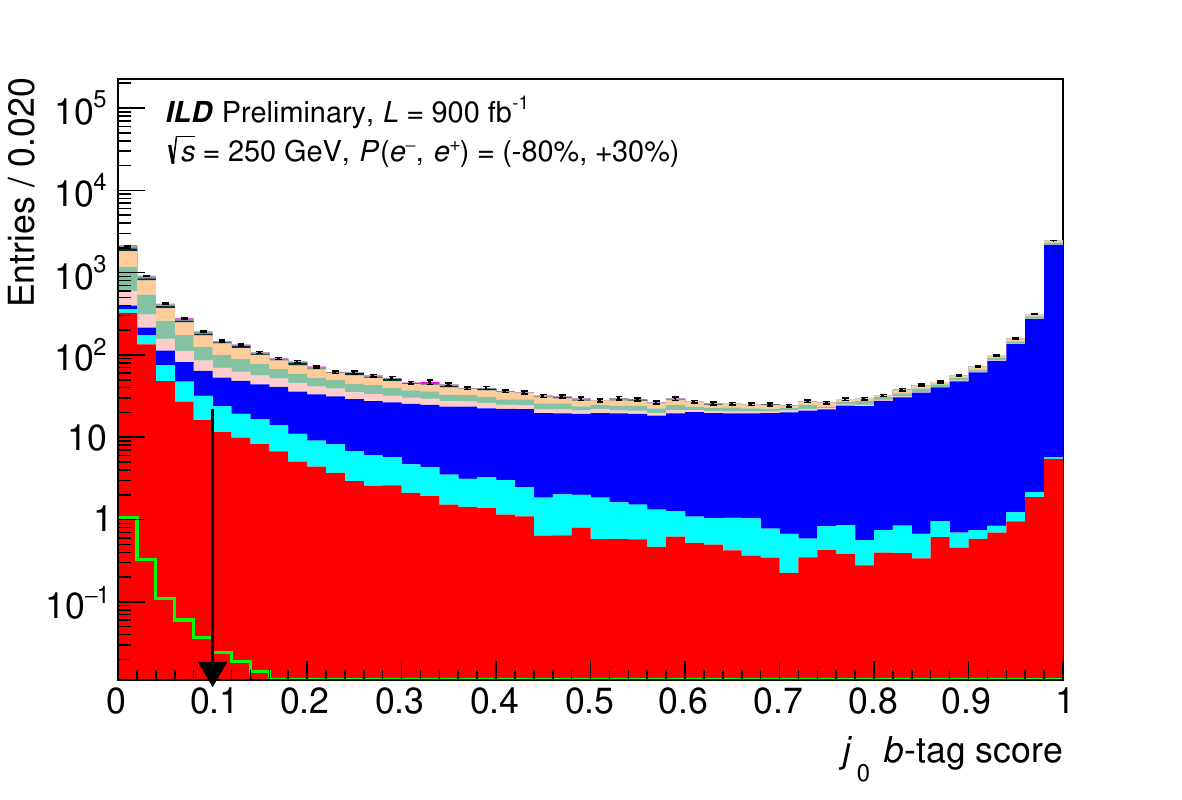}
        \caption{Leading jet BTag score}
    \end{subfigure}
    \hfill
    \begin{subfigure}{0.49\textwidth}
        \centering
        \includegraphics[width=1.\textwidth]{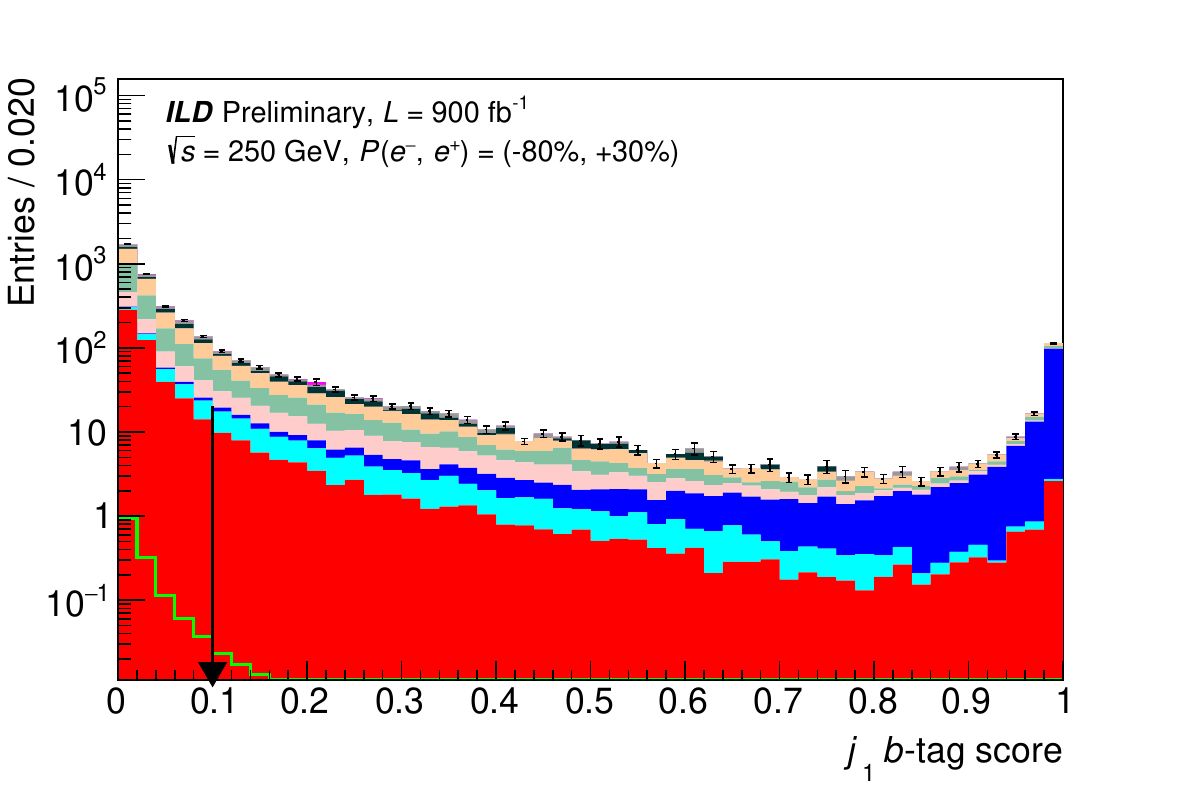}
        \caption{Subleading jet BTag score}
    \end{subfigure} \\
    \begin{subfigure}{0.49\textwidth}
        \centering
        \includegraphics[width=1.\textwidth]{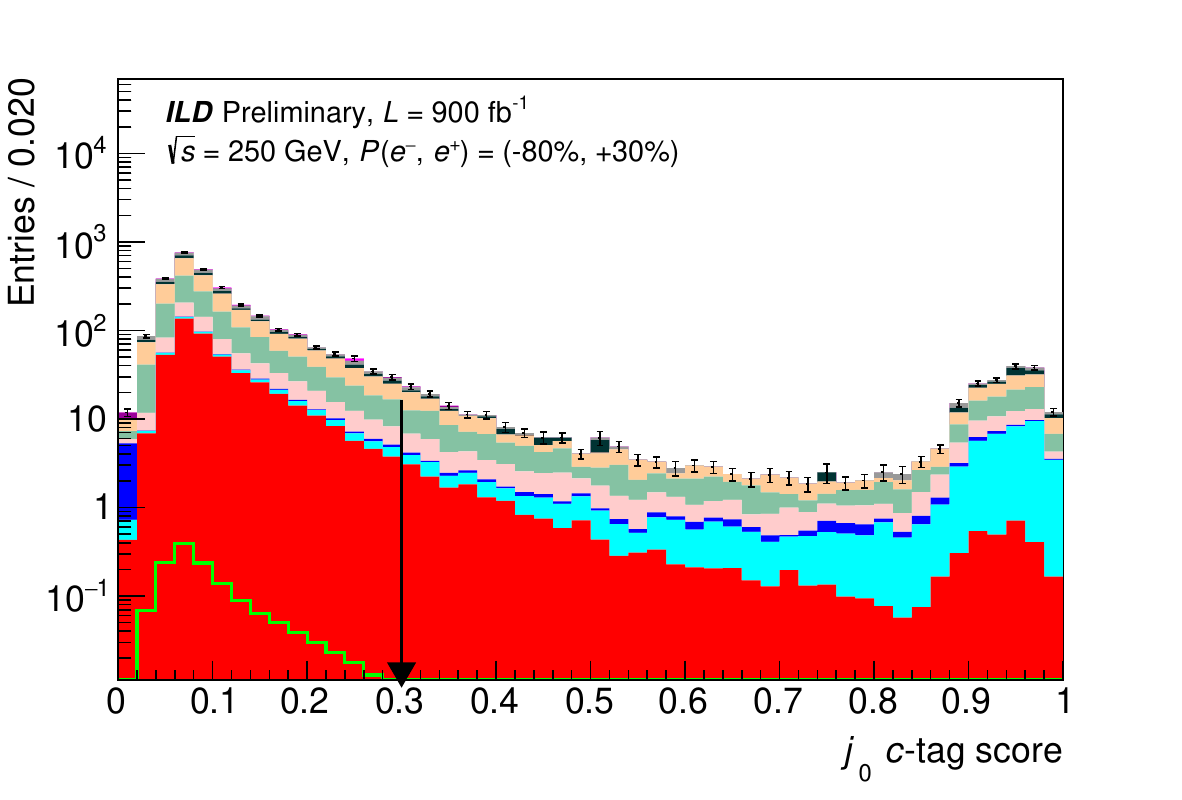}
        \caption{Leading jet CTag score}
    \end{subfigure}
    \hfill
    \begin{subfigure}{0.49\textwidth}
        \centering
        \includegraphics[width=1.\textwidth]{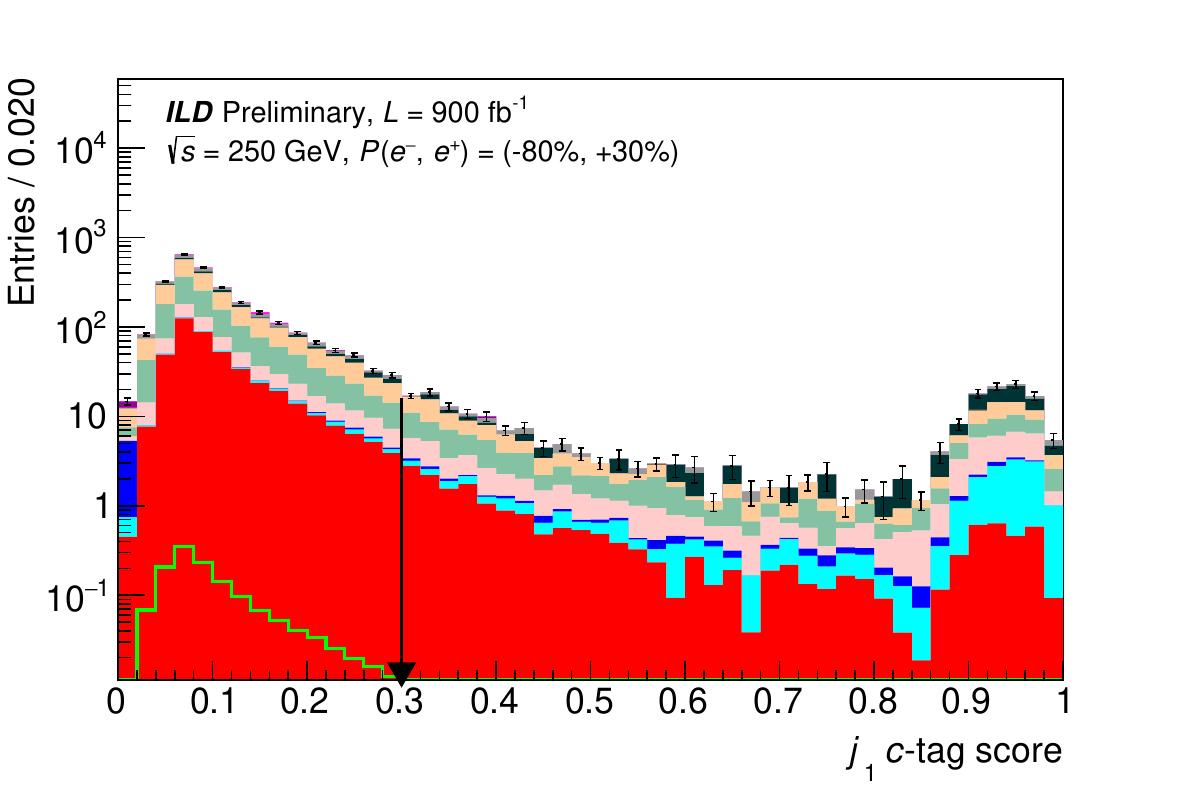}
        \caption{Subleading jet CTag score}
    \end{subfigure} \\
    }
    \begin{subfigure}{0.49\textwidth}
        \centering
        \includegraphics[width=1.\textwidth]{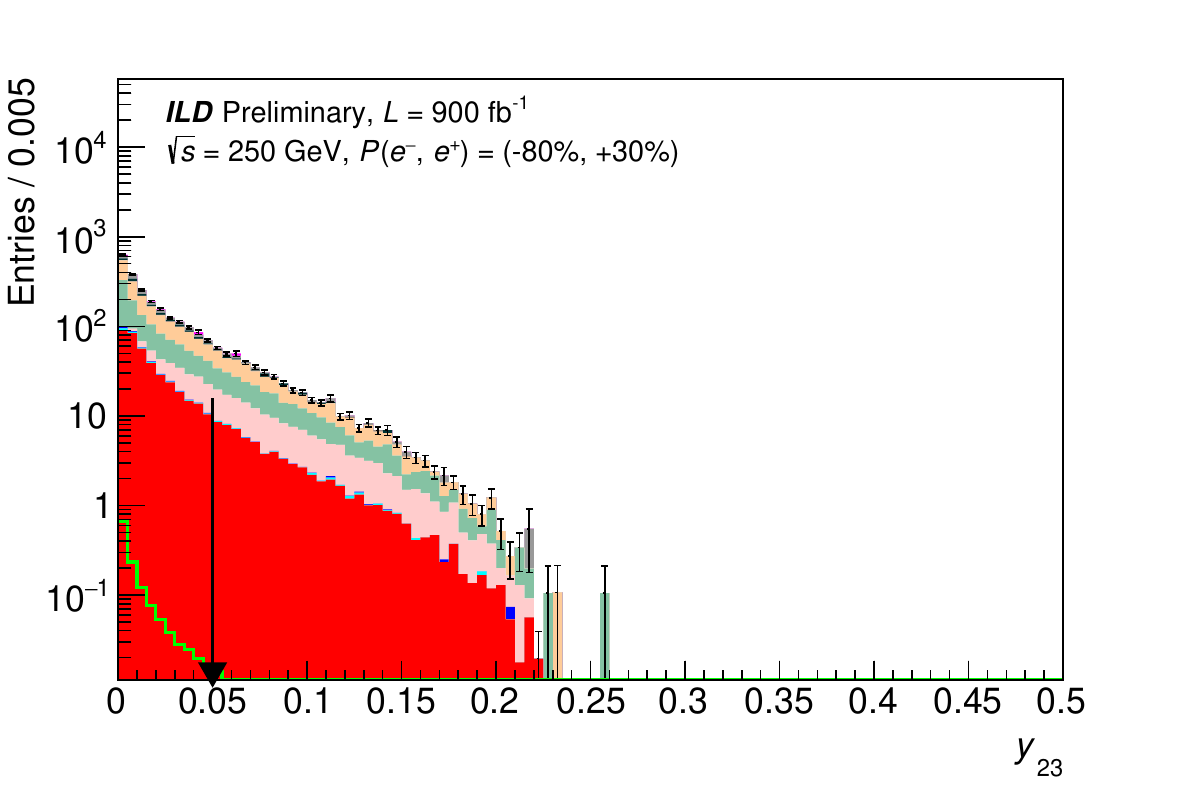}
        \caption{$2\rightarrow3$ jet transition variable $y_{23}$}
    \end{subfigure}
    \begin{subfigure}{0.23\textwidth}
        \centering
        \includegraphics[width=1.\textwidth]{figures/analysis_Zinv/legend.pdf}
    \end{subfigure} \\
    \caption{Histograms of the variables used in the kinematic selections of the \Zll channel, as described in Table~\ref{tab:selections}. Each histogram is given at the level of its corresponding selection but \emph{before} that selection is applied. The arrows represent the placement of the selection cuts, and the error bars represent the MC statistical uncertainties. The sum-of-weights per process is normalised to the SM cross section. N.B. the $h(\rightarrow s\bar{s})Z(\rightarrow\ell\bar{\ell}/\nu\bar{\nu})$ signal is unstacked. A continuation of Fig.~\ref{fig:histograms_Zll_2}.}
    \label{fig:histograms_Zll_3}
\end{figure}

\begin{figure}[htbp]
    {\centering
    \begin{subfigure}{0.49\textwidth}
        \centering
        \includegraphics[width=1.\textwidth]{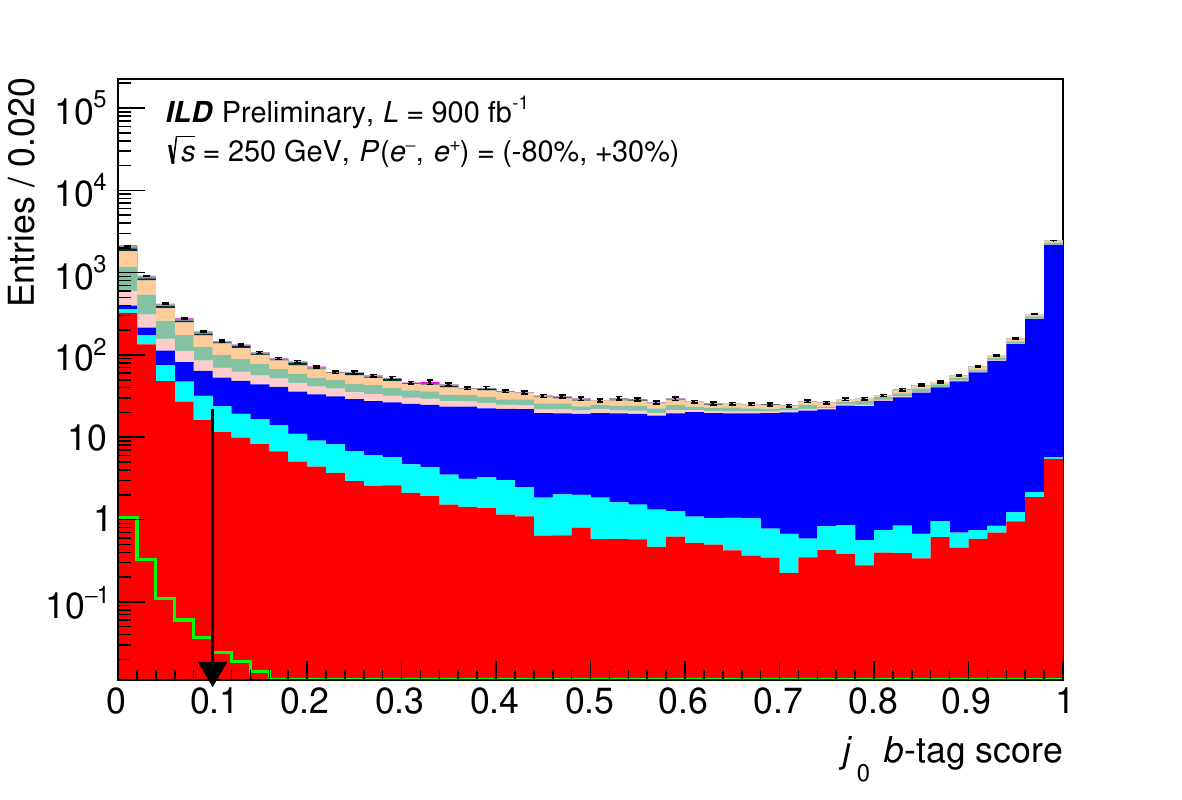}
        \caption{Leading jet BTag score}
    \end{subfigure}
    \hfill
    \begin{subfigure}{0.49\textwidth}
        \centering
        \includegraphics[width=1.\textwidth]{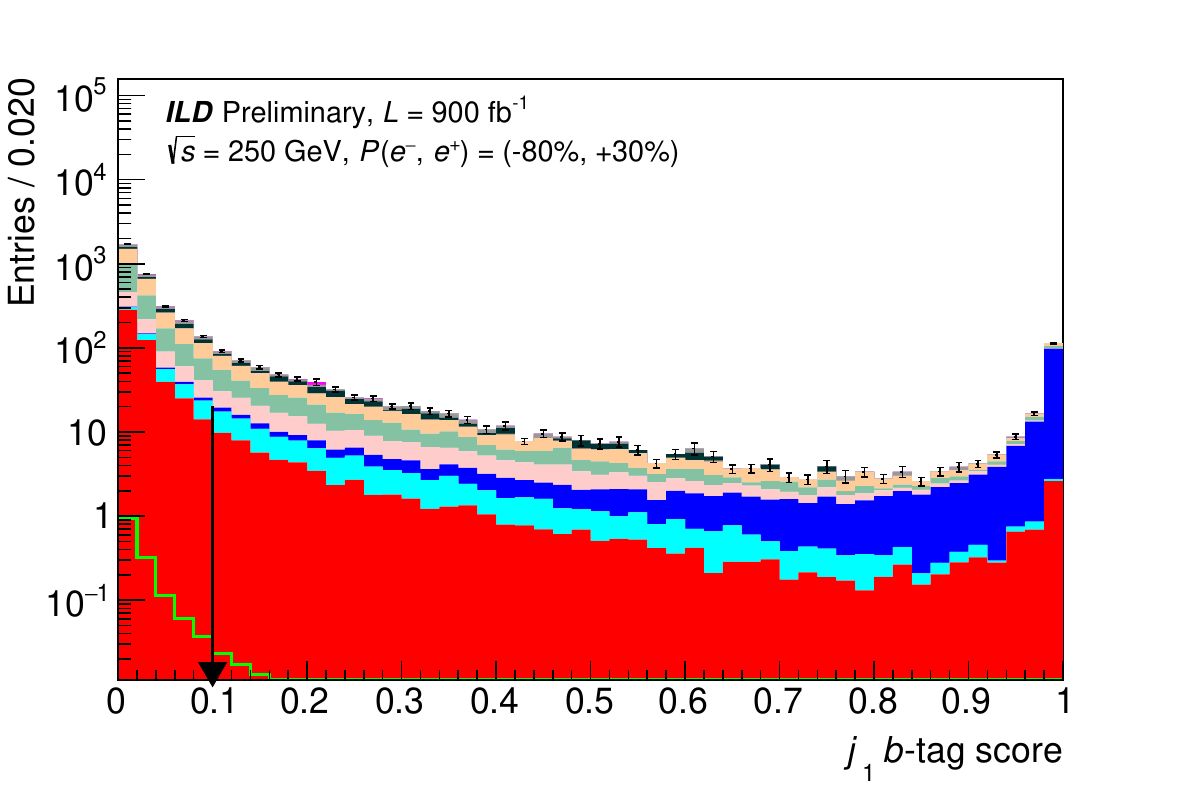}
        \caption{Subleading jet BTag score}
    \end{subfigure} \\
    \begin{subfigure}{0.49\textwidth}
        \centering
        \includegraphics[width=1.\textwidth]{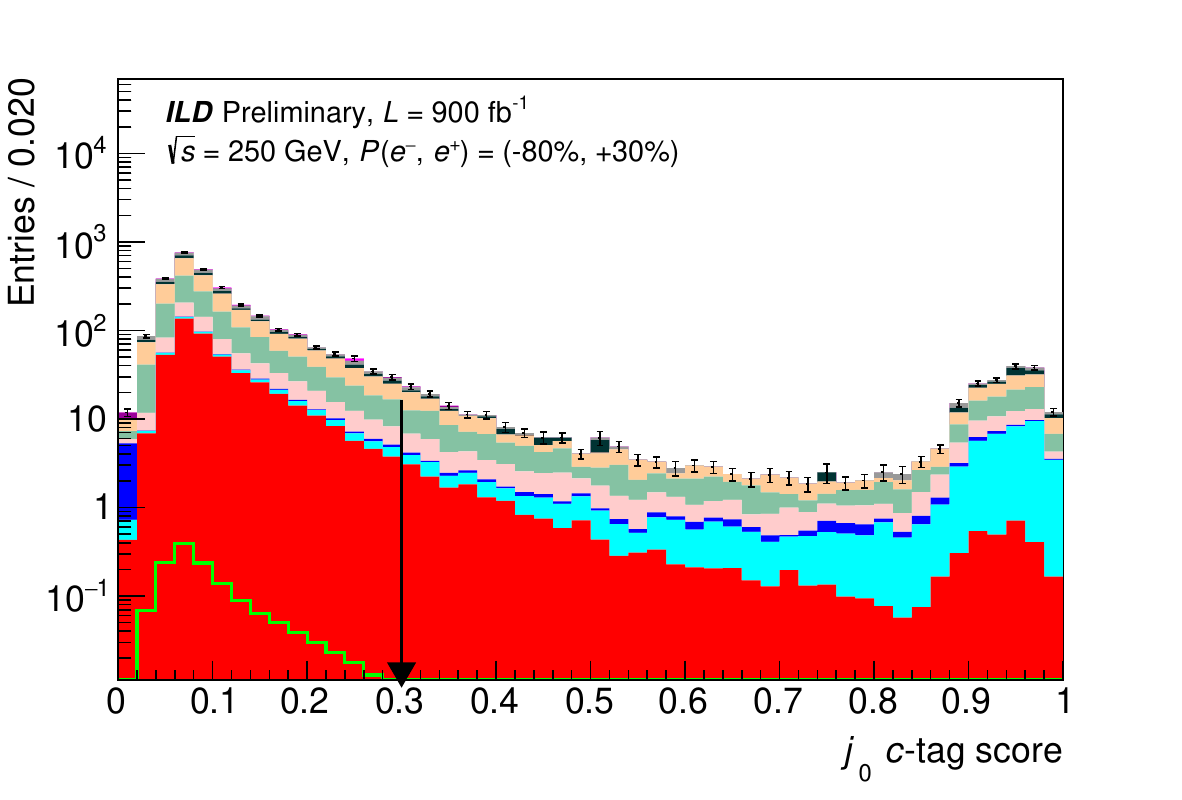}
        \caption{Leading jet CTag score}
    \end{subfigure}
    \hfill
    \begin{subfigure}{0.49\textwidth}
        \centering
        \includegraphics[width=1.\textwidth]{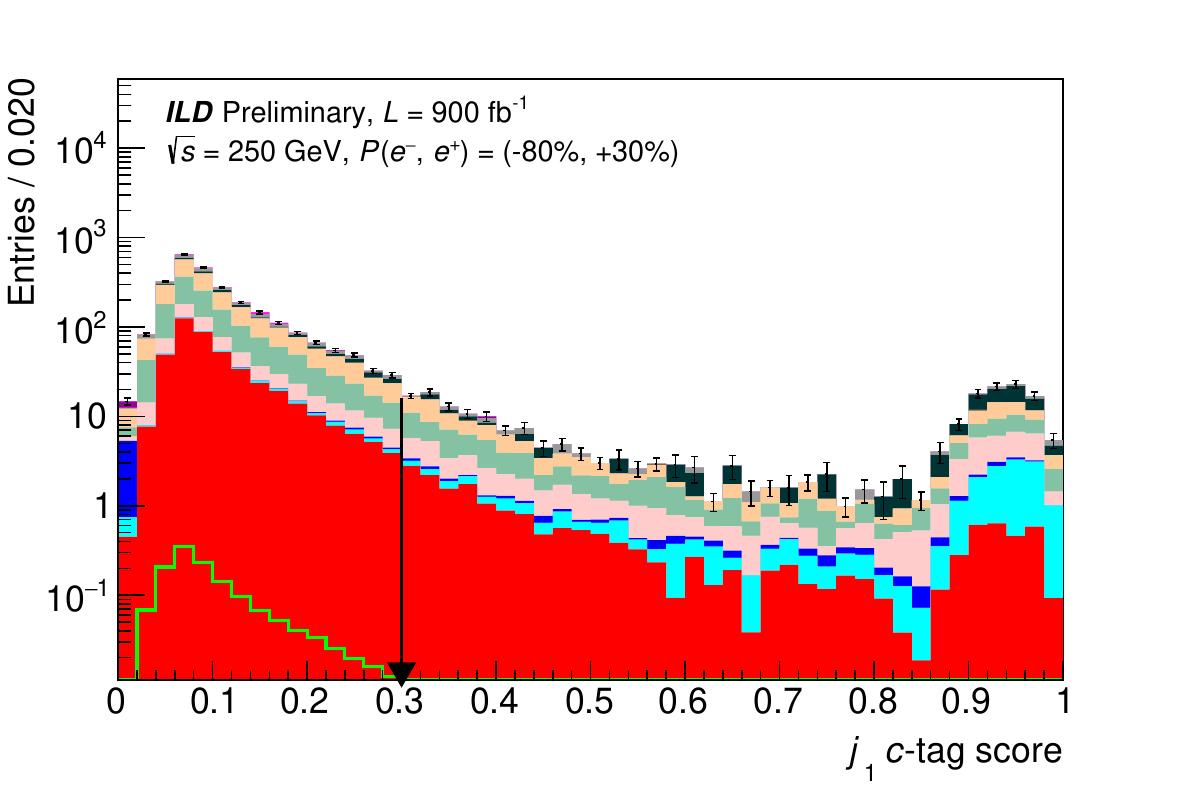}
        \caption{Subleading jet CTag score}
    \end{subfigure} \\
    }
    \begin{subfigure}{0.49\textwidth}
        \centering
        \includegraphics[width=1.\textwidth]{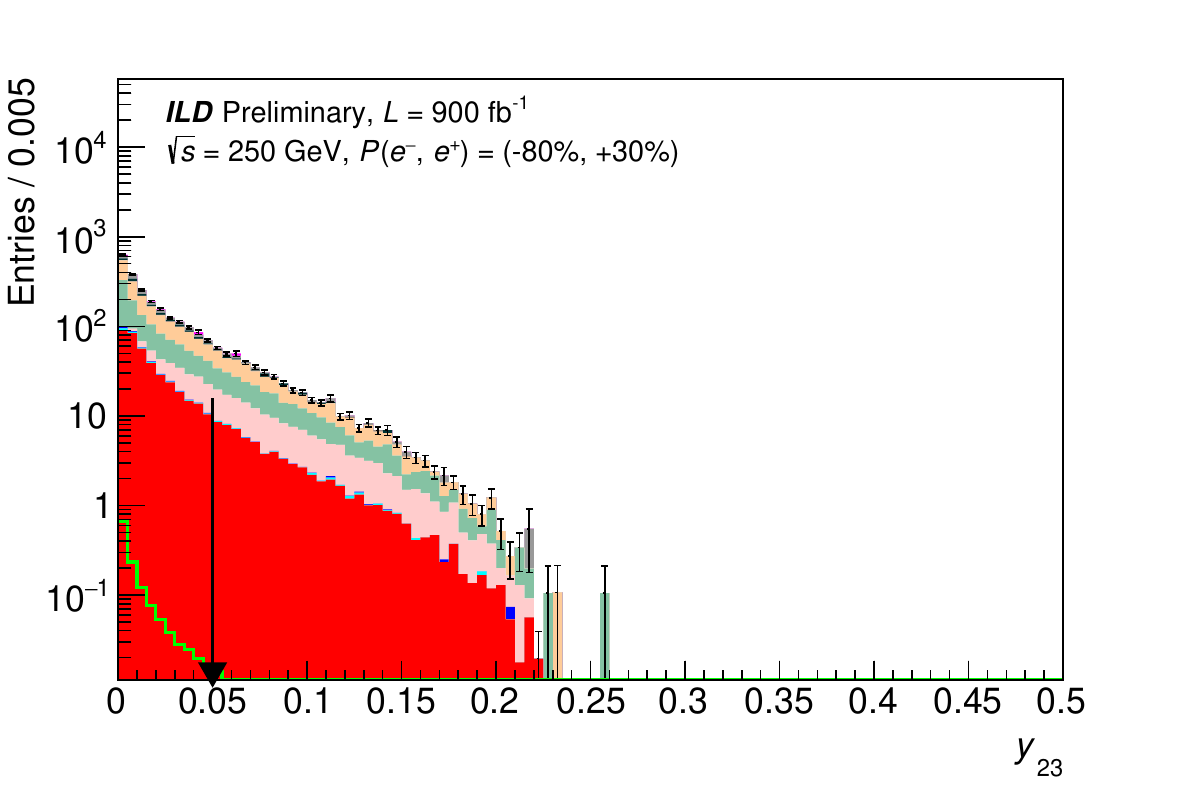}
        \caption{$2\rightarrow3$ jet transition variable $y_{23}$}
    \end{subfigure}
    \begin{subfigure}{0.23\textwidth}
        \centering
        \includegraphics[width=1.\textwidth]{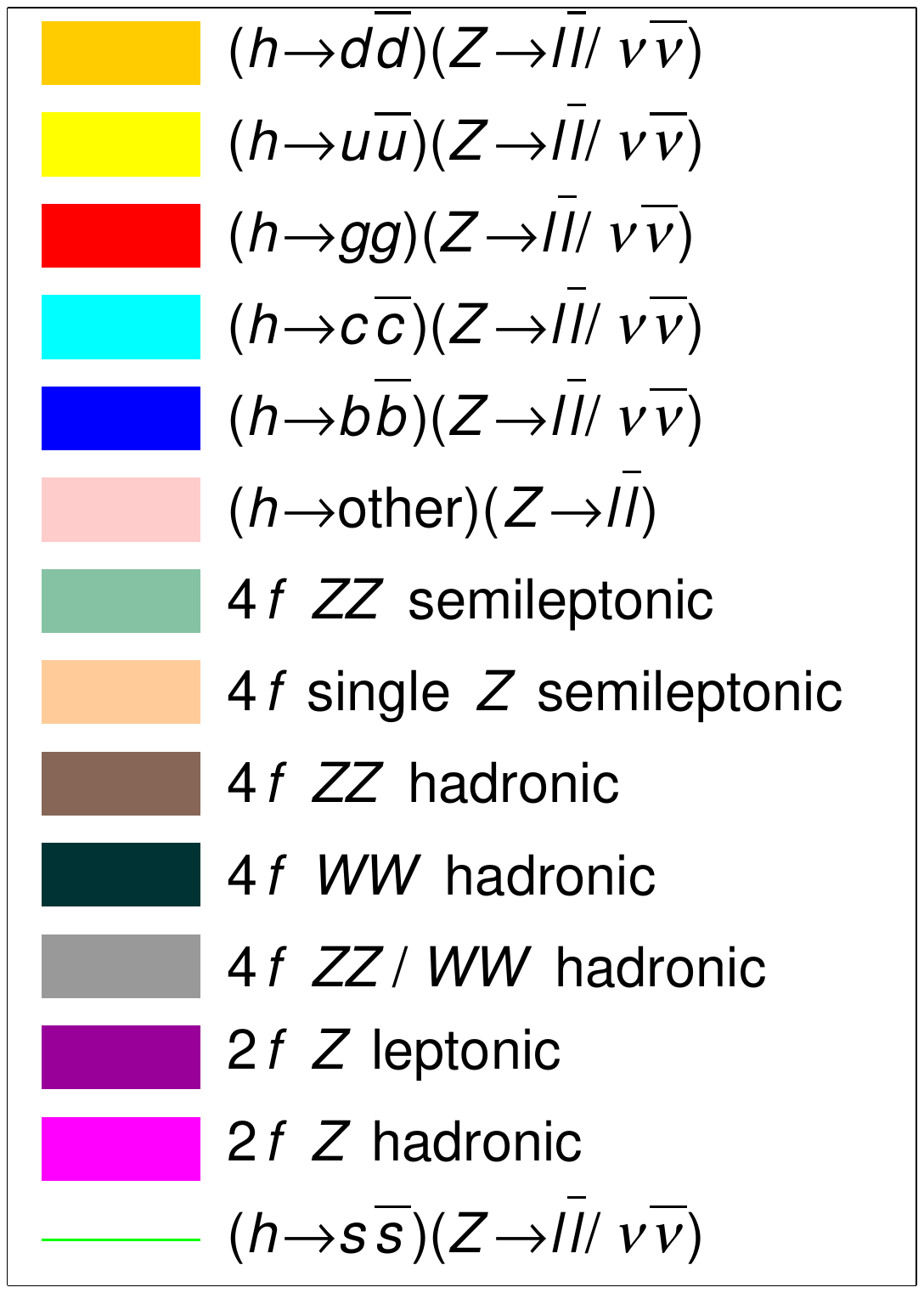}
    \end{subfigure} \\
    \caption{Histograms of the variables used in the kinematic selections of the \Zll channel, as described in Table~\ref{tab:selections}. Each histogram is given at the level of its corresponding selection but \emph{before} that selection is applied. The arrows represent the placement of the selection cuts, and the error bars represent the MC statistical uncertainties. The sum-of-weights per process is normalised to the SM cross section. N.B. the $h(\rightarrow s\bar{s})Z(\rightarrow\ell\bar{\ell}/\nu\bar{\nu})$ signal is unstacked. A continuation of Fig.~\ref{fig:histograms_Zll_3}.}
    \label{fig:histograms_Zll_4}
\end{figure}

\begin{figure}[htbp]
    {\centering
    \begin{subfigure}{0.49\textwidth}
        \centering
        \includegraphics[width=1.\textwidth]{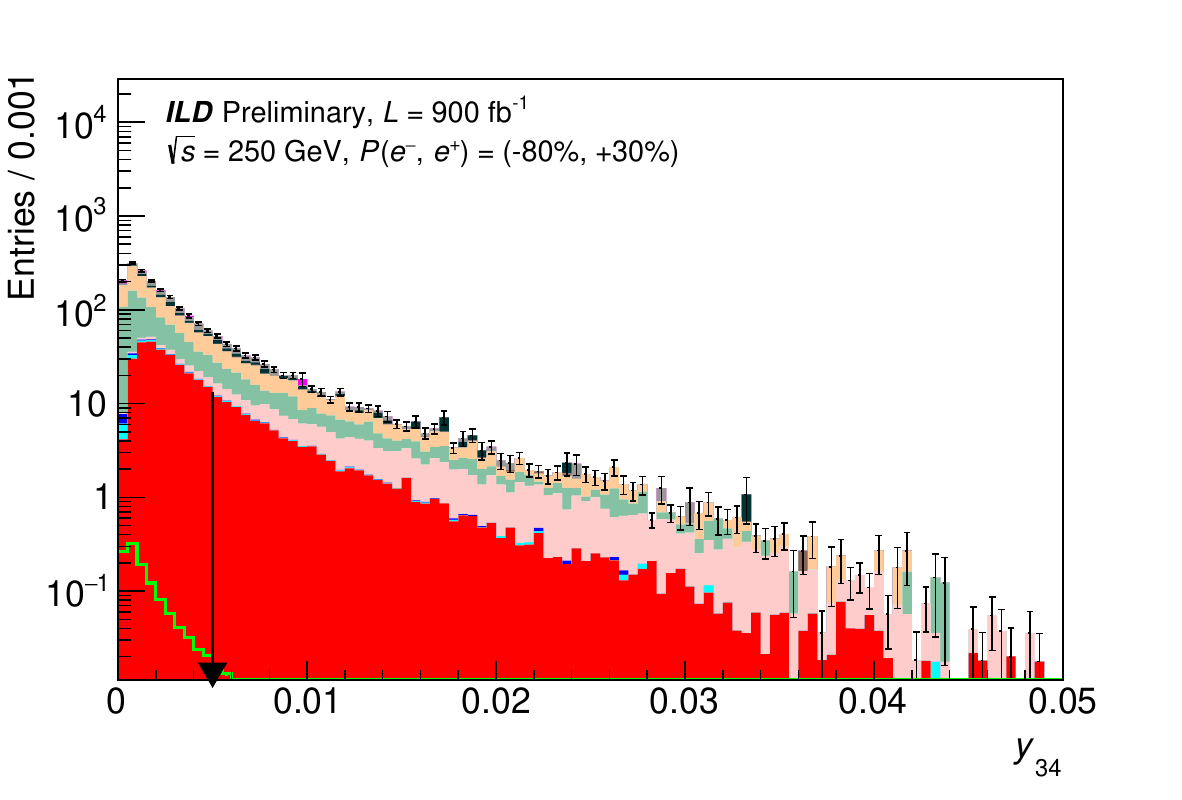}
        \caption{$3\rightarrow4$ jet transition variable $y_{34}$}
    \end{subfigure}
    \hfill
    \begin{subfigure}{0.49\textwidth}
        \centering
        \includegraphics[width=1.\textwidth]{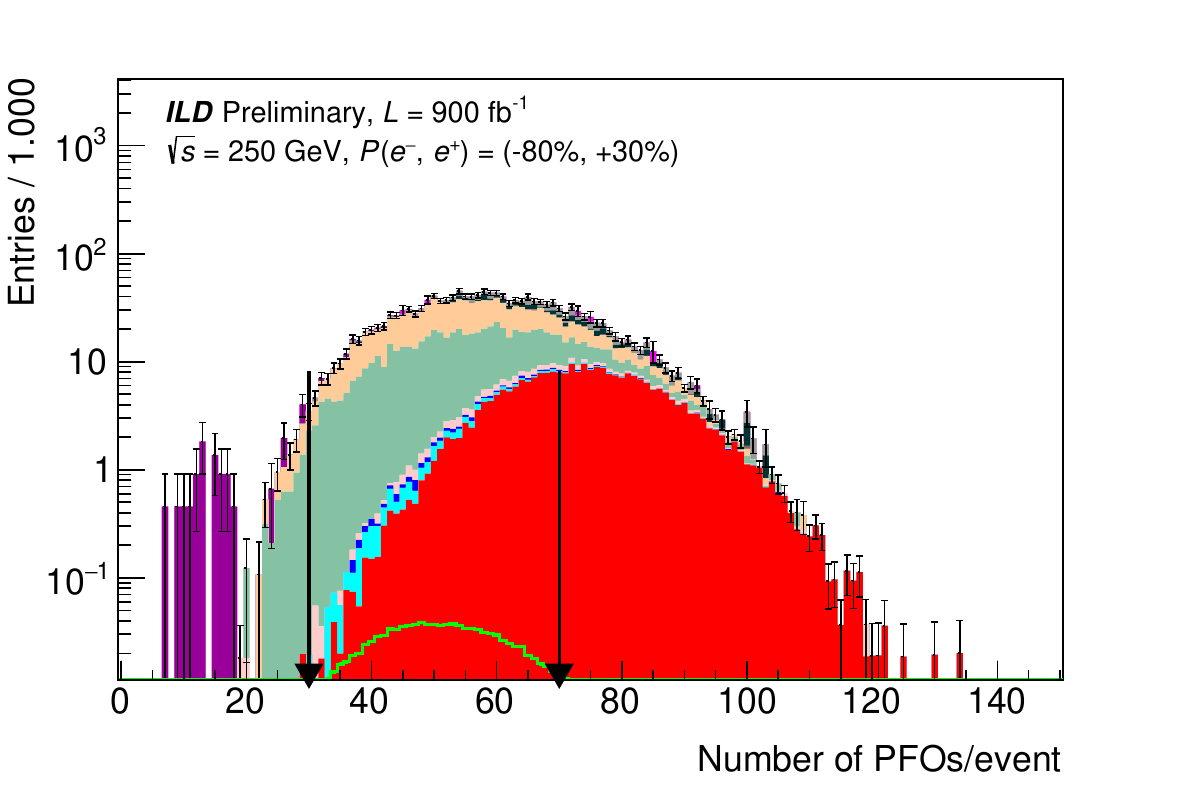}
        \caption{Number of PFOs in event}
    \end{subfigure} \\
    }
    \begin{subfigure}{0.49\textwidth}
        \centering
        \includegraphics[width=1.\textwidth]{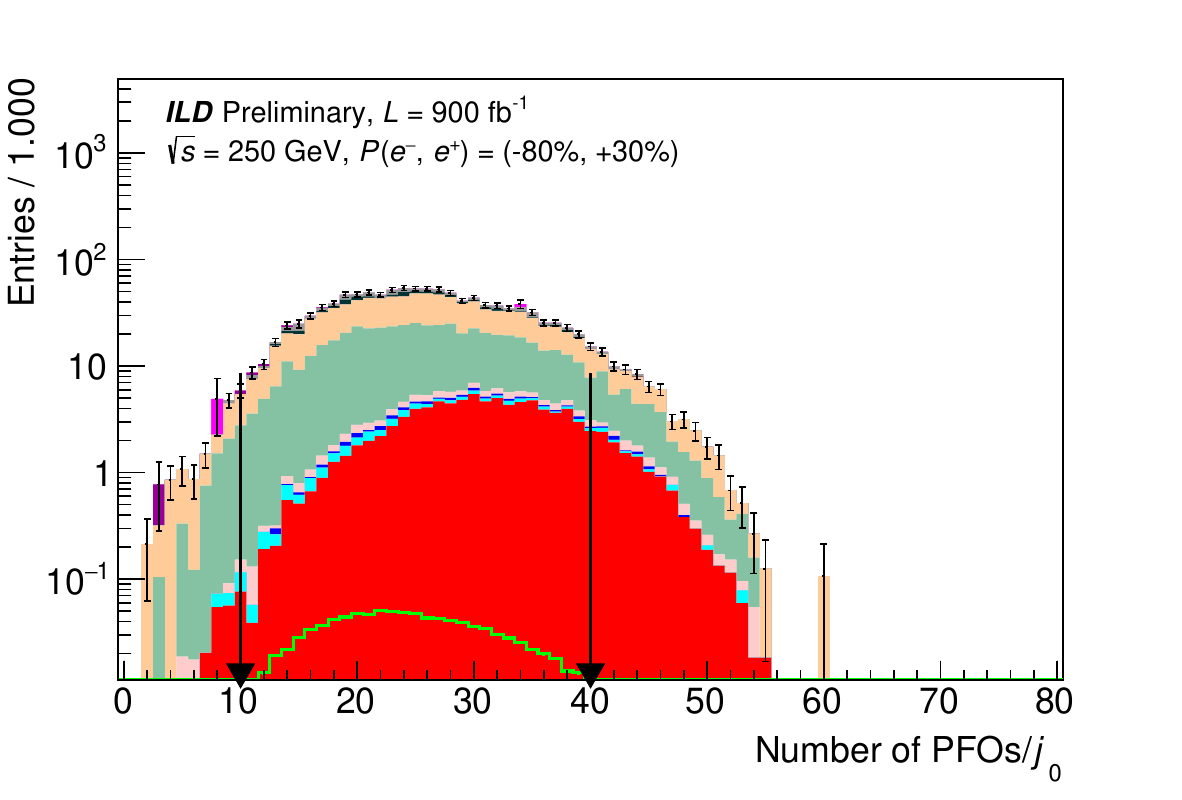}
        \caption{Number of PFOs in leading jet}
    \end{subfigure}
    \begin{subfigure}{0.23\textwidth}
        \centering
        \includegraphics[width=1.\textwidth]{figures/analysis_Zinv/legend.pdf}
    \end{subfigure} \\
    \begin{subfigure}{0.49\textwidth}
        \centering
        \includegraphics[width=1.\textwidth]{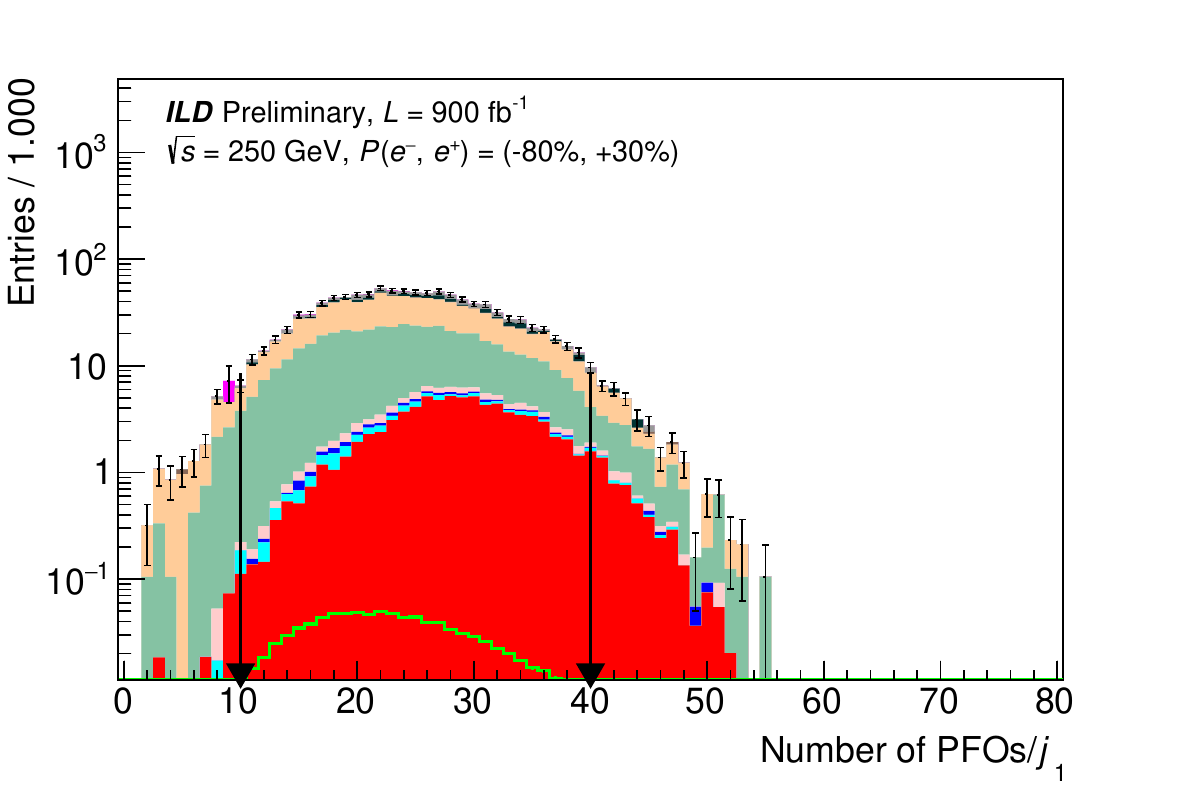}
        \caption{Number of PFOs in subleading jet}
    \end{subfigure} \\
    \caption{Histograms of the variables used in the kinematic selections of the \Zll channel, as described in Table~\ref{tab:selections}. Each histogram is given at the level of its corresponding selection but \emph{before} that selection is applied. The arrows represent the placement of the selection cuts, and the error bars represent the MC statistical uncertainties. The sum-of-weights per process is normalised to the SM cross section. N.B. the $h(\rightarrow s\bar{s})Z(\rightarrow\ell\bar{\ell}/\nu\bar{\nu})$ signal is unstacked. A continuation of Fig.~\ref{fig:histograms_Zll_4}.}
    \label{fig:histograms_Zll_5}
\end{figure}

The cutflow for the \Zll channel is plotted in Fig.~\ref{fig:cutflow_Zll} -- the full cutflow table is given in Table~\ref{tab:cutflow_Zll} in Appendix~\ref{app:cutflows}. Histograms of the variables included as part of this channel's selections (showing the evolution of the yields as each selection is applied) are shown in Figs.~\ref{fig:histograms_Zll_1} through \ref{fig:histograms_Zll_5}. From Table~\ref{tab:cutflow_Zll}, the hadronic backgrounds are almost entirely removed by cutting on the number of leptons. The signal efficiency for our selections is 5\% while our background efficiency is 0.001\%. The $4f$ single $Z$ and $ZZ$ backgrounds are the dominant backgrounds, with $\sim$730~events compared to the $\sim$1~events expected for \Hss. As with the \Zinv channel, the $\Hgg$ process is the dominant Higgs background with $\sim$76~events.

\FloatBarrier

\FloatBarrier

\begin{figure}[htbp]
    \centering
    \begin{subfigure}{\textwidth}
        \centering
        \includegraphics[width=\textwidth]{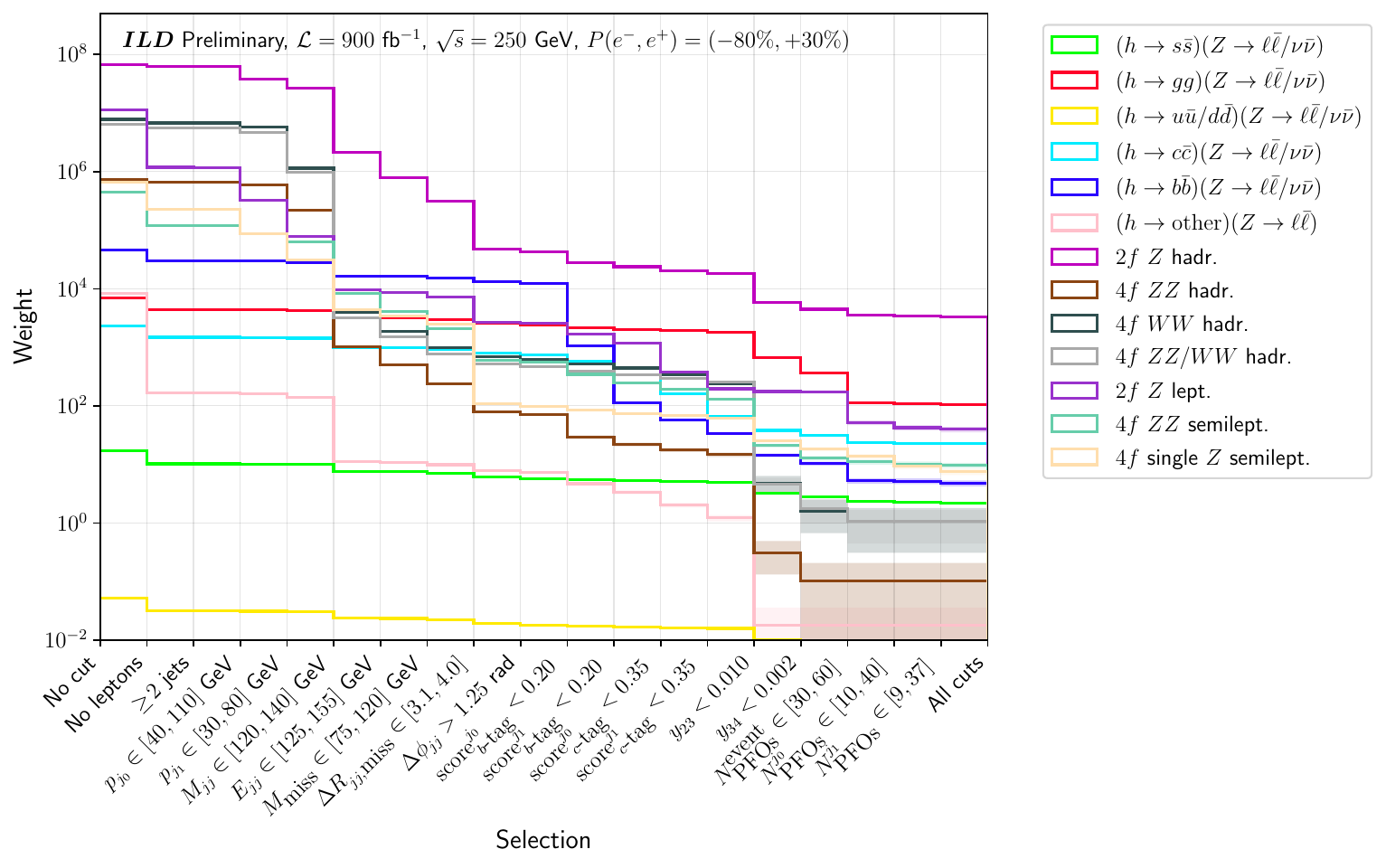}
        \caption{\Zinv channel}
        \label{fig:cutflow_Zinv}
    \end{subfigure} \\
    \vspace{2em}
    \begin{subfigure}{\textwidth}
        \centering
        \includegraphics[width=\textwidth]{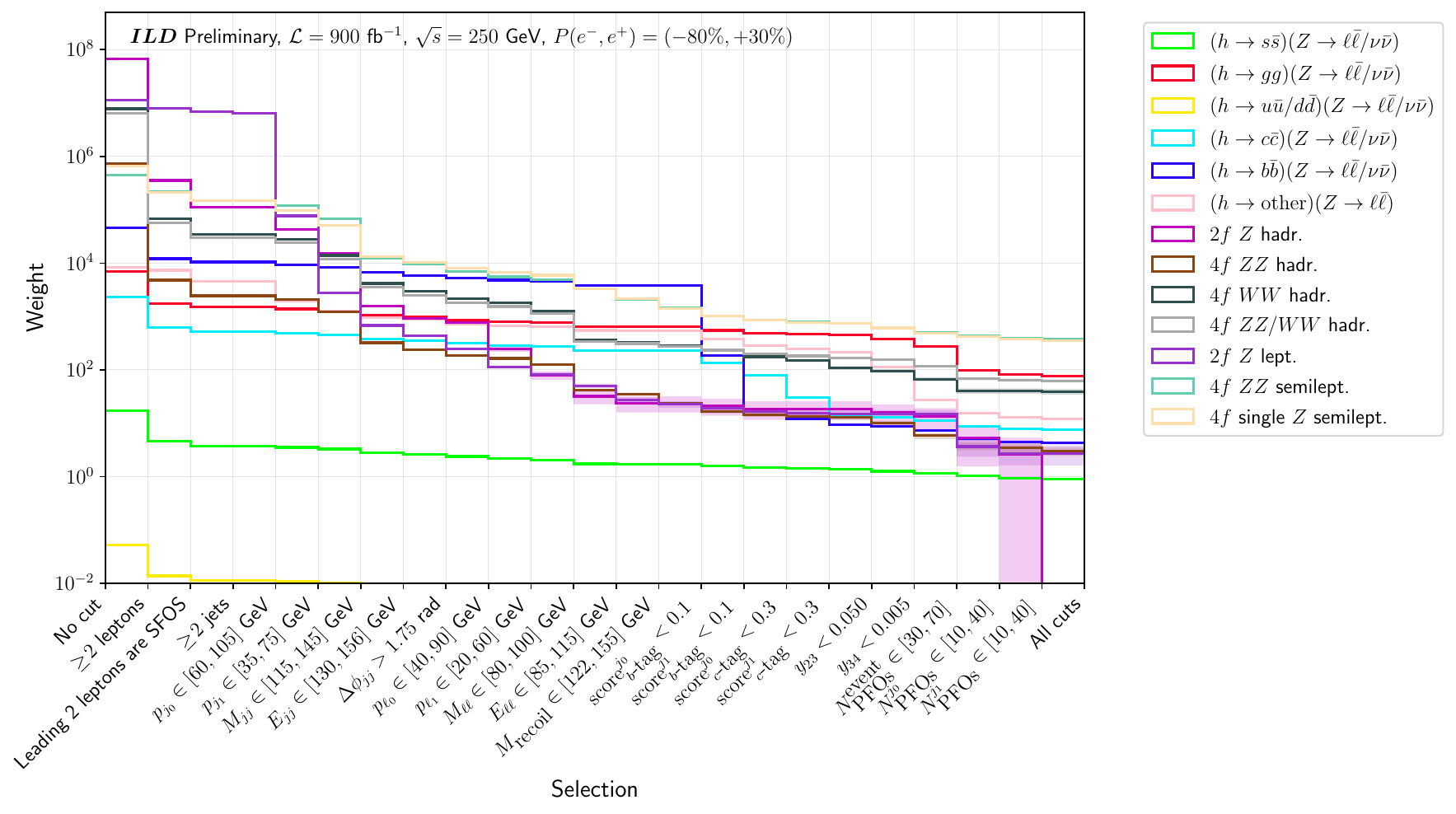}
        \caption{\Zll channel}
        \label{fig:cutflow_Zll}
    \end{subfigure} \\
    \caption{Visualisations of the cutflows in (a) Tables~\ref{tab:cutflow_Zinv} and (b) \ref{tab:cutflow_Zll}. The error bars correspond to MC statistical uncertainties.}
    \label{fig:cutflows}
\end{figure}

\FloatBarrier

\subsection{Limits on Higgs-strange coupling strength modifier}

The estimated significance of discovery, $Z_0 \approx s/\sqrt{s+b}$ (valid for $s/b \ll 1$), using the signal and background yields at the level of the last selection in Table~\ref{tab:selections} is $\sim$0.1. Therefore, a discovery measurement of \Hss is \emph{unlikely}, given the use of a best-case jet flavour tagger. However, limits on $\kappa_s$, and accordingly $\BR[\Hss]$, may be set instead, allowing us to reduce the phase space for BSM enhancements to the \Hss rate. 

\begin{figure}[htbp]
    \centering
    \begin{subfigure}{\textwidth}
        \centering
        \includegraphics[width=0.9\textwidth]{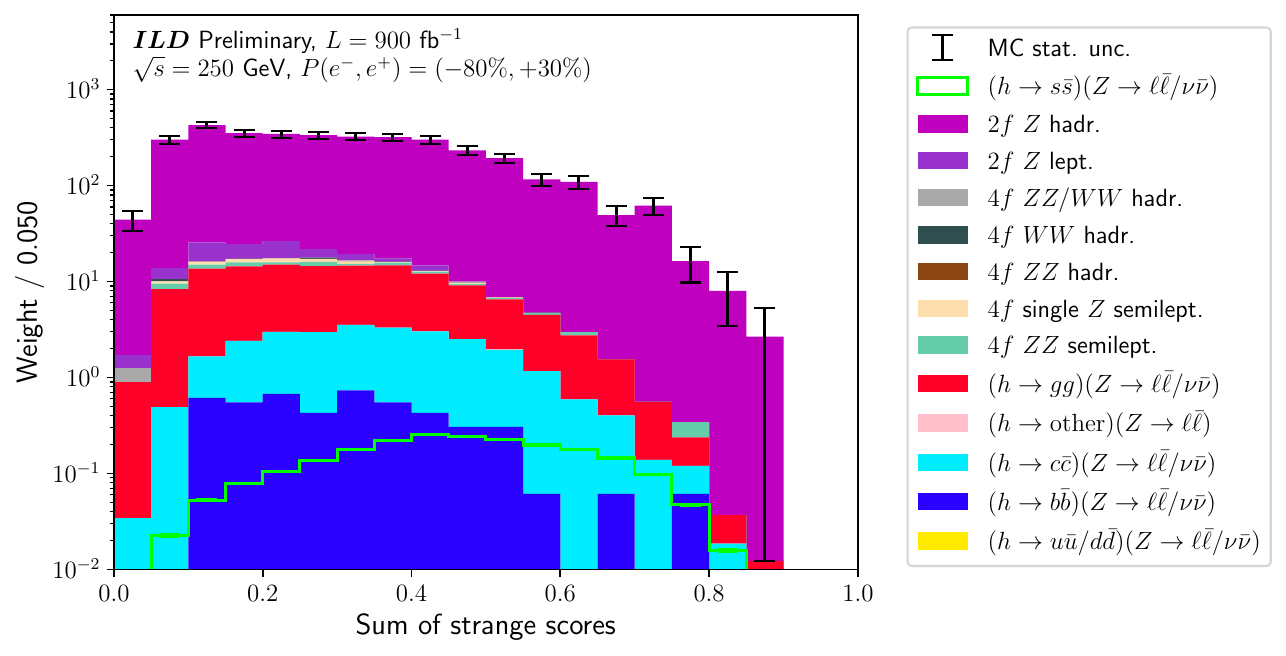}
        \caption{\Zinv channel}
    \end{subfigure} \\
    \vspace{0.5em}
    \begin{subfigure}{\textwidth}
        \centering
        \includegraphics[width=0.9\textwidth]{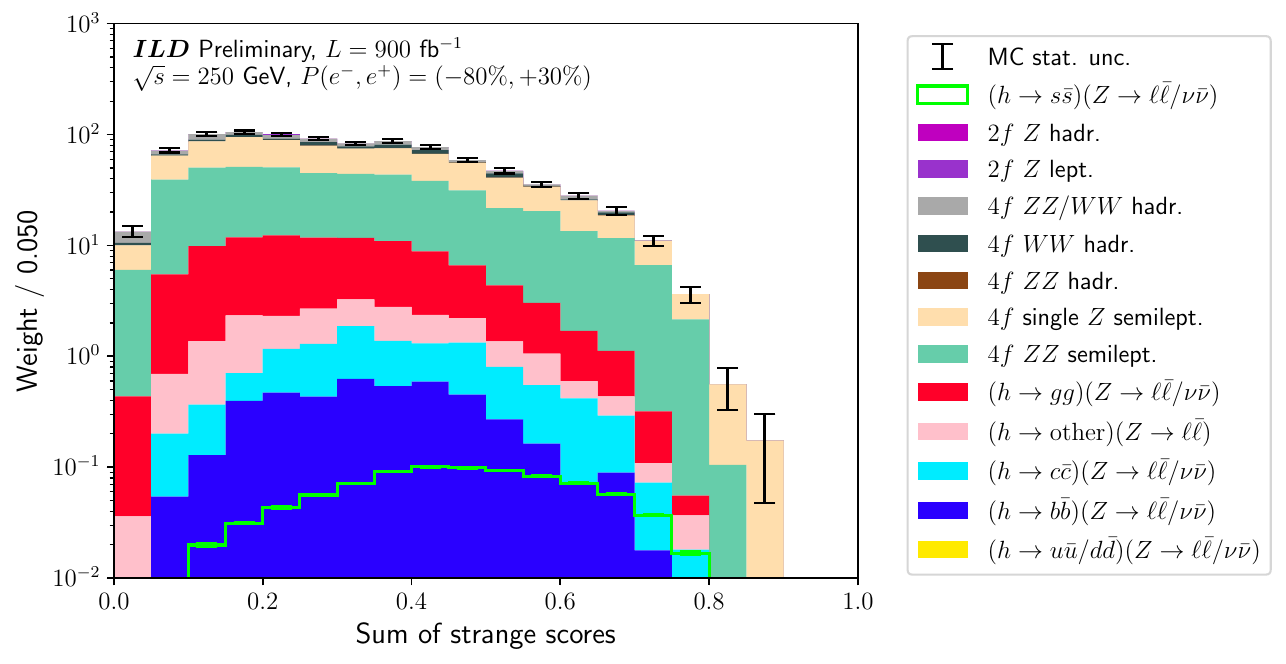}
        \caption{\Zll channel}
    \end{subfigure} \\
    \caption{Fit discriminants for each channel of the SM \Hss analysis, Eq.~\ref{eqn:discriminant}. Each histogram is produced at the level of the last selection of their respective channel in Table~\ref{tab:selections}. The error bars represent the MC statistical uncertainties. The sum-of-weights per process is normalised to the SM cross section. N.B. the $h(\rightarrow s\bar{s})Z(\rightarrow\ell\bar{\ell}/\nu\bar{\nu})$ signal is unstacked.}
    \label{fig:discriminant}
\end{figure}

The chosen fit discriminant for \Hss is the sum of the strange scores for the leading and subleading momentum jets, using the jet flavour tagger described in Section~\ref{sec:tagger}. Mathematically written, the discriminant, $\mathcal{D}$, is given by:

\begin{equation}
    \mathcal{D}(\vec{x}_{j_0}, \vec{x}_{j_1}) = \frac{1}{2} \times \left( [\vec{F}(\vec{x}_{j_0})]_s + [\vec{F}(\vec{x}_{j_1})]_s \right) \,,
    \label{eqn:discriminant}
\end{equation}

\noindent where $\vec{x}_{j_{0/1}}$ correspond to the flavour tagger's inputs for jets 0 and 1 and the subscript $s$ indicates the $s$-jet output node. The discriminant is shown for both channels in Fig.~\ref{fig:discriminant}. A higher sum of scores corresponds to a higher probability of an event containing an $s\bar{s}$ system.

The fitted likelihood is a product of Poisson probability density functions (PDFs)\footnote{The likelihood should match what is specified by \texttt{HistFactory}~\cite{HistFactory} for uncorrelated, counting-like uncertainties on the background yields.}:

\begin{equation}
    L(\kappa_s, \vec{\gamma}_b \,|\, \vec{n}, \vec{s}, \vec{b}, \vec{\sigma}_b) = \prod_{i=1}^{N} \left( \textrm{Pois}(n_i \,|\, \mu(\kappa_s) s_i + \gamma_{b,i} b_i) \times \textrm{Pois}(\sigma_{b,i}^{-2} \,|\, \gamma_{b,i} \sigma_{b,i}^{-2}) \right) \,,
    \label{eqn:likelihood}
\end{equation}

\noindent where $\textrm{Pois}(r \,|\, \rho) \equiv \rho^r\exp(-\rho)/r!$ is a Poisson PDF with the number of occurrences given by $r$ and the expectation value given by $\rho$. In the above:

\begin{itemize}
    \item $\mu(\kappa_s)$ is our signal strength modifier as a function of our POI, $\kappa_s$, given by Eq.~\ref{eqn:mu};
    \item $\vec{s}$ is the vector of expected signal yields (in $N$~regions or bins -- $s_i \equiv [\vec{s}]_i$ is the expected signal yield in $i$-th bin);
    \item $\vec{b}$ is the vector of expected background yields;
    \item $\vec{\sigma}_b$ is the vector of (uncorrelated) relative uncertainties on the background yields;
    \item $\vec{\gamma}_b$ is the vector of (uncorrelated) shape parameters for the background yields;
    \item $\vec{n}$ is the vector of observed yields.
\end{itemize}

\noindent As observed yields are unavailable, ``Asimov''~\cite{Asimov} data is assumed. The Python package \texttt{pyhf}~\cite{pyhf, pyhf_joss} is used to set the limits. Assuming the measurements performed by ILD are limited by the availability of data statistics, have well-constrained experimental systematics, and have excellent MC statistics, the background uncertainty is taken as the Poisson counting uncertainty for expected background yield in each bin, $\sigma_{b,i} = \sqrt{b_i}/b_i \,\forall\, i = 1,\ldots,N$.

\begin{figure}[htbp]
    \centering
    \begin{subfigure}{0.75\textwidth}
        \centering
        \includegraphics[width=\textwidth]{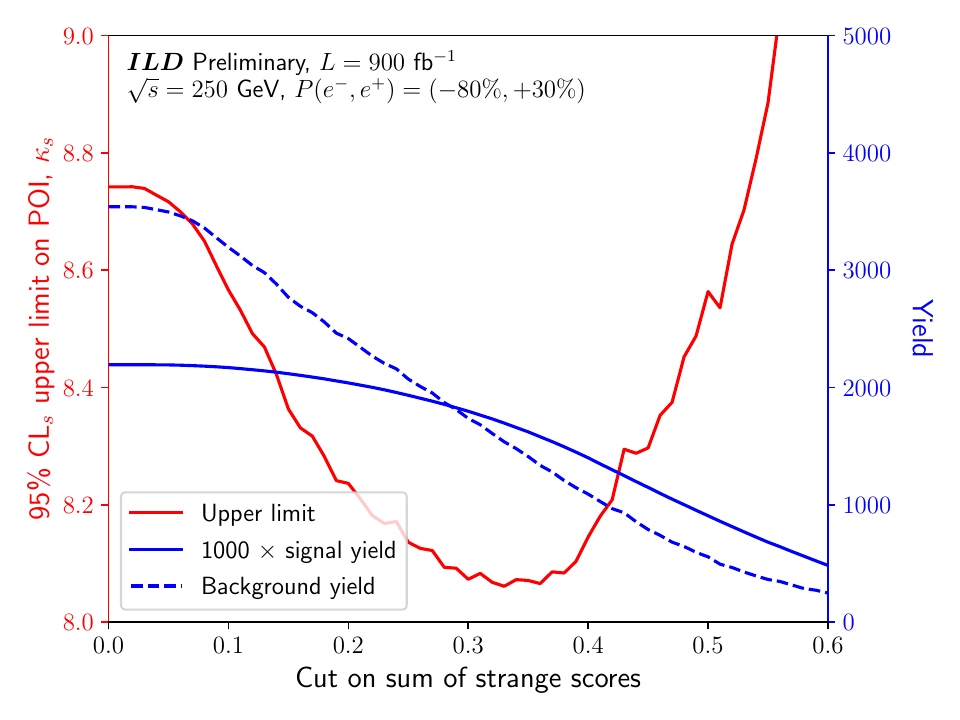}
        \caption{\Zinv channel}
    \end{subfigure} \\
    \begin{subfigure}{0.75\textwidth}
        \centering
        \includegraphics[width=\textwidth]{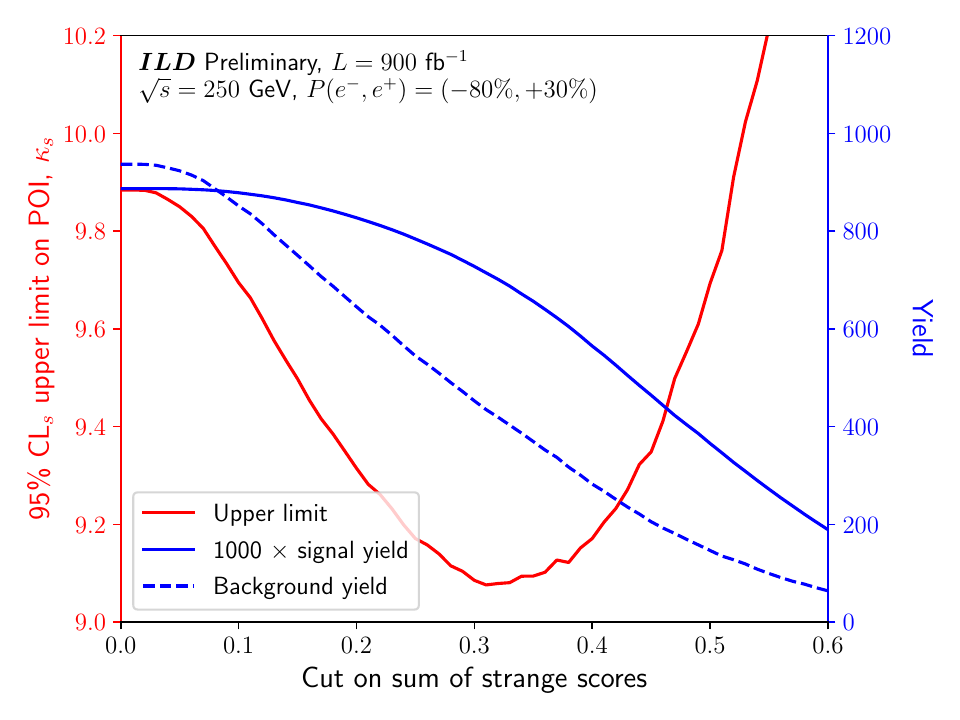}
        \caption{\Zll channel}
    \end{subfigure} \\
    \caption{Scans of the 95\% \CLs upper limit for the Higgs-strange coupling strength modifier, $\kappa_s$, obtained by varying the choice of the lower thresholds on the discriminants shown in Fig.~\ref{fig:discriminant}. Also shown are the signal (i.e., $h(\rightarrow s\bar{s})Z(\rightarrow\ell\bar{\ell}/\nu\bar{\nu})$) and background (i.e., non-$h(\rightarrow s\bar{s})Z(\rightarrow\ell\bar{\ell}/\nu\bar{\nu})$) yields in the resulting regions.}
    \label{fig:limit_scan}
\end{figure}

\begin{figure}[htbp]
    \centering
    \includegraphics[width=0.9\textwidth]{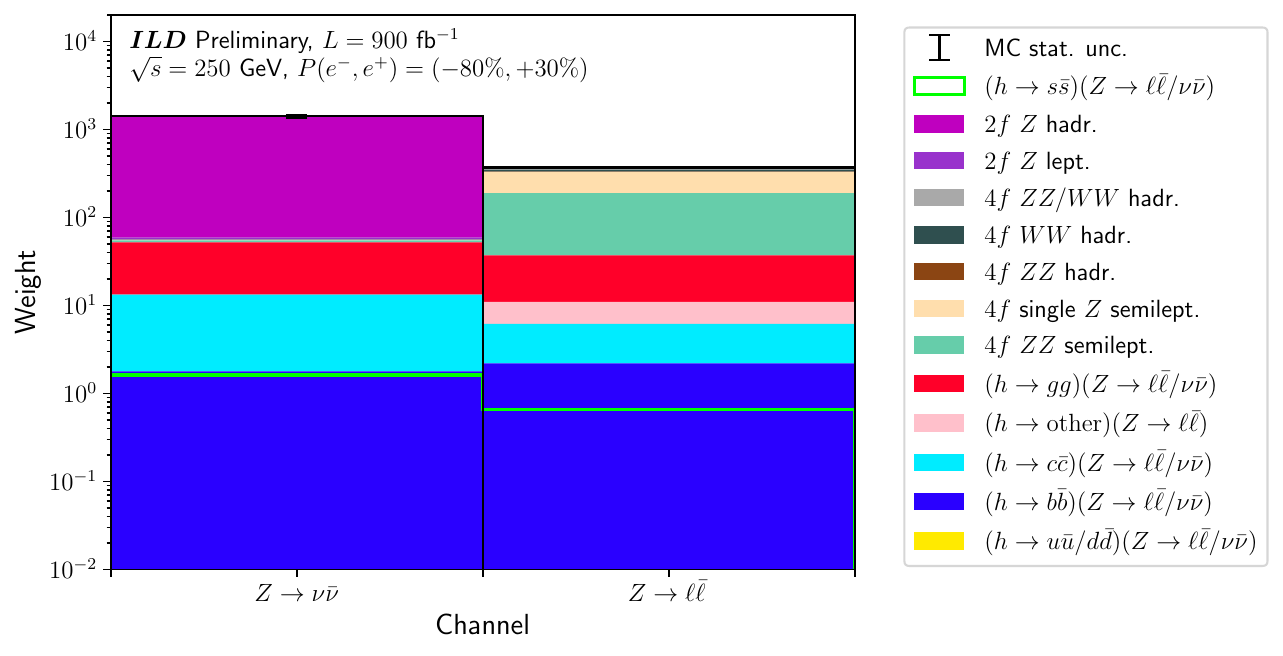}
    \caption{Yields in the signal regions for the \Zinv and \Zll channels, obtained by applying selections of $>$0.35 on the respective discriminants shown in Fig.~\ref{fig:discriminant}. The error bars represent the MC statistical uncertainties, and the sum-of-weights per process is normalised to the SM cross section. N.B. the $h(\rightarrow s\bar{s})Z(\rightarrow\ell\bar{\ell}/\nu\bar{\nu})$ signal is unstacked.}
    \label{fig:fit_inputs}
\end{figure}

\begin{figure}[htbp]
    \centering
    \begin{subfigure}{\textwidth}
        \centering
        \includegraphics[width=0.53\textwidth]{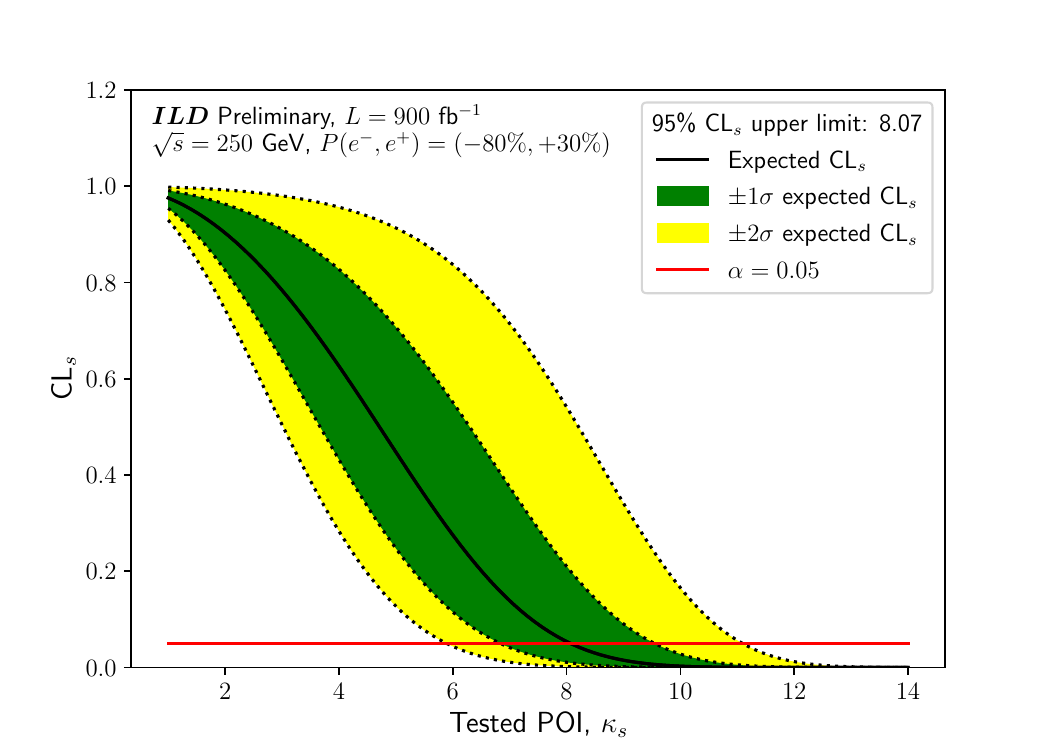}
        \caption{\Zinv channel}
    \end{subfigure} \\
    \begin{subfigure}{\textwidth}
        \centering
        \includegraphics[width=0.53\textwidth]{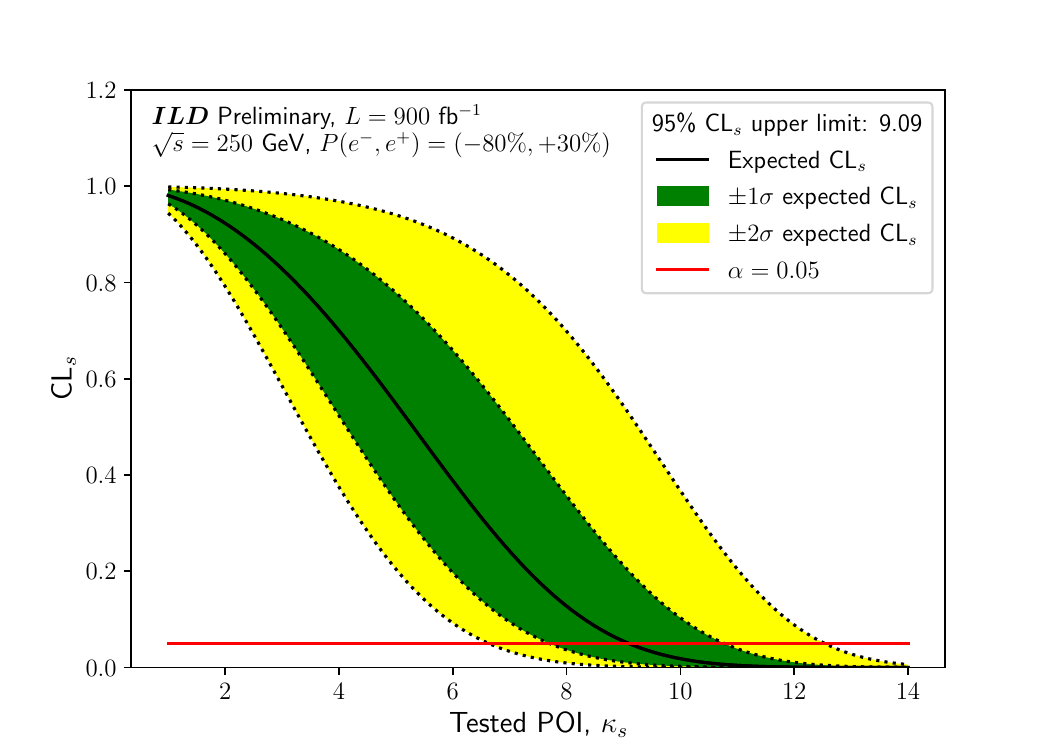}
        \caption{\Zll channel}
    \end{subfigure} \\
    \begin{subfigure}{\textwidth}
        \centering
        \includegraphics[width=0.53\textwidth]{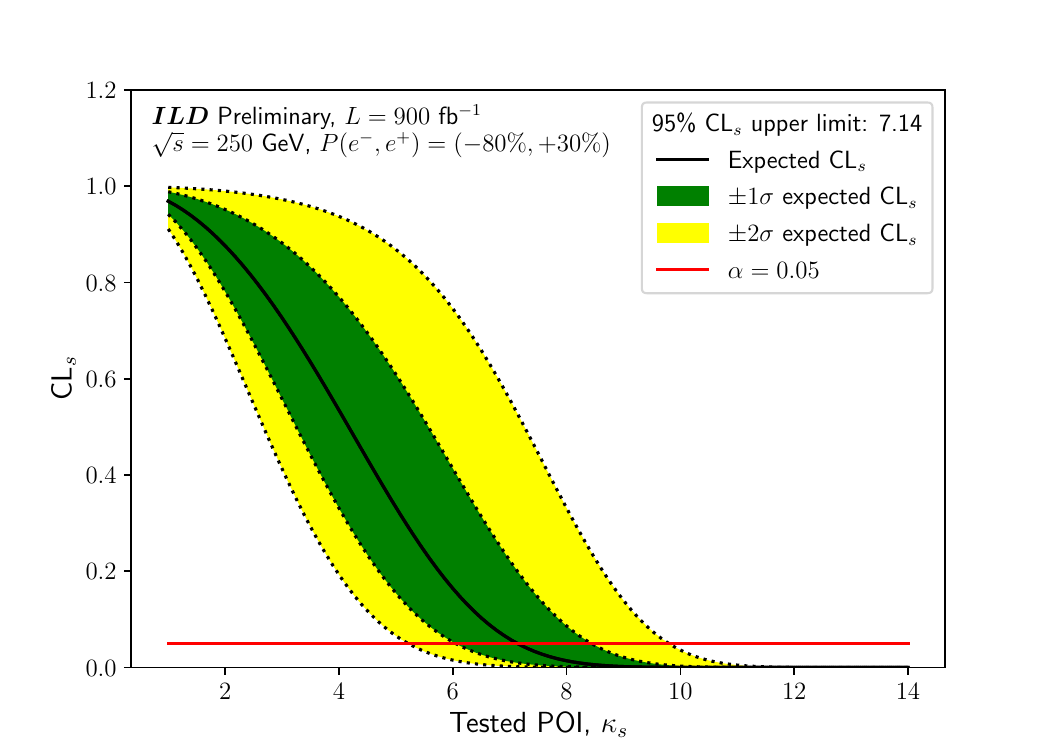}
        \caption{Combined}
    \end{subfigure} \\
    \caption{\CLs upper limit plots for the Higgs-strange coupling strength modifier, $\kappa_s$, obtained from fitting the yields in the \Zinv and \Zll signal regions shown in Fig.~\ref{fig:fit_inputs}. The combination fit using both channels is also shown. The crossing of the black and red lines indicates the 95\% confidence level.}
    \label{fig:limits}
\end{figure}

Signal regions are built by requiring the discriminants to be greater than some threshold -- these thresholds are chosen such that the best (i.e., strongest) 95\% \CLs upper limits~\cite{CLs} are obtained for the $\Zinv$ and $\Zll$ channels independently. Scans on the choice of threshold are shown in Fig.~\ref{fig:limit_scan}, which is found by eye to be 0.35 for both the $\Zinv$ and $\Zll$ channels. The choice of threshold is a trade-off between reducing the background while retaining signal and the finiteness of MC statistics. The yields in the signal regions for these particular thresholds are shown in Fig.~\ref{fig:fit_inputs} and the resulting limit plots for $\kappa_s$ are shown in Fig.~\ref{fig:limits}, including both the single-channel and combined results.\footnote{The single-channel analyses are both ``cut-and-count'' analyses, as each uses flat cuts to generate a single signal region bin. Accordingly, each channel fits Eq.~\ref{eqn:likelihood} with $N=1$. The combined fit utilises both of these bins, and therefore fits Eq.~\ref{eqn:likelihood} with $N=2$. In the future, the fit may be performed using the shape of the \Hss discriminant, enhancing the sensitivity.}

From Fig.~\ref{fig:limits}, the 95\% upper confidence bound on $\kappa_s$ is found to be 8.07 for the \Zinv channel and 9.09 for the \Zll channel, leading to a combined limit of 7.14. This number is comparable to what has been estimated for the ILC (all data, \sqrts = \unit[250]{GeV} as well as \sqrts = \unit[500]{GeV}) from other studies~\cite{FutureHiggs} using indirect measurements, $\kappa_s < 7.5$ at the 95\% confidence level (CL). However, the study here includes only two measurement channels and approximately 50\% of the expected dataset for ILD at \unit[250]{GeV}. The limits are therefore expected to improve even more.

\FloatBarrier

\subsubsection{Implications on BSM models}

We may study the implications of our expected results on extended Higgs sector models. A particular class of 2HDMs, a spontaneous flavour violating (SFV) 2HDM allows for large couplings of additional Higgs to strange/light quarks while suppressing flavour-changing neutral currents. The SFV 2HDM has been studied extensively in Refs.~\cite{Egana-Ugrinovic:2019, Egana-Ugrinovic:2021}. Ega{\~n}a-Ugrinovic \emph{et al.} consider two cases:

\begin{enumerate}[label=(\alph*)]
    \item the up-type SFV 2HDM, where the up-type quark Yukawa matrix for the second Higgs doublet, $H_2$, is required to be proportional to the SM up-type quark Yukawa matrix for the first Higgs doublet, $H_1$, while the down-type quark Yukawa matrix is left free;
    \item and the down-type SFV 2HDM, where the down-type quark Yukawa matrix for $H_2$ is required to be proportional to the SM down-type quark Yukawa matrix for $H_1$, while the up-type quark Yukawa matrix is left free.
\end{enumerate}

\noindent Assuming non-zero mixing between $H_1$ and $H_2$, an up-type SFV 2HDM may manifest itself as an enhancement to the \Hss coupling, $\lambda_{hs\bar{s}}$.

\begin{figure}[ht]
    \centering
    \includegraphics[width=0.8\textwidth]{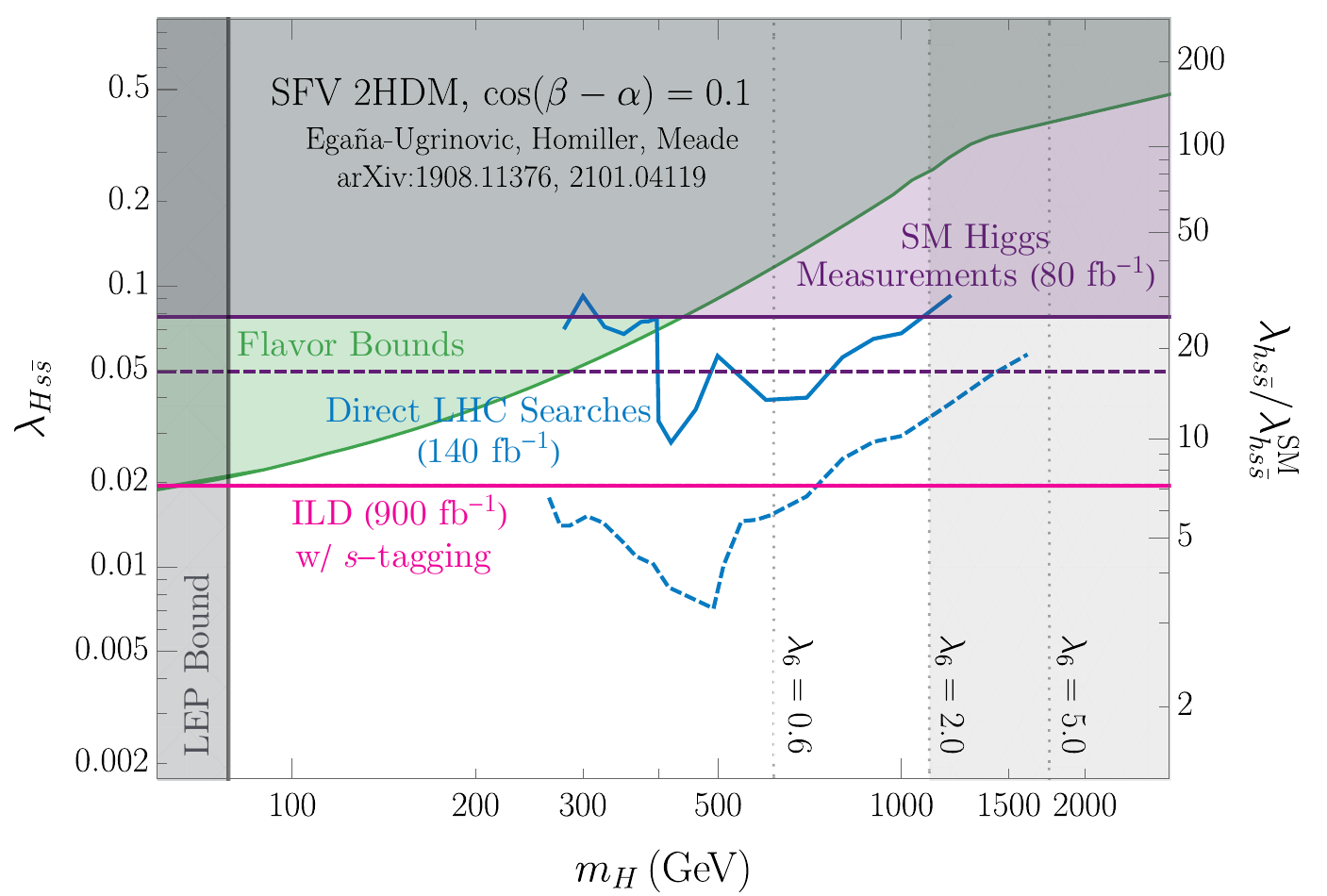}
    \caption{95\% CL bounds on the CP-even Higgs-strange Yukawa coupling $\lambda_{Hs\bar{s}}$ as well as on \unit[125]{GeV} SM Higgs-strange Yukawa coupling $\lambda_{hs\bar{s}}/\lambda_{hs\bar{s}}^\textrm{SM}$ (i.e., $\kappa_s$) for the SFV 2HDM described in Refs.~\cite{Egana-Ugrinovic:2019, Egana-Ugrinovic:2021}. The limits are shown as a function of the mass of the CP-even Higgs, $m_H$. The model assumes the CP-even Higgs $H$, the CP-odd Higgs $A$, and the charged Higgs $H^\pm$ are all degenerate (i.e., $m_H = m_A = m_{H^\pm}$) -- additionally, an alignment parameter of $\cos(\beta-\alpha) = 0.1$ is used for the $h$ -- $H$ mixing. The green line shows the bounds obtained from $D$ -- $\bar{D}$ mixing as described in Ref.~\cite{Egana-Ugrinovic:2019}; the purple lines show the bounds obtained by requiring the inclusive gluon-gluon fusion cross section to be consistent with combination measurements from ATLAS~\cite{ATLAS:2019}; the blue lines show the bounds obtained $H\rightarrow hh$ and $A\rightarrow Zh$ measurements from ATLAS and CMS~\cite{ATLAS:2019-Hhh, CMS:2018, ATLAS:2022enb}; and the pink line shows the bounds obtained from the \Hss analysis presented in this paper. The dashed lines correspond to bounds expected from the HL-LHC. Also shown are bounds from charged Higgs searches performed at LEP~\cite{LEP:2001}. Drawn as dotted lines are the contours for the 2HDM's quartic coupling $\lambda_6$: $\mathcal{L} \supset (\lambda_6 H_1^\dagger H_1 H_1^\dagger H_2 + \textrm{h.c.})$.}
    \label{fig:exclusion_plot_2HDM}
\end{figure}

We show the limits placed on the Yukawa couplings of such a model in Fig.~\ref{fig:exclusion_plot_2HDM}, including the limits obtained from the \Hss analysis presented in this paper. We find that the limits from the \Hss analysis are the strongest throughout the parameter space considered, exceeding even those expected from measurements performed at the High Luminosity LHC (HL-LHC) except for a small range of parameters. Therefore, tests of SFV 2HDMs are expected to be highly competitive at future lepton colliders like the ILC.

\FloatBarrier

\section{Proposal for an alternative detector layout}
\label{sec:alt_detector}

We have made a preliminary investigation of a possible Ring Imaging Cerenkov system (RICH) detector capable of $\pi$/$K$ separation up to \unit[25]{GeV} at the SiD or ILD detectors. It has been discussed many times that a gaseous RICH detector is the only way to reach $\pi$/$K$ separation up to \unit[30--40]{GeV} -- see Appendix~\ref{app:PID_reach}.

\subsection{Overall concept}

\begin{figure}
    \centering
   \begin{subfigure}{0.8\textwidth}
        \centering
        \includegraphics[width=1.\textwidth]{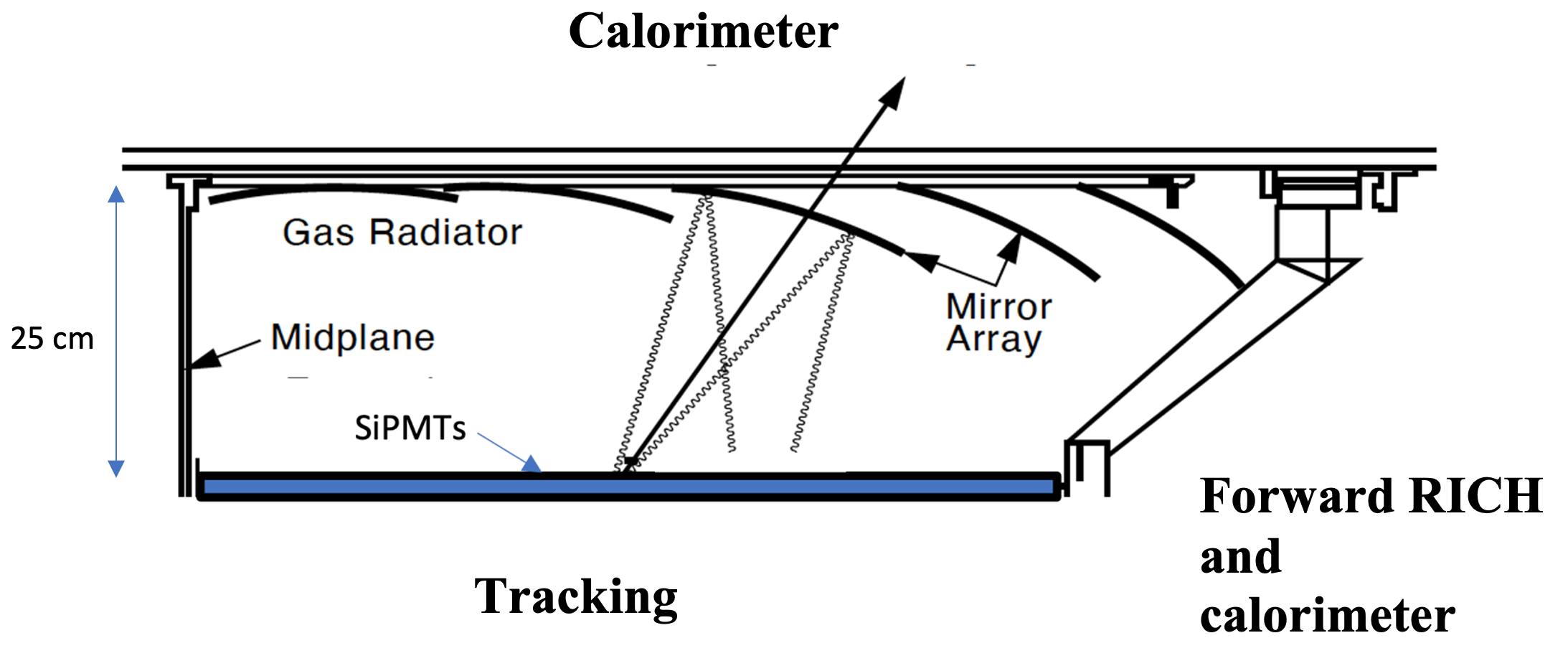}
        \caption{Side view of overall layout}
    \end{subfigure} \\
    \begin{subfigure}[b]{0.65\textwidth}
        \centering
        \includegraphics[width=1.\textwidth]{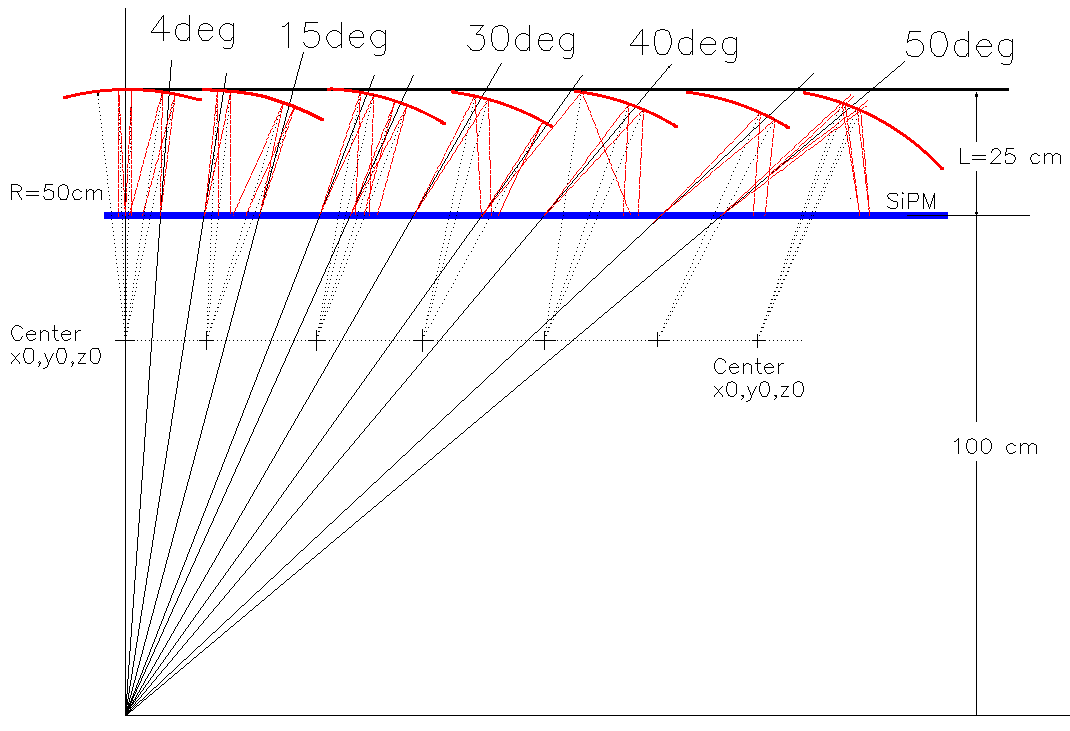}
        \caption{Side view with tracks}
    \end{subfigure}
    \hfill
    \begin{subfigure}[b]{0.30\textwidth}
        \centering
        \includegraphics[width=1.\textwidth]{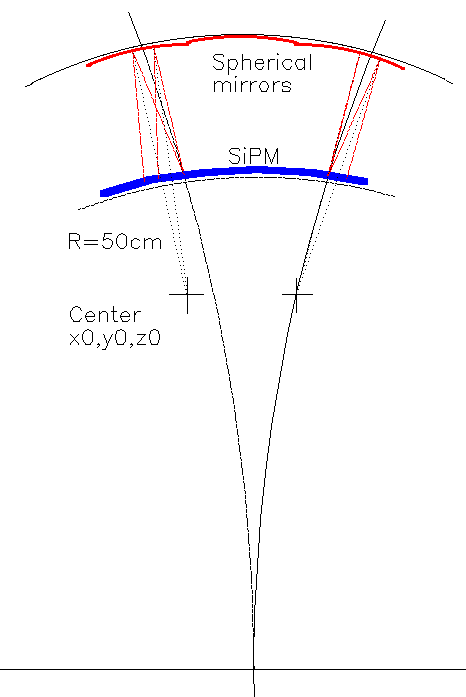}
        \caption{Front view with tracks}
    \end{subfigure}
    \caption{Proposed gaseous RICH detector at SiD/ILD. (a) The relative placement of the tracking, calorimetry, and forward instrumentation is indicated. (b) Side view and (c) front view of the proposed detector, with tracks. All of the mirrors have a radius of \unit[50]{cm}. This optical design is preliminary as further tuning of the mirror positions is required.}
    \label{fig:proposed_RICH}
\end{figure}

The detector concept is shown in Fig.~\ref{fig:proposed_RICH}. The initial choice for the RICH detector thickness was \unit[25]{cm} active length; however, we also looked at a \unit[10]{cm} active length to minimise the magnetic field smearing effect.\footnote{The Cherenkov ring is smeared in the focal plane due to the helical motion of the particle in a large magnetic field -- see Section~\ref{sec:magentic_smearing} for more details.} The RICH detector is designed using spherical mirrors and Silicon Photomultipliers (SiPMs -- also referred to as SiPMTs) as photon detectors.\footnote{The present design with SiPM detectors requires that the total neutron dose at RICH's location is less than \unit[$\sim\!5 \times 10^{10}$]{$n_\textrm{eq}$/10 years}, for which the SiPM damage is expected to be low.} Fig.~\ref{fig:proposed_RICH} resembles the gaseous RICH detector of the SLAC Large Detector's (SLD's) Cherenkov Ring Imaging Detector (CRID)~\cite{Vavra:1999}; however, introducing SiPM-based design improves the PID performance by a factor of two compared to the SLD's and DELPHI's gaseous RICH detectors, who pioneered this type of PID concept. Although we have selected a specific type of SiPM in this paper in order to do the calculation (a commercially-available Hamamatsu SiPM), we believe that the photon  technology will improve over the next 15 years in terms of noise performance, timing capability, pixel size, and detection efficiency. The overall aim is to make this RICH detector with as low mass as possible because we do not want to degrade the calorimeter. This speaks for mirrors made of beryllium~\cite{Barber:2006} and the structure made of low mass carbon-composite material. Another important aspect is to make the RICH detector depth as thin as possible in order to reduce the cost of the calorimeter. Our initial choice of \unit[25]{cm} could be reduced further if the detection efficiency of future photon detectors improve. For example, if the detection efficiency improves by $\sim$50\%, the radial depth can be reduced to \unit[10--15]{cm}, in turn reducing the magnetic smearing contribution to Cherenkov angle resolution.

\subsubsection{Gas choices}

\begin{enumerate}[label=(\alph*)]
    \item Pure C$_5$F$_{12}$ gas at \unit[1]{bar} requires a detector temperature of \unit[40]{\oC} since the boiling point of this gas is \unit[31]{\oC} at \unit[1]{bar}. That could prove to be difficult since SiPMs need to be cooled.
    \item A gas choice of pure C$_4$F$_{10}$ at \unit[1]{bar} allows detector operation at a few degrees Celsius since boiling point of this gas is \unit[-1.9]{\oC} at \unit[1]{bar}. This is presently our \emph{preferred} choice.
    \item A choice of C$_2$F$_6$ gas at \unit[1]{bar} would allow detector operation even below \unit[0]{\oC} since the boiling point of this gas is \unit[-70.2]{\oC} at \unit[1]{bar}. However, this gas would deliver insufficient number of photoelectrons in the geometry shown in Fig.~\ref{fig:proposed_RICH} and therefore it was not considered.
    \item A choice of C$_3$F$_8$ gas at \unit[1]{bar} would allow detector operation at \unit[-30]{\oC} since the boiling point of C$_3$F$_8$ is \unit[-37]{\oC}. The detector's PID performance will be between C$_2$F$_6$ and C$_4$F$_{10}$. It is certainly worthwhile to look into this solution.
    \item Among non-freon-based gases, one could consider either C$_3$H$_8$ or C$_3$H$_6$, each of which has a reasonably high refraction index; however, these gases are flammable.
\end{enumerate}

\subsubsection{Number of photoelectrons per ring}

\begin{figure}[htbp]
    \centering
    \begin{subfigure}[b]{0.44\textwidth}
        \centering
        \includegraphics[width=1.\textwidth,valign=b]{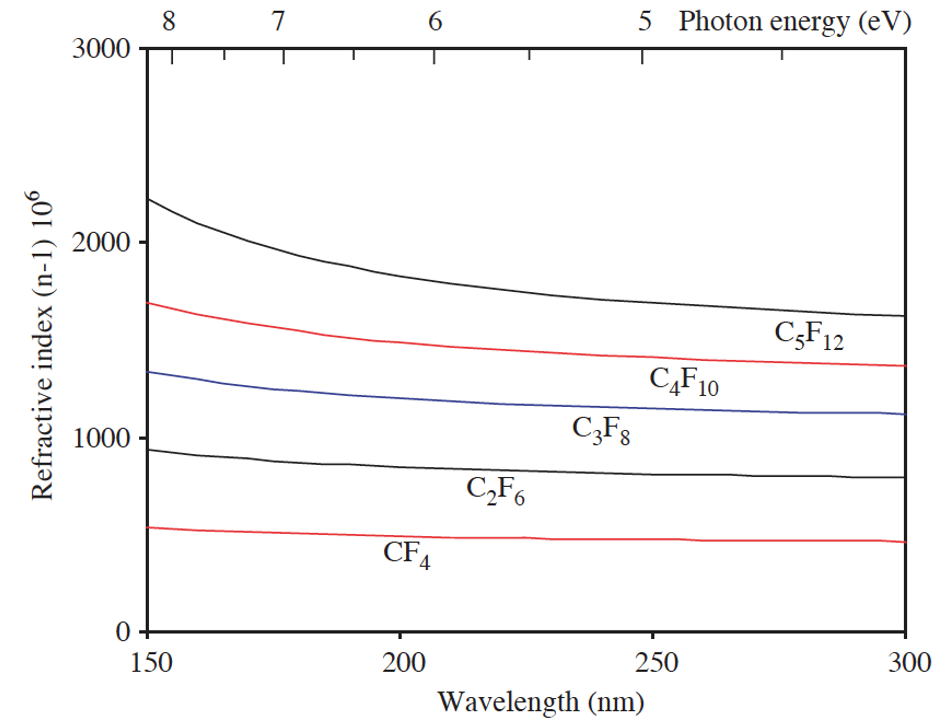}
        \caption{}
        \label{fig:efficiency_refractive_index}
    \end{subfigure}
    \hfill
    \begin{subfigure}[b]{0.54\textwidth}
        \centering
        \includegraphics[width=1.\textwidth,valign=b]{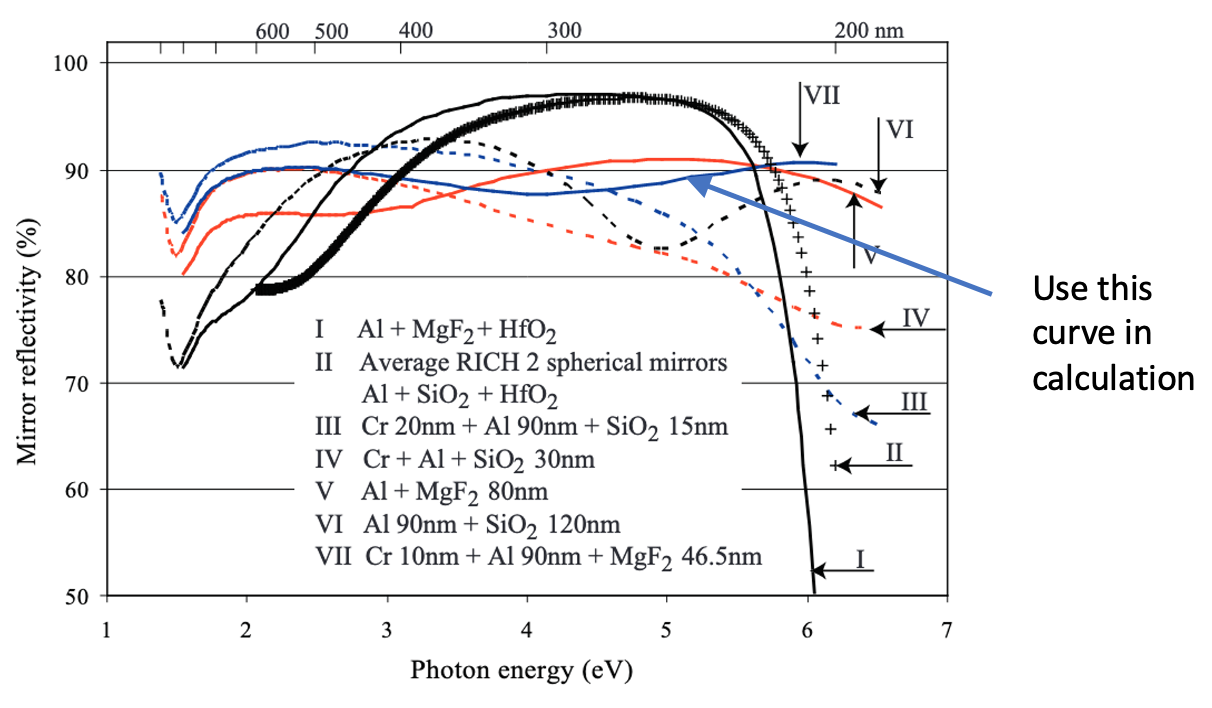}
        \caption{}
        \label{fig:efficiency_reflectivity}
    \end{subfigure} \\
    \vspace{0.5em}
    \begin{subfigure}[b]{0.54\textwidth}
        \centering
        \includegraphics[width=1.\textwidth,valign=b]{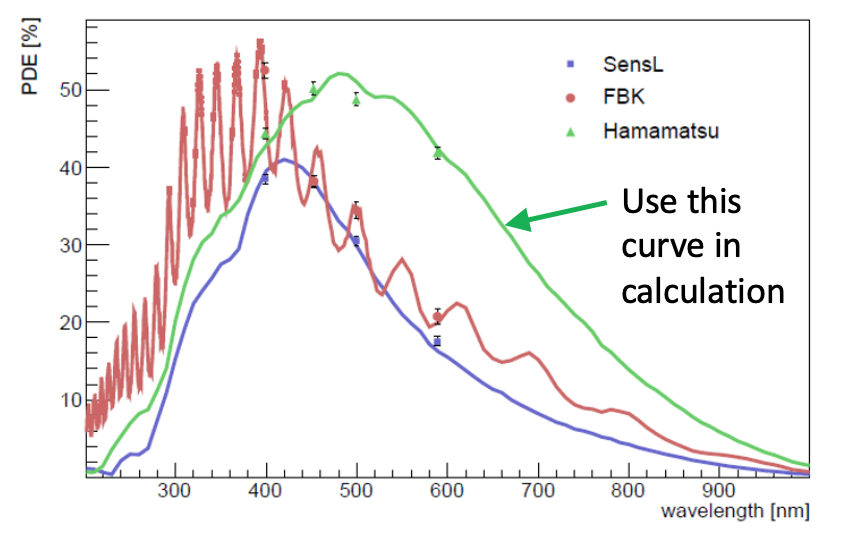}
        \caption{}
        \label{fig:efficiency_PDE}
    \end{subfigure}
    \hfill
    \begin{subfigure}[b]{0.44\textwidth}
        \centering
        \includegraphics[width=1.\textwidth,valign=b]{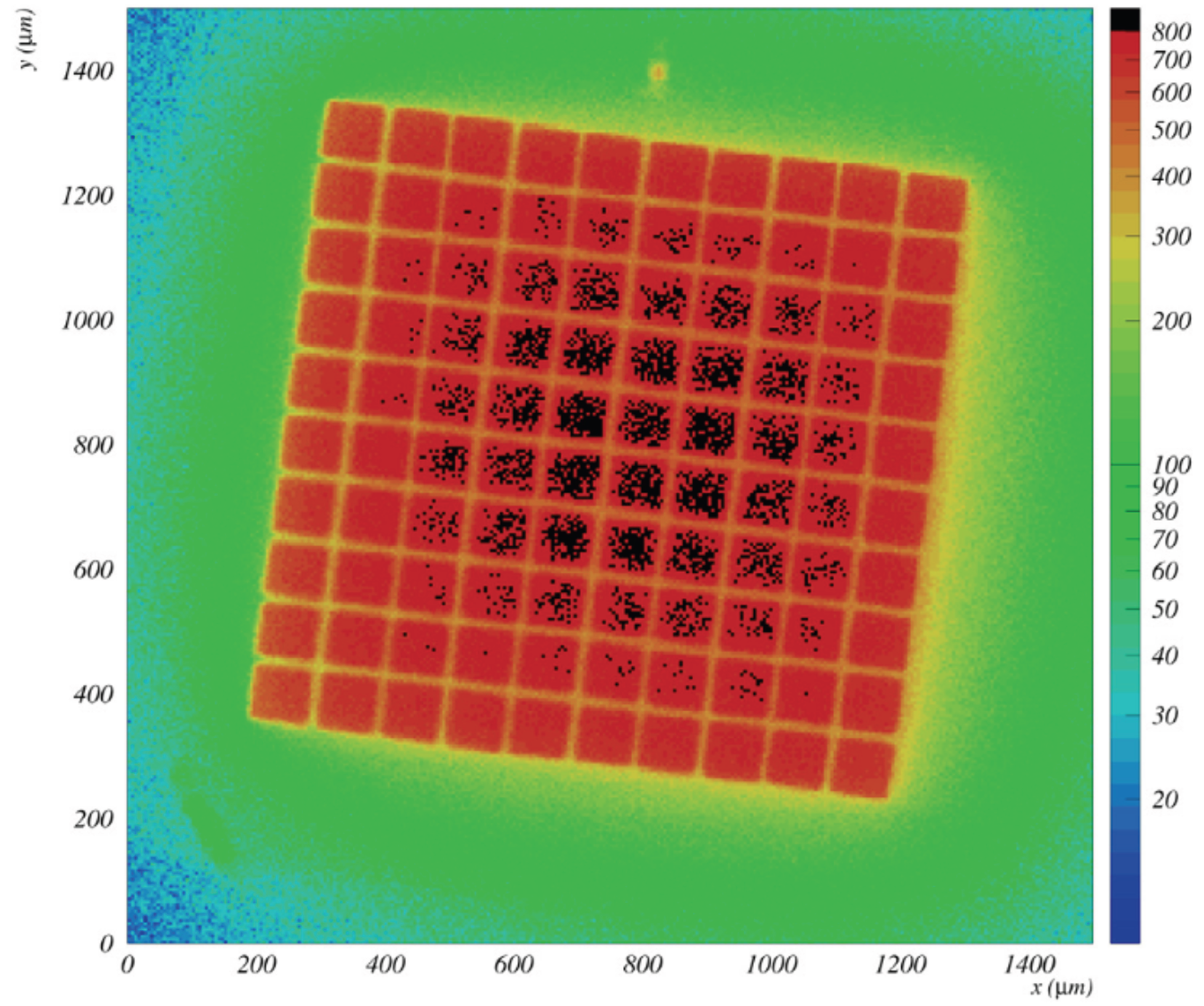}
        \caption{}
        \label{fig:efficiency_PMT}
    \end{subfigure} \\
    \vspace{0.5em}
    \begin{subfigure}[b]{0.49\textwidth}
        \centering
        \includegraphics[width=1.\textwidth,valign=b]{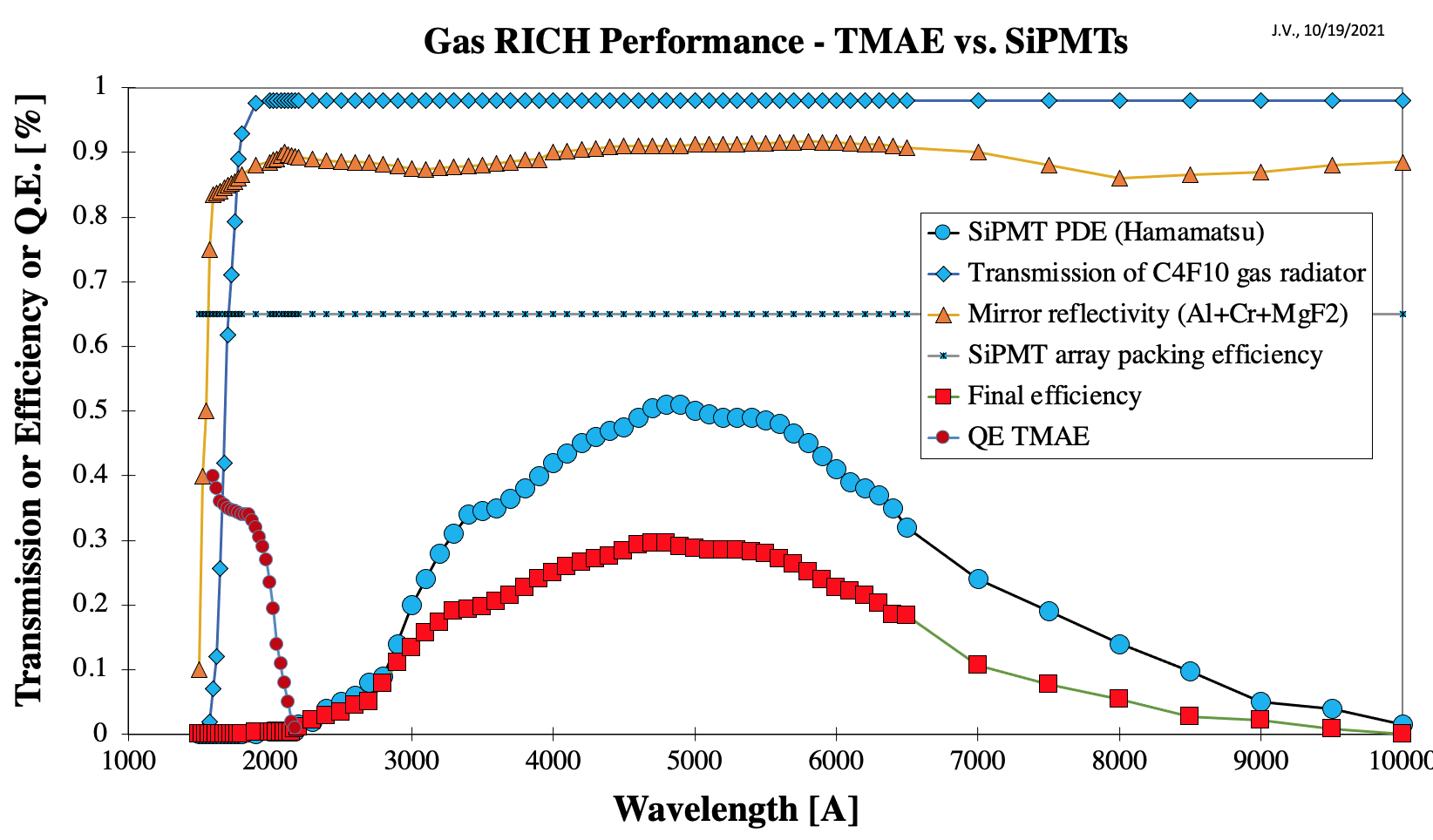}
        \caption{}
        \label{fig:efficiency_separate}
    \end{subfigure}
    \hfill
    \begin{subfigure}[b]{0.49\textwidth}
        \centering
        \includegraphics[width=1.\textwidth,valign=b]{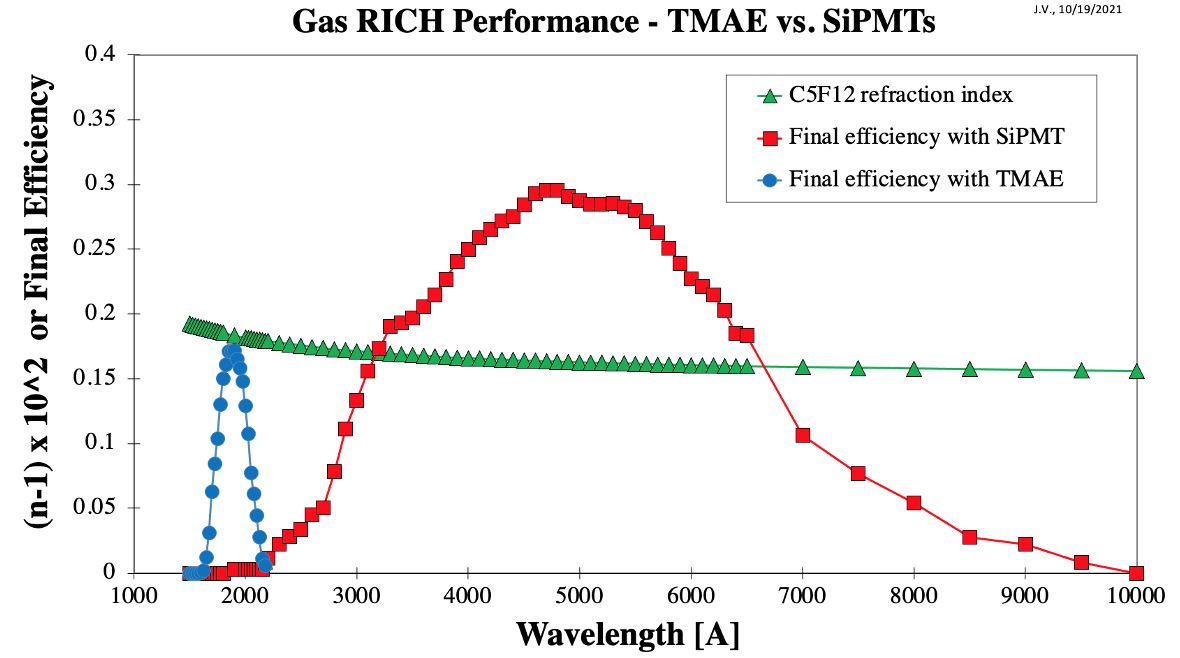}
        \caption{}
        \label{fig:efficiency_combined}
    \end{subfigure} \\
    \caption{(a) Refraction index for the gases considered~\cite{Ullaland:2005, Vavra:2014}. (b) Reflectivity of various mirror coatings~\cite{LHCb:2008}; we used Cr/Al/MgF$_2$ coating in the calculation. (c) Photon detection efficiency (PDE) of a single SiPM from several sources~\cite{NepomukOtte:2016}. We used the Hamamatsu PDE in the calculation. (d) A SiPM array has additional losses due to gaps between pixel elements~\cite{Korpar:2020}, the so called ``packing efficiency''. In this paper, we assume an additional loss of 65\% due to this effect. (e) The various efficiencies, including packing efficiency, gas transmission, mirror reflectivity, and the SiPM PDE, used in our calculation. (f) Final efficiency of the SiPM compared to the final efficiency of TMAE used by the SLD CRID, as calculated in this work. Also shown is the C$_5$F$_{12}$ refraction index to indicate chromaticity in the present detector proposal.}
    \label{fig:efficiency}
\end{figure}

The number of photoelectrons, $N_\textrm{pe}$, is calculated using:

\begin{equation}
    N_\textrm{pe} = N_0 L \sin^2\!\left(\langle\theta_c\rangle\right) \,,
\end{equation}

\noindent where $L$ is the length of the radiator, $\langle\theta_c\rangle$ is the mean Cherenkov angle, and:

\begin{equation}
    N_0 = \frac{\alpha}{hc}\int \frac{\textrm{Eff}(E)\sin^2\!\theta_c}{\sin^2\!\left(\langle\theta_c\rangle\right)} dE \,,
\end{equation}

\noindent where $\alpha$ is the fine-structure constant, $h$ is Planck's constant, $c$ is the speed of light, and $E$ is the energy of the photon. The Cherenkov angle, $\theta_c$, is given by:

\begin{equation}
    \cos\theta_c(\lambda) = \frac{1}{n(\lambda)\beta} \,,
\end{equation}

\noindent where $\lambda$ is the wavelength of the photon, $n$ is the refractive index of the medium, and $\beta$ is the Lorentz factor. To calculate $N_0$, one also needs to calculate $\textrm{Eff}(E)$, which is the product of all of the efficiencies in the problem, and to determine the refraction index as a function of wavelength to calculate the Cherenkov angle. Fig.~\ref{fig:efficiency_refractive_index} shows the refraction index for all gases considered. These gases wee measured by Ullaland~\cite{Ullaland:2005}. Fig.~\ref{fig:efficiency_reflectivity} shows reflectivity of various mirror coatings~\cite{LHCb:2008}. We chose the reflectivity of Cr/Al/MgF$_2$ coating in the calculation, as indicated on the graph. Fig.~\ref{fig:efficiency_PDE} shows photon detection efficiency (PDE) of a single SiPM~\cite{NepomukOtte:2016}. We have chosen the Hamamatsu PDE for our calculation. Fig.~\ref{fig:efficiency_PMT} shows that a SiPM array has additional losses due to gaps between the pixel elements of the array~\cite{Korpar:2020}, the so called ``packing efficiency''. We have chosen a packing efficiency of 65\% in our calculation. Fig.~\ref{fig:efficiency_separate} shows the various efficiencies used in our calculation, and Fig.~\ref{fig:efficiency_combined} shows the final efficiency of the SiPM-based and the TMAE-based\footnote{``TMAE'' $\coloneqq$ ``tetrakis(dimethylamine)ethylene''.} detector solutions used by the SLD CRID and the DELPHI RICH. Also shown is the C$_4$F$_{10}$ refraction index to indicate chromaticity\footnote{By chromaticity, we mean the variation of the refraction index as a function of wavelength. This effect causes an increase of the Cherenkov angle resolution, referred to as ``chromatic'' broadening.} in the problem. The SiPM solution is vastly better than TMAE solution in terms of overall efficiency, as one can see from Fig.~\ref{fig:efficiency_combined}.

\begin{figure}[htbp]
    \centering
    \begin{subfigure}[b]{0.49\textwidth}
        \centering
        \includegraphics[width=1.\textwidth]{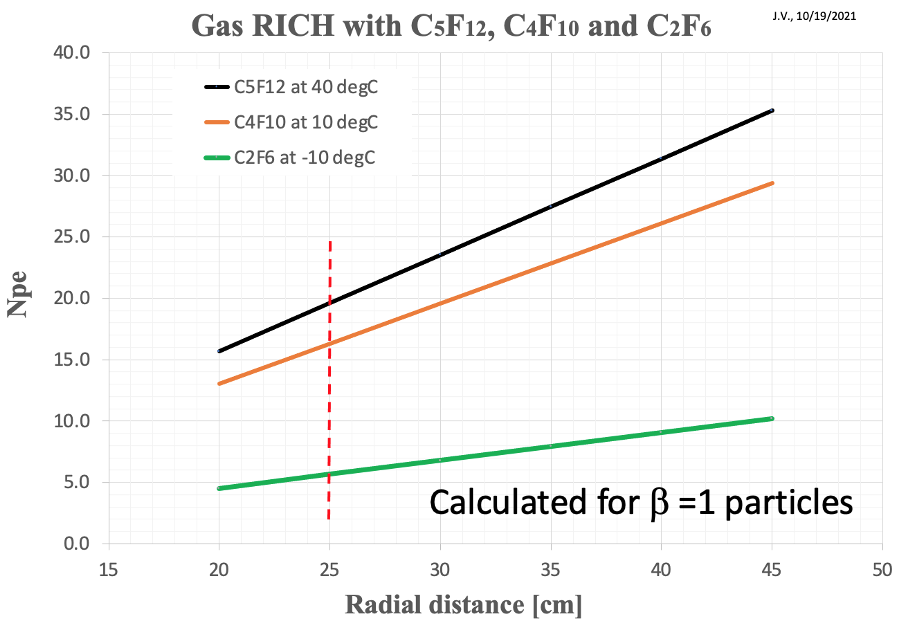}
        \caption{}
        \label{fig:number_of_photoelectrons_L}
    \end{subfigure}
    \hfill
    \begin{subfigure}[b]{0.49\textwidth}
        \centering
        \includegraphics[width=1.\textwidth]{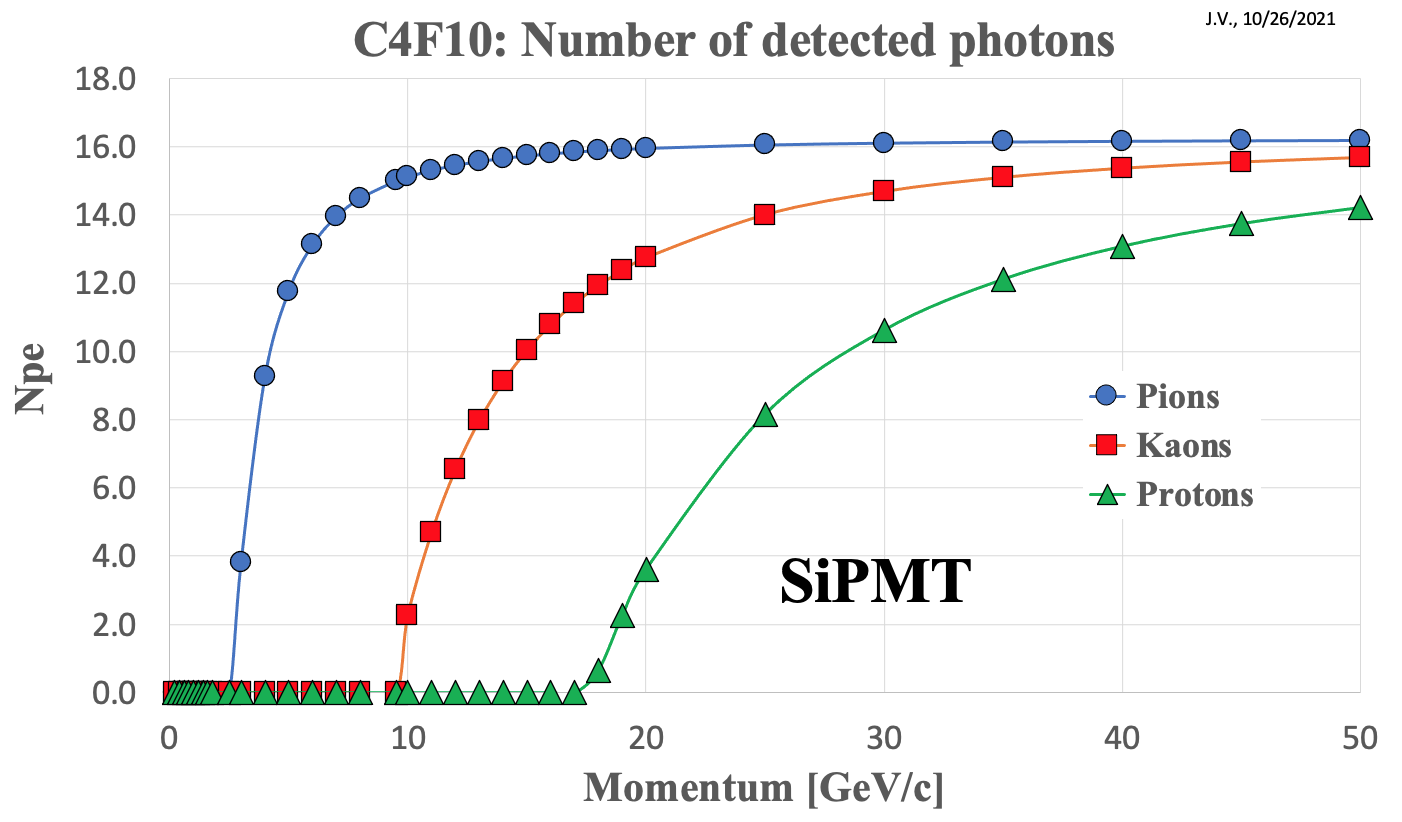}
        \caption{}
        \label{fig:number_of_photoelectrons_p}
    \end{subfigure} \\
    \begin{subfigure}{0.49\textwidth}
        \centering
        \includegraphics[width=1.\textwidth]{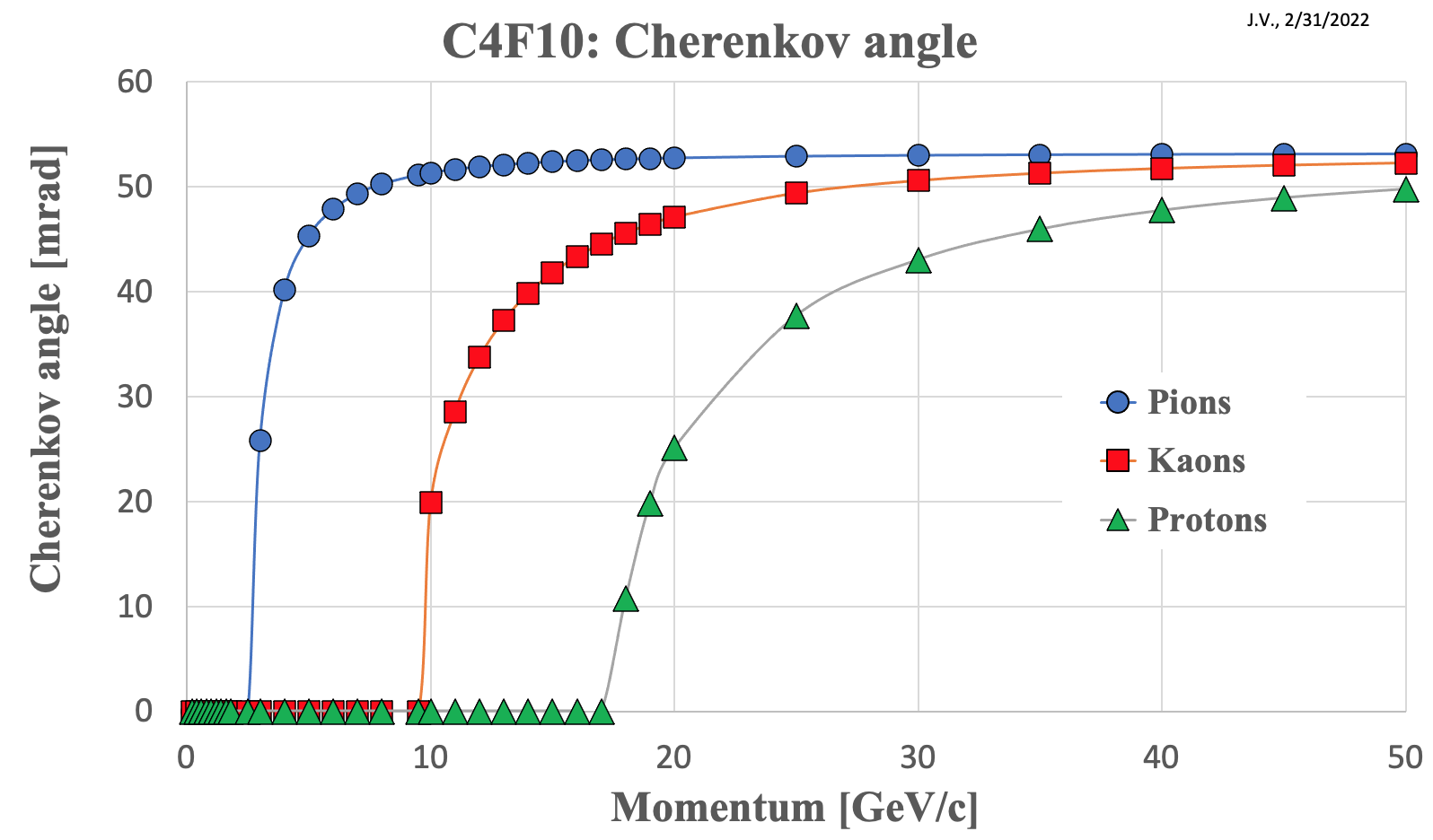}
        \caption{}
        \label{fig:cherenkov_angle_p}
    \end{subfigure} \\
    \caption{(a) Calculated number of photoelectrons per ring as a function of radiator length $L$. (b) Calculated number of photoelectrons and (c) Cherenkov angle as a function of momentum for pions, kaons, and protons. One can see that the kaon threshold is \unit[$\sim$10]{GeV} for C$_4$F$_{10}$ gas and the expected number of photoelectrons per ring is about 16 for $L = \unit[25]{cm}$ and $\beta \sim 1$.}
    \label{fig:number_of_photoelectrons}
\end{figure}

Fig.~\ref{fig:number_of_photoelectrons} shows the calculated number of photoelectrons per ring as a function of radiator length $L$ and as a function of momentum. One can see that the kaon threshold is at \unit[$\sim$10]{GeV} for C$_4$F$_{10}$ gas and that the expected number of photoelectrons per ring is about 16 for $L = \unit[25]{cm}$ and $\beta \sim 1$. For comparison, the SLD CRID's gaseous RICH had $\sim$10 photoelectrons per ring for 80\%~C$_5$F$_{12}$/20\%~N$_2$ mix and $L = \unit[45]{cm}$~\cite{Vavra:1999}.

\subsubsection{PID performance as a function of Cherenkov angle resolution}

\begin{figure}
    \centering
    \includegraphics[width=0.8\textwidth]{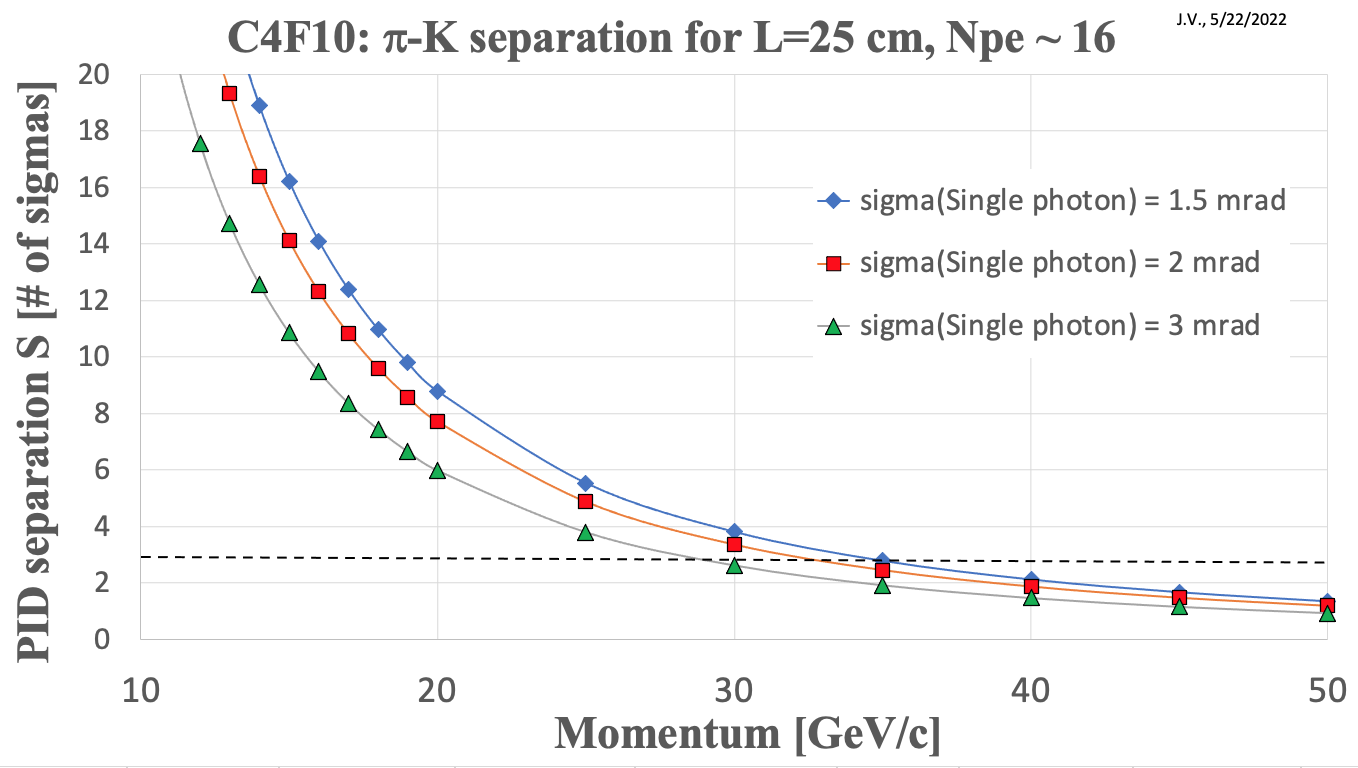}
    \caption{Expected PID performance as a function of momentum and (total) single-photon Cherenkov angle resolution. A resolution higher than \unit[4]{mrad} starts severely affecting the performance. Here, we assume 16 photoelectrons per ring and a tracking error of \unit[0.5]{mrad}, which does not scale with $N_\textrm{pe}$.}
    \label{fig:expected_sigmas}
\end{figure}

The RICH detector performance can be divided into a threshold region, where one can identify particles based on threshold, ring size, and number of photoelectrons per ring (see Figs.~\ref{fig:number_of_photoelectrons_p} and \ref{fig:cherenkov_angle_p}), and a high momentum region, where one can use the following formula to the determine particle separation $S$ (in number of sigmas):

\begin{equation}
    S = \frac{|\theta_\pi - \theta_K|}{\sigma_{\theta_c} \sqrt{N_\textrm{pe}}} \,,
\end{equation}

\noindent where $\theta_\pi$ is the Cherenkov angle for pions, $\theta_K$ is the Cherenkov angle for kaons, $\sigma_{\theta_c}$ is the single-photon Cherenkov angle resolution, and $N_\textrm{pe}$ is number of photoelectrons per ring.\footnote{We took $N_\textrm{pe} = (N_\textrm{pions}+N_\textrm{kaons})/2$.} Fig.~\ref{fig:expected_sigmas} shows the PID performance of the proposed detector for a C$_4$F$_{10}$ gas as a function of Cherenkov angle resolution, where we haved added in quadrature the tracking error of \unit[0.5]{mrad}. The conclusion is clear: going over \unit[4]{mrad} will severely impact the performance, as the requirement of $S > 3\sigma$ reduces the momentum window where PID works.

\subsection{Resolution contributions to the Cherenkov angle measurement}

In the following section, we will discuss the various contributions to the Cherenkov angle resolution. We will see that the largest contribution is a smearing error in the large magnetic field of \unit[5]{T}.

\subsubsection{Chromatic error}

\begin{figure}
    \centering
    \includegraphics[width=0.8\textwidth]{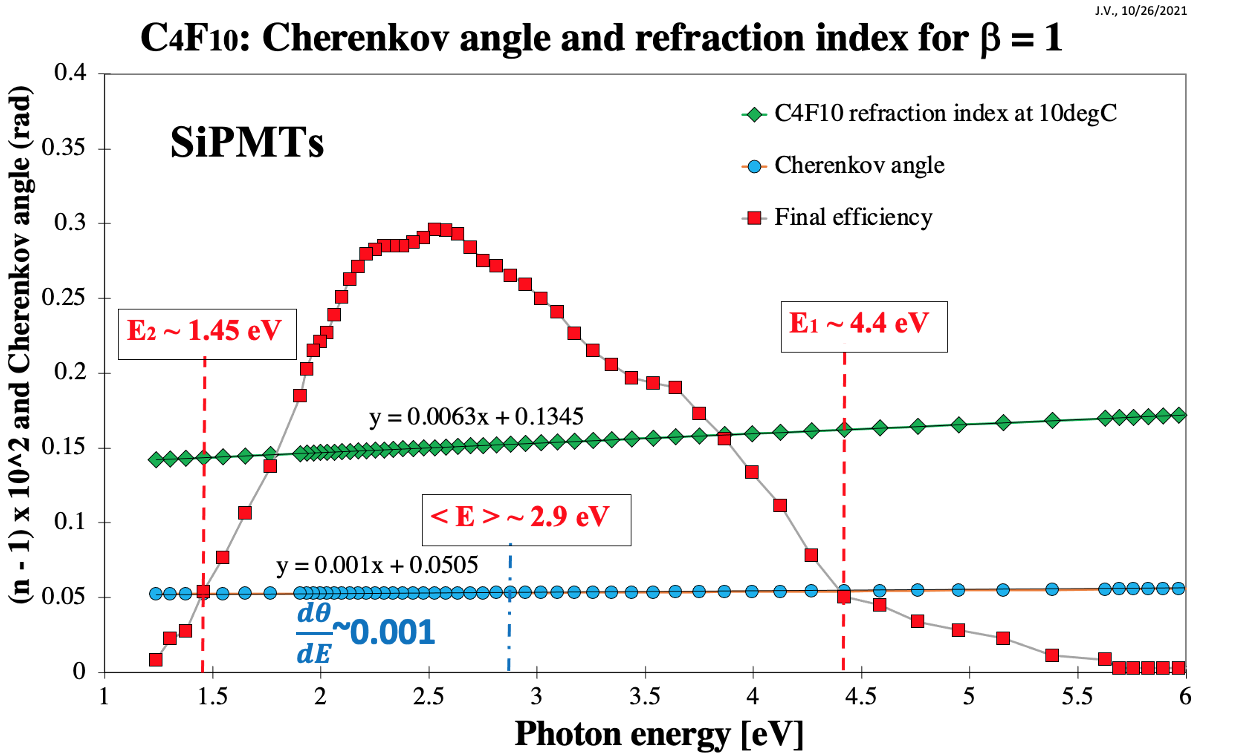}
    \caption{Final efficiency for the SiPM design: the C$_4$F$_{10}$ refraction index plotted as a function of photon energy.}
    \label{fig:final_efficiency}
\end{figure}

The chromatic effect may affect the RICH performance significantly. Although the SLD CRID, using TMAE, operated in a region where the refraction index changed more rapidly, its wavelength acceptance was very narrow and therefore the chromatic error was smaller than that of a SiPM-based detector. From Fig.~\ref{fig:efficiency_combined}, we determine the average wavelength to be \unit[$\sim$500]{nm}, which corresponds to average refraction index of $n \sim 1.001415$ -- see Fig.~\ref{fig:efficiency_refractive_index}. For SiD/ILD, we determine from Fig.~\ref{fig:final_efficiency} that the chromatic error contribution for the SiD/ILD RICH is $\sigma_{\theta_c} \sim (d\theta_c/dE)(E_1 - E_2)/\sqrt{12} \sim \unit[0.85]{mrad}$, which is twice as a large as that of the SLD CRID, which was \unit[$\sim$0.4]{mrad}, determined using the same method. This large chromatic error is due to a very broad wavelength acceptance provided by the SiPM-based design.

\subsubsection{Error due to a finite SiPM pixel size}

We assume that SiPMs will have \unit[3$\times$3]{mm$^2$} pixels. The Cherenkov error contribution due to pixel size effect is $\sigma_{\theta_c} \sim (\unit[0.3]{cm} / \sqrt{12}) / (1.5 \times \unit[25]{cm}) \sim \unit[2.3]{mrad}$. This is relatively large contribution to the final error, and one could argue that one should use smaller pixels to reduce this error. For example, \unit[2$\times$2]{mm$^2$} pixels would reduce this error to \unit[$\sim$1.5]{mrad} and \unit[1$\times$1]{mm$^2$} pixels would reduce this error to \unit[$\sim$0.8]{mrad}.

\subsubsection{Alignment errors and other systematic effects}

There are several errors which should be minimised as much as possible:

\begin{enumerate}[label=(\alph*)]
    \item mirror misalignment contribution goal: \unit[$<0.5$]{mrad};
    \item tracking direction error goal: \unit[$<0.5$]{mrad};
\end{enumerate}

\subsubsection{Cherenkov angle smearing error due to a large magnetic field}
\label{sec:magentic_smearing}

\begin{figure}
    \centering
    \includegraphics[width=0.5\textwidth]{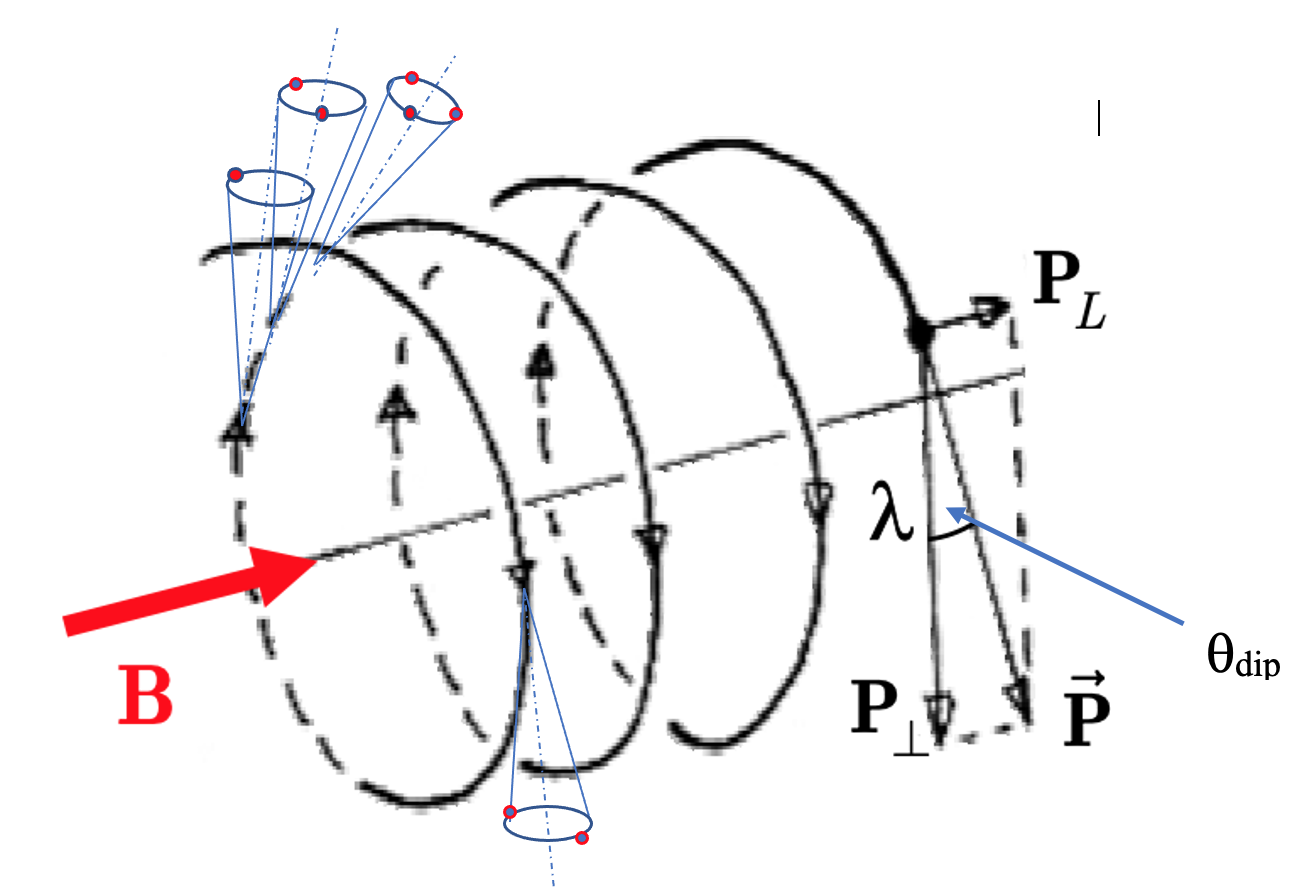}
    \caption{Schematic diagram of the helix trajectory and Cherenkov cones. Notice that the Cherenkov cones move in 3D. This contributes to smearing of the detected image at large magnetic field.}
    \label{fig:helix_trajectory}
\end{figure}

Running this type of RICH detector at \unit[5]{T} has some consequences: there is a considerable contribution to the Cherenkov angle error due to a magnetic field smearing effect. Fig.~\ref{fig:helix_trajectory} shows that the Cherenkov cone rotates in 3D as particle trajectory follows helix. This contributes to the smearing of the image. This smearing affects detected points around the Cherenkov azimuth angle $\phi_c$ differently, and is generally larger for larger magnetic fields, larger dip angles, and smaller momenta. In this section, we will try to estimate the size of this effect.

\paragraph{Estimate using analytical formula}

\begin{figure}[htbp]
    \centering
    \begin{subfigure}{\textwidth}
        \centering
        \includegraphics[width=0.8\textwidth]{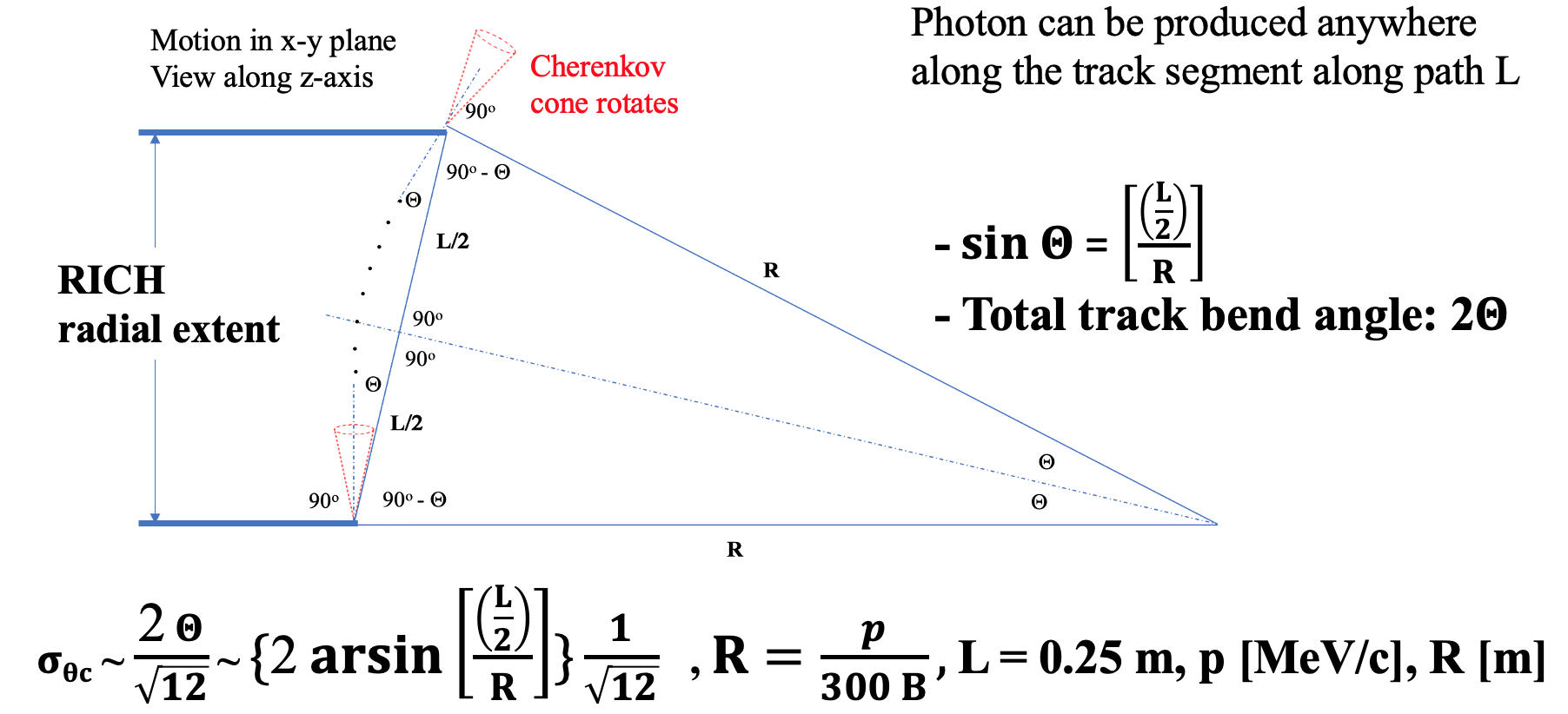}
        \caption{}
        \label{fig:smearing_ana_calc}
    \end{subfigure} \\
    \begin{subfigure}{\textwidth}
        \centering
        \includegraphics[width=1.\textwidth]{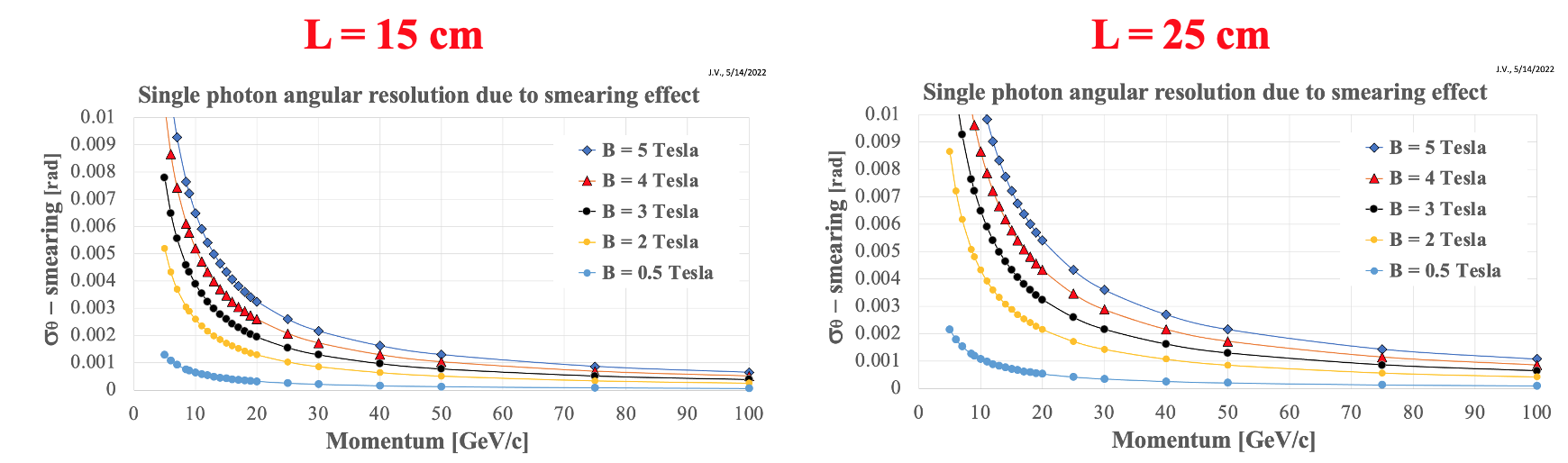}
        \caption{}
        \label{fig:smearing_ana_resolution}
    \end{subfigure} \\
    \caption{(a) Principle of the analytical calculation for the magnetic field smearing effect. (b) Angular resolution for two radiator lengths: $L = \unit[15]{cm}$ and \unit[25]{cm}.}
    \label{fig:smearing_ana}
\end{figure}

We used an analytical solution first -- this is described in Fig.~\ref{fig:smearing_ana_calc}. Fig.~\ref{fig:smearing_ana_resolution} shows the prediction for two different radiator lengths and several values of the magnetic field. The simple model predicts a larger error for a radiator length of \unit[25]{cm} and a larger magnetic field. The smearing error applies to single photons and therefore its final contribution is divided by $\sqrt{N_\textrm{pe}}$.

\paragraph{Estimate using simulation code}

\begin{figure}[htbp]
    \centering
    \begin{subfigure}[b]{0.49\textwidth}
        \centering
        \includegraphics[width=1.\textwidth]{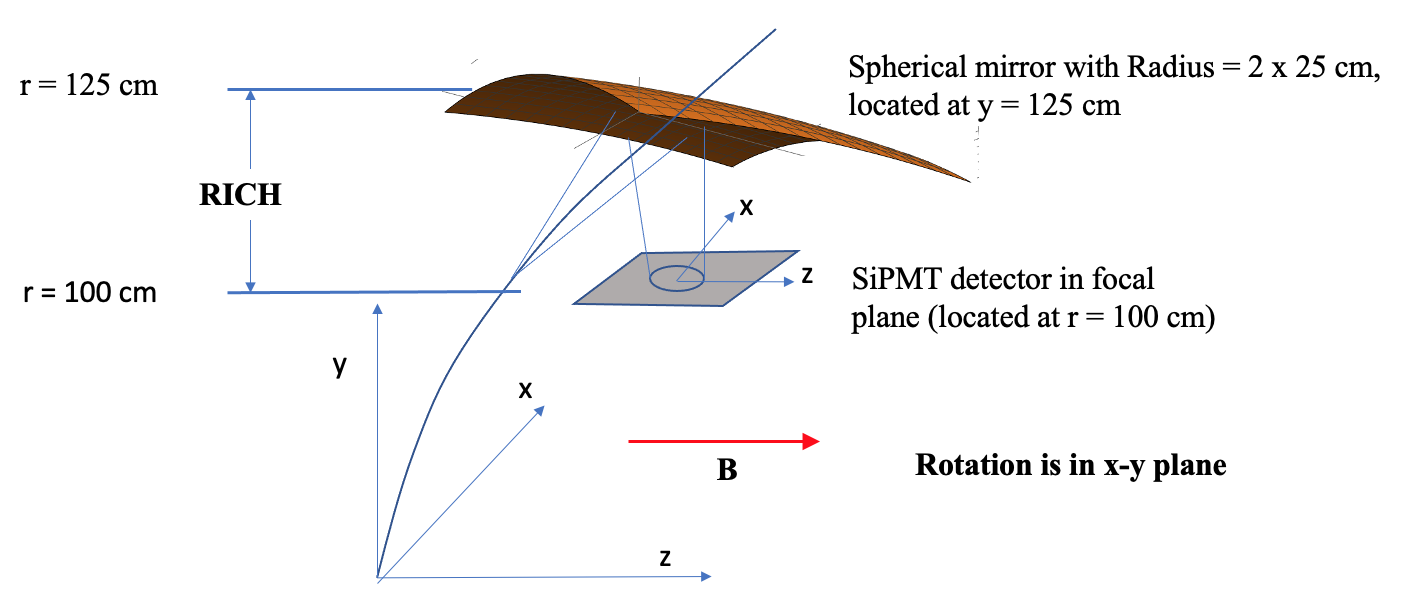}
        \caption{}
        \label{fig:simulation_schematic}
    \end{subfigure}
    \hfill
    \begin{subfigure}[b]{0.49\textwidth}
        \centering
        \includegraphics[width=1.\textwidth]{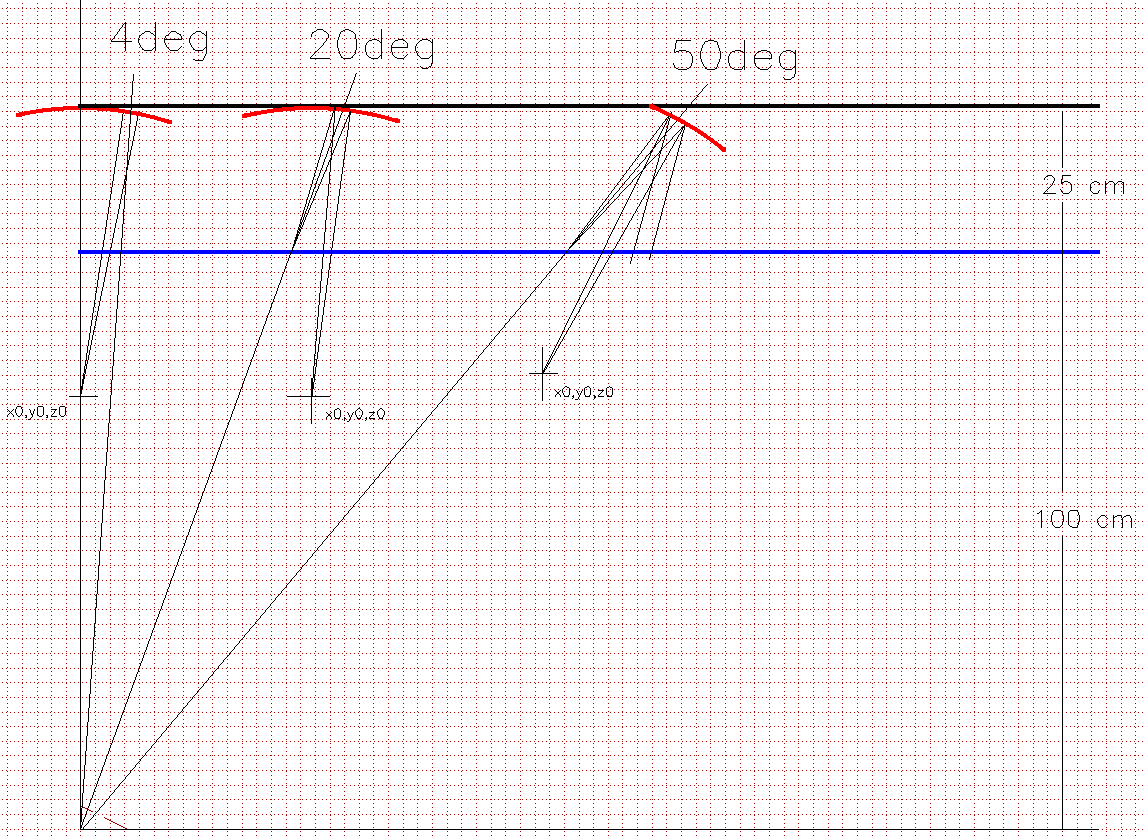}
        \caption{}
        \label{fig:simulation_ray_trace}
    \end{subfigure} \\
    \caption{(a) A schematic diagram of the helix trajectory and Cherenkov cone. Notice that cones move in 3D. A simple program was implemented: step through the magnetic field, radiate Cherenkov photons when $100 < r < \unit[125]{cm}$, reflect them from a spherical mirror, and find their intersection with a detector plane. (b) Ray tracing model for the simulation of three dip angles.}
    \label{fig:simulation}
\end{figure}

We have created a code which steps charged particles in a magnetic field following a helix. Fig.~\ref{fig:simulation} shows schematically the simulation model. Once in the radiator region ($100 < r < \unit[125]{cm}$), the particle radiates Cherenkov photons. Photons reflect from a spherical mirror and are imaged on a plane of SiPMs. We will discuss in this paper only case where SiPM detector plane is horizontal at $y = \unit[100]{cm}$. This is a simplified model, which stops working for a certain choice of parameters. For example, the analysis gets more complicated for dip angles less than 70$^\circ$ because the particles are spiraling -- the SiPM detector plane should be replaced by a segmented cylinder (this part of the analysis was not done in this paper). Nevertheless, our simple model provides useful insight. N.B. the alignment of mirror centers and the detector plane orientation must be within a fraction of a millimeter to get sharp images.

\begin{figure}[htbp]
    \centering
    \begin{subfigure}[b]{0.49\textwidth}
        \centering
        \includegraphics[width=1.\textwidth]{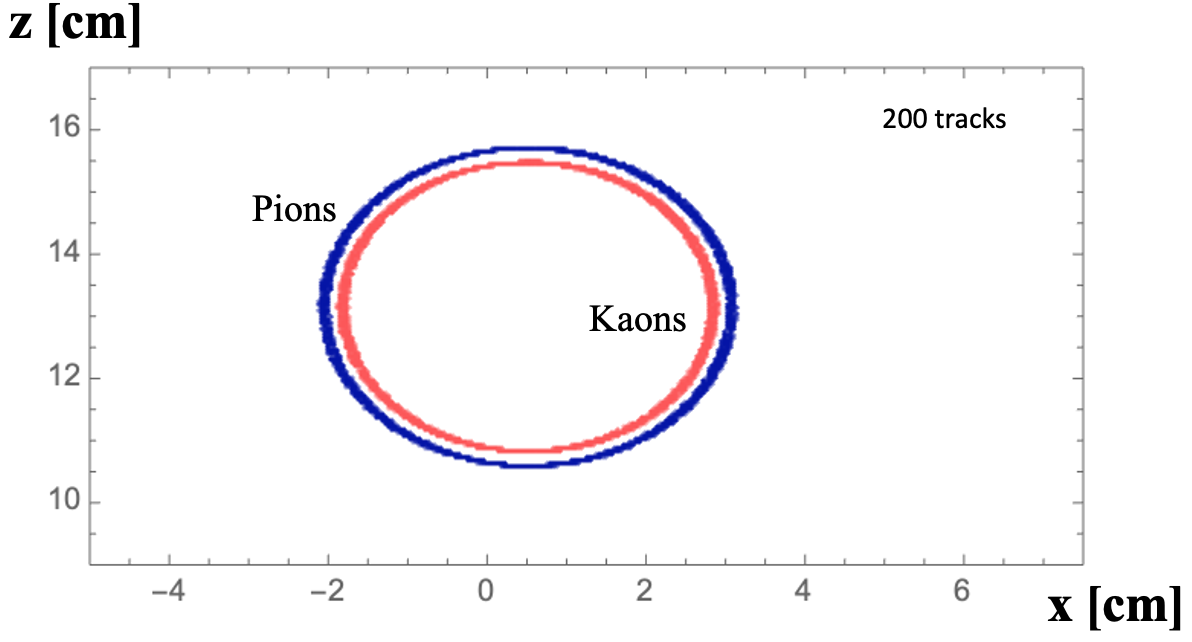}
        \caption{}
        \label{fig:SLD_CRID_rings}
    \end{subfigure}
    \hfill
    \begin{subfigure}[b]{0.49\textwidth}
        \centering
        \includegraphics[width=1.\textwidth]{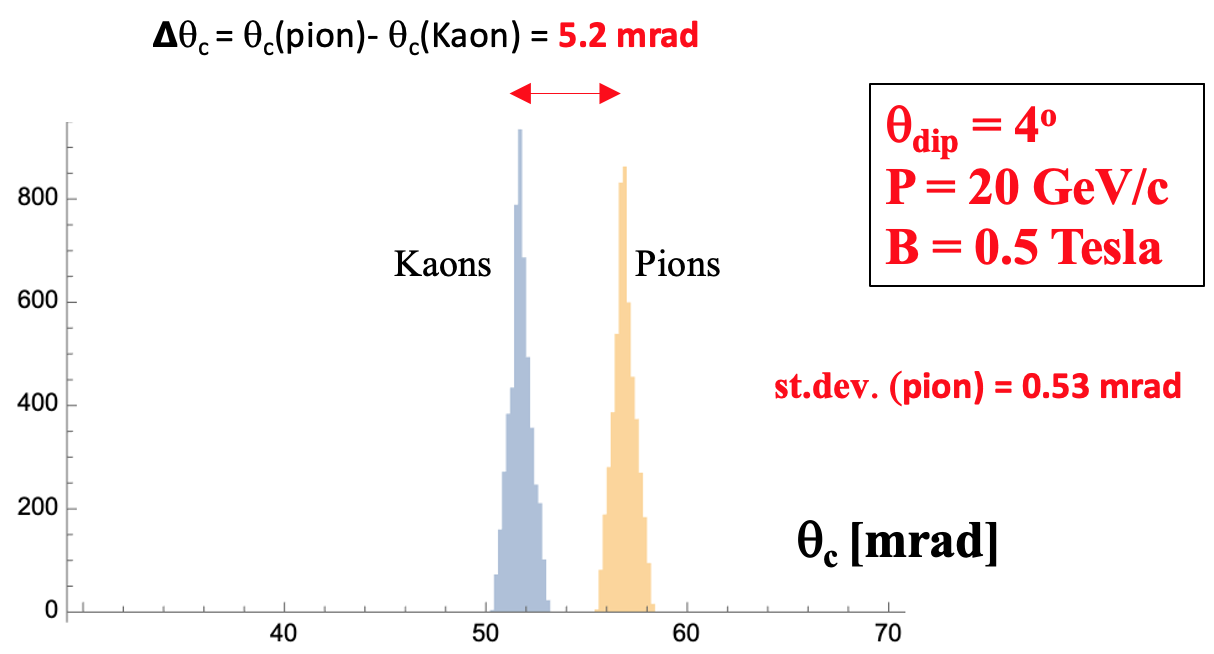}
        \caption{}
        \label{fig:SLD_CRID_angle}
    \end{subfigure} \\
    \caption{Smearing effect in the SLD CRID gaseous RICH for $\theta_\textrm{dip} = 4^\circ$, $p = \unit[20]{GeV}$, $B = \unit[0.5]{T}$, and $L = \unit[45]{cm}$. (a) Cherenkov rings imaged as 2D-hits, $\{x_\textrm{final}[i],\, z_\textrm{final}[i]\}$, in the SiPM detector plane with no cuts and no fitting, showing the smearing effect alone. (b) A plot of the Cherenkov angle $\theta_c = (\textrm{Cherenkov radius})\,/\, (\textrm{focal length})$, where the Cherenkov radius $ = \sqrt{(x_\textrm{final}[i] - x_0)^2 + (z_\textrm{final}[i] - z_0)^2}$. We tune $x_0$ and $z_0$ to obtain the smallest possible standard deviation.}
    \label{fig:SLD_CRID}
\end{figure}

We have decided to test the program on the SLD CRID gaseous RICH first. The SLD CRID operated at \unit[0.5]{T}, and so we do not expect a large smearing effect. It used an 80\%~C$_5$F$_{12}$/20\%~N$_2$ gas mixture with a \unit[45]{cm} long gaseous radiator length. Fig.~\ref{fig:SLD_CRID_rings} shows a clear separation of $\pi$/$K$ rings and Cherenkov angle distributions at \unit[20]{GeV}. From Fig.~\ref{fig:SLD_CRID_angle}, we estimate the smearing error contribution to be \unit[$\sim$0.75]{mrad}.

To obtain the estimate of the Cherenkov angle resolution in Fig.~\ref{fig:SLD_CRID_angle}, we used a simple method of calculating $R[i] = \sqrt{(x_\textrm{final}[i] - x_0)^2 + (z_\textrm{final}[i] - z_0)^2}$ from all hits in detector plane. For each 2D entry, we then calculated $\theta_c[i] = R[i]\,/\,(\textrm{focal length})$ and plotted histograms without any cuts (focal length = \unit[45]{cm}, in this case). This algorithm assumes that the ring is circular, and the procedure requires tuning of the circle center $x_0$ and $z_0$. To get the correct distributions, the center of the ring has to be known to fraction of a millimeter. Similarly, the alignment of all optical elements is critical in this type of RICH.

\begin{figure}[htbp]
    \centering
    \begin{subfigure}{\textwidth}
        \centering
        \includegraphics[width=0.49\textwidth,valign=t]{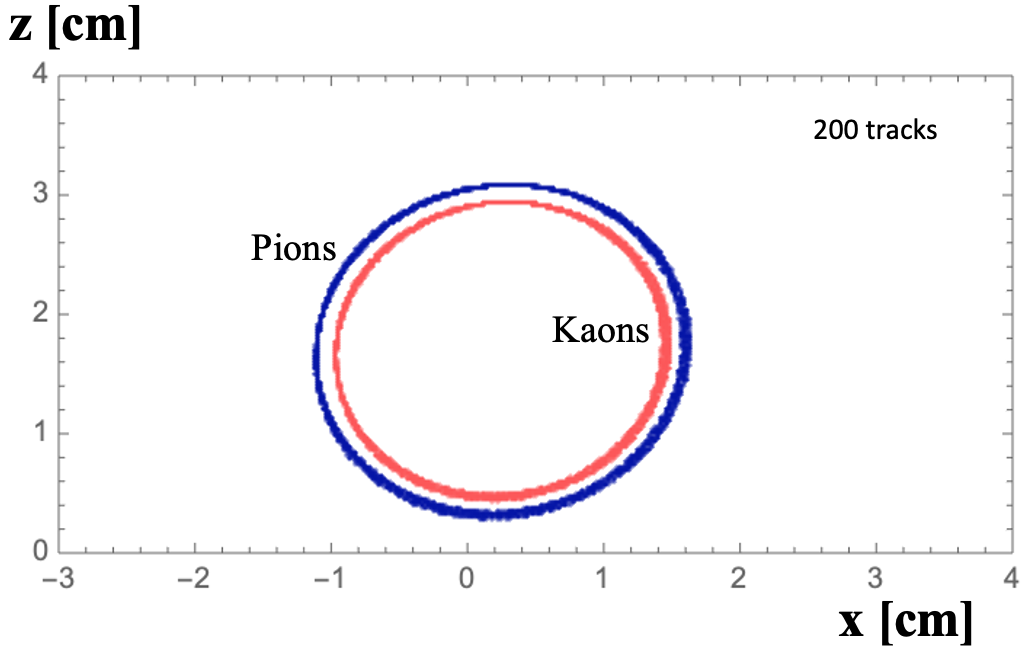}
        \hfill
        \includegraphics[width=0.49\textwidth,valign=t]{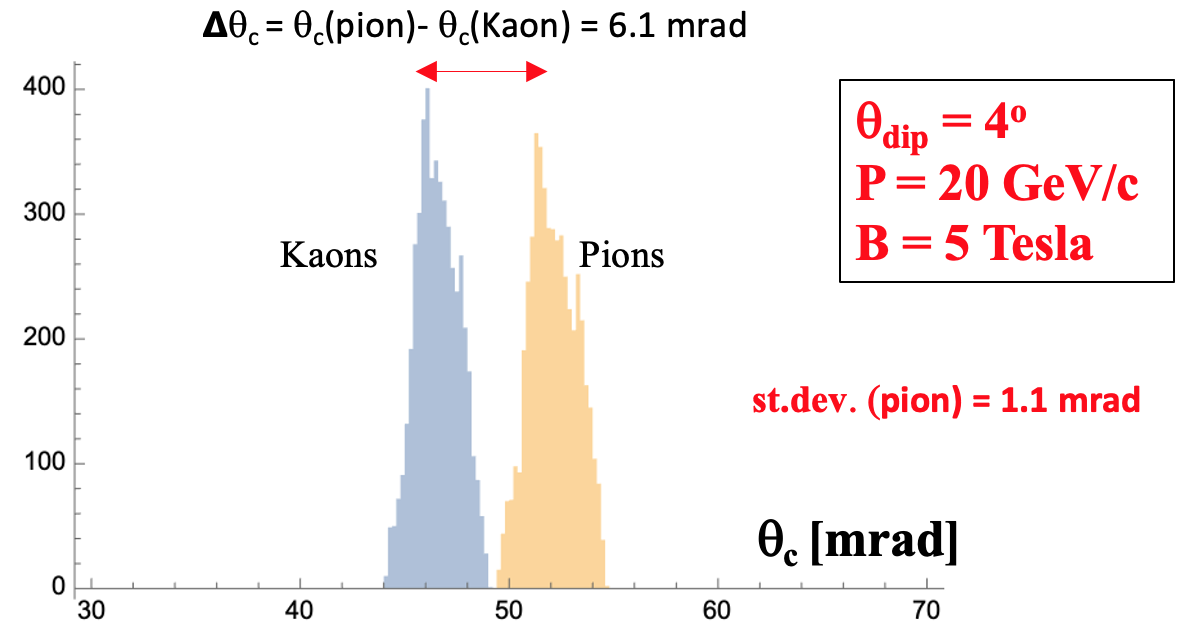}
        \caption{$\theta_\textrm{dip} = 4^\circ$}
        \label{fig:SiD_CRID_dip_4}
    \end{subfigure} \\
    \vspace{0.5em}
    \begin{subfigure}{\textwidth}
        \centering
        \includegraphics[width=0.49\textwidth,valign=t]{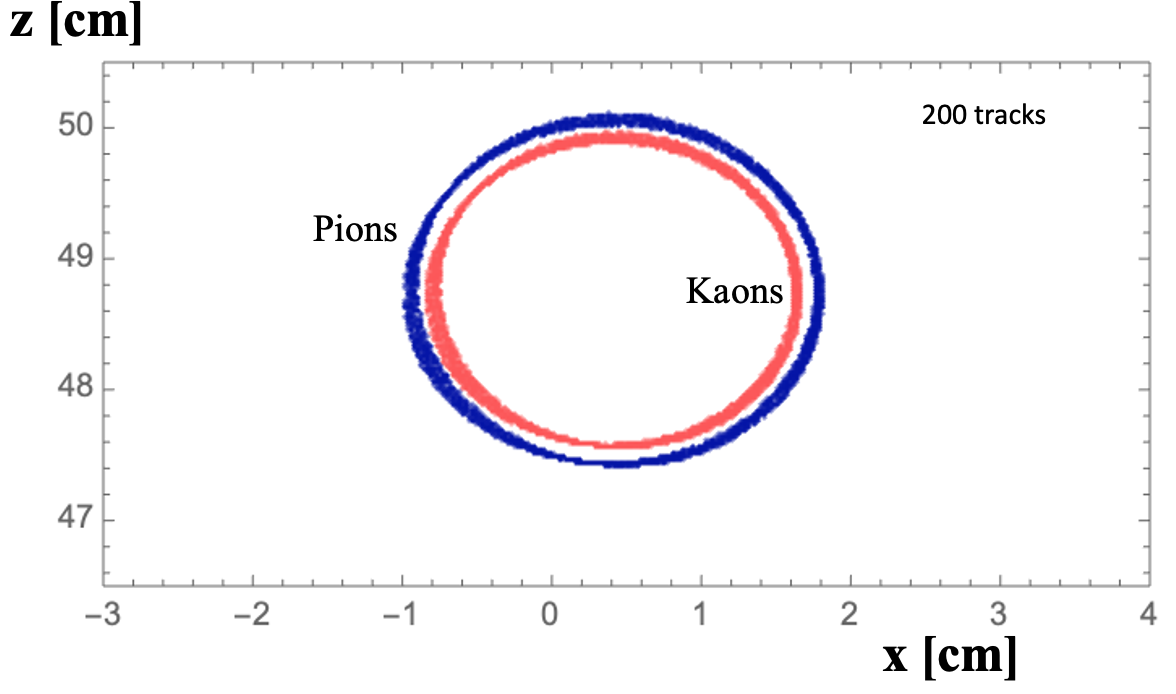}
        \hfill
        \includegraphics[width=0.49\textwidth,valign=t]{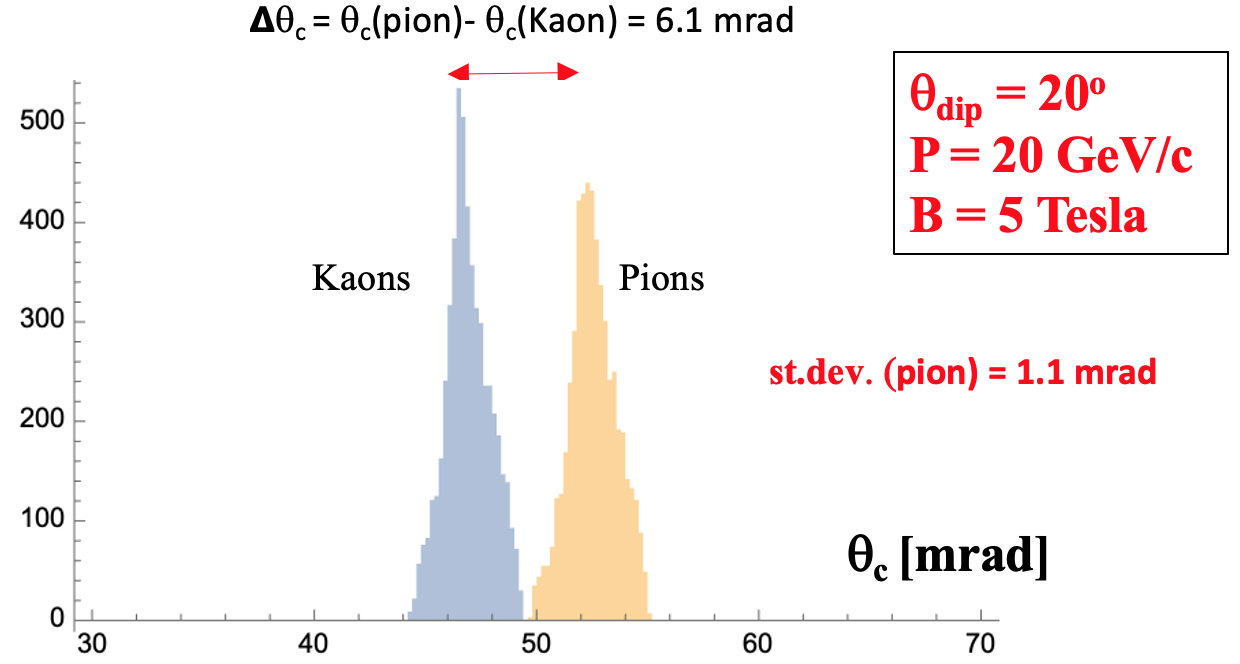}
        \caption{$\theta_\textrm{dip} = 20^\circ$}
        \label{fig:SiD_CRID_dip_20}
    \end{subfigure} \\
    \vspace{0.5em}
    \begin{subfigure}{\textwidth}
        \centering
        \includegraphics[width=0.49\textwidth,valign=t]{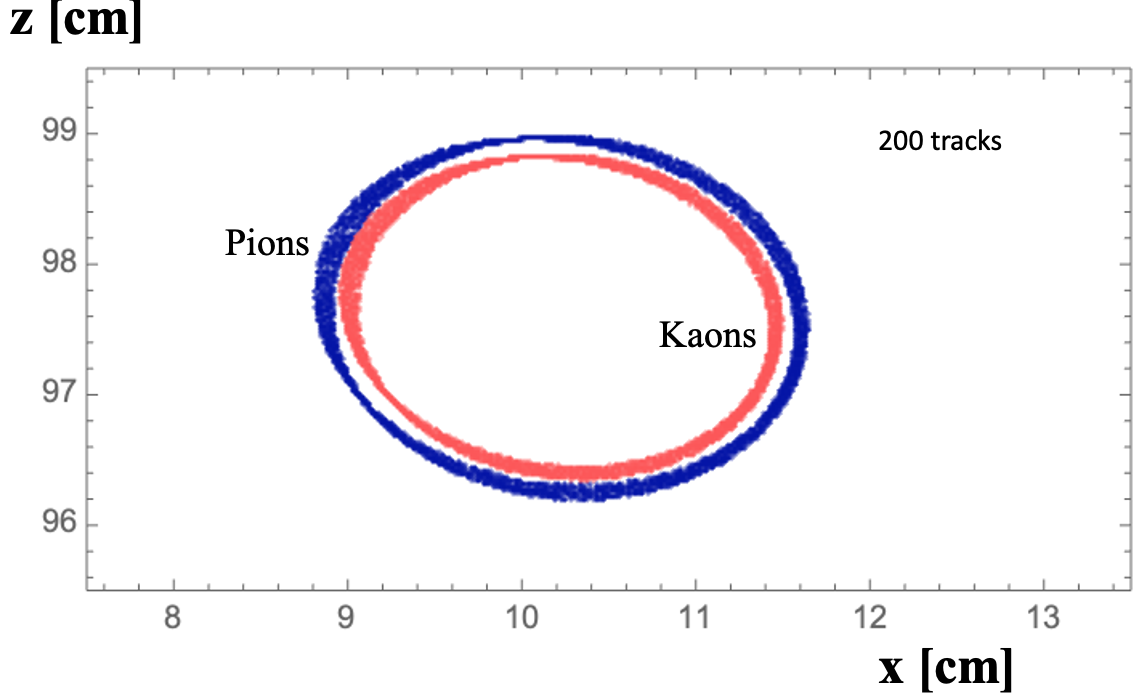}
        \hfill
        \includegraphics[width=0.49\textwidth,valign=t]{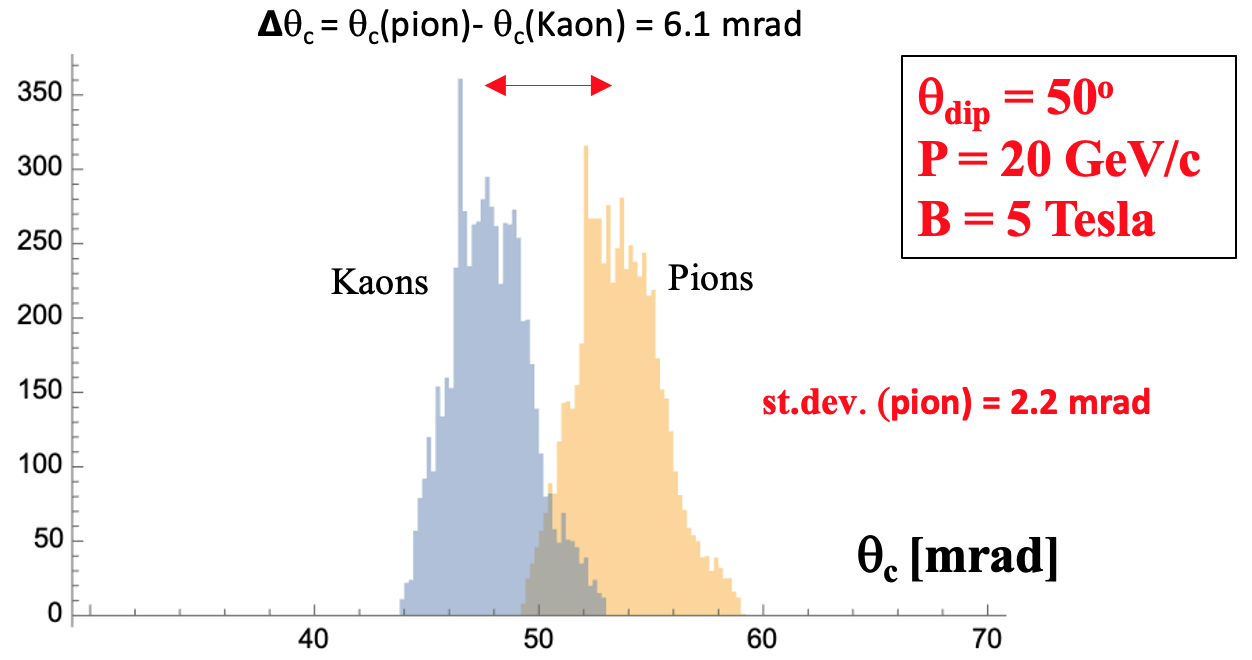}
        \caption{$\theta_\textrm{dip} = 50^\circ$}
        \label{fig:SiD_CRID_dip_50}
    \end{subfigure} \\
    \caption{Smearing effect in the SiD/ILD RICH for $p = \unit[20]{GeV}$, $B = \unit[5]{T}$, and $L = \unit[25]{cm}$. Cherenkov rings are imaged in the detector plane and plots of all 2D-hits, $\{x_\textrm{final}[i],\, z_\textrm{final}[i]\}$, with no cuts and no fitting are shown for (a) $\theta_\textrm{dip} = 4^\circ$, (b) $\theta_\textrm{dip} = 20^\circ$, and (c) $\theta_\textrm{dip} = 50^\circ$.}
    \label{fig:SiD_CRID_dip}
\end{figure}

Now, we turn to a RICH design for SiD/ILD where the focal length of the spherical mirror is \unit[25]{cm}, the magnetic field is \unit[5]{T}, $L = \unit[25]{cm}$, and a C$_4$F$_{10}$ gas at normal pressure is used. Fig.~\ref{fig:SiD_CRID_dip} shows Cherenkov rings and resolutions for three dip angles: 4$^\circ$, 20$^\circ$, and 50$^\circ$.

\begin{figure}[htbp]
    \centering
    \begin{subfigure}{\textwidth}
        \centering
        \includegraphics[width=0.95\textwidth]{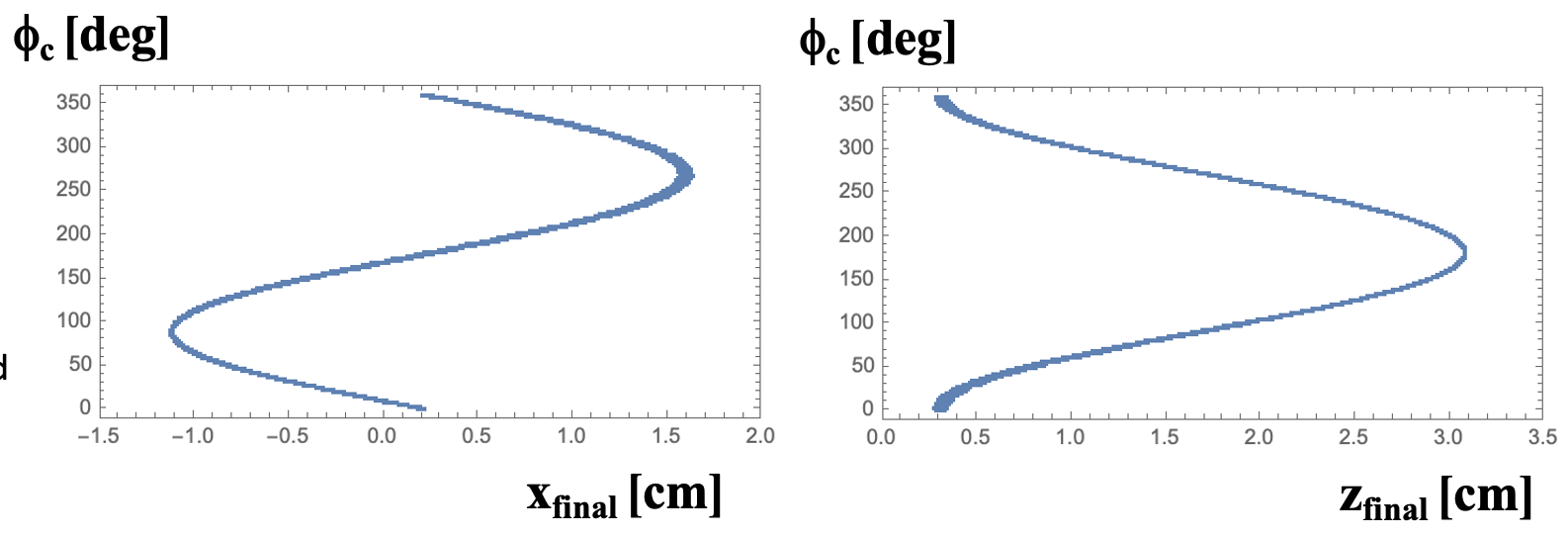}
        \caption{$\theta_\textrm{dip} = 4^\circ$}
        \label{fig:SiD_CRID_smearing_dip_4}
    \end{subfigure} \\
    \vspace{0.5em}
    \begin{subfigure}{\textwidth}
        \centering
        \includegraphics[width=0.95\textwidth]{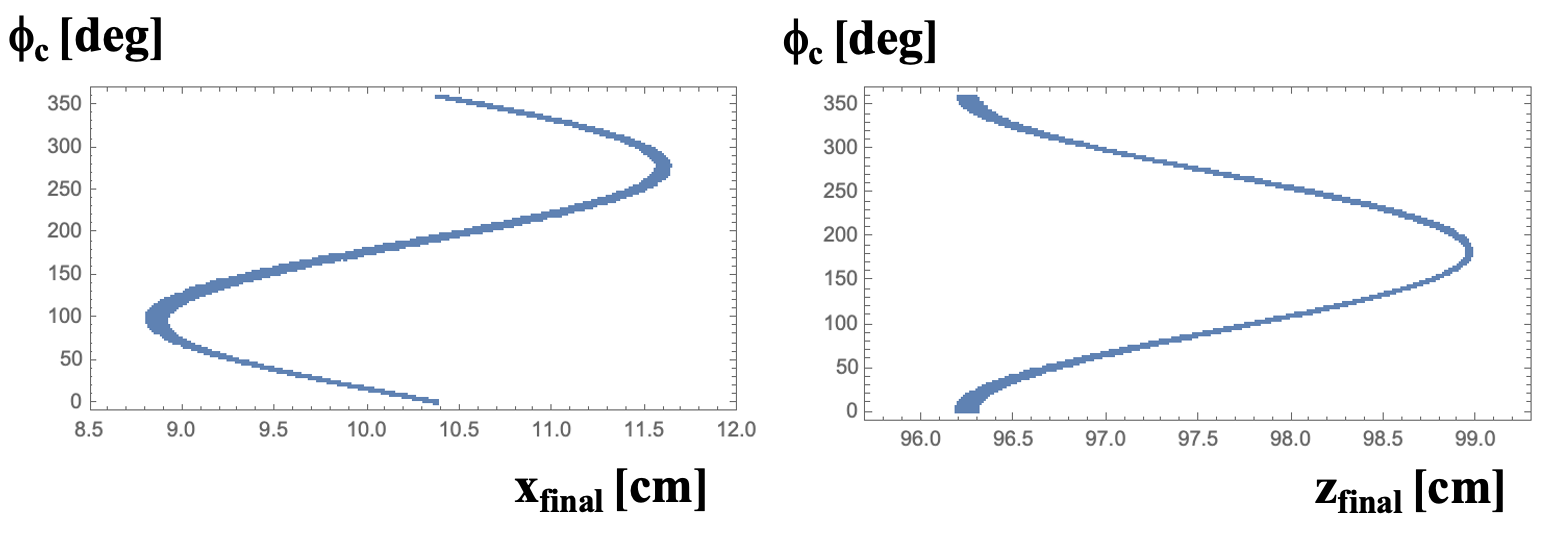}
        \caption{$\theta_\textrm{dip} = 50^\circ$}
        \label{fig:SiD_CRID_smearing_dip_50}
    \end{subfigure} \\
    \caption{Smearing effect in the SiD/ILD RICH for $p = \unit[20]{GeV}$, $B = \unit[5]{T}$, and $L = \unit[25]{cm}$. The effects manifests itself as a variation in the Cherenkov angle resolution in $x$ and $z$ final positions as a function of Cherenkov azimuthal angle $\phi_c$ for (a) $\theta_\textrm{dip} = 4^\circ$ and (b) $\theta_\textrm{dip} = 50^\circ$.}
    \label{fig:SiD_CRID_smearing_dip}
\end{figure}

The smearing effect due to the large magnetic field has two consequences: (a) there is a hint that ring images might have slightly elliptical shapes, especially in the very forward direction, and (b) there is a clear variation of errors in both $x$ and $z$ as a function of the Cherenkov angle azimuth $\phi_c$ and the dip angle $\theta_\textrm{dip}$ -- see Fig.~\ref{fig:SiD_CRID_smearing_dip}. Both effects were not removed in the Cherenkov angle resolution algorithm described in this paper. However, the final analysis can take care of these two effects by a proper weighted fitting, which may include weights as a function of $\phi_c$ and $\theta_\textrm{dip}$ and by possibly fitting a rotated ellipse rather than a circle. We clearly observe that the Cherenkov angle resolution contribution from the smearing effect at \unit[5]{T} is larger than that of the SLD CRID. However, it is not as large of an effect as initially feared, especially if more sophisticated analysis will be performed in future, and this gives hope that this type of RICH is doable.

\begin{figure}[H]
    \centering
    \begin{subfigure}{\textwidth}
        \centering
        \includegraphics[width=0.49\textwidth,valign=t]{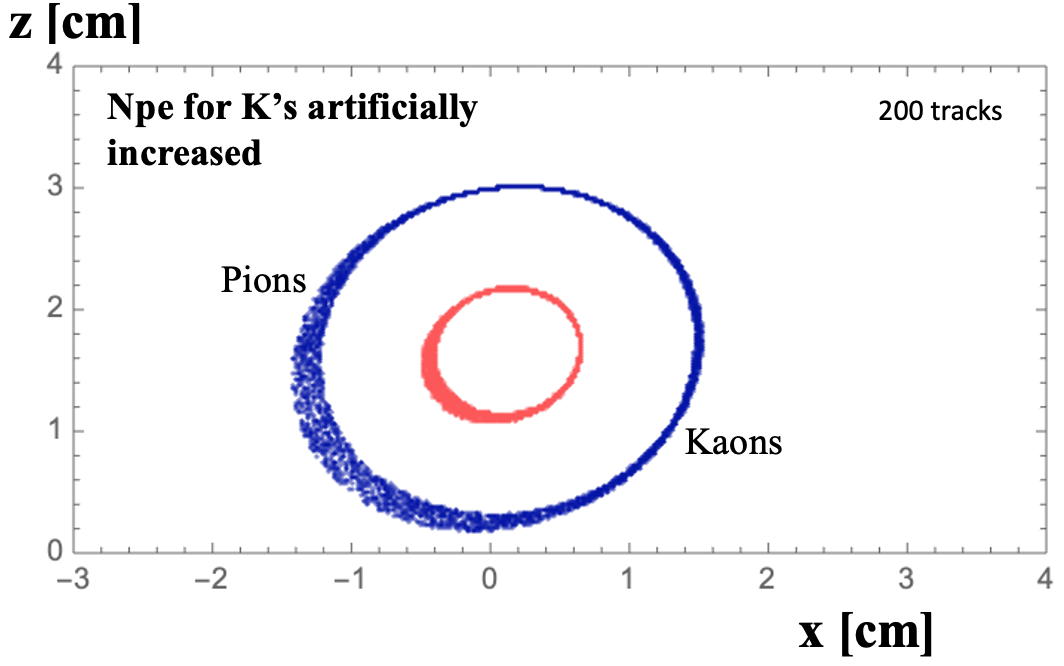}
        \hfill
        \includegraphics[width=0.49\textwidth,valign=t]{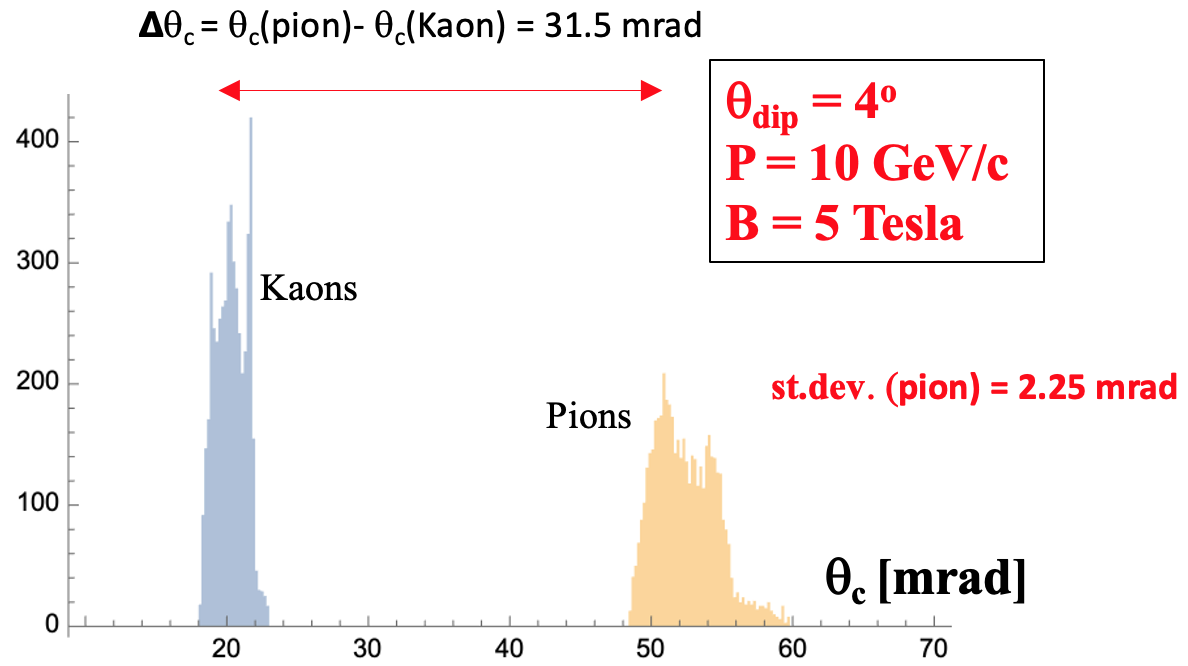}
        \caption{$p = \unit[10]{GeV}$}
        \label{fig:SiD_CRID_p_10}
    \end{subfigure} \\
    \vspace{0.5em}
    \begin{subfigure}{\textwidth}
        \centering
        \includegraphics[width=0.49\textwidth,valign=t]{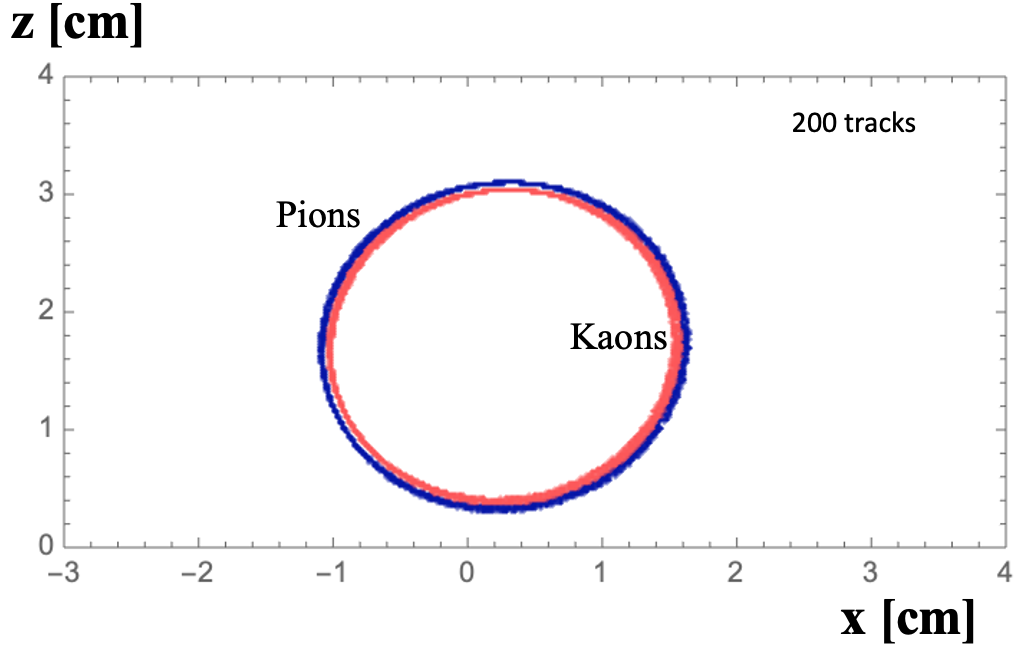}
        \hfill
        \includegraphics[width=0.49\textwidth,valign=t]{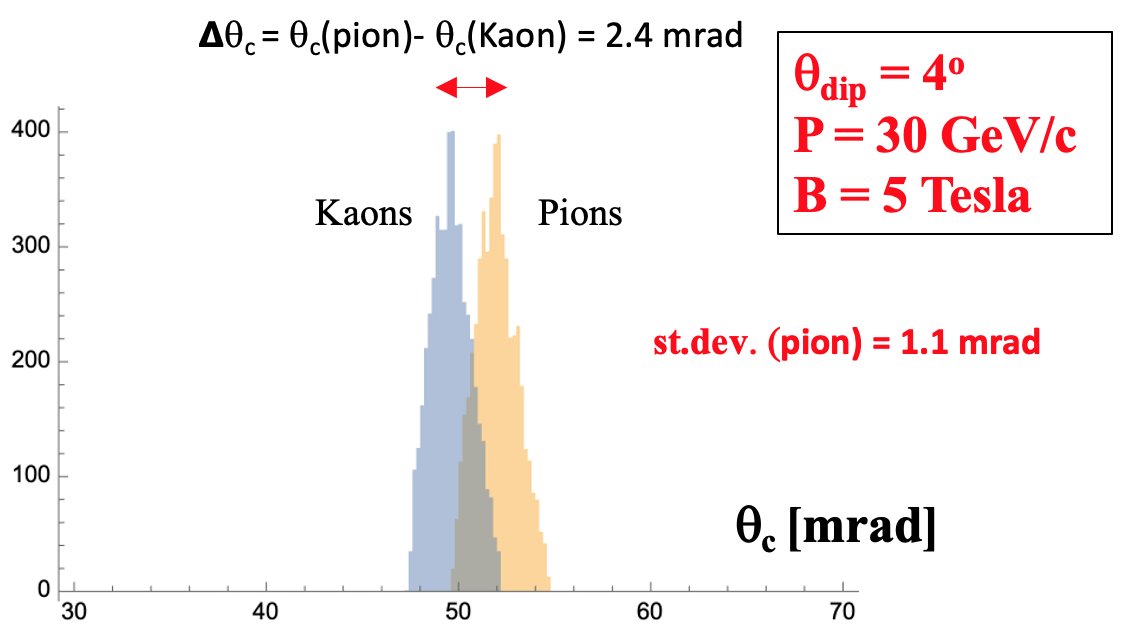}
        \caption{$p = \unit[30]{GeV}$}
        \label{fig:SiD_CRID_p_30}
    \end{subfigure} \\
    \caption{Smearing effect in the SiD/ILD RICH for $\theta_\textrm{dip} = 4^\circ$, $B = \unit[5]{T}$, and $L = \unit[25]{cm}$. Cherenkov rings are imaged in the detector plane and plots of all 2D-hits, $\{x_\textrm{final}[i],\, z_\textrm{final}[i]\}$, with no cuts and no fitting are shown for (a) $p = \unit[10]{GeV}$ and (b) $p = \unit[30]{GeV}$.}
    \label{fig:SiD_CRID_p}
\end{figure}

Fig.~\ref{fig:SiD_CRID_p} shows two other extreme conditions for PID: Fig.~\ref{fig:SiD_CRID_p_10} for \unit[10]{GeV}, near the kaon threshold, and Fig.~\ref{fig:SiD_CRID_p_30} for \unit[30]{GeV}. Although the $\pi$/$K$ separation is very clear at \unit[10]{GeV}, the number of kaon photoelectrons per ring is only 2--3, so the SiPM noise could be an issue in this region -- see Appendix~\ref{app:SiPM_noise}. Here is where the timing of SiPM hits relative to the crossing signal is critical to eliminate the random noise. Fig.~\ref{fig:SiD_CRID_p_30} shows that the smearing effect is a significant contribution to PID at \unit[30]{GeV}. It is clear that in this region one will have to work very hard on all contributions to the resolution.

\subsubsection{Summary of resolution study}

Table~\ref{tab:resolution} shows a summary of various error contributions to the Cherenkov angle resolution. The SiD/ILD RICH design is compared with the SLD CRID gaseous RICH design. The SLD CRID had a local resolution of \unit[$\sim$3.8]{mrad}, determined by fitting rings alone; however, the final overall single-photon resolution was quoted at a level of \unit[$\sim$4.3]{mrad} due to additional overall systematic errors. These systematic errors include effects such as: (a) angular track resolution, (b) electron path and drift velocity in the TPC, (c) TPC position and orientation, (d) mirror position, orientation and radius, (e) refraction index variation, (f) radiator gas stability (i.e., mix and pressure), and (g) electronics gain. These effects made the CRID analysis difficult but successful~\cite{Muller, SLD:1998} -- see Appendix~\ref{app:SLD_CRID}.

The SiD/ILD RICH has a larger chromatic error and much larger smearing effect due to the magnetic field of \unit[5]{T}. Not much can be done about the chromatic effect except possibly filters at the expense of the number of photoelectrons. The smearing error at \unit[5]{T} can be reduced in the future by clever fitting strategies. The pixel-based error depends on the choice of the pixel size, and this really depends on the future technology developments. Another critical contribution is the tracking angular resolution, which needs to be below \unit[1]{mrad} if one wants to achieve PID at \unit[30]{GeV}. For comparison, the SLD drift chamber provided the CRID with a tracking angular resolution of \unit[$\sim$0.8]{mrad}~\cite{Markiewicz, Hildreth:1995}. Many of the other systematic effects will not exist in the SiD/ILD design thanks to its solid-state photodetector choice. However, some resolution effects will remain similar, such items (a), (d), (f) and (e) in the above list.

\begin{figure}[htbp]
    \centering
    \begin{subfigure}{\textwidth}
        \centering
        \includegraphics[width=\textwidth]{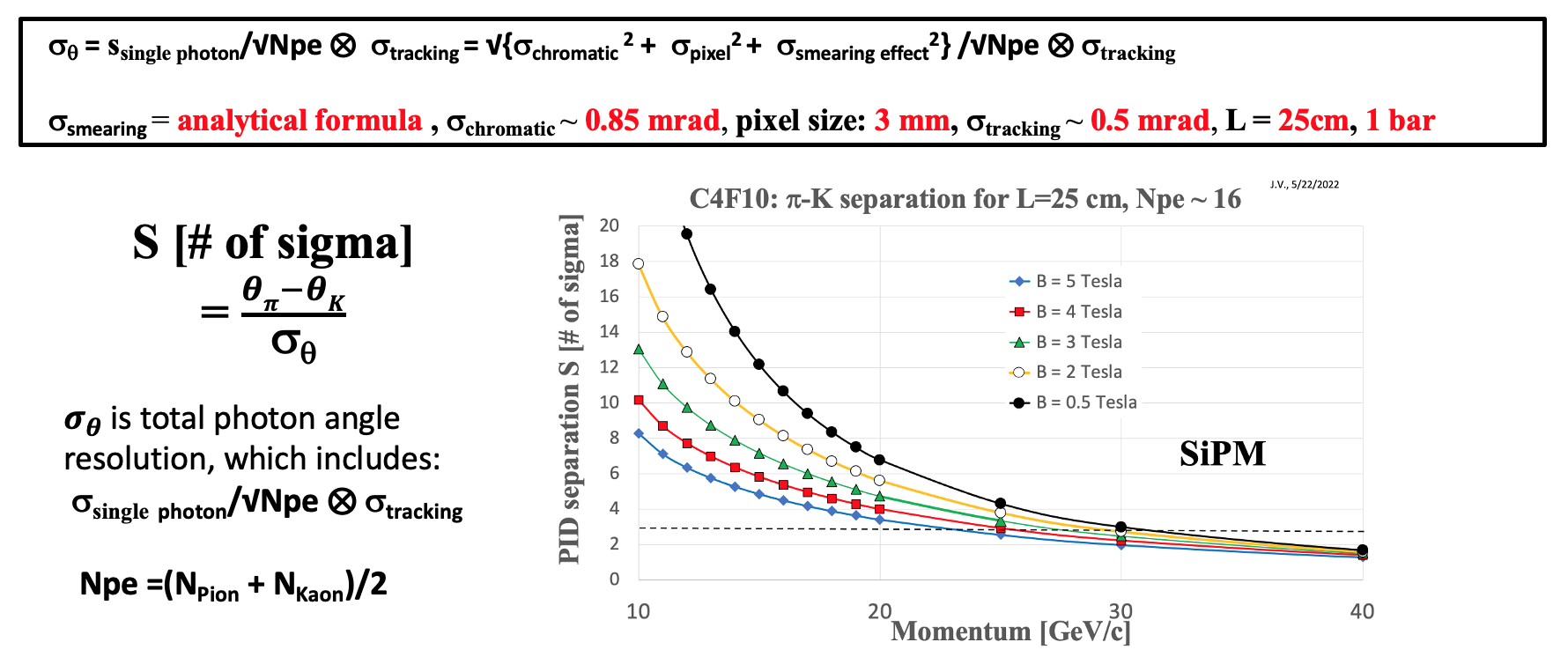}
        \caption{Nominal design}
        \label{fig:design_perf_nom}
    \end{subfigure} \\
    \vspace{0.5em}
    \begin{subfigure}{\textwidth}
        \centering
        \includegraphics[width=\textwidth]{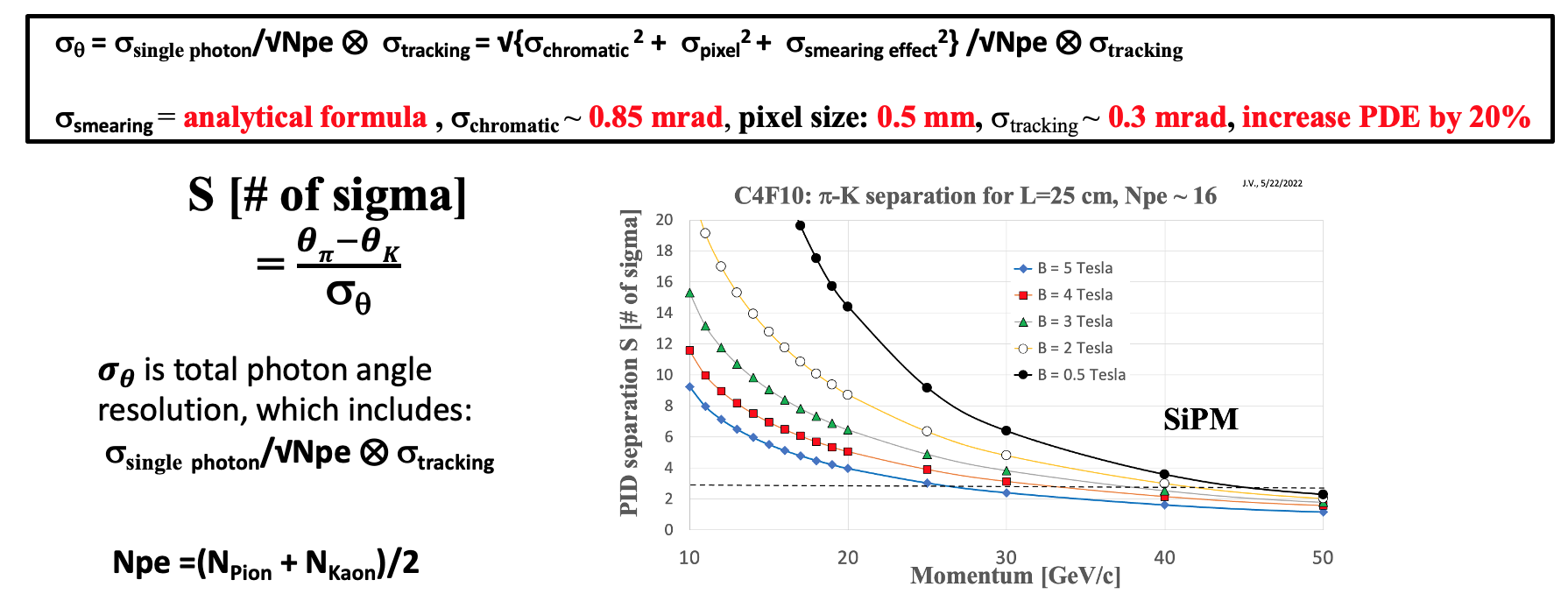}
        \caption{Design with improved performance}
        \label{fig:design_perf_improved}
    \end{subfigure} \\
    \caption{(a) Expected PID performance as a function of momentum for the nominal design: the 3-sigma limit is \unit[23--31]{GeV}, depending on the magnetic field. (b) Performance improvement from a smaller pixel size, a smaller tracking angular error, and a 20\% improvement in the SiPM PDE: the 3-sigma limit is \unit[30--45]{GeV}, depending on the magnetic field. This, we believe, is a limit of our design at \unit[1]{bar}.}
    \label{fig:design_perf}
\end{figure}

Table~\ref{tab:resolution} and Fig.~\ref{fig:design_perf} show the predicted PID performance for two designs. The only way to improve this performance is to increase the gas pressure and to reduce the radial length, as shown in Ref.~\cite{FortyConf}. However, the price for this improvement is significant: one needs to deal with a pressure vessel holding \unit[3.5]{bar} and the increase in detector mass ($X/X_0 \sim 10\%$). We believe that our design can be built with $X/X_0 \sim \textrm{3--4\%}$.

\begin{table}[htbp]
    \centering
    \caption{Various contributions to the Cherenkov angle resolution.}
    \label{tab:resolution}
    \begin{tabular}{l|c|c}
        \toprule
        Single-photon error source & SiD/ILD RICH detector & SLD CRID detector \\
        & @ \unit[5]{T} [mrad] & @ \unit[0.5]{T} [mrad] \\
        \midrule
        Chromatic error & $\sim$0.85 & $\sim$0.4 \\
        Pixel size error (0.5$\times$0.5 -- \unit[3$\times$3]{mm$^2$}) & 0.4--2.3 & $\sim$0.5 \\
        Smearing effect due to magnetic field & 1.5--2.5 & $\sim$0.5 \\
        Mirror alignment & $\ll1$ & $\sim$1 (?) \\
        Other systematic errors & $\ll1$ & a few mrad \\
        \midrule
        Total single-photon error & 1.8--3.5 & $\sim$3.4 \\
        Total error including systematic effects & -- & $\sim$4.3 \\
        \midrule
        Tracking angular error & $\sim$0.5 & $\sim$0.8~\cite{Markiewicz, Hildreth:1995} \\
        \bottomrule
    \end{tabular}
\end{table}

\subsection{Summary}

This simple study indicates that there is a hope for PID using this type of RICH design at the SiD or ILD detectors operating at \unit[5]{T}. The final performance, shown in Table~\ref{tab:resolution} and Fig.~\ref{fig:design_perf_nom}, critically depends on the Cherenkov angle resolution contributions. Results from this work justify a full \GeantFour simulation.

Although we have discussed SiPM as the only photodetector option in this paper, one can safely assume that the photon detector technology will improve significantly by the time at which the SiD or ILD detectors are seriously considered. For example, if the PDE will improve significantly, one could consider reducing the radial thickness of the RICH, which would in turn reduce the smearing effect

\section{Conclusion}
\label{sec:conclusion}

This paper presented a novel algorithm for strange tagging developed using the simulated data of the ILD at the ILC. It also described the first application of such a strange tagger to a \Hss analysis with the $P(e^-,e^+) = (-80\%,+30\%)$ polarisation scenario corresponding to \unit[900]{\ifb} of the initial proposed \unit[2000]{\ifb} of data which will be collected by ILD during its first 10~years of data taking at \sqrts = \unit[250]{GeV}. Upper limits on the Standard Model Higgs-strange coupling strength modifier, $\kappa_s$, were derived at the 95\% confidence level to be 7.14 and the implications on models predicting an extended Higgs sector were investigated. In the context of SFV 2HDMs, the limits on the strange Yukawa coupling presented in this paper are the strongest throughout much of the parameter space considered, exceeding those expected from measurements performed at the HL-LHC, competing directly with searches for the new states, and confirming the potential of future $e^+e^-$ colliders in probing extended Higgs sectors. It must be noted that the presented results only focus on a small fraction of the foreseen ILC dataset and will be updated in the near future to include larger statistics and polarisation scenarios.

Particle identification at high momenta has been proven to boost strange tagging capabilities as well as the analysis sensitivity in constraining the available phase space for new physics. A preliminary study of a RICH system was also carried out. The results show that in a compact RICH with a radial extension of \unit[25]{cm}, the Cherenkov angle resolution can be maintained at the level of \unit[$\sim$5]{mrad} in magnetic fields up to \unit[5]{T}. This leads to a discrimination power of $3\sigma$ between kaons and pions up to momenta of approximately \unit[25]{GeV}.

This work is largely independent from the specific accelerator and experiment considered. The conclusion strongly motivates further explorations of dedicated analysis techniques and detector technologies enhancing strange tagging performance and, in turn, allowing to better constrain strange/light Yukawa couplings and new physics models at any future Higgs factory.

Additional improvements in the analysis sensitivity could arise from the usage of more complex neural networks for flavour tagging and machine learning approaches for the event selection. We also plan to reinterpret the analysis and perform a search for charged Higgs bosons decaying into a charm- and a strange-initiated jet. It will then be of paramount importance to perform a full simulation study and understand the impact that the introduction of a RICH system would have more broadly on object reconstruction, such as particle flow jets, and on other physics benchmarks, when used in conjunction with silicon or gaseous tracking detectors.

\section*{Acknowledgements}
\addcontentsline{toc}{section}{Acknowledgements}

The authors would like to thank the broader ILD community for their input and assistance in the study and appreciated very much the thorough and thoughtful review of the paper by Alberto~Ruiz (U.~Cantabria), Daniel~Jeans (KEK), and Kiyotomo~Kawagoe (Kyushu~U.). The authors are also very grateful to Jenny~List (DESY) and Markus~Elsing (CERN) for their support and very helpful insights throughout the development of these results. This paper benefits from hand-made Feynman diagrams by Federica~Cairo, whom the authors thank deeply.

\FloatBarrier

\Urlmuskip=0mu plus 1mu\relax
\bibliographystyle{spphys} 
\bibliography{bibliography}

\newpage
\appendix

\numberwithin{equation}{section}
\setcounter{equation}{0}
\renewcommand{\theequation}{\thesection\arabic{equation}}

\numberwithin{figure}{section}
\setcounter{figure}{0}
\renewcommand{\thefigure}{\thesection\arabic{figure}}

\FloatBarrier

\section{Additional jet flavour tagger training plots}
\label{app:training_plots}

\begin{figure}[htbp]
    \centering
    \begin{subfigure}{0.49\textwidth}
        \centering
        \includegraphics[width=1.\textwidth]{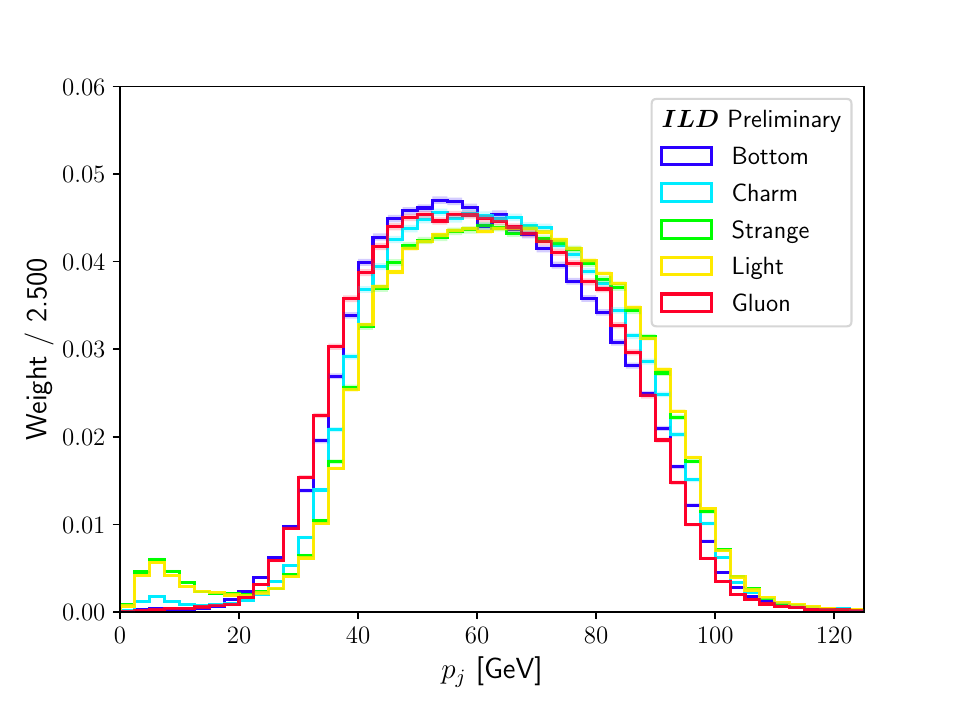}
        \caption{Momentum $p_j$}
    \end{subfigure}
    \hfill
    \begin{subfigure}{0.49\textwidth}
        \centering
        \includegraphics[width=1.\textwidth]{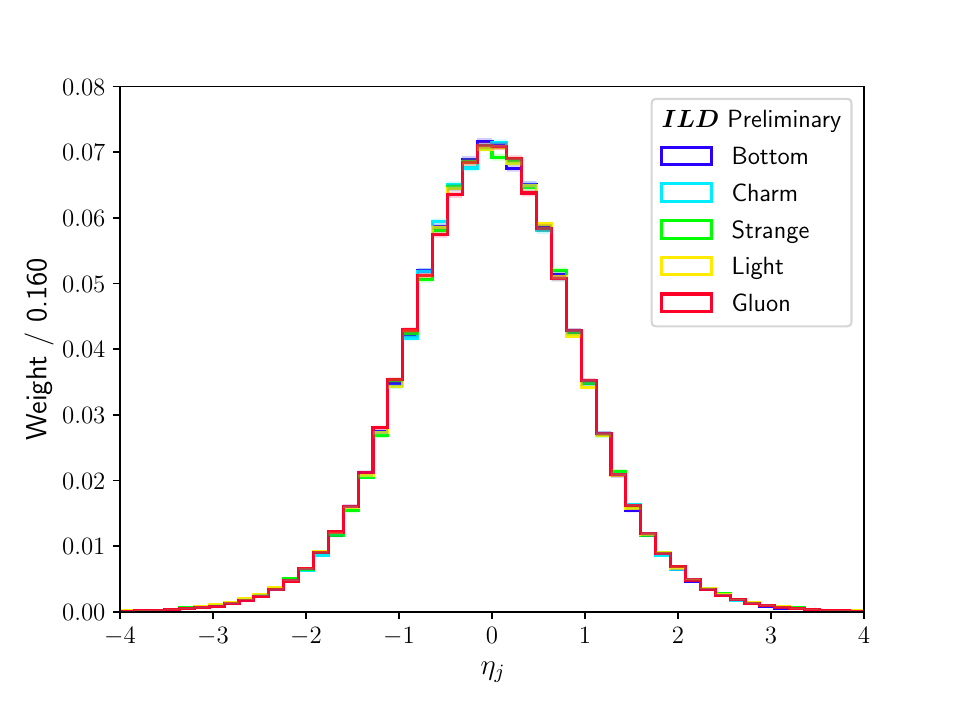}
        \caption{Pseudorapidity $\eta_j$}
    \end{subfigure} \\
    \begin{subfigure}{0.49\textwidth}
        \centering
        \includegraphics[width=1.\textwidth]{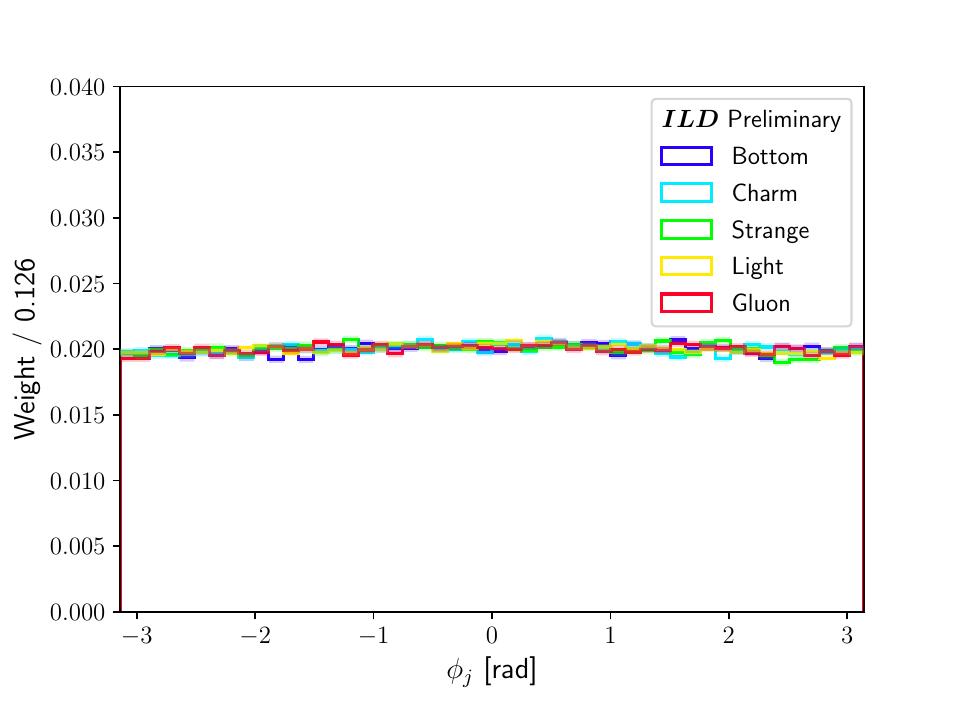}
        \caption{Azimuthal angle $\phi_j$}
    \end{subfigure}
    \hfill
    \begin{subfigure}{0.49\textwidth}
        \centering
        \includegraphics[width=1.\textwidth]{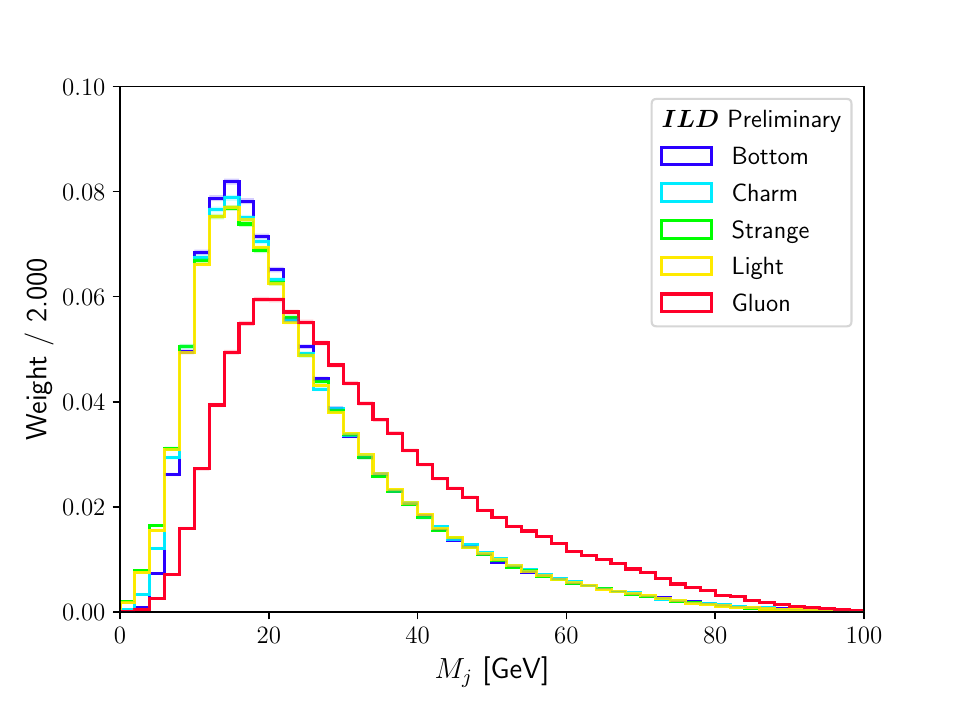}
        \caption{Mass $M$}
    \end{subfigure} \\
    \caption{Distributions of the jet-level inputs for the ANN described in Section~\ref{sec:tagger}. The sum-of-weights for each class is normalised to 1. The error bars correspond to MC statistical uncertainties.}
    \label{fig:inputs_jet_1}
\end{figure}

\begin{figure}[htbp]
    \centering
    \begin{subfigure}{0.49\textwidth}
        \centering
        \includegraphics[width=1.\textwidth]{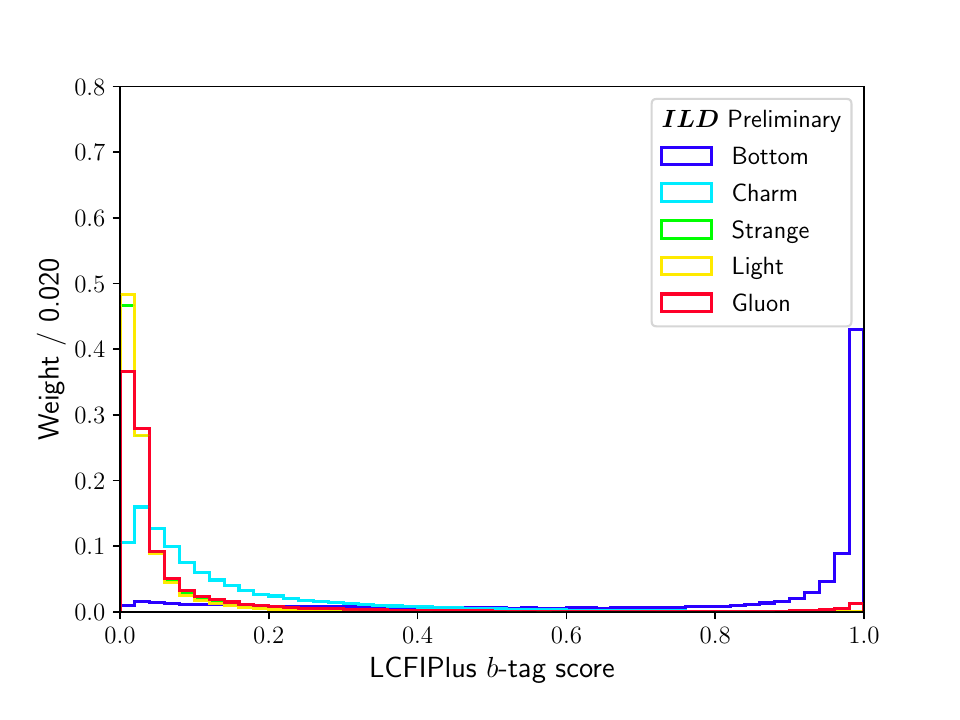}
        \caption{LCFIPlus $b$-tag score}
    \end{subfigure}
    \hfill
    \begin{subfigure}{0.49\textwidth}
        \centering
        \includegraphics[width=1.\textwidth]{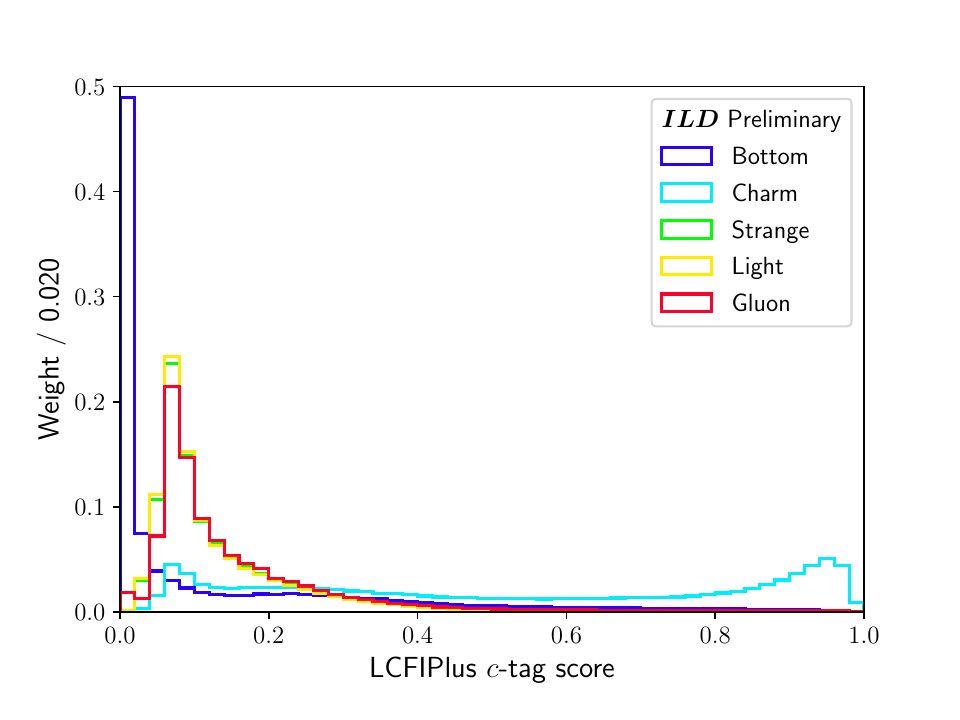}
        \caption{LCFIPlus $c$-tag score}
    \end{subfigure} \\
    \begin{subfigure}{0.49\textwidth}
        \centering
        \includegraphics[width=1.\textwidth]{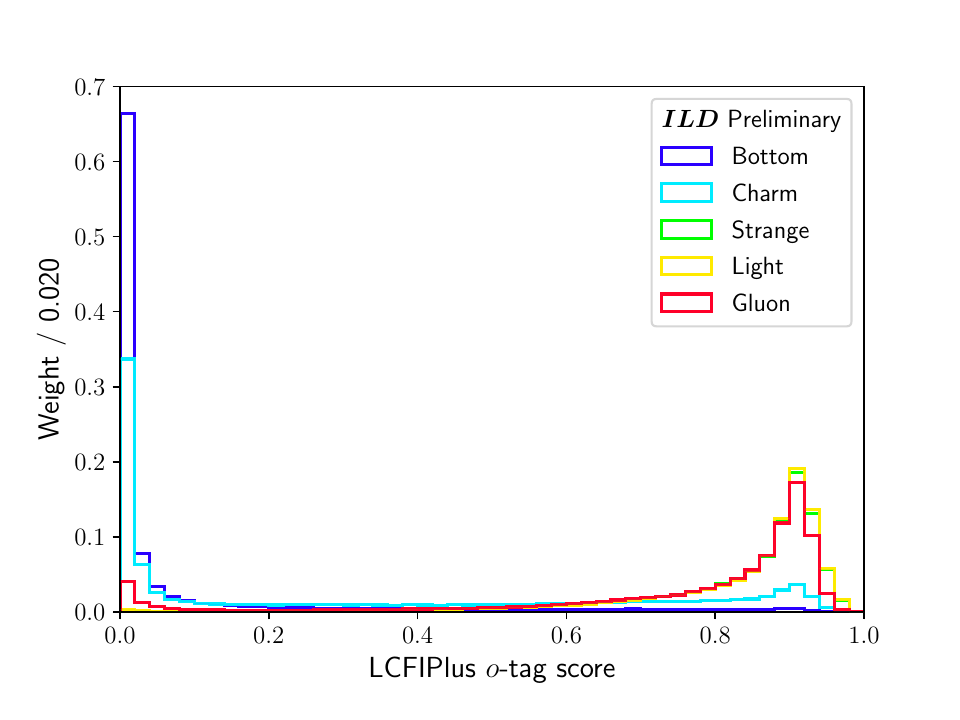}
        \caption{LCFIPlus $o$-tag score}
    \end{subfigure}
    \hfill
    \begin{subfigure}{0.49\textwidth}
        \centering
        \includegraphics[width=1.\textwidth]{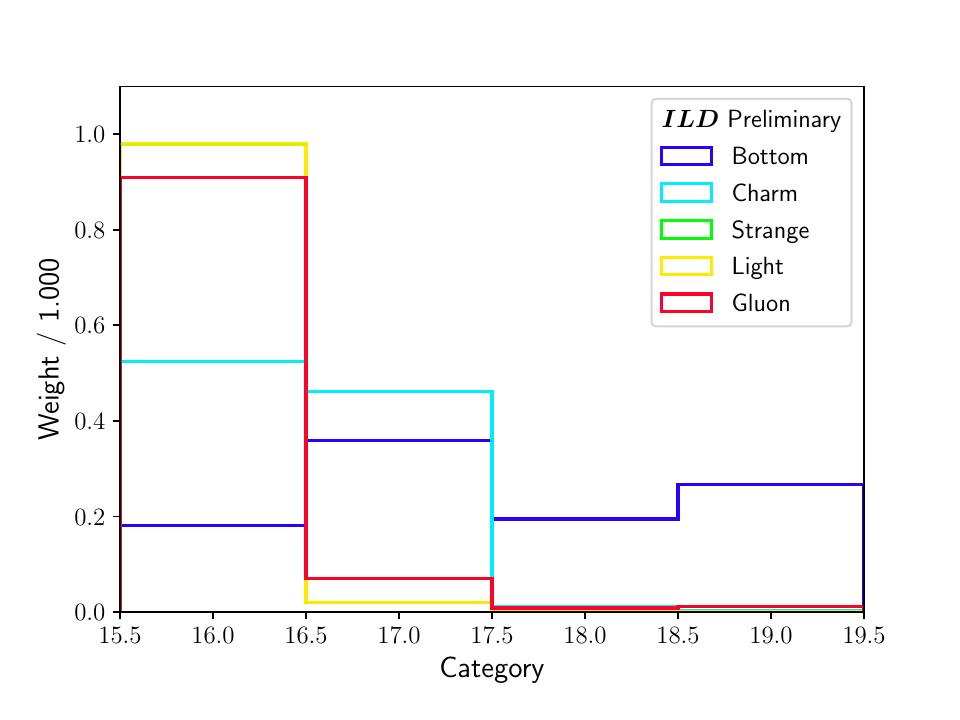}
        \caption{Category}
    \end{subfigure} \\
    \begin{subfigure}{0.49\textwidth}
        \centering
        \includegraphics[width=1.\textwidth]{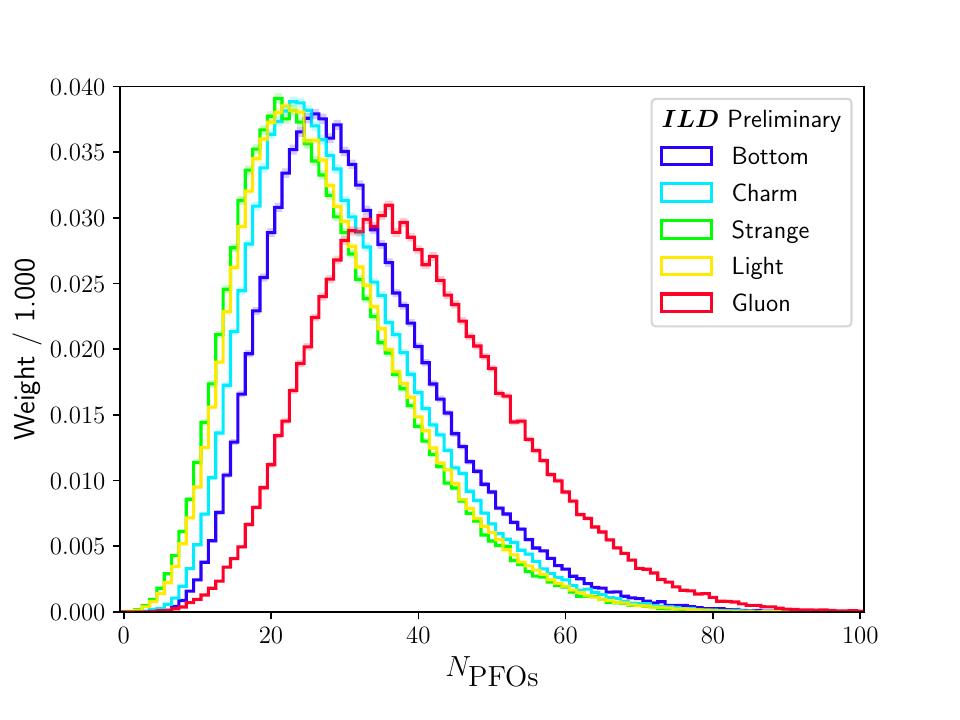}
        \caption{Number of PFOs $N_\textrm{PFOs}$}
    \end{subfigure} \\
    \caption{Distributions of the jet-level inputs for the ANN described in Section~\ref{sec:tagger}. The sum-of-weights for each class is normalised to 1. The error bars correspond to MC statistical uncertainties. A continuation of Fig.~\ref{fig:inputs_jet_1}.}
    \label{fig:inputs_jet_2}
\end{figure}

\begin{figure}[htbp]
    \centering
    \begin{subfigure}{0.49\textwidth}
        \centering
        \includegraphics[width=1.\textwidth]{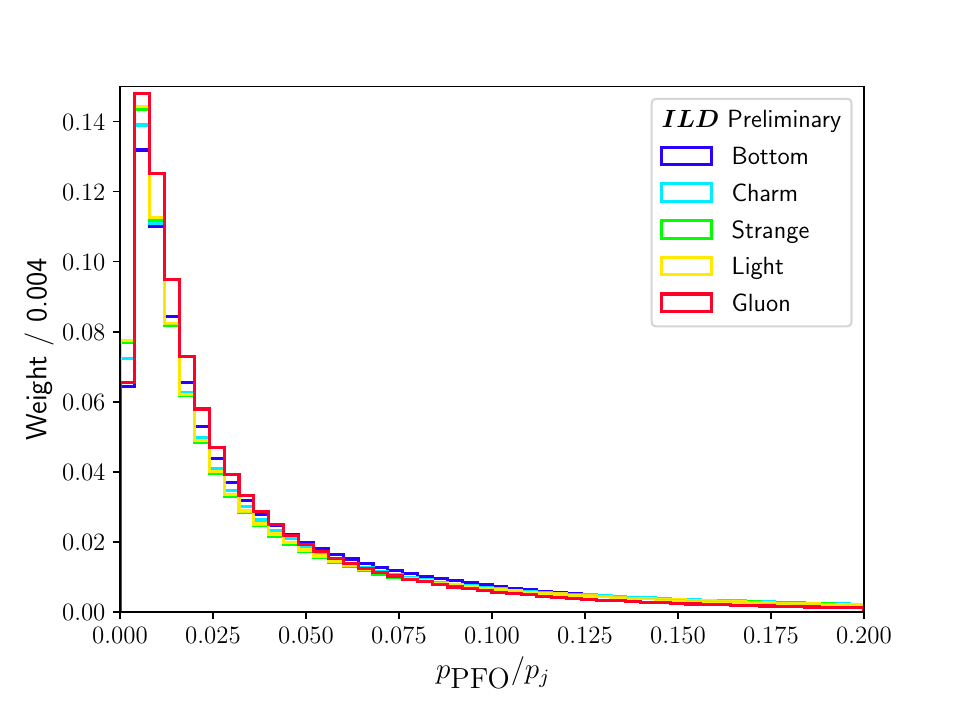}
        \caption{Momentum $p_\textrm{PFO}$}
    \end{subfigure}
    \hfill
    \begin{subfigure}{0.49\textwidth}
        \centering
        \includegraphics[width=1.\textwidth]{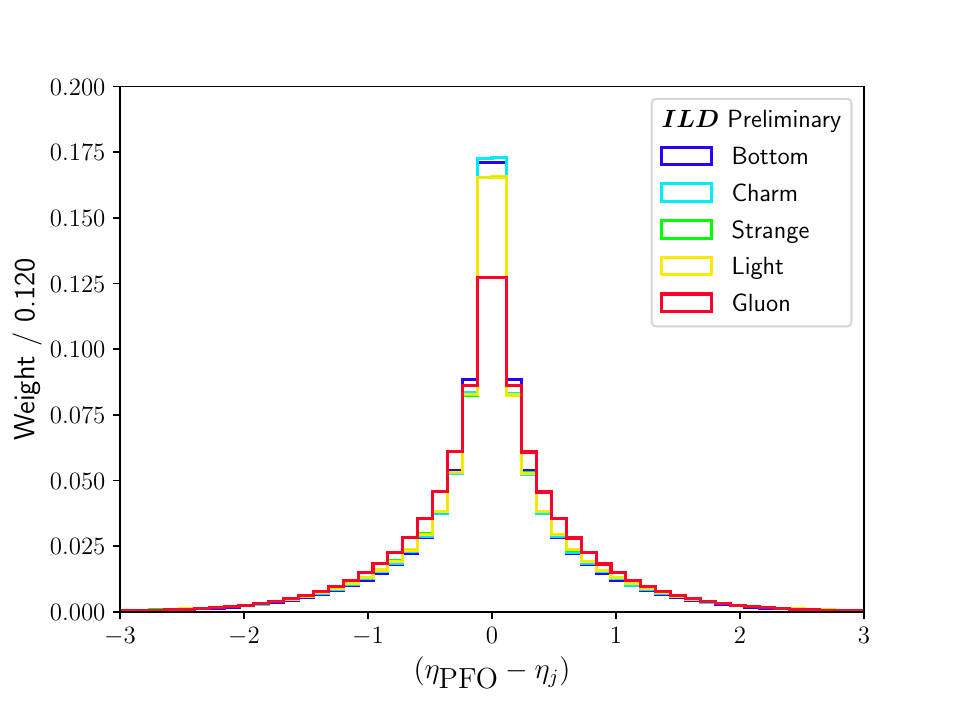}
        \caption{Pseudorapidity $\eta_\textrm{PFO}$}
    \end{subfigure} \\
    \begin{subfigure}{0.49\textwidth}
        \centering
        \includegraphics[width=1.\textwidth]{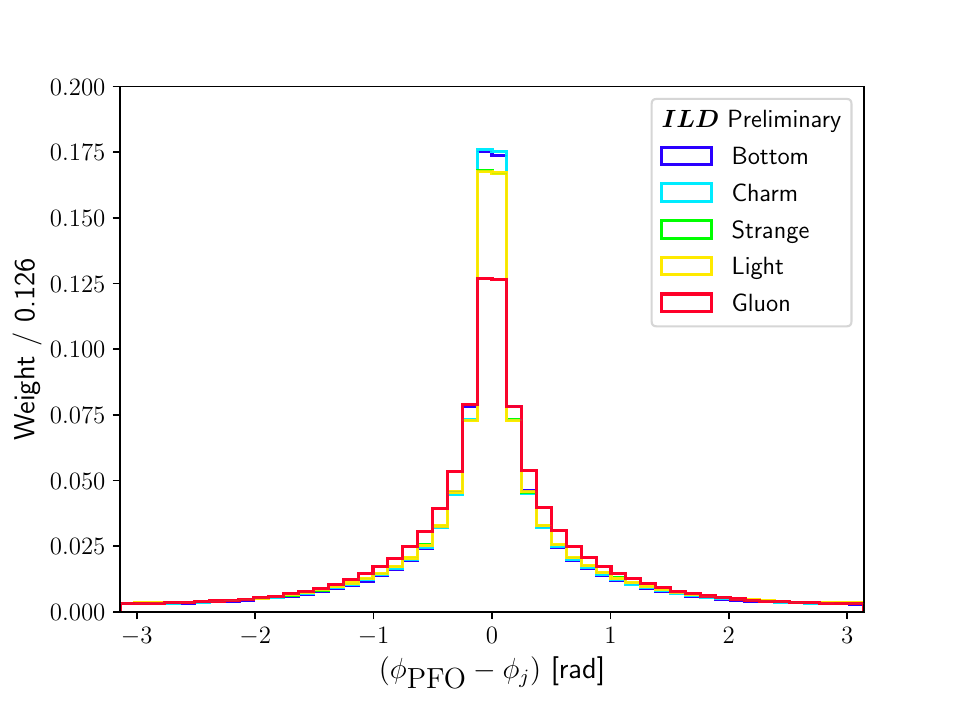}
        \caption{Azimuthal angle $\phi_\textrm{PFO}$}
    \end{subfigure}
    \hfill
    \begin{subfigure}{0.49\textwidth}
        \centering
        \includegraphics[width=1.\textwidth]{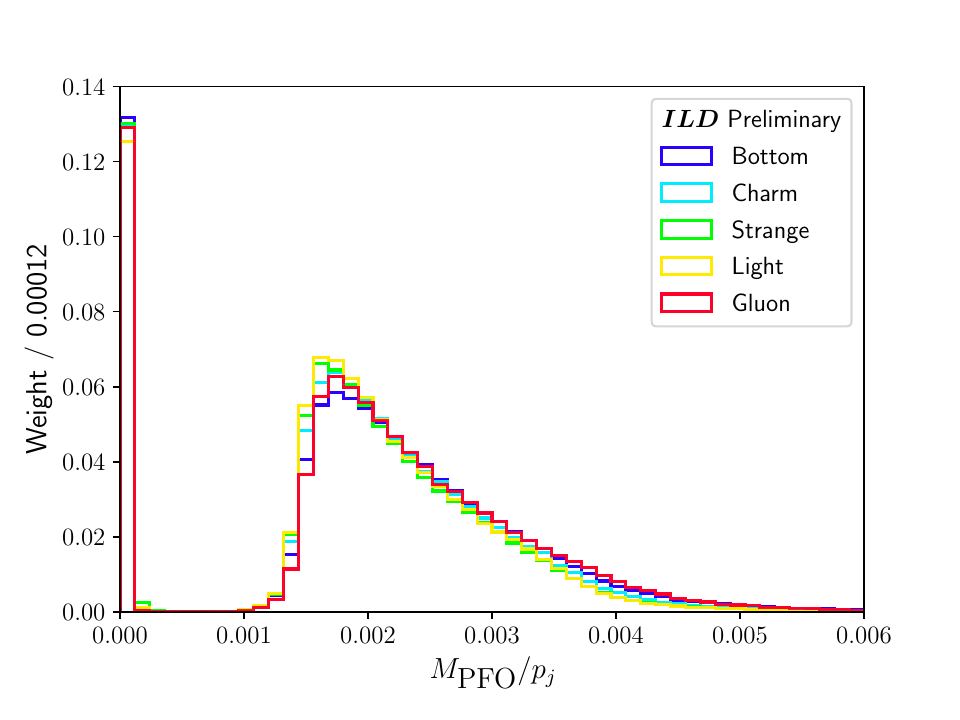}
        \caption{Mass $M_\textrm{PFO}$}
    \end{subfigure} \\
    \begin{subfigure}{0.49\textwidth}
        \centering
        \includegraphics[width=1.\textwidth]{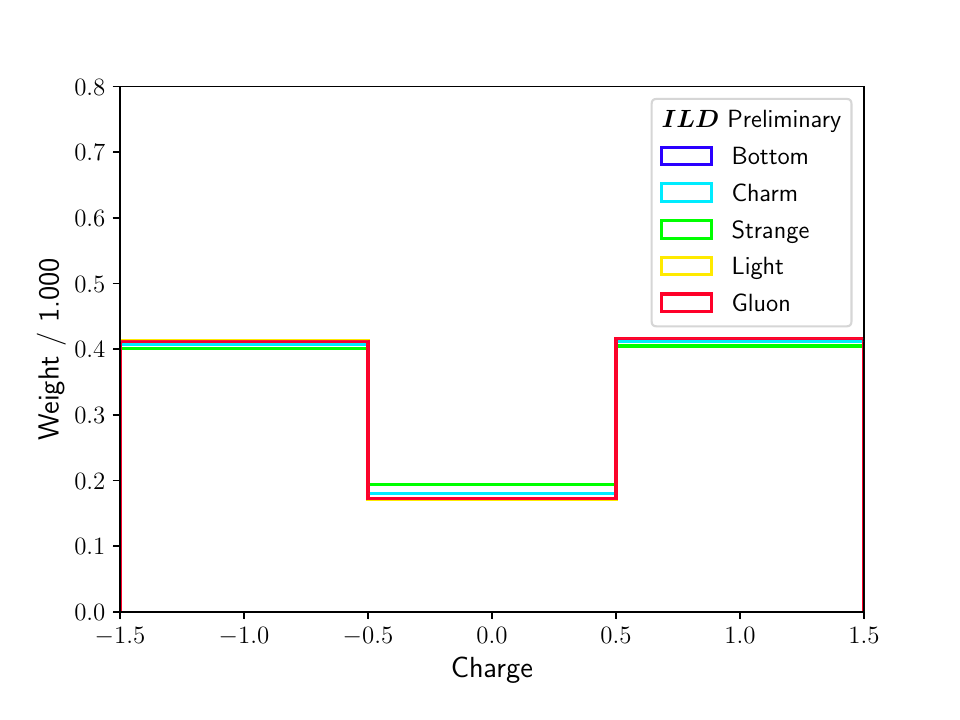}
        \caption{Charge}
    \end{subfigure}
    \hfill
    \begin{subfigure}{0.49\textwidth}
        \centering
        \includegraphics[width=1.\textwidth]{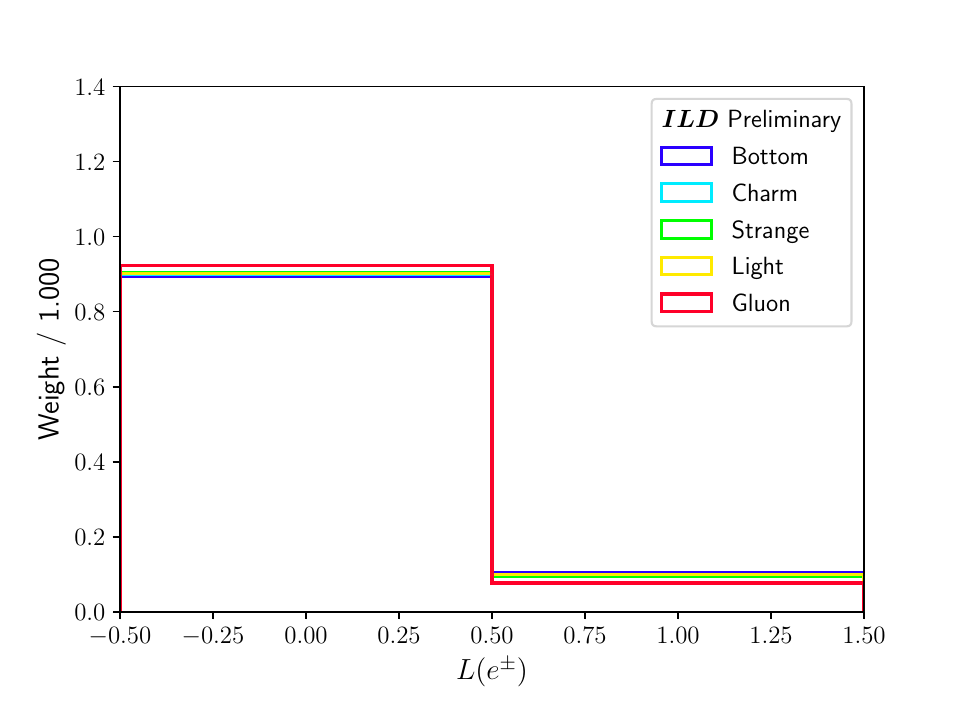}
        \caption{Electron truth likelihood $L(e^\pm)$}
    \end{subfigure} \\
    \caption{Distributions of the PFO-level inputs for the ANN described in Section~\ref{sec:tagger}. The sum-of-weights for each class is normalised to 1. The error bars correspond to MC statistical uncertainties.}
    \label{fig:inputs_PFO_1}
\end{figure}

\begin{figure}[htbp]
    \begin{subfigure}{0.49\textwidth}
        \centering
        \includegraphics[width=1.\textwidth]{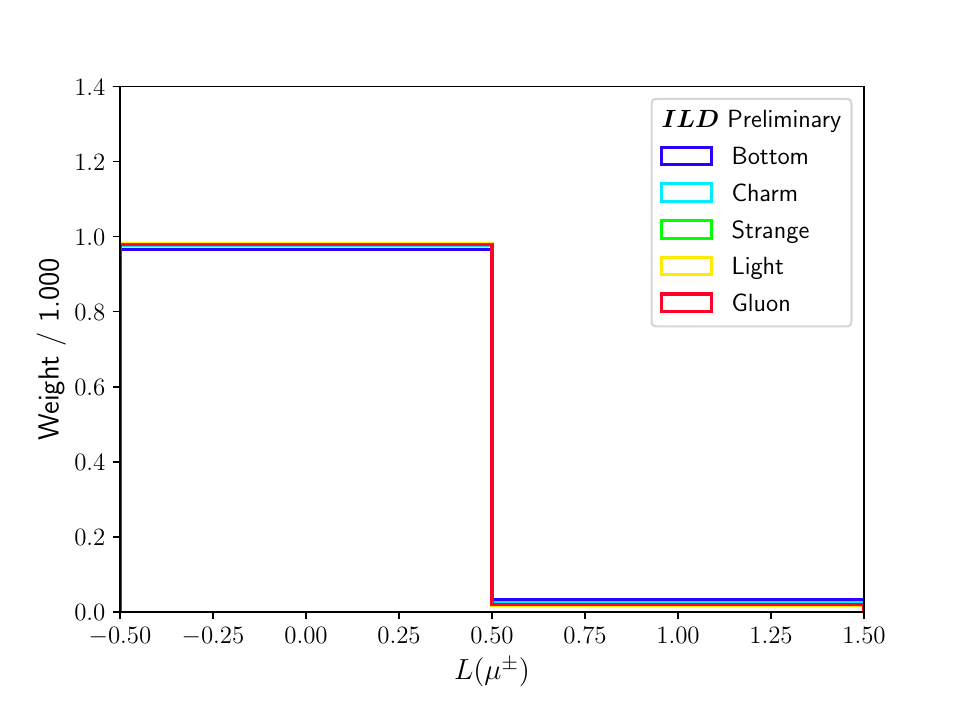}
        \caption{Muon truth likelihood $L(\mu^\pm)$}
    \end{subfigure}
    \hfill
    \begin{subfigure}{0.49\textwidth}
        \centering
        \includegraphics[width=1.\textwidth]{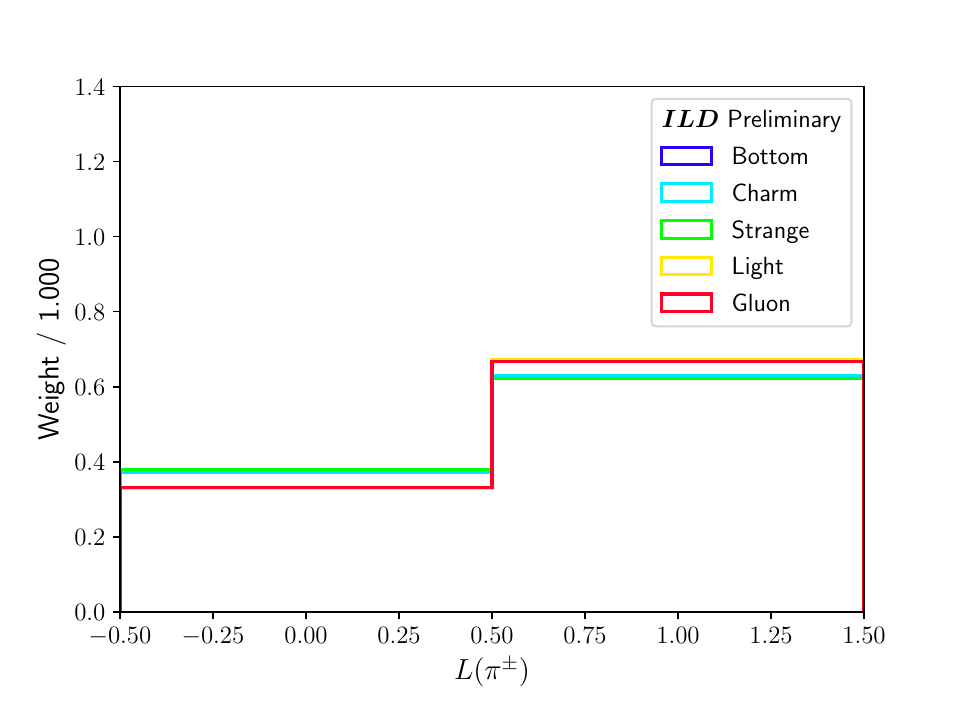}
        \caption{Pion truth likelihood $L(\pi^\pm)$}
    \end{subfigure} \\
    \begin{subfigure}{0.49\textwidth}
        \centering
        \includegraphics[width=1.\textwidth]{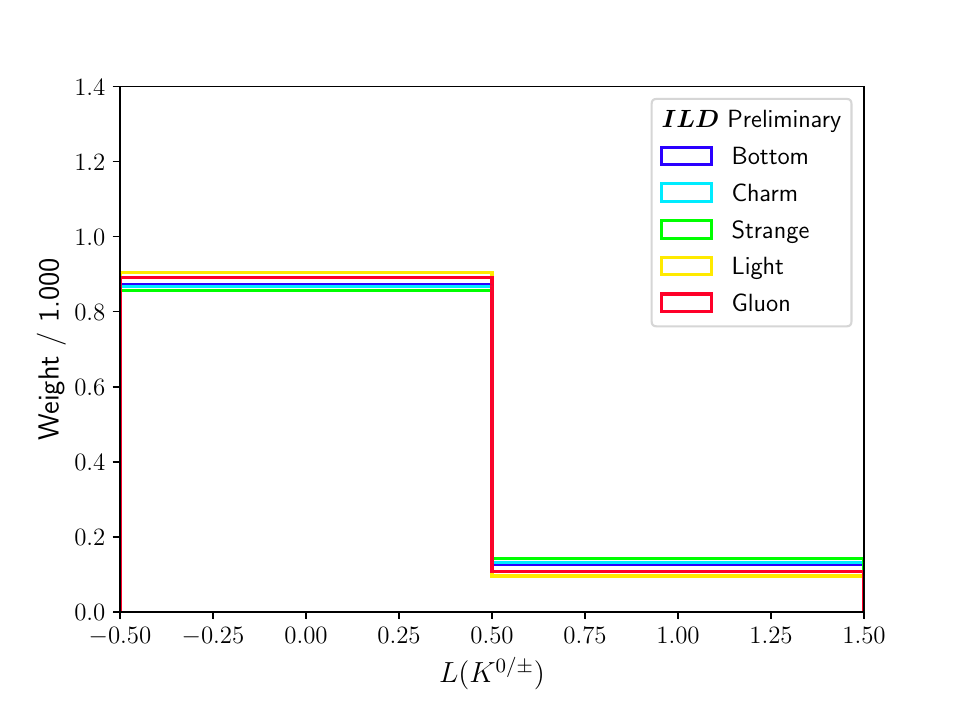}
        \caption{Kaon/strange hadron truth likelihood $L(K^{0/\pm})$}
    \end{subfigure}
    \hfill
    \begin{subfigure}{0.49\textwidth}
        \centering
        \includegraphics[width=1.\textwidth]{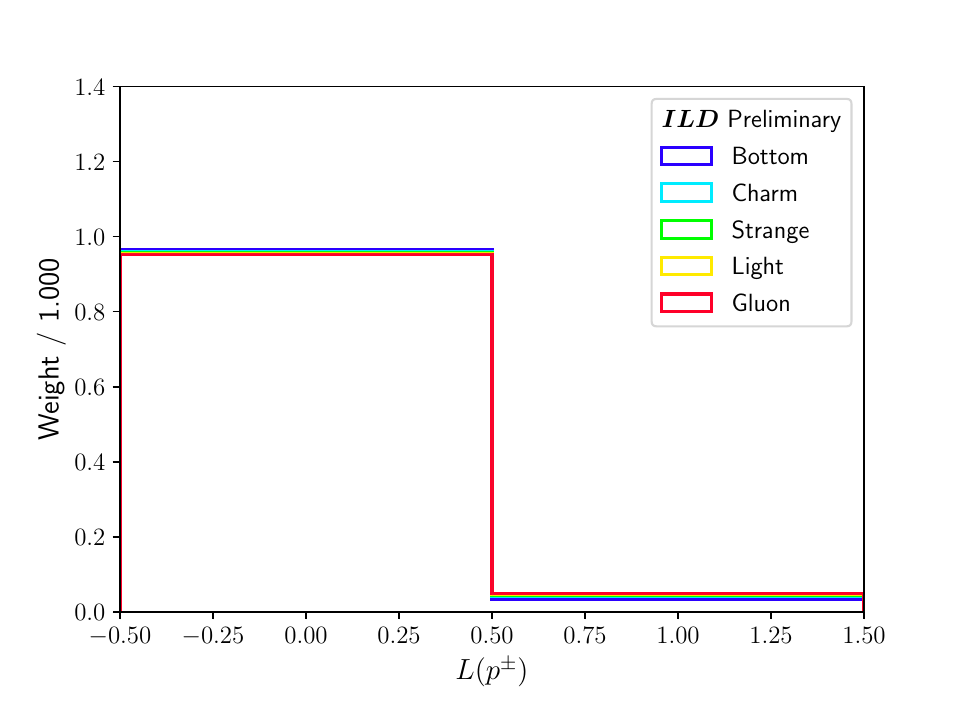}
        \caption{Proton truth likelihood $L(p^\pm)$}
    \end{subfigure} \\
    \caption{Distributions of the PFO-level inputs for the ANN described in Section~\ref{sec:tagger}. The sum-of-weights for each class is normalised to 1. The error bars correspond to MC statistical uncertainties. A continuation of Fig.~\ref{fig:inputs_PFO_1}.}
    \label{fig:inputs_PFO_2}
\end{figure}

\begin{figure}[htbp]
    \centering
    \begin{subfigure}{0.49\textwidth}
        \centering
        \includegraphics[width=1.\textwidth]{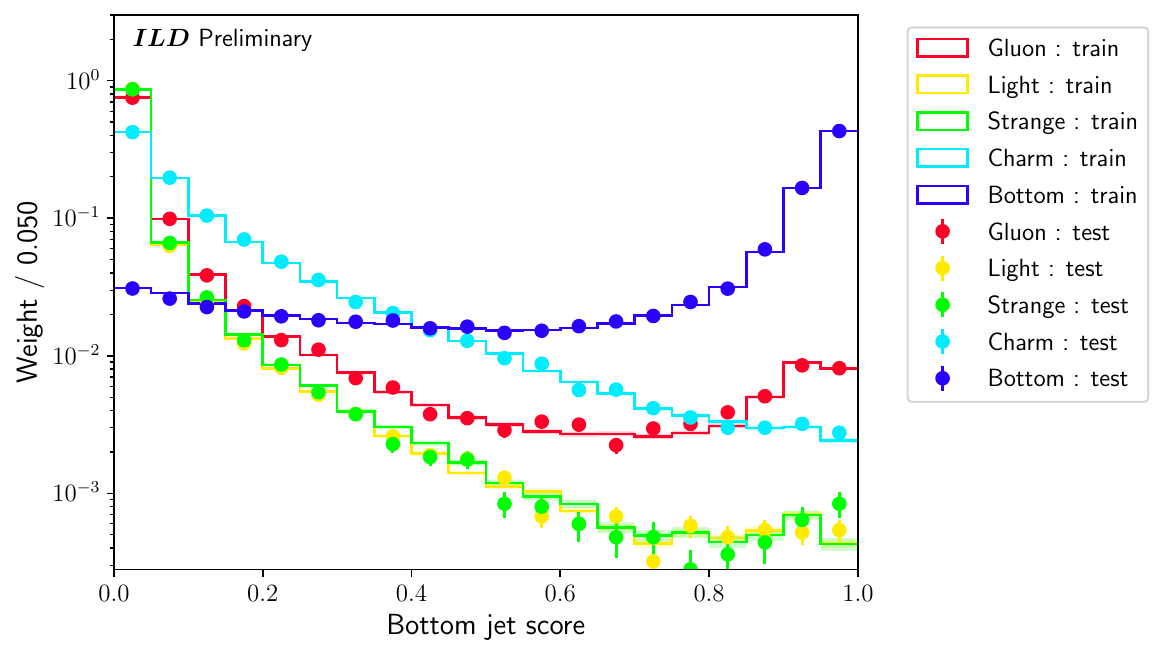}
        \caption{$b$-jet score}
    \end{subfigure}
    \hfill
    \begin{subfigure}{0.49\textwidth}
        \centering
        \includegraphics[width=1.\textwidth]{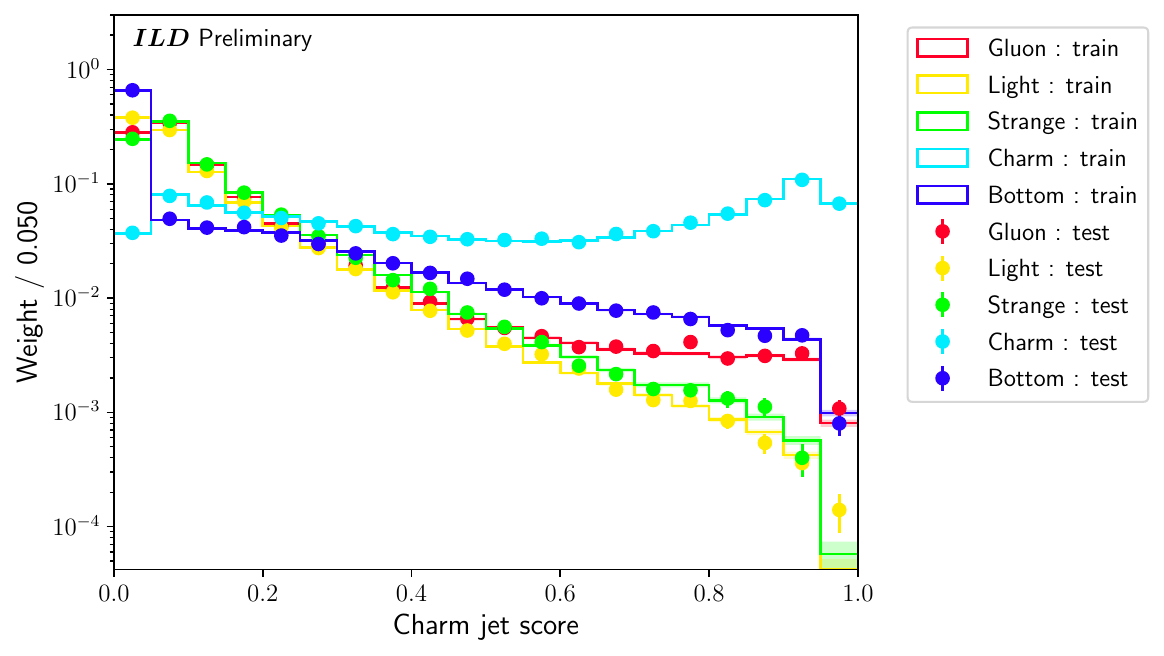}
        \caption{$c$-jet score}
    \end{subfigure} \\
    \begin{subfigure}{0.49\textwidth}
        \centering
        \includegraphics[width=1.\textwidth]{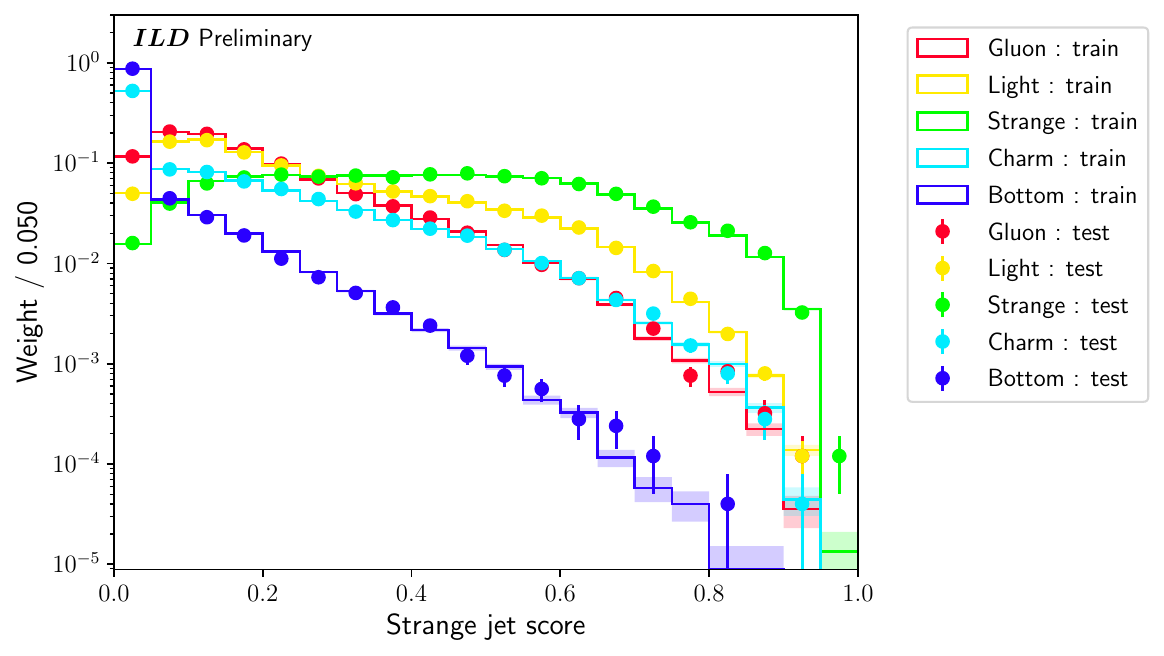}
        \caption{$s$-jet score}
    \end{subfigure}
    \hfill
    \begin{subfigure}{0.49\textwidth}
        \centering
        \includegraphics[width=1.\textwidth]{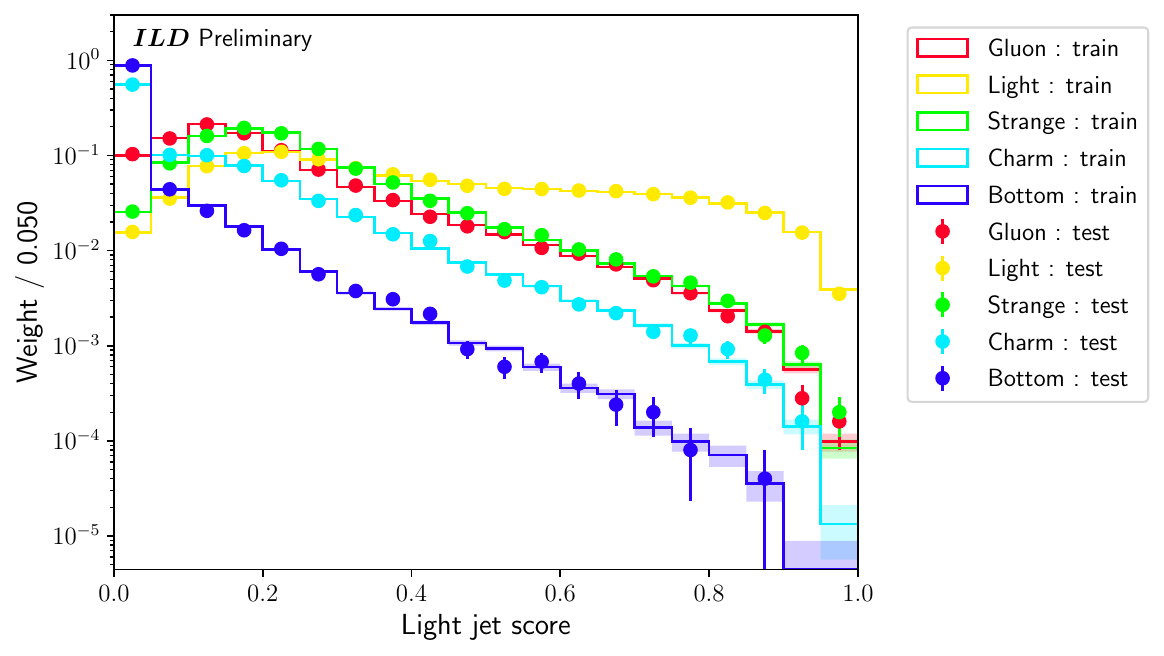}
        \caption{Light-jet score}
    \end{subfigure} \\
    \begin{subfigure}{0.49\textwidth}
        \centering
        \includegraphics[width=1.\textwidth]{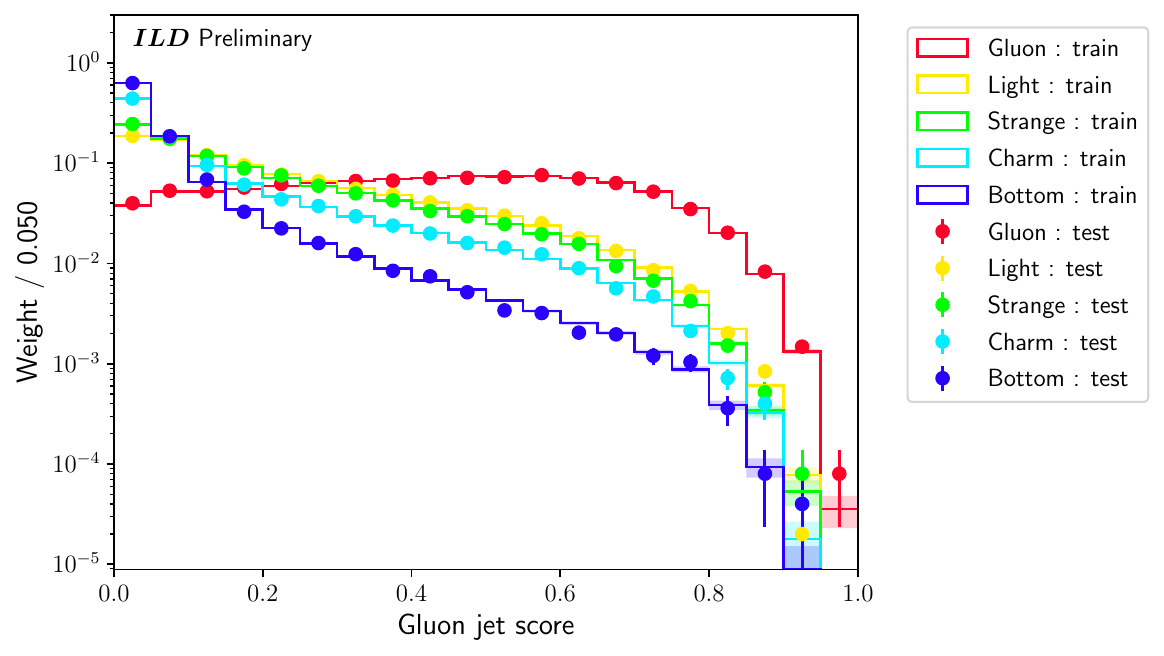}
        \caption{$g$-jet score}
    \end{subfigure} \\
    \caption{Distributions of the ANN's output nodes for the training and testing slices of \kfold~0 tagger described in Section~\ref{sec:tagger}. The sum-of-weights for each class of each slice is normalised to 1 and logarithmic $y$-axis scales are used. The error bars correspond to MC statistical uncertainties.}
    \label{fig:traintest0}
\end{figure}

\begin{figure}[htbp]
    \centering
    \begin{subfigure}{0.49\textwidth}
        \centering
        \includegraphics[width=1.\textwidth]{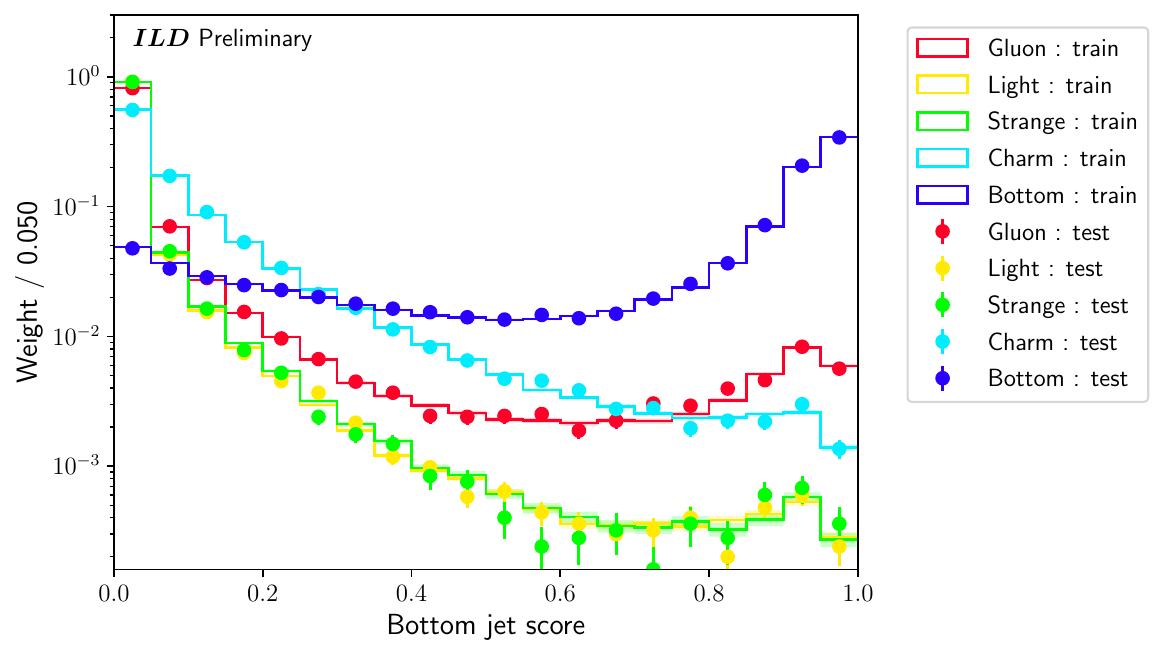}
        \caption{$b$-jet score}
    \end{subfigure}
    \hfill
    \begin{subfigure}{0.49\textwidth}
        \centering
        \includegraphics[width=1.\textwidth]{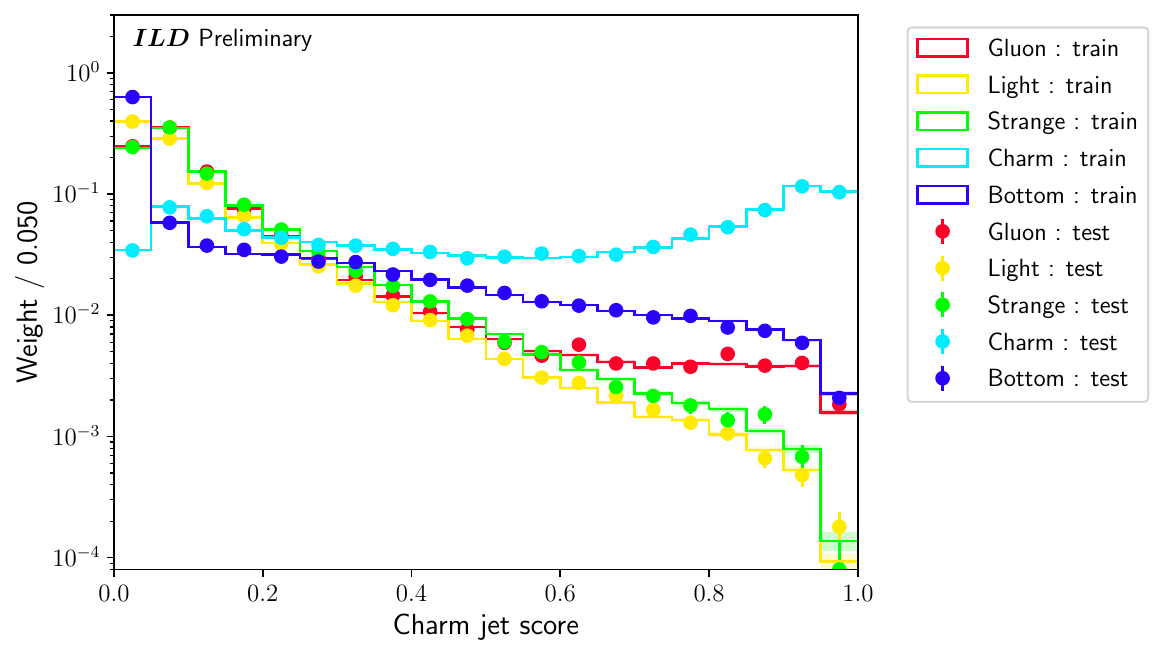}
        \caption{$c$-jet score}
    \end{subfigure} \\
    \begin{subfigure}{0.49\textwidth}
        \centering
        \includegraphics[width=1.\textwidth]{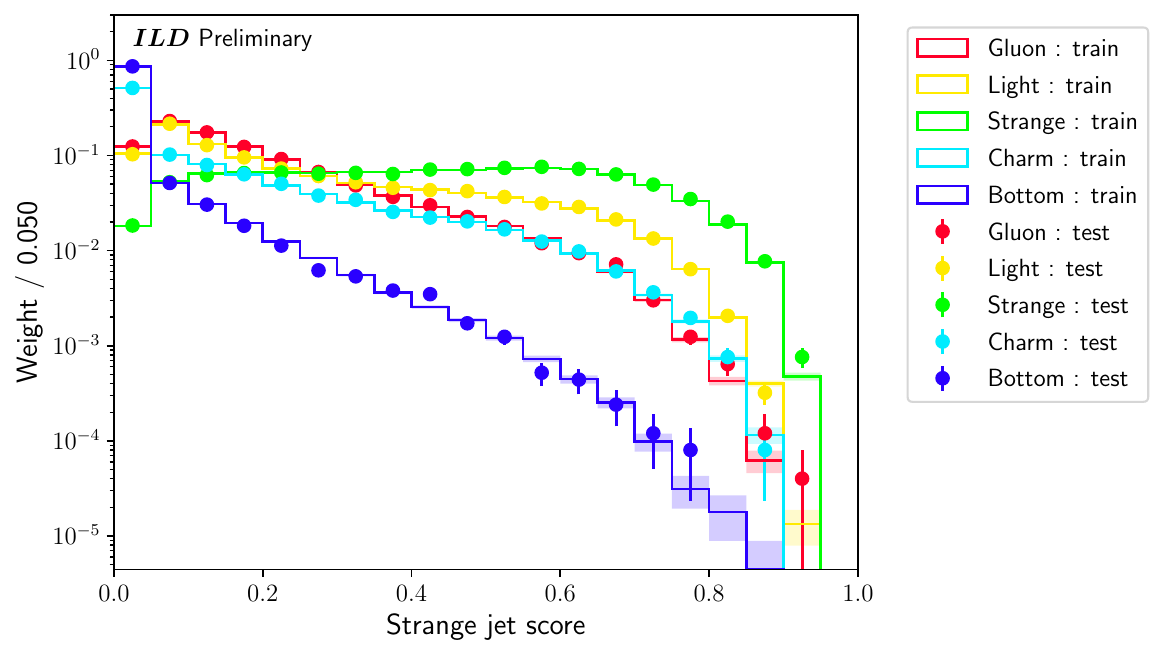}
        \caption{$s$-jet score}
    \end{subfigure}
    \hfill
    \begin{subfigure}{0.49\textwidth}
        \centering
        \includegraphics[width=1.\textwidth]{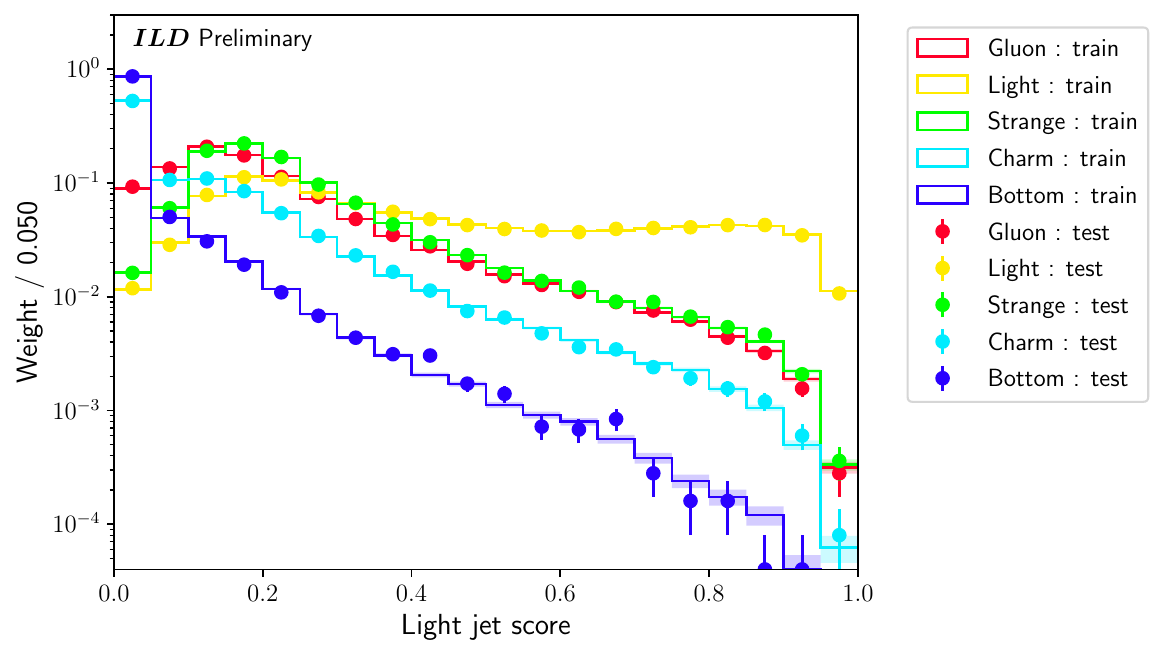}
        \caption{Light-jet score}
    \end{subfigure} \\
    \begin{subfigure}{0.49\textwidth}
        \centering
        \includegraphics[width=1.\textwidth]{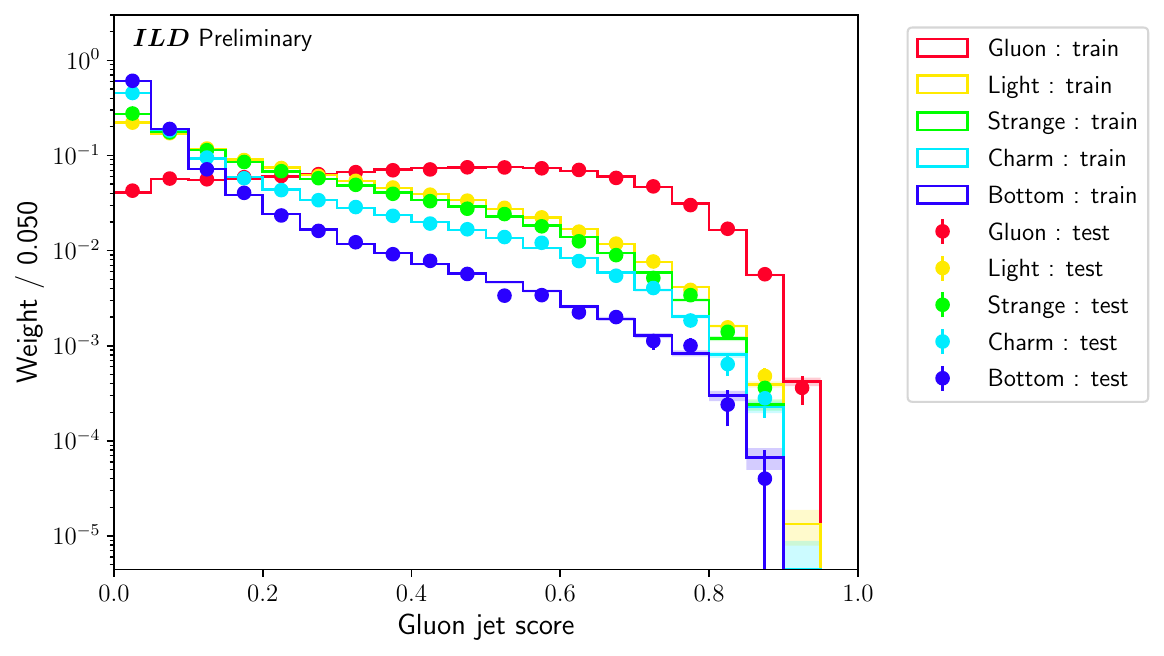}
        \caption{$g$-jet score}
    \end{subfigure} \\
    \caption{Distributions of the ANN's output nodes for the training and testing slices of \kfold~1 tagger described in Section~\ref{sec:tagger}. The sum-of-weights for each class of each slice is normalised to 1 and logarithmic $y$-axis scales are used. The error bars correspond to MC statistical uncertainties.}
    \label{fig:traintest1}
\end{figure}

This appendix contains additional plots related to the training of the jet flavour tagger described in Section~\ref{sec:tagger}. In particular, Figs.~\ref{fig:inputs_jet_1} and \ref{fig:inputs_jet_2} show the shapes of the jet-level inputs for training, and Figs.~\ref{fig:inputs_PFO_1} and \ref{fig:inputs_PFO_2} show the shapes of the PFO-level inputs for training. Figs.~\ref{fig:traintest0} and \ref{fig:traintest1} show the train-test agreement for the \kfold~0 and 1 networks.

\FloatBarrier

\section{Jet flavour tagger without PID and with partial PID}
\label{app:no_PID_tagger}

To study the extent to which measurements of $\kappa_s$ depend on PID, we re-trained a jet flavour tagger using the same architecture as described in Section~\ref{sec:tagger}. The truth likelihood information per jet constituent PFO was \emph{not} provided; otherwise, all of the same inputs were used. This is equivalent to the tagger having no PID for any of the input particles. N.B. the tagger was verified to have good train-test agreement following training. For brevity, metrics used to monitor over-training are excluded from the following sections. The tagger is then applied to the same SM \Hss analysis presented in Section~\ref{sec:analysis} and limits on $\kappa_s$ are calculated.

We also re-trained the jet flavour tagger using the same architecture as described in Section~\ref{sec:tagger} but with partial PID, motivated by the fact that PID becomes less powerful for momentum above $\unit[\mathcal{O}(10)]{GeV}$. The partial PID is applied by modifying the truth likelihoods, $L(\zeta)$, as:

\begin{equation}
    L^\prime(\zeta) = \left\{
        \begin{array}{ll}
            L(\zeta) \,, & p_\textrm{PFO} < p_\textrm{cut} \\
            0.5      \,, & p_\textrm{PFO} > p_\textrm{cut} \\
        \end{array}
    \right. \,\forall\, \zeta \in [e^\pm,\mu^\pm,\pi^\pm,K^{0/\pm},p^\pm]\,,
\end{equation}

\noindent where $L^\prime(\zeta)$ is the modified truth likelihood. In words: each PFO in an input jet has its truth likelihoods set to 0.5 if its momentum, $p_\textrm{PFO}$, is above some threshold, $p_\textrm{cut}$. The choice of 0.5 is made to represent ``maximal confusion'' being the two extremes of each likelihood (i.e., ``$L(\zeta) = 0$'' $\coloneqq$ ``PFO is \emph{not} of type $\zeta$ with 100\% certainty'' and ``$L(\zeta) = 1$'' $\coloneqq$ ``PFO is of type $\zeta$ with 100\% certainty''). Different values of the momentum threshold are tested, $p_\textrm{cut} \in [10, 20, 30]\,\textrm{GeV}$, and a tagger is trained for each choice of threshold. The taggers with partial PID are compared to the taggers with full PID and with no PID in the following sections; however, they are not applied to the \Hss analysis and used to set limits on $\kappa_s$ in the same way as the taggers with full PID and without PID.

\subsection{Tagger performance}
\label{app:no_PID_tagger_perf}

\begin{figure}[htbp]
    \centering
    \begin{subfigure}{0.49\textwidth}
        \centering
        \includegraphics[width=1.\textwidth]{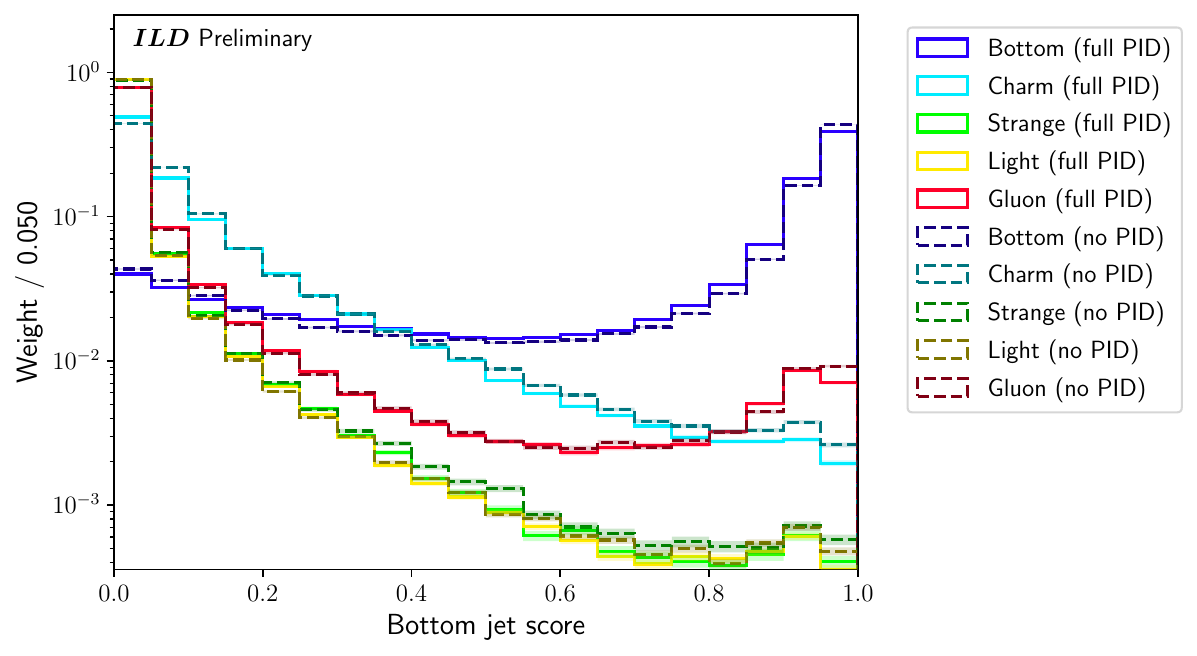}
        \caption{$b$-jet score}
    \end{subfigure}
    \hfill
    \begin{subfigure}{0.49\textwidth}
        \centering
        \includegraphics[width=1.\textwidth]{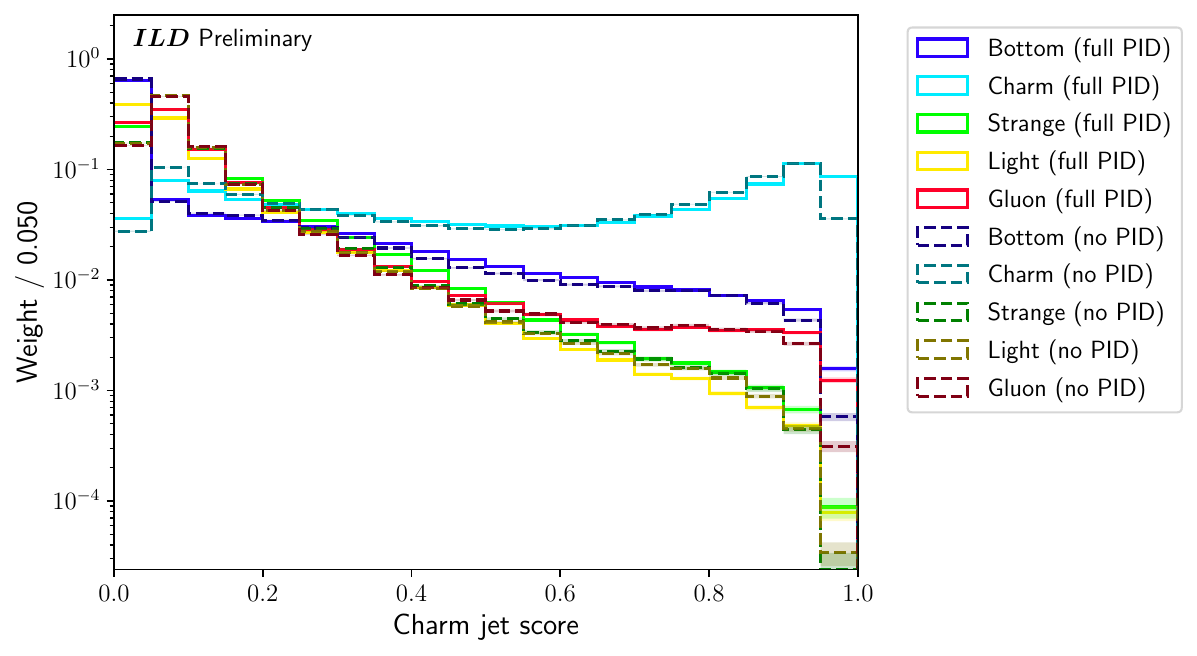}
        \caption{$c$-jet score}
    \end{subfigure} \\
    \begin{subfigure}{0.49\textwidth}
        \centering
        \includegraphics[width=1.\textwidth]{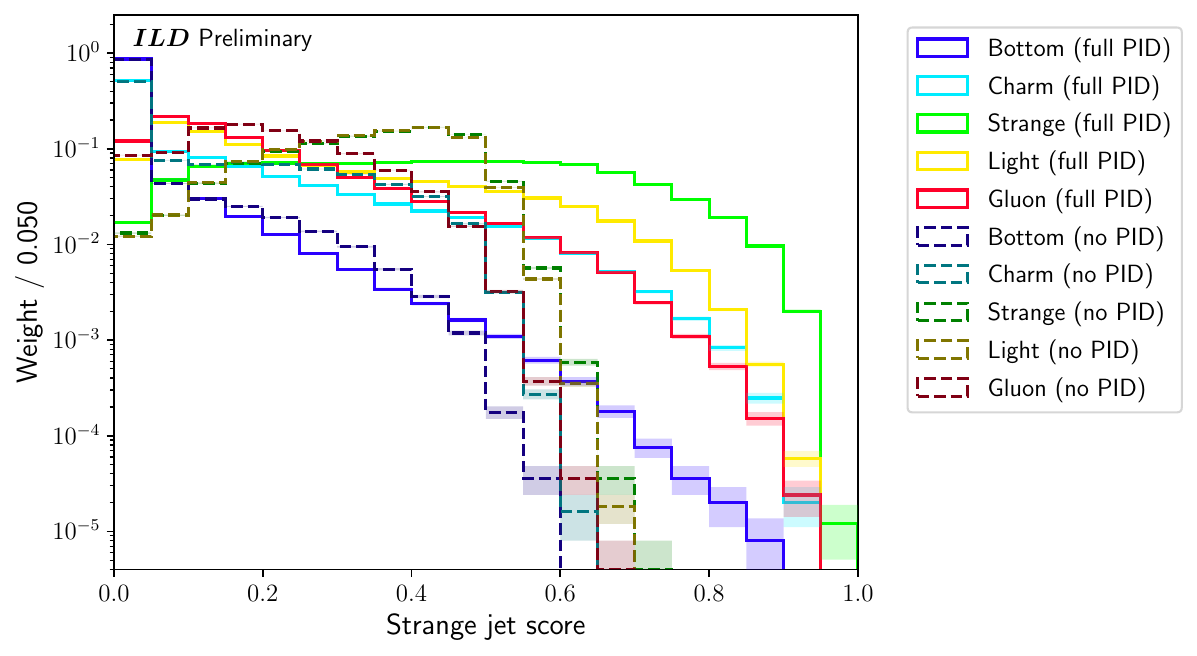}
        \caption{$s$-jet score}
        \label{fig:output_no_PID_s}
    \end{subfigure}
    \hfill
    \begin{subfigure}{0.49\textwidth}
        \centering
        \includegraphics[width=1.\textwidth]{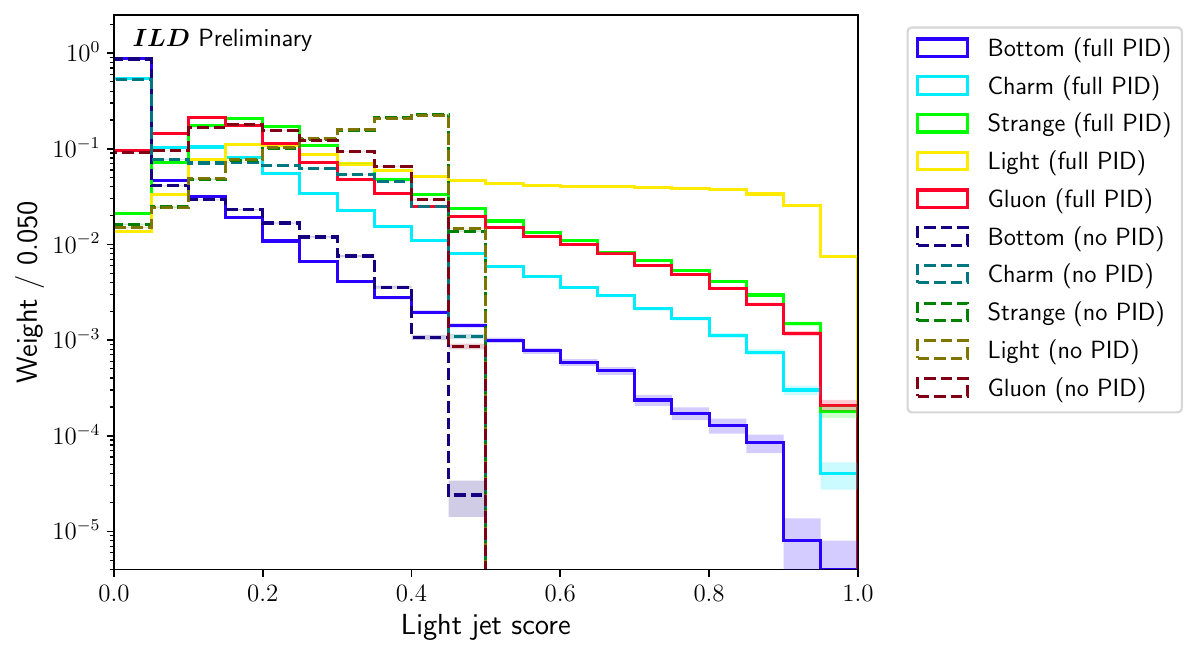}
        \caption{Light-jet score}
        \label{fig:output_no_PID_ud}
    \end{subfigure} \\
    \begin{subfigure}{0.49\textwidth}
        \centering
        \includegraphics[width=1.\textwidth]{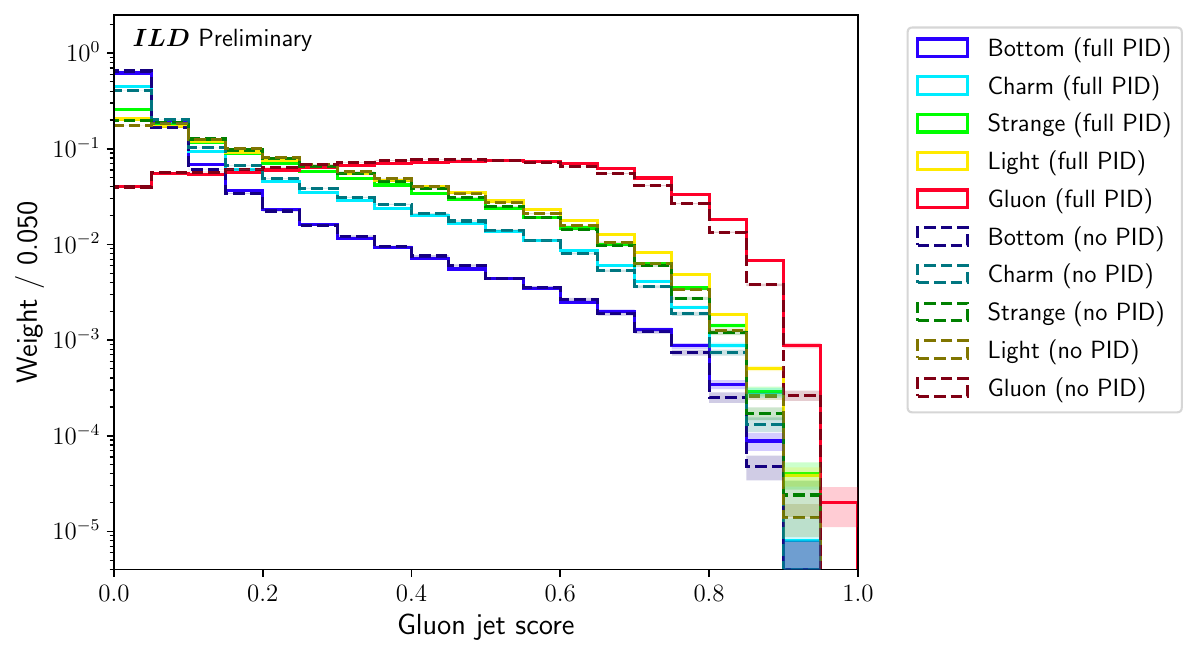}
        \caption{$g$-jet score}
    \end{subfigure} \\
    \caption{Distributions for each output node of the jet flavour tagger with full PID (``full PID''), as described in Section~\ref{sec:tagger}, as well as for the jet flavour tagger without PID (``no PID''), as described in Appendix~\ref{app:no_PID_tagger}. The sum-of-weights for each class is normalised to 1 and logarithmic $y$-axis scales are used. The error bars correspond to MC statistical uncertainties.}
    \label{fig:output_no_PID}
\end{figure}

Eq.~\ref{eqn:tagger} is plotted in Fig.~\ref{fig:output_no_PID} for the taggers with and without PID, showing the output scores for each class and each output node. We see that the output shapes for the taggers with and without PID are identical for each class of the $b$-, $c$- and $g$-jet output nodes. The output shapes between the taggers for the $s$- and light-jet output nodes are very \emph{different} for all classes, however. From Fig.~\ref{fig:output_no_PID_s}, we see that the $s$-jet score for strange jets falls off sharply at $\sim$0.5 rather than occupying the full output range of 0 to 1. A similar remark can be made about the light-jet score for light jets from Fig.~\ref{fig:output_no_PID_ud}. This indicates that there is much greater confusion when classifying the jet as a strange or light jet -- much more so than when the classifying the jet as a bottom, charm, or gluon jet.

\begin{figure}
    \centering
    \includegraphics[width=0.9\textwidth]{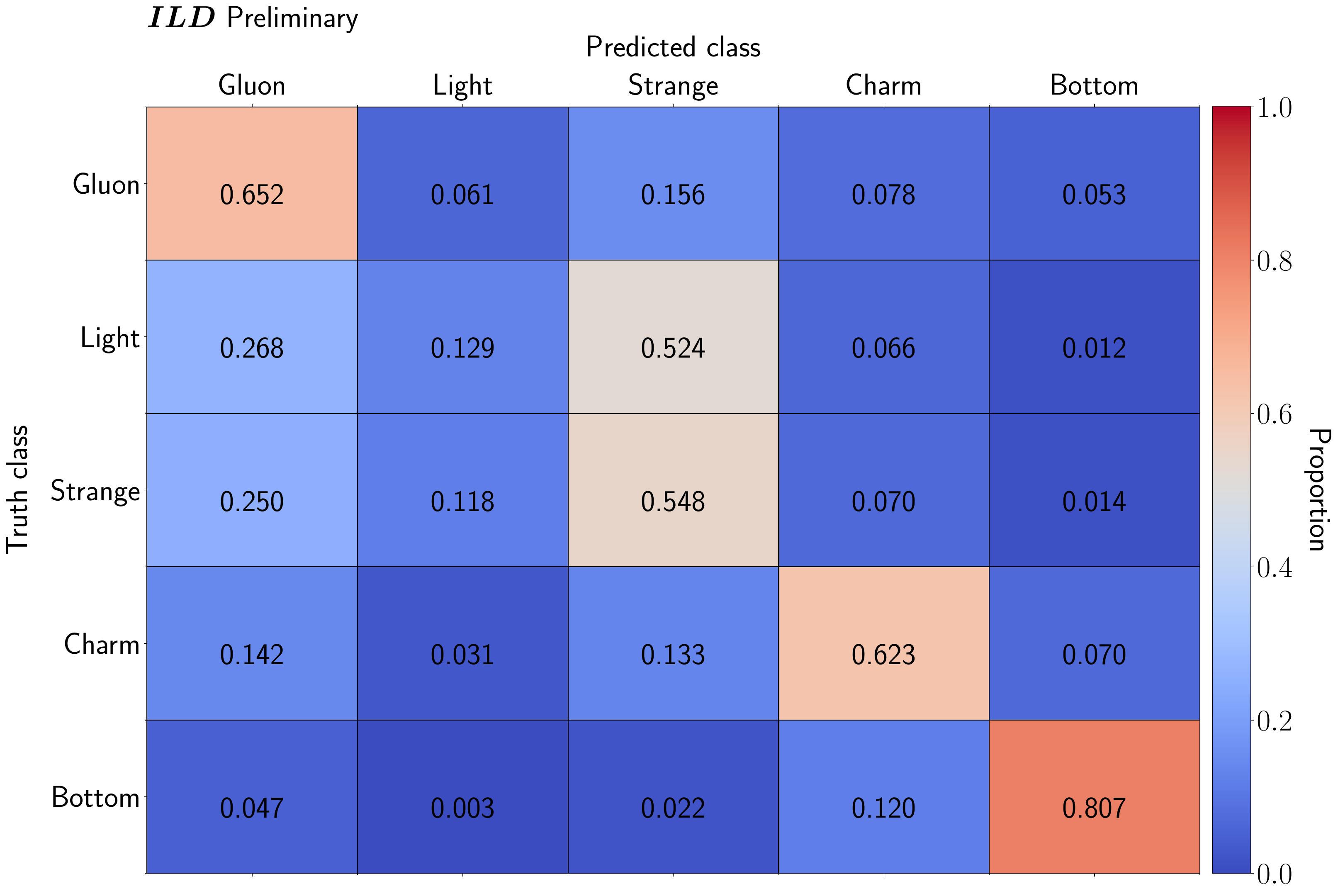}
    \caption{Confusion matrix for the output of the described jet flavour tagger without PID, as described in Appendix~\ref{app:no_PID_tagger}. Each truth class (i.e., row) is normalised to 1.}
    \label{fig:confusion_matrix_no_PID}
\end{figure}

Our conclusions are further supported by the confusion matrix for the tagger without PID, shown in Fig.~\ref{fig:confusion_matrix_no_PID}. From the confusion matrix, we see that ground truth light jets are more often classified as strange jets (52.4\%) than light jets (12.9\%). This is disparate from the confusion matrix for the tagger with PID, Fig.~\ref{fig:confusion_matrix}, where ground truth light jets are most often classified as light jets (47.1\%). For ground truth strange jets, the classification using the tagger without PID is degraded as compared to the classification using the tagger with PID (54.8\%, previously 58.1\%). This seemingly comes from the higher rate of classification of ground truth strange jets as gluon jets when using the tagger without PID (25.0\%, previously 19.4\%). N.B. the reason for the higher rate of classification of ground truth light jets as strange jets than light jets is likely due to the $s$-jet scores being marginally higher than the light-jet scores (but otherwise very similar) for light jets. When deciding which flavour the tagger classifies a jet as (for the purposes of building a confusion matrix), the highest score is taken. This means that if a ground truth light jet has an $s$-jet score of 0.501 and a light-jet score of 0.499, the jet is classified as a ``strange'' jet.

\begin{figure}[htbp]
    \centering
    \begin{subfigure}{0.49\textwidth}
        \centering
        \includegraphics[width=1.\textwidth]{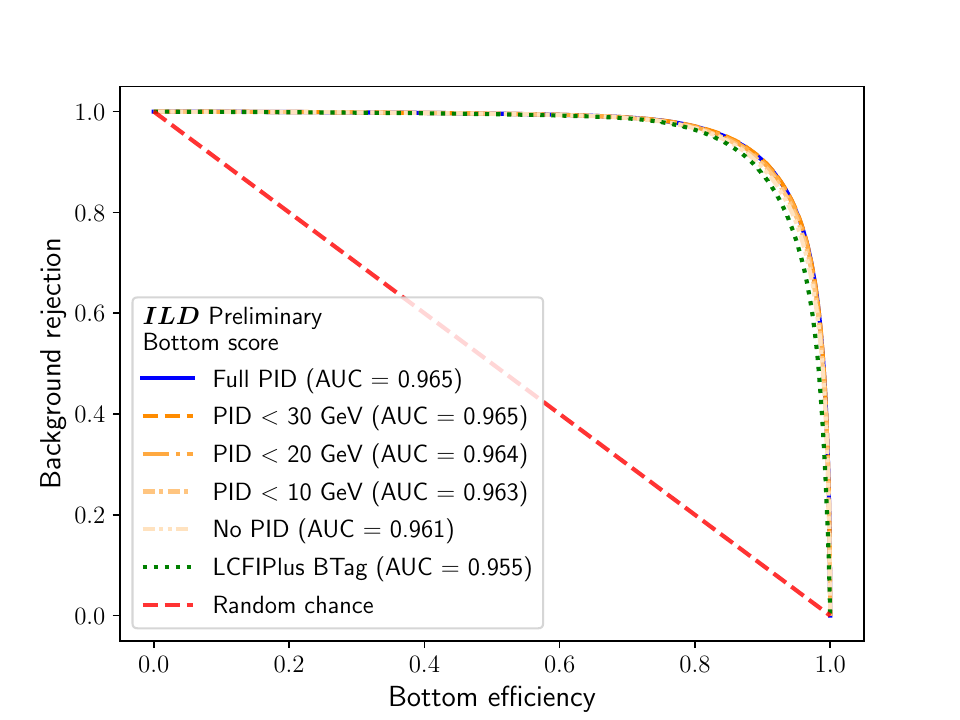}
        \caption{$b$-jet score}
    \end{subfigure}
    \hfill
    \begin{subfigure}{0.49\textwidth}
        \centering
        \includegraphics[width=1.\textwidth]{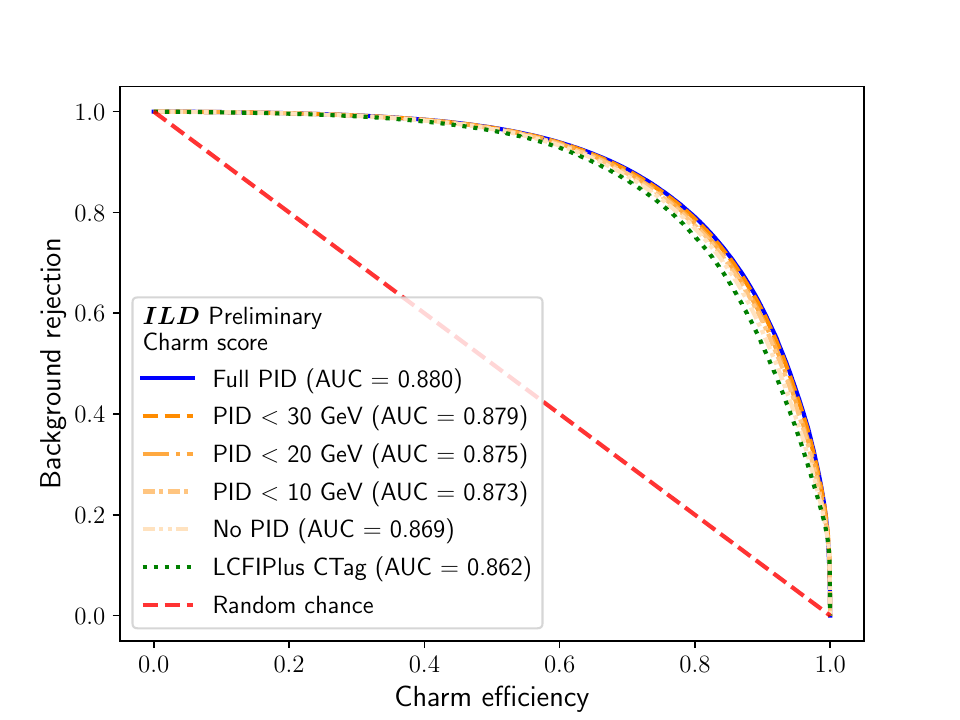}
        \caption{$c$-jet score}
    \end{subfigure} \\
    \begin{subfigure}{0.49\textwidth}
        \centering
        \includegraphics[width=1.\textwidth]{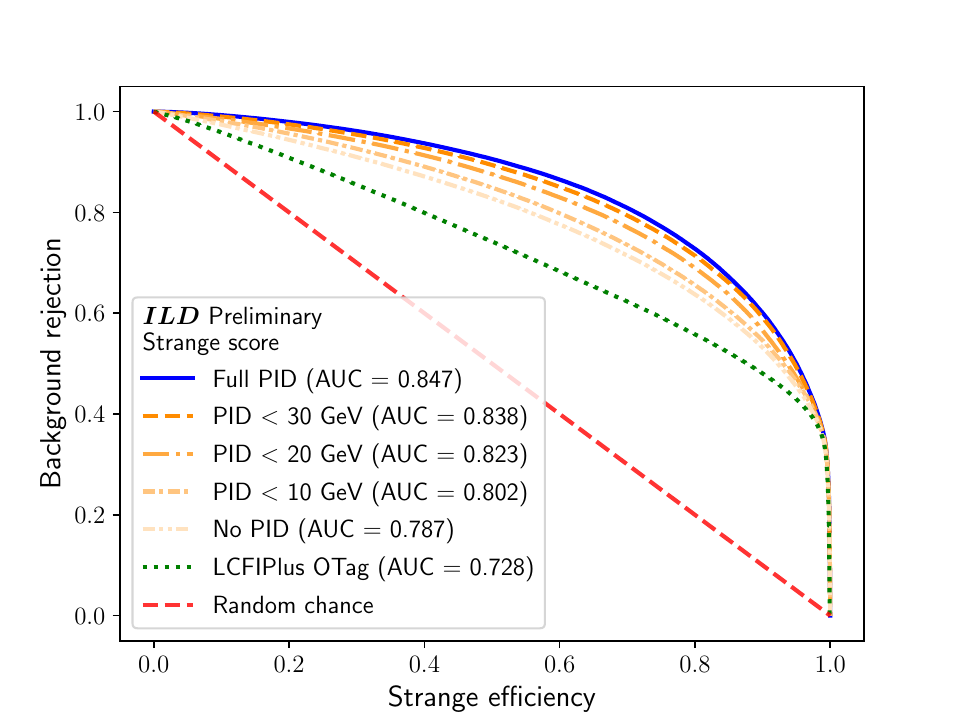}
        \caption{$s$-jet score}
    \end{subfigure}
    \hfill
    \begin{subfigure}{0.49\textwidth}
        \centering
        \includegraphics[width=1.\textwidth]{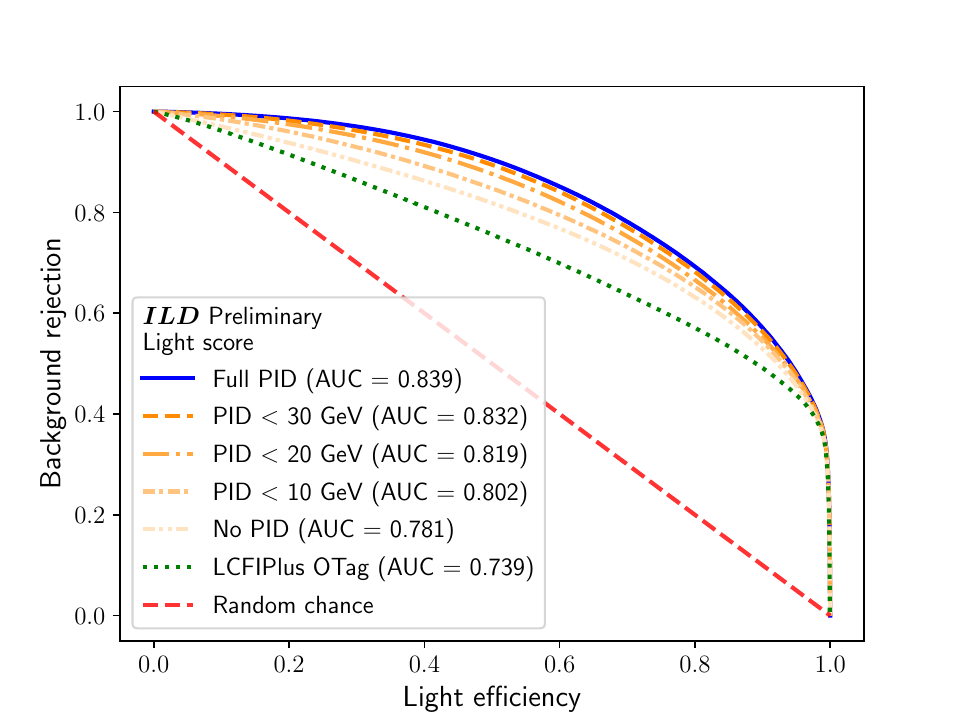}
        \caption{Light-jet score}
    \end{subfigure} \\
    \begin{subfigure}{0.49\textwidth}
        \centering
        \includegraphics[width=1.\textwidth]{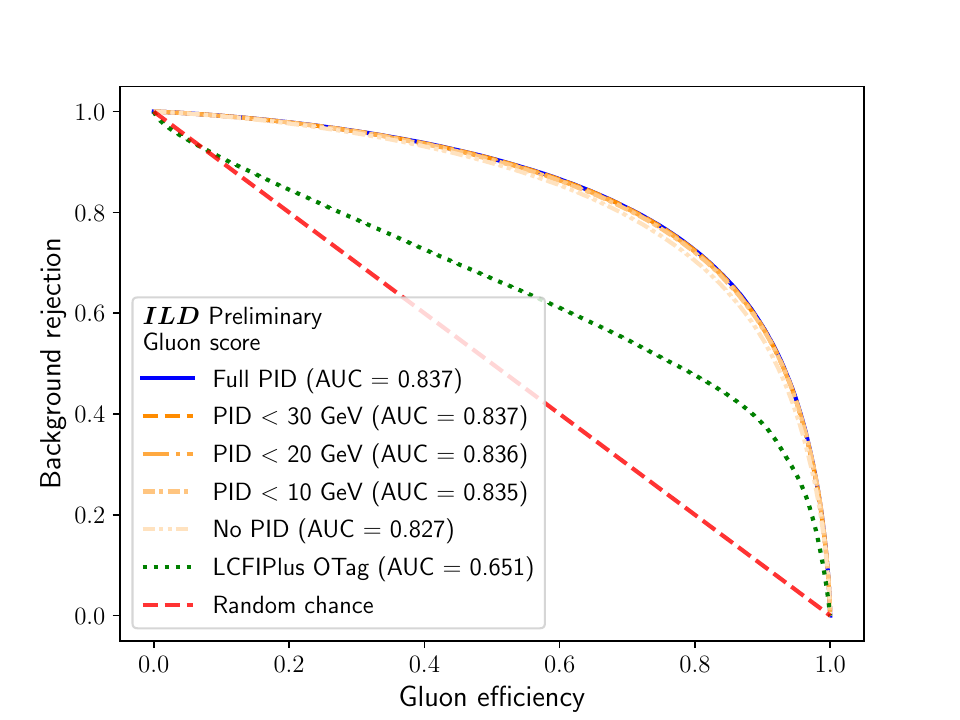}
        \caption{$g$-jet score}
    \end{subfigure} \\
    \caption{ROC curves for each output node of the jet flavour tagger with full PID (``Full PID''), as described in Section~\ref{sec:tagger}, as well as for the jet flavour taggers without PID (``No PID'') and with partial PID (``PID $< \unit[X]{GeV}$''), as described in Appendix~\ref{app:no_PID_tagger}. The sum-of-weights for each class is normalised to 1. The ``Background'' in a given plot corresponds to all classes not targeted by that node of the tagger. N.B. the blue and orange curves lie nearly on top of one another in (a), (b), and (e).}
    \label{fig:roc_no_PID}
\end{figure}

Shown in Fig.~\ref{fig:roc_no_PID} are the ROC curves for the taggers with PID, without PID, and with partial PID. As expected, there is a significant improvement in the separation of strange and light jets going from the tagger without PID to the tagger with PID: e.g., for 80\% background (i.e, bottom, charm, light, and gluon) rejection, the strange tagging efficiency improves from 60\% with no PID to 80\% with full PID. The improvement is marginal for bottom, charm, and gluon jets. It is also worth noting that the multiclassifier taggers have equal or better performance than LCFIPlus for all jet flavours, as expected. 

\begin{figure}[htbp]
    \centering
    \begin{subfigure}{0.49\textwidth}
        \centering
        \includegraphics[width=1.\textwidth]{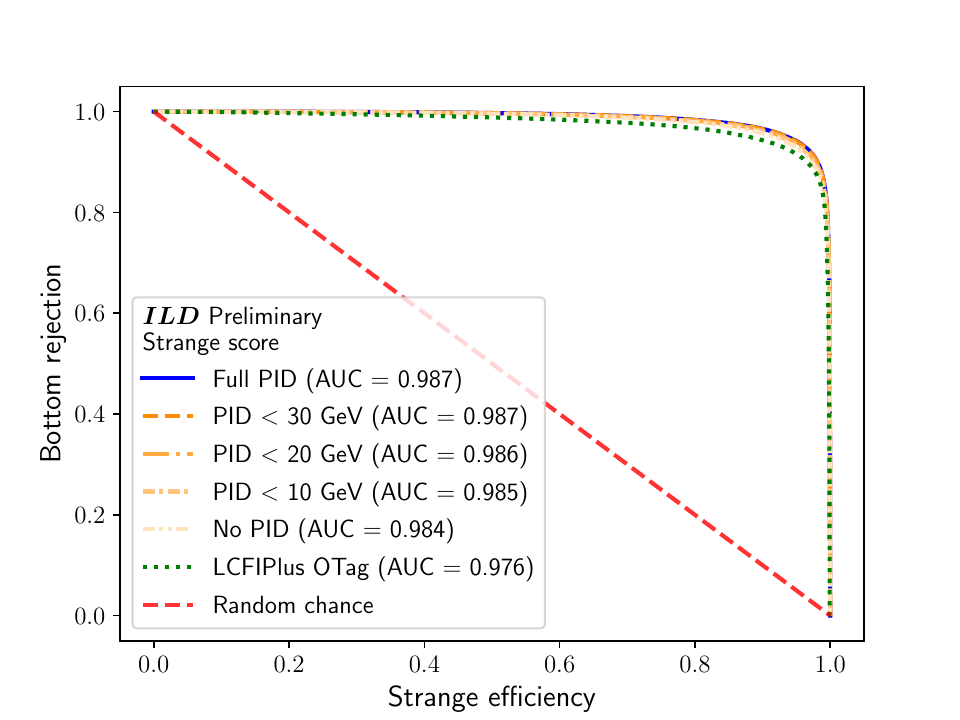}
        \caption{Strange vs. bottom, using $s$-jet score}
        \label{fig:roc_no_PID_binary_s_vs_b}
    \end{subfigure}
    \hfill
    \begin{subfigure}{0.49\textwidth}
        \centering
        \includegraphics[width=1.\textwidth]{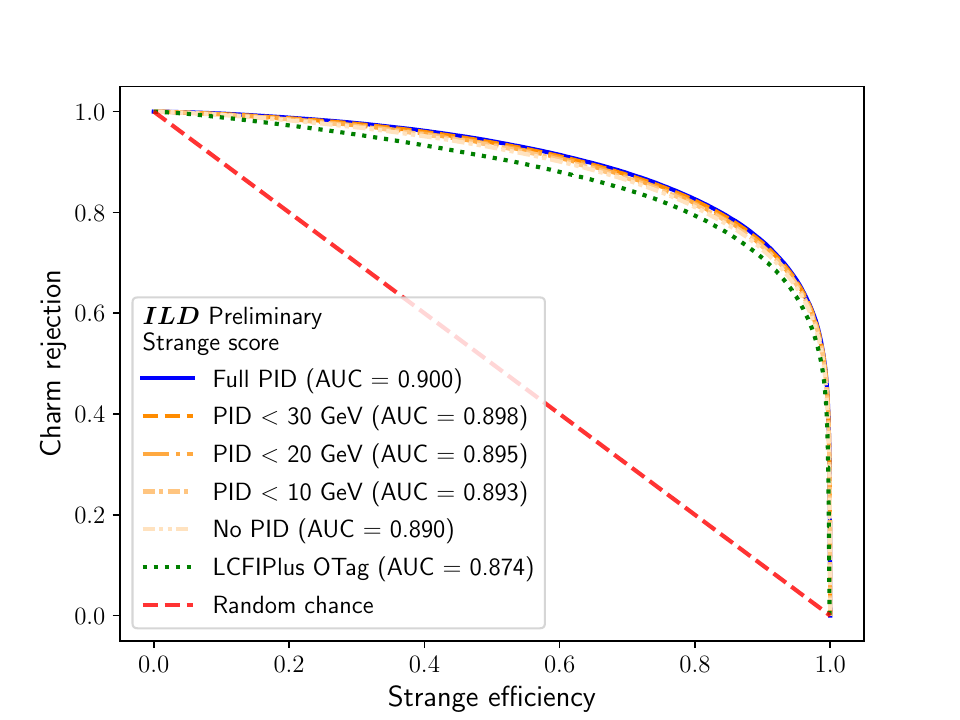}
        \caption{Strange vs. charm, using $s$-jet score}
        \label{fig:roc_no_PID_binary_s_vs_c}
    \end{subfigure} \\
    \begin{subfigure}{0.49\textwidth}
        \centering
        \includegraphics[width=1.\textwidth]{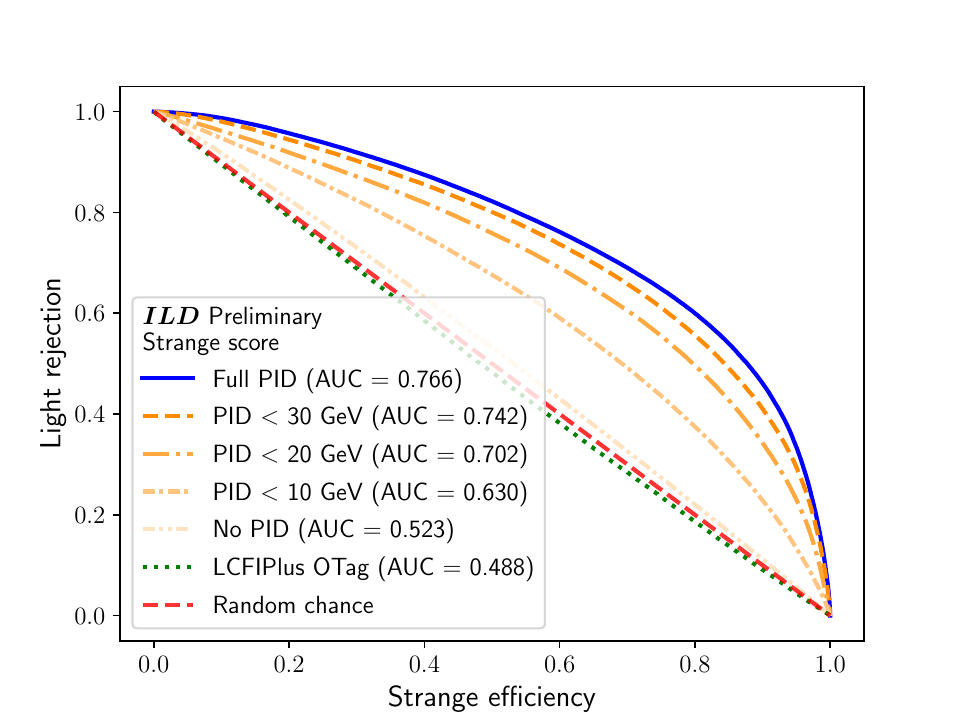}
        \caption{Strange vs. light, using $s$-jet score}
        \label{fig:roc_no_PID_binary_s_vs_ud}
    \end{subfigure}
    \hfill
    \begin{subfigure}{0.49\textwidth}
        \centering
        \includegraphics[width=1.\textwidth]{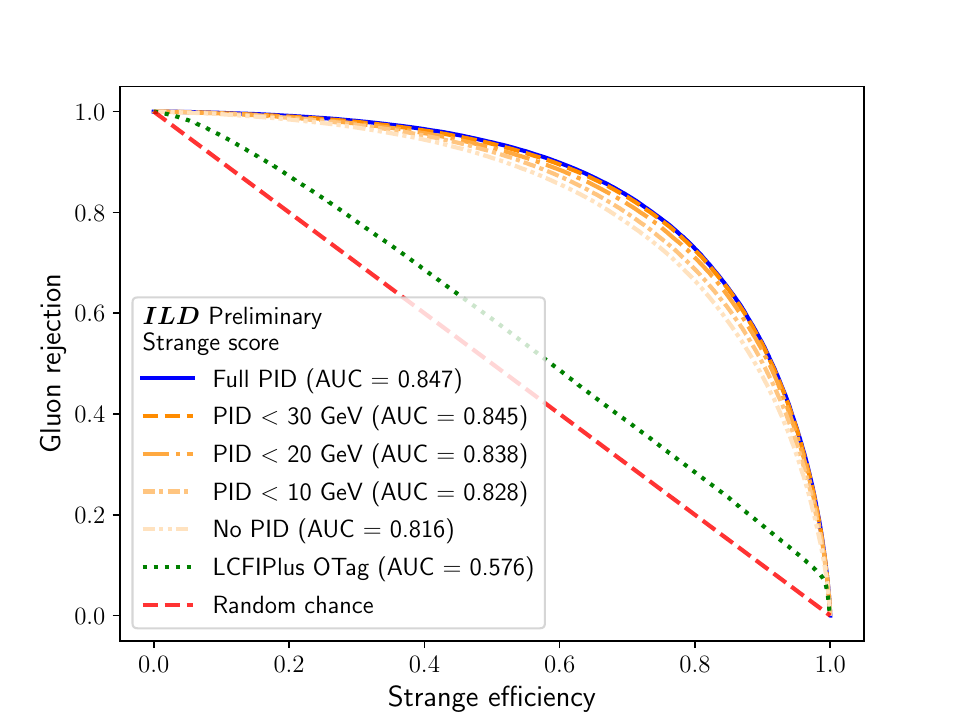}
        \caption{Strange vs. gluon, using $s$-jet score}
        \label{fig:roc_no_PID_binary_s_vs_g}
    \end{subfigure} \\
    \begin{subfigure}{0.49\textwidth}
        \centering
        \includegraphics[width=1.\textwidth]{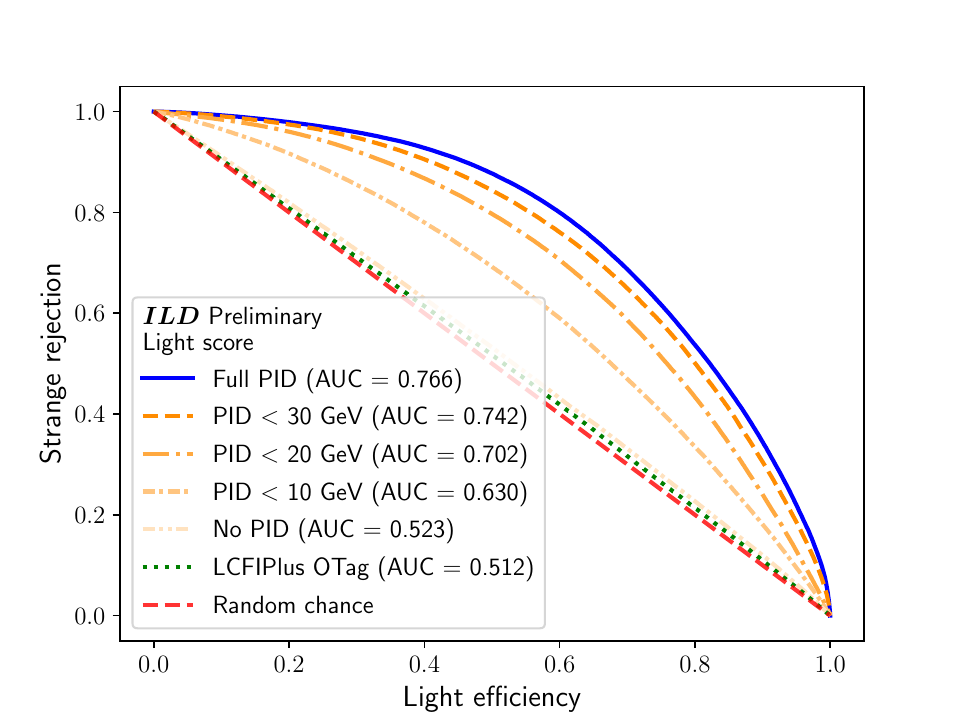}
        \caption{Light vs. strange, using light-jet score}
        \label{fig:roc_no_PID_binary_ud_vs_s}
    \end{subfigure} \\
    \caption{Pairwise ROC curves for various output nodes of the jet flavour tagger with full PID (``Full PID''), as described in Section~\ref{sec:tagger}, as well as for the jet flavour taggers without PID (``No PID'') and with partial PID (``PID $< \unit[X]{GeV}$''), as described in Appendix~\ref{app:no_PID_tagger}. The sum-of-weights for each class is normalised to 1. N.B. the blue and orange curves lie nearly on top of one another in (a) and (b).}
    \label{fig:roc_no_PID_binary}
\end{figure}

We have also included pairwise ROC curves where only two classes are considered at a time, one as signal (for which we are interested in efficiency) and the other as background (for which we are interested in rejection) -- we show 5 of the 120 possible combinations in Fig.~\ref{fig:roc_no_PID_binary}. The tagger scores used to generate each plot are modified as:

\begin{equation}
    f_\textrm{sig}(\vec{x}) = \frac{[\vec{F}(\vec{x})]_{i_\textrm{sig}}}{[\vec{F}(\vec{x})]_{i_\textrm{sig}} + [\vec{F}(\vec{x})]_{i_\textrm{bkg}}} \,,
\end{equation}

\noindent the so-called ``pairwise couplings of probabilities'', where $f_\textrm{sig}(\vec{x})$ is the binary classifier score, $\vec{F}(\vec{x})$ is the multiclassifier score (i.e., Eq.~\ref{eqn:tagger}), and $i_\textrm{sig}$ ($i_\textrm{bkg}$) is the index of the signal (background) class in the multiclassifier output vector. Effectively a renormalisation, these modified scores are more optimal than the multiclassifier outputs on their own. Figs.~\ref{fig:roc_no_PID_binary_s_vs_ud} and \ref{fig:roc_no_PID_binary_ud_vs_s} corroborate our conclusions on the effect of PID, where the tagger without PID and the LCFIPlus OTag do not perform better than random chance. For 80\% light rejection, the strange efficiency improves from 20\% with no PID to 60\% with full PID. As expected, the inclusion of PID does not significantly affect our ability to separate strange jets from bottom, charm, and gluon jets, as per Figs.~\ref{fig:roc_no_PID_binary_s_vs_b}, \ref{fig:roc_no_PID_binary_s_vs_c}, and \ref{fig:roc_no_PID_binary_s_vs_g}, respectively.

\subsection{Effect on Higgs to strange measurement}
\label{app:no_PID_tagger_ana}

A SM \Hss analysis following the same procedure as outlined in Section~\ref{sec:analysis} has been performed using the tagger trained without PID inputs. The same input samples, Table~\ref{tab:samples}, and analysis cuts, Table~\ref{tab:selections}, have been used. Accordingly, the cutflows for the \Zinv and \Zll channels are the same as Tables~\ref{tab:cutflow_Zinv} and \ref{tab:cutflow_Zll}, respectively.

\begin{figure}[htbp]
    \centering
    \begin{subfigure}{0.75\textwidth}
        \centering
        \includegraphics[width=\textwidth]{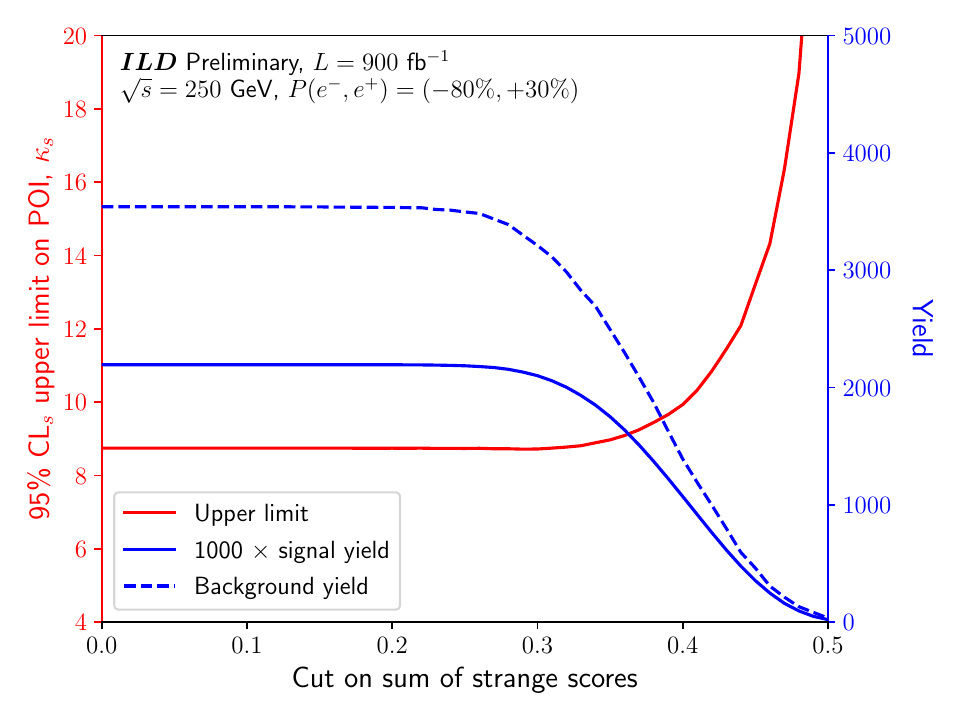}
        \caption{\Zinv channel}
    \end{subfigure} \\
    \begin{subfigure}{0.75\textwidth}
        \centering
        \includegraphics[width=\textwidth]{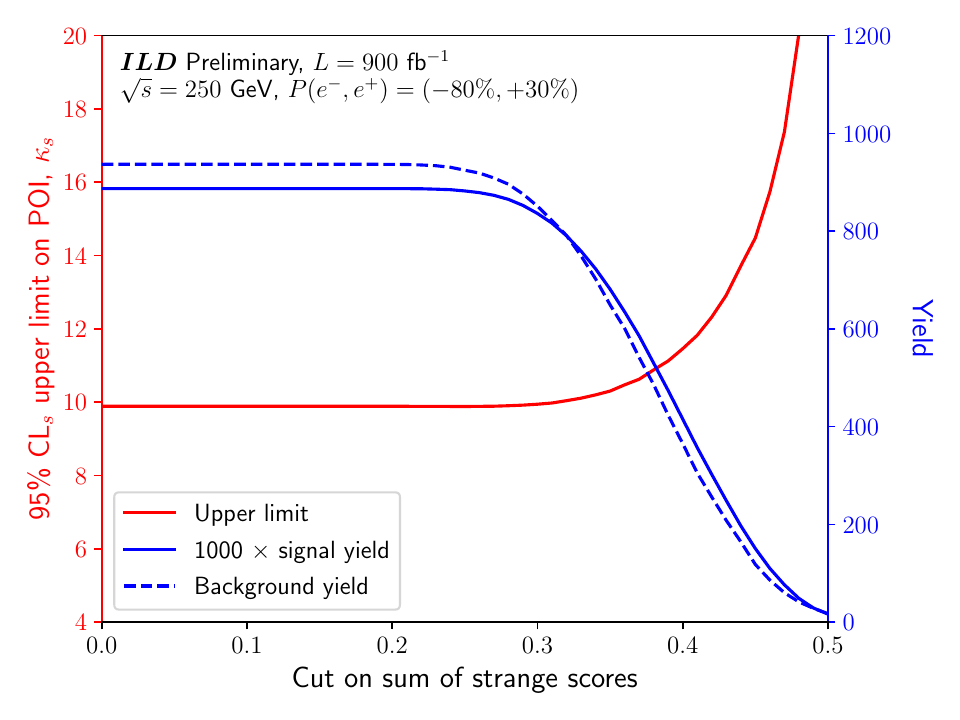}
        \caption{\Zll channel}
    \end{subfigure} \\
    \caption{Scans of the 95\% \CLs upper limit for the Higgs-strange coupling strength modifier, $\kappa_s$, obtained by varying the choice of the lower thresholds on the discriminant, Eq.~\ref{eqn:discriminant}, using the jet flavour tagger trained without PID, for both the \Zinv and \Zll channels. Also shown are the signal (i.e., $h(\rightarrow s\bar{s})Z(\rightarrow\ell\bar{\ell}/\nu\bar{\nu})$) and background (i.e., non-$h(\rightarrow s\bar{s})Z(\rightarrow\ell\bar{\ell}/\nu\bar{\nu})$) yields in the resulting regions.}
    \label{fig:limit_scan_no_PID}
\end{figure}

\begin{figure}[htbp]
    \centering
    \begin{subfigure}{\textwidth}
        \centering
        \includegraphics[width=0.53\textwidth]{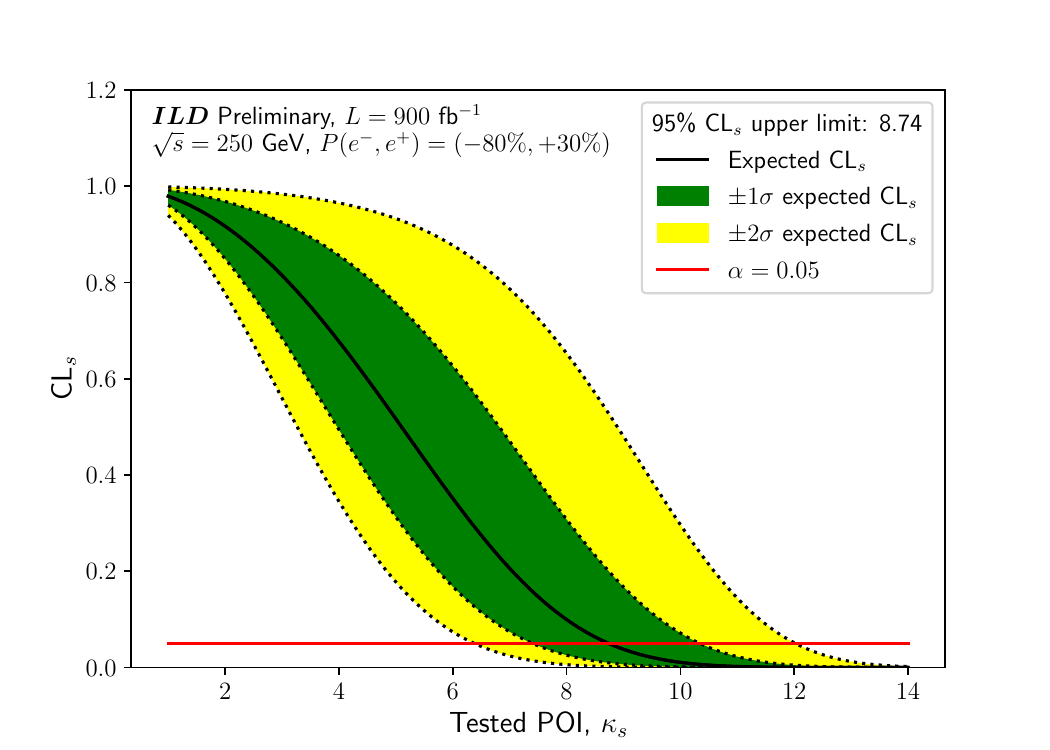}
        \caption{\Zinv channel}
    \end{subfigure} \\
    \begin{subfigure}{\textwidth}
        \centering
        \includegraphics[width=0.53\textwidth]{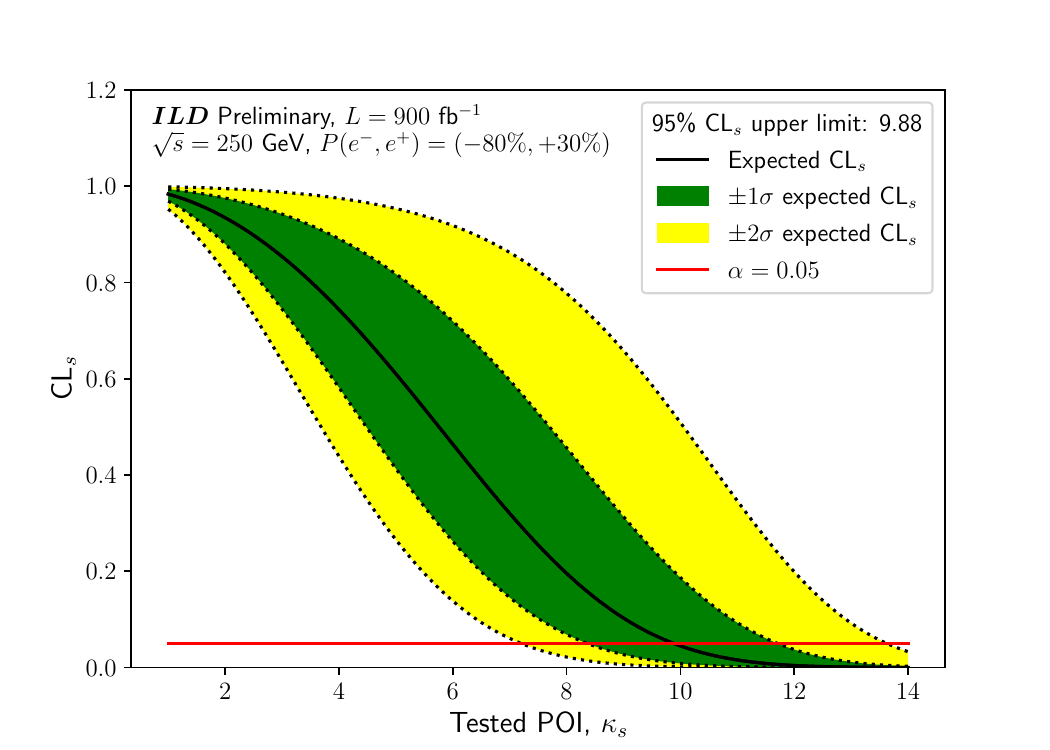}
        \caption{\Zll channel}
    \end{subfigure} \\
    \begin{subfigure}{\textwidth}
        \centering
        \includegraphics[width=0.53\textwidth]{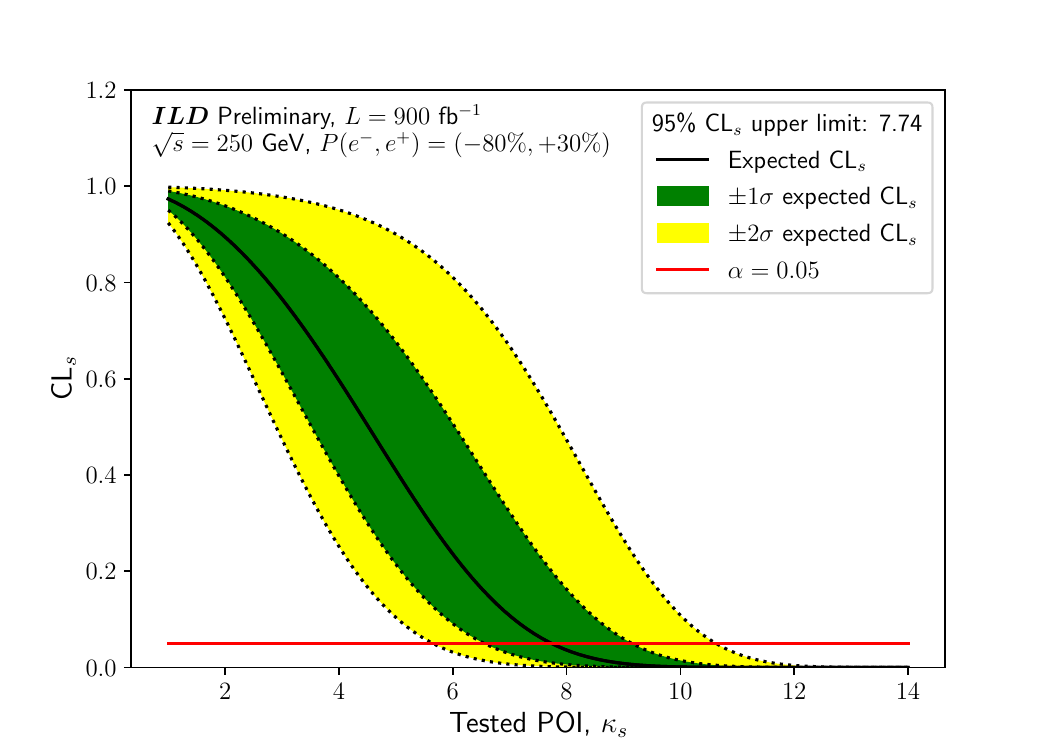}
        \caption{Combined}
    \end{subfigure} \\
    \caption{\CLs upper limit plots for the Higgs-strange coupling strength modifier, $\kappa_s$, obtained from fitting the signal regions described in Table~\ref{tab:selections} \emph{without} any additional cuts on the jet flavour tagger. Only a single bin is used for each channel -- the combination fit using both of these bins is also shown. The crossing of the black and red lines indicates the 95\% confidence level.}
    \label{fig:limits_no_PID}
\end{figure}

The same discriminant as before, the sum of the leading and subleading jet strange scores, is used to produce signal regions for each channel. Scans of the choice of lower threshold on this discriminant are performed for both channels and shown in Fig.~\ref{fig:limit_scan_no_PID}. We note that there is no additional discriminating power for \Hss using the jet flavour tagger trained without PID -- cutting more tightly on the discriminant does not improve the limits on $\kappa_s$ any more than \emph{not} cutting on it at all. Accordingly, we have performed single bin fits \textrm{without} any cuts on the discriminant for the \Zinv and \Zll channels as well as performed a combined fit using both channels -- the resulting limit plots for $\kappa_s$ are shown in Fig.~\ref{fig:limits_no_PID}.

From Fig.~\ref{fig:limits_no_PID}, we find the 95\% upper confidence bound on $\kappa_s$ is found to be 8.74 for the \Zinv channel and 9.88 for the \Zll channel, leading to a combined limit of 7.74. Compared to the combined limited achieved using a jet flavour tagger \emph{with} PID, 7.14, there is a $\sim$8\% degradation in the limit achieved using a jet flavour tagger \emph{without} PID. We conclude that jet tagging utilising PID offers gains in analyses targeting strange jets -- while the gains are small in the SM \Hss analysis performed in this paper, we expect the effect to be more pronounced in analyses with stronger expected signals and/or more luminosity.

\FloatBarrier

\section{Full cutflows for the Higgs to strange analysis}
\label{app:cutflows}

This appendix contains the full cutflow tables for the SM \Hss analysis described in Section~\ref{sec:analysis}. In particular, Table~\ref{tab:cutflow_Zinv} shows the cutflow for the \Zinv channel and Table~\ref{tab:cutflow_Zll} shows the cutflow for the \Zll channel.

\begin{sidewaystable}[htbp]
    \centering
    \caption{Cutflow for the \Zinv channel selections described in Table~\ref{tab:selections}. Shown are the central values in the yields as well as MC statistical uncertainties on those yields. The signal (background) efficiency is defined as the ratio of the $h(\rightarrow s\bar{s})Z(\rightarrow\ell\bar{\ell}/\nu\bar{\nu})$ (all non-$h(\rightarrow s\bar{s})Z(\rightarrow\ell\bar{\ell}/\nu\bar{\nu})$) yields at a given cut value to the yields at ``No cut''. The sum-of-weights per process is normalised to the SM cross section. N.B. ``hadr.'' $\coloneqq$ ``hadronic'' and ``lept.'' $\coloneqq$ ``leptonic'', and decimals are suppressed for yields $>$1000.}
    \label{tab:cutflow_Zinv}
    \resizebox{\textwidth}{!}{
        \begin{tabular}{ r | r r r r r r r r r r r r r | r | r r }
    \multicolumn{16}{l}{\textit{\textbf{ILD}} Preliminary, $\mathcal{L} = 900$~fb$^{-1}$, $\sqrt{s} = 250$~GeV, $P(e^-,e^+) = (-80\%,+30\%)$} \\
    \toprule
    & $(h \rightarrow s\bar{s})(Z \rightarrow \ell\bar{\ell}/\nu\bar{\nu})$ & $(h \rightarrow gg)(Z \rightarrow \ell\bar{\ell}/\nu\bar{\nu})$ & $(h \rightarrow u\bar{u}/d\bar{d})(Z \rightarrow \ell\bar{\ell}/\nu\bar{\nu})$ & $(h \rightarrow c\bar{c})(Z \rightarrow \ell\bar{\ell}/\nu\bar{\nu})$ & $(h \rightarrow b\bar{b})(Z \rightarrow \ell\bar{\ell}/\nu\bar{\nu})$ & $(h \rightarrow \mathrm{other})(Z \rightarrow \ell\bar{\ell})$ & $2f$ $Z$ hadr. & $4f$ $ZZ$ hadr. & $4f$ $WW$ hadr. & $4f$ $ZZ/WW$ hadr. & $2f$ $Z$ lept. & $4f$ $ZZ$ semilept. & $4f$ single $Z$ semilept. & Total bkg. & Sig. eff. & Bkg. eff. \\
    \midrule
    No cut & $16.95 \pm 0.02$ & $6870 \pm 9$ & $0.05 \pm 0.00$ & $2333 \pm 4$ & $46252 \pm 46$ & $8364 \pm 12$ & $67373852 \pm 13380$ & $739764 \pm 278$ & $7827170 \pm 2035$ & $6364669 \pm 1498$ & $11169172 \pm 2257$ & $441249 \pm 215$ & $668607 \pm 268$ & $94648302 \pm 13810$ & 1.00e+00 & 1.00e+00 \\
    No leptons & $10.24 \pm 0.02$ & $4404 \pm 6$ & $0.03 \pm 0.00$ & $1487 \pm 2$ & $29877 \pm 42$ & $168.66 \pm 1.74$ & $62055159 \pm 12841$ & $657885 \pm 262$ & $6768357 \pm 1893$ & $5506699 \pm 1394$ & $1188400 \pm 736$ & $119742 \pm 112$ & $225518 \pm 155$ & $76557698 \pm 13079$ & 6.04e-01 & 8.09e-01 \\
    $\geq\!2$ jets & $10.24 \pm 0.02$ & $4404 \pm 6$ & $0.03 \pm 0.00$ & $1487 \pm 2$ & $29877 \pm 42$ & $168.66 \pm 1.74$ & $62055154 \pm 12841$ & $657885 \pm 262$ & $6768357 \pm 1893$ & $5506699 \pm 1394$ & $1158631 \pm 727$ & $119742 \pm 112$ & $225518 \pm 155$ & $76527923 \pm 13079$ & 6.04e-01 & 8.09e-01 \\
    $p_{j_0} \in [40, 110]$~GeV & $10.13 \pm 0.02$ & $4370 \pm 6$ & $0.03 \pm 0.00$ & $1472 \pm 2$ & $29591 \pm 42$ & $161.65 \pm 1.70$ & $38383674 \pm 10099$ & $587745 \pm 247$ & $5759260 \pm 1746$ & $4704439 \pm 1288$ & $324262 \pm 384$ & $86634 \pm 95$ & $88308 \pm 97$ & $49969916 \pm 10341$ & 5.98e-01 & 5.28e-01 \\
    $p_{j_1} \in [30, 80]$~GeV & $9.91 \pm 0.01$ & $4188 \pm 6$ & $0.03 \pm 0.00$ & $1431 \pm 2$ & $28071 \pm 41$ & $137.98 \pm 1.57$ & $26985146 \pm 8468$ & $220115 \pm 151$ & $1143105 \pm 778$ & $989273 \pm 591$ & $77615 \pm 188$ & $63744 \pm 82$ & $31441 \pm 58$ & $29544267 \pm 8528$ & 5.85e-01 & 3.12e-01 \\
    $M_{jj} \in [120, 140]$~GeV & $7.59 \pm 0.01$ & $3187 \pm 5$ & $0.02 \pm 0.00$ & $992.58 \pm 1.76$ & $16366 \pm 32$ & $11.30 \pm 0.45$ & $2143270 \pm 2386$ & $1008 \pm 10$ & $3932 \pm 46$ & $3229 \pm 34$ & $9577 \pm 66$ & $8331 \pm 30$ & $4356 \pm 22$ & $2194261 \pm 2389$ & 4.48e-01 & 2.32e-02 \\
    $E_{jj} \in [125, 155]$~GeV & $7.49 \pm 0.01$ & $3148 \pm 5$ & $0.02 \pm 0.00$ & $979.57 \pm 1.74$ & $16133 \pm 31$ & $10.82 \pm 0.44$ & $777141 \pm 1437$ & $495.82 \pm 7.19$ & $1863 \pm 31$ & $1505 \pm 23$ & $8664 \pm 63$ & $4045 \pm 21$ & $3438 \pm 19$ & $817423 \pm 1440$ & 4.42e-01 & 8.64e-03 \\
    $M_\textrm{miss} \in [75, 120]$~GeV & $7.04 \pm 0.01$ & $2974 \pm 5$ & $0.02 \pm 0.00$ & $924.69 \pm 1.69$ & $15285 \pm 31$ & $9.87 \pm 0.42$ & $313724 \pm 913$ & $237.28 \pm 4.97$ & $981.66 \pm 22.79$ & $781.59 \pm 16.60$ & $7226 \pm 57$ & $2056 \pm 15$ & $2517 \pm 16$ & $346718 \pm 916$ & 4.15e-01 & 3.66e-03 \\
    ${\Delta}R_{jj,\textrm{miss}} \in [3.1, 4.0]$ & $6.06 \pm 0.01$ & $2560 \pm 5$ & $0.02 \pm 0.00$ & $796.17 \pm 1.57$ & $13161 \pm 28$ & $7.92 \pm 0.38$ & $47148 \pm 354$ & $79.41 \pm 2.88$ & $687.96 \pm 19.08$ & $514.59 \pm 13.47$ & $2662 \pm 35$ & $604.99 \pm 7.97$ & $109.63 \pm 3.43$ & $68332 \pm 358$ & 3.57e-01 & 7.22e-04 \\
    ${\Delta}\phi_{jj} > 1.25$~rad & $5.70 \pm 0.01$ & $2376 \pm 5$ & $0.02 \pm 0.00$ & $749.37 \pm 1.53$ & $12379 \pm 27$ & $7.29 \pm 0.36$ & $43176 \pm 339$ & $71.59 \pm 2.73$ & $626.57 \pm 18.21$ & $460.98 \pm 12.75$ & $2559 \pm 34$ & $563.80 \pm 7.70$ & $96.57 \pm 3.22$ & $63066 \pm 342$ & 3.37e-01 & 6.66e-04 \\
    $\textrm{score}_{b\textrm{-tag}}^{j_0} < 0.20$ & $5.51 \pm 0.01$ & $2169 \pm 4$ & $0.02 \pm 0.00$ & $583.88 \pm 1.35$ & $1064 \pm 8$ & $4.70 \pm 0.29$ & $27713 \pm 271$ & $29.59 \pm 1.76$ & $528.67 \pm 16.73$ & $394.67 \pm 11.80$ & $1671 \pm 28$ & $345.68 \pm 6.03$ & $84.36 \pm 3.01$ & $34589 \pm 274$ & 3.25e-01 & 3.65e-04 \\
    $\textrm{score}_{b\textrm{-tag}}^{j_1} < 0.20$ & $5.32 \pm 0.01$ & $2038 \pm 4$ & $0.02 \pm 0.00$ & $458.26 \pm 1.19$ & $114.19 \pm 2.64$ & $3.35 \pm 0.25$ & $23761 \pm 251$ & $21.99 \pm 1.51$ & $444.00 \pm 15.33$ & $342.12 \pm 10.98$ & $1163 \pm 23$ & $248.59 \pm 5.11$ & $73.12 \pm 2.80$ & $28669 \pm 253$ & 3.14e-01 & 3.03e-04 \\
    $\textrm{score}_{c\textrm{-tag}}^{j_0} < 0.35$ & $5.13 \pm 0.01$ & $1929 \pm 4$ & $0.02 \pm 0.00$ & $162.70 \pm 0.71$ & $58.22 \pm 1.89$ & $2.02 \pm 0.19$ & $20177 \pm 232$ & $17.72 \pm 1.36$ & $341.33 \pm 13.44$ & $292.04 \pm 10.15$ & $382.49 \pm 13.20$ & $189.86 \pm 4.47$ & $67.77 \pm 2.69$ & $23620 \pm 233$ & 3.03e-01 & 2.50e-04 \\
    $\textrm{score}_{c\textrm{-tag}}^{j_1} < 0.35$ & $4.96 \pm 0.01$ & $1835 \pm 4$ & $0.02 \pm 0.00$ & $67.23 \pm 0.46$ & $33.41 \pm 1.43$ & $1.22 \pm 0.15$ & $18250 \pm 220$ & $14.69 \pm 1.24$ & $242.37 \pm 11.33$ & $253.95 \pm 9.46$ & $195.57 \pm 9.44$ & $130.08 \pm 3.70$ & $61.88 \pm 2.57$ & $21085 \pm 221$ & 2.93e-01 & 2.23e-04 \\
    $y_{23} < 0.010$ & $3.23 \pm 0.01$ & $668.48 \pm 2.47$ & $0.01 \pm 0.00$ & $38.16 \pm 0.34$ & $14.39 \pm 0.94$ & $0.02 \pm 0.02$ & $5793 \pm 124$ & $0.31 \pm 0.18$ & $4.76 \pm 1.59$ & $4.59 \pm 1.27$ & $175.97 \pm 8.96$ & $21.01 \pm 1.49$ & $25.48 \pm 1.65$ & $6746 \pm 124$ & 1.90e-01 & 7.13e-05 \\
    $y_{34} < 0.002$ & $2.78 \pm 0.01$ & $362.10 \pm 1.81$ & $0.01 \pm 0.00$ & $31.04 \pm 0.31$ & $10.34 \pm 0.80$ & $0.02 \pm 0.02$ & $4475 \pm 109$ & $0.10 \pm 0.10$ & $1.59 \pm 0.92$ & $1.76 \pm 0.79$ & $172.32 \pm 8.86$ & $12.82 \pm 1.16$ & $18.20 \pm 1.40$ & $5085 \pm 109$ & 1.64e-01 & 5.37e-05 \\
    $N_\textrm{PFOs}^\textrm{event} \in [30, 60]$ & $2.34 \pm 0.01$ & $113.43 \pm 1.02$ & $0.01 \pm 0.00$ & $23.65 \pm 0.27$ & $5.26 \pm 0.57$ & $0.02 \pm 0.02$ & $3558 \pm 97$ & $0.10 \pm 0.10$ & $1.06 \pm 0.75$ & $1.06 \pm 0.61$ & $51.97 \pm 4.87$ & $11.03 \pm 1.08$ & $13.70 \pm 1.21$ & $3779 \pm 97$ & 1.38e-01 & 3.99e-05 \\
    $N_\textrm{PFOs}^{j_0} \in [10, 40]$ & $2.27 \pm 0.01$ & $109.57 \pm 1.00$ & $0.01 \pm 0.00$ & $23.14 \pm 0.27$ & $5.20 \pm 0.56$ & $0.02 \pm 0.02$ & $3460 \pm 96$ & $0.10 \pm 0.10$ & $1.06 \pm 0.75$ & $1.06 \pm 0.61$ & $42.40 \pm 4.40$ & $10.19 \pm 1.03$ & $9.31 \pm 1.00$ & $3662 \pm 96$ & 1.34e-01 & 3.87e-05 \\
    $N_\textrm{PFOs}^{j_1} \in [9, 37]$ & $2.19 \pm 0.01$ & $106.17 \pm 0.98$ & $0.01 \pm 0.00$ & $22.55 \pm 0.26$ & $4.77 \pm 0.54$ & $0.02 \pm 0.02$ & $3348 \pm 94$ & $0.10 \pm 0.10$ & $1.06 \pm 0.75$ & $1.06 \pm 0.61$ & $39.66 \pm 4.25$ & $9.67 \pm 1.01$ & $7.49 \pm 0.90$ & $3541 \pm 94$ & 1.29e-01 & 3.74e-05 \\
    \bottomrule
\end{tabular}

    }
\end{sidewaystable}

\begin{sidewaystable}[htbp]
    \centering
    \caption{Cutflow for the \Zll channel selections described in Table~\ref{tab:selections}. Shown are the central values in the yields as well as MC statistical uncertainties on those yields. The signal (background) efficiency is defined as the ratio of the $h(\rightarrow s\bar{s})Z(\rightarrow\ell\bar{\ell}/\nu\bar{\nu})$ (all non-$h(\rightarrow s\bar{s})Z(\rightarrow\ell\bar{\ell}/\nu\bar{\nu})$) yields at a given cut value to the yields at ``No cut''. The sum-of-weights per process is normalised to the SM cross section. N.B. ``hadr.'' $\coloneqq$ ``hadronic'' and ``lept.'' $\coloneqq$ ``leptonic'', and decimals are suppressed for yields $>$1000.}
    \label{tab:cutflow_Zll}
    \resizebox{\textwidth}{!}{
        \begin{tabular}{ r | r r r r r r r r r r r r r | r | r r }
    \multicolumn{16}{l}{\textit{\textbf{ILD}} Preliminary, $\mathcal{L} = 900$~fb$^{-1}$, $\sqrt{s} = 250$~GeV, $P(e^-,e^+) = (-80\%,+30\%)$} \\
    \toprule
    & $(h \rightarrow s\bar{s})(Z \rightarrow \ell\bar{\ell}/\nu\bar{\nu})$ & $(h \rightarrow gg)(Z \rightarrow \ell\bar{\ell}/\nu\bar{\nu})$ & $(h \rightarrow u\bar{u}/d\bar{d})(Z \rightarrow \ell\bar{\ell}/\nu\bar{\nu})$ & $(h \rightarrow c\bar{c})(Z \rightarrow \ell\bar{\ell}/\nu\bar{\nu})$ & $(h \rightarrow b\bar{b})(Z \rightarrow \ell\bar{\ell}/\nu\bar{\nu})$ & $(h \rightarrow \mathrm{other})(Z \rightarrow \ell\bar{\ell})$ & $2f$ $Z$ hadr. & $4f$ $ZZ$ hadr. & $4f$ $WW$ hadr. & $4f$ $ZZ/WW$ hadr. & $2f$ $Z$ lept. & $4f$ $ZZ$ semilept. & $4f$ single $Z$ semilept. & Total bkg. & Sig. eff. & Bkg. eff. \\
    \midrule
    No cut & $16.95 \pm 0.02$ & $6870 \pm 9$ & $0.05 \pm 0.00$ & $2333 \pm 4$ & $46252 \pm 46$ & $8364 \pm 12$ & $67373852 \pm 13380$ & $739764 \pm 278$ & $7827170 \pm 2035$ & $6364669 \pm 1498$ & $11169172 \pm 2257$ & $441249 \pm 215$ & $668607 \pm 268$ & $94648302 \pm 13810$ & 1.00e+00 & 1.00e+00 \\
    $\geq\!2$ leptons & $4.53 \pm 0.01$ & $1755 \pm 6$ & $0.01 \pm 0.00$ & $615.40 \pm 3.35$ & $12105 \pm 15$ & $7351 \pm 12$ & $354596 \pm 971$ & $4806 \pm 22$ & $67568 \pm 189$ & $56529 \pm 141$ & $7903040 \pm 1898$ & $216573 \pm 151$ & $212173 \pm 151$ & $8837113 \pm 2156$ & 2.68e-01 & 9.34e-02 \\
    Leading 2 leptons are SFOS & $3.73 \pm 0.01$ & $1527 \pm 5$ & $0.01 \pm 0.00$ & $526.32 \pm 3.10$ & $10498 \pm 14$ & $4563 \pm 9$ & $111672 \pm 545$ & $2444 \pm 16$ & $34712 \pm 136$ & $29967 \pm 103$ & $6850848 \pm 1767$ & $148515 \pm 125$ & $149319 \pm 126$ & $7344593 \pm 1866$ & 2.20e-01 & 7.76e-02 \\
    $\geq\!2$ jets & $3.73 \pm 0.01$ & $1527 \pm 5$ & $0.01 \pm 0.00$ & $526.32 \pm 3.10$ & $10498 \pm 14$ & $4539 \pm 9$ & $111664 \pm 545$ & $2444 \pm 16$ & $34712 \pm 136$ & $29967 \pm 103$ & $6410375 \pm 1710$ & $148515 \pm 125$ & $149319 \pm 126$ & $6904088 \pm 1811$ & 2.20e-01 & 7.29e-02 \\
    $p_{j_0} \in [60, 105]$~GeV & $3.52 \pm 0.01$ & $1385 \pm 5$ & $0.01 \pm 0.00$ & $491.64 \pm 3.00$ & $9313 \pm 13$ & $1966 \pm 6$ & $43192 \pm 339$ & $2071 \pm 15$ & $27602 \pm 121$ & $23835 \pm 92$ & $76723 \pm 187$ & $118578 \pm 112$ & $96574 \pm 102$ & $401730 \pm 443$ & 2.08e-01 & 4.24e-03 \\
    $p_{j_1} \in [35, 75]$~GeV & $3.29 \pm 0.01$ & $1239 \pm 5$ & $0.01 \pm 0.00$ & $457.18 \pm 2.89$ & $8302 \pm 12$ & $1286 \pm 5$ & $15245 \pm 201$ & $1204 \pm 11$ & $13930 \pm 86$ & $12011 \pm 65$ & $2759 \pm 35$ & $67182 \pm 84$ & $50622 \pm 74$ & $174236 \pm 257$ & 1.94e-01 & 1.84e-03 \\
    $M_{jj} \in [115, 145]$~GeV & $2.82 \pm 0.01$ & $1052 \pm 4$ & $0.01 \pm 0.00$ & $381.75 \pm 2.64$ & $6644 \pm 11$ & $950.69 \pm 4.16$ & $1555 \pm 64$ & $321.89 \pm 5.79$ & $4165 \pm 47$ & $3525 \pm 35$ & $679.72 \pm 17.60$ & $12207 \pm 36$ & $13390 \pm 38$ & $44872 \pm 104$ & 1.66e-01 & 4.74e-04 \\
    $E_{jj} \in [130, 156]$~GeV & $2.65 \pm 0.01$ & $987.92 \pm 4.33$ & $0.01 \pm 0.00$ & $349.66 \pm 2.53$ & $5855 \pm 10$ & $880.97 \pm 4.00$ & $930.05 \pm 49.71$ & $237.49 \pm 4.97$ & $2974 \pm 40$ & $2519 \pm 30$ & $442.21 \pm 14.20$ & $9463 \pm 32$ & $10374 \pm 33$ & $35014 \pm 86$ & 1.56e-01 & 3.70e-04 \\
    ${\Delta}\phi_{jj} > 1.75$~rad & $2.38 \pm 0.01$ & $865.86 \pm 4.05$ & $0.01 \pm 0.00$ & $313.90 \pm 2.40$ & $5240 \pm 10$ & $731.30 \pm 3.65$ & $781.24 \pm 45.56$ & $183.40 \pm 4.37$ & $2138 \pm 34$ & $1818 \pm 25$ & $249.82 \pm 10.67$ & $7059 \pm 27$ & $8110 \pm 29$ & $27492 \pm 76$ & 1.41e-01 & 2.90e-04 \\
    $p_{\ell_0} \in [40, 90]$~GeV & $2.16 \pm 0.01$ & $795.57 \pm 3.89$ & $0.01 \pm 0.00$ & $286.55 \pm 2.29$ & $4792 \pm 9$ & $673.76 \pm 3.50$ & $244.47 \pm 25.49$ & $164.13 \pm 4.14$ & $1814 \pm 31$ & $1541 \pm 23$ & $113.06 \pm 7.18$ & $5539 \pm 24$ & $6795 \pm 27$ & $22758 \pm 60$ & 1.28e-01 & 2.40e-04 \\
    $p_{\ell_1} \in [20, 60]$~GeV & $2.05 \pm 0.01$ & $759.39 \pm 3.80$ & $0.01 \pm 0.00$ & $272.03 \pm 2.23$ & $4553 \pm 9$ & $644.16 \pm 3.43$ & $79.72 \pm 14.55$ & $127.44 \pm 3.64$ & $1285 \pm 26$ & $1124 \pm 20$ & $82.97 \pm 6.15$ & $4846 \pm 23$ & $5953 \pm 25$ & $19726 \pm 51$ & 1.21e-01 & 2.08e-04 \\
    $M_{\ell\ell} \in [80, 100]$~GeV & $1.74 \pm 0.01$ & $647.65 \pm 3.51$ & $0.01 \pm 0.00$ & $230.93 \pm 2.06$ & $3864 \pm 8$ & $549.67 \pm 3.16$ & $31.89 \pm 9.21$ & $41.27 \pm 2.07$ & $361.44 \pm 13.83$ & $341.77 \pm 10.98$ & $50.60 \pm 4.80$ & $3325 \pm 19$ & $3344 \pm 19$ & $12789 \pm 35$ & 1.02e-01 & 1.35e-04 \\
    $E_{\ell\ell} \in [85, 115]$~GeV & $1.72 \pm 0.01$ & $643.17 \pm 3.49$ & $0.01 \pm 0.00$ & $229.34 \pm 2.05$ & $3839 \pm 8$ & $545.63 \pm 3.15$ & $23.92 \pm 7.97$ & $35.43 \pm 1.92$ & $322.28 \pm 13.06$ & $311.08 \pm 10.47$ & $26.90 \pm 3.50$ & $2055 \pm 15$ & $2172 \pm 15$ & $10204 \pm 30$ & 1.02e-01 & 1.08e-04 \\
    $M_\textrm{recoil} \in [122, 155]$~GeV & $1.71 \pm 0.01$ & $638.53 \pm 3.48$ & $0.01 \pm 0.00$ & $227.30 \pm 2.04$ & $3813 \pm 8$ & $542.27 \pm 3.14$ & $23.92 \pm 7.97$ & $23.34 \pm 1.56$ & $279.95 \pm 12.17$ & $273.70 \pm 9.83$ & $22.79 \pm 3.22$ & $1447 \pm 12$ & $1429 \pm 12$ & $8721 \pm 27$ & 1.01e-01 & 9.21e-05 \\
    $\textrm{score}_{b\textrm{-tag}}^{j_0} < 0.1$ & $1.58 \pm 0.01$ & $549.82 \pm 3.23$ & $0.00 \pm 0.00$ & $135.09 \pm 1.57$ & $187.68 \pm 1.84$ & $381.73 \pm 2.64$ & $21.26 \pm 7.52$ & $16.67 \pm 1.32$ & $233.91 \pm 11.13$ & $234.19 \pm 9.09$ & $19.15 \pm 2.95$ & $1015 \pm 10$ & $1011 \pm 10$ & $3806 \pm 23$ & 9.31e-02 & 4.02e-05 \\
    $\textrm{score}_{b\textrm{-tag}}^{j_1} < 0.1$ & $1.46 \pm 0.01$ & $484.97 \pm 3.03$ & $0.00 \pm 0.00$ & $79.06 \pm 1.20$ & $17.55 \pm 0.56$ & $287.75 \pm 2.29$ & $18.60 \pm 7.03$ & $14.38 \pm 1.22$ & $175.69 \pm 9.64$ & $198.22 \pm 8.36$ & $16.41 \pm 2.74$ & $868.40 \pm 9.55$ & $854.44 \pm 9.56$ & $3015 \pm 20$ & 8.59e-02 & 3.19e-05 \\
    $\textrm{score}_{c\textrm{-tag}}^{j_0} < 0.3$ & $1.41 \pm 0.01$ & $464.67 \pm 2.97$ & $0.00 \pm 0.00$ & $30.51 \pm 0.75$ & $12.24 \pm 0.47$ & $245.95 \pm 2.12$ & $18.60 \pm 7.03$ & $13.55 \pm 1.19$ & $151.35 \pm 8.95$ & $182.70 \pm 8.03$ & $15.50 \pm 2.66$ & $782.87 \pm 9.07$ & $781.53 \pm 9.15$ & $2699 \pm 20$ & 8.31e-02 & 2.85e-05 \\
    $\textrm{score}_{c\textrm{-tag}}^{j_1} < 0.3$ & $1.36 \pm 0.01$ & $446.04 \pm 2.91$ & $0.00 \pm 0.00$ & $14.12 \pm 0.51$ & $9.31 \pm 0.41$ & $211.20 \pm 1.96$ & $18.60 \pm 7.03$ & $12.71 \pm 1.15$ & $109.01 \pm 7.60$ & $169.30 \pm 7.73$ & $15.04 \pm 2.62$ & $734.64 \pm 8.79$ & $736.35 \pm 8.88$ & $2476 \pm 19$ & 8.02e-02 & 2.62e-05 \\
    $y_{23} < 0.050$ & $1.26 \pm 0.01$ & $377.94 \pm 2.68$ & $0.00 \pm 0.00$ & $12.76 \pm 0.48$ & $8.63 \pm 0.39$ & $111.48 \pm 1.42$ & $15.94 \pm 6.51$ & $10.00 \pm 1.02$ & $95.26 \pm 7.10$ & $153.07 \pm 7.35$ & $15.04 \pm 2.62$ & $608.98 \pm 8.00$ & $610.88 \pm 8.09$ & $2020 \pm 17$ & 7.44e-02 & 2.13e-05 \\
    $y_{34} < 0.005$ & $1.15 \pm 0.01$ & $275.00 \pm 2.28$ & $0.00 \pm 0.00$ & $11.07 \pm 0.45$ & $7.34 \pm 0.36$ & $27.33 \pm 0.71$ & $13.29 \pm 5.94$ & $5.84 \pm 0.78$ & $67.21 \pm 5.96$ & $115.33 \pm 6.38$ & $15.04 \pm 2.62$ & $509.59 \pm 7.32$ & $489.37 \pm 7.24$ & $1536 \pm 15$ & 6.79e-02 & 1.62e-05 \\
    $N_\textrm{PFOs}^\textrm{event} \in [30, 70]$ & $1.02 \pm 0.01$ & $96.85 \pm 1.35$ & $0.00 \pm 0.00$ & $8.79 \pm 0.40$ & $5.06 \pm 0.30$ & $15.40 \pm 0.53$ & $5.31 \pm 3.76$ & $3.75 \pm 0.63$ & $40.22 \pm 4.61$ & $67.72 \pm 4.89$ & $3.65 \pm 1.29$ & $435.62 \pm 6.77$ & $415.50 \pm 6.67$ & $1098 \pm 12$ & 6.02e-02 & 1.16e-05 \\
    $N_\textrm{PFOs}^{j_0} \in [10, 40]$ & $0.94 \pm 0.00$ & $83.13 \pm 1.26$ & $0.00 \pm 0.00$ & $7.93 \pm 0.38$ & $4.44 \pm 0.28$ & $12.93 \pm 0.48$ & $2.66 \pm 2.66$ & $3.44 \pm 0.60$ & $40.22 \pm 4.61$ & $64.54 \pm 4.77$ & $2.74 \pm 1.12$ & $397.79 \pm 6.46$ & $378.88 \pm 6.37$ & $998.71 \pm 11.71$ & 5.56e-02 & 1.06e-05 \\
    $N_\textrm{PFOs}^{j_1} \in [10, 40]$ & $0.89 \pm 0.00$ & $76.67 \pm 1.21$ & $0.00 \pm 0.00$ & $7.47 \pm 0.37$ & $4.30 \pm 0.28$ & $11.89 \pm 0.47$ & $0.00 \pm 0.00$ & $3.02 \pm 0.56$ & $39.16 \pm 4.55$ & $62.43 \pm 4.69$ & $2.74 \pm 1.12$ & $375.10 \pm 6.28$ & $354.05 \pm 6.16$ & $936.82 \pm 11.11$ & 5.23e-02 & 9.90e-06 \\
    \bottomrule
\end{tabular}

    }
\end{sidewaystable}

\FloatBarrier

\section{Limits on light quark Yukawa couplings}
\label{app:light_couplings}

\begin{figure}[htbp]
    \centering
    \begin{subfigure}{\textwidth}
        \centering
        \includegraphics[width=0.9\textwidth]{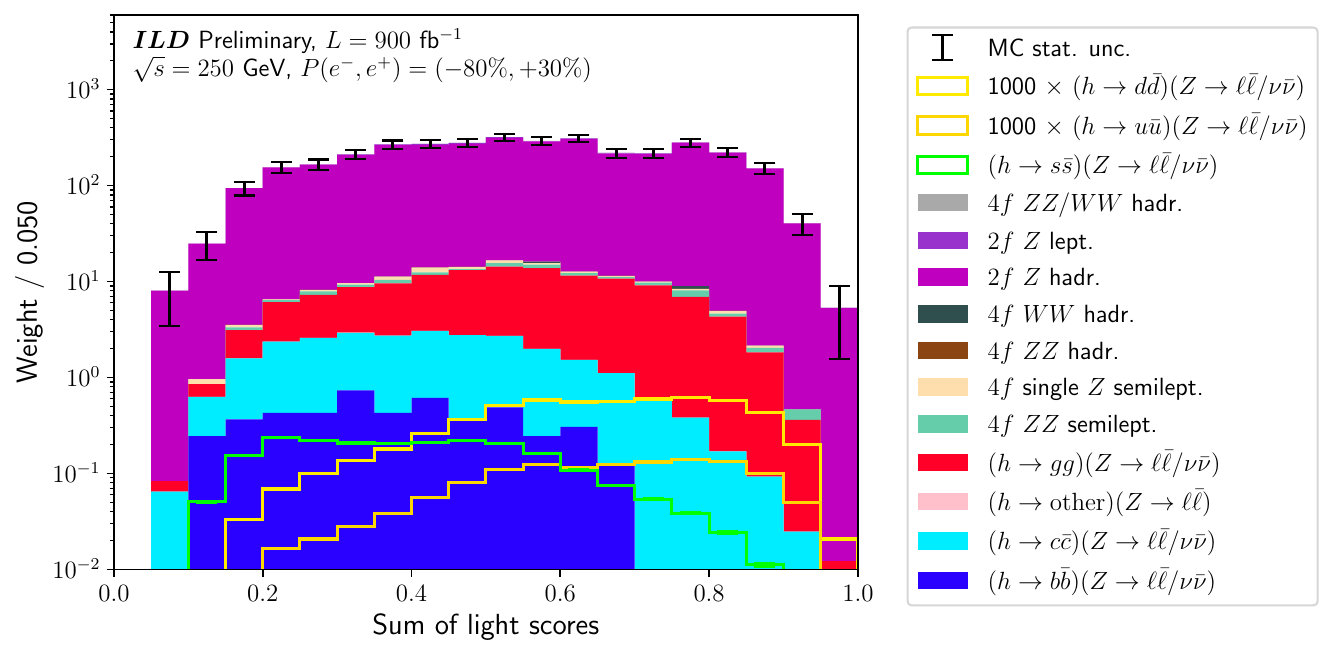}
        \caption{\Zinv channel}
    \end{subfigure} \\
    \vspace{0.5em}
    \begin{subfigure}{\textwidth}
        \centering
        \includegraphics[width=0.9\textwidth]{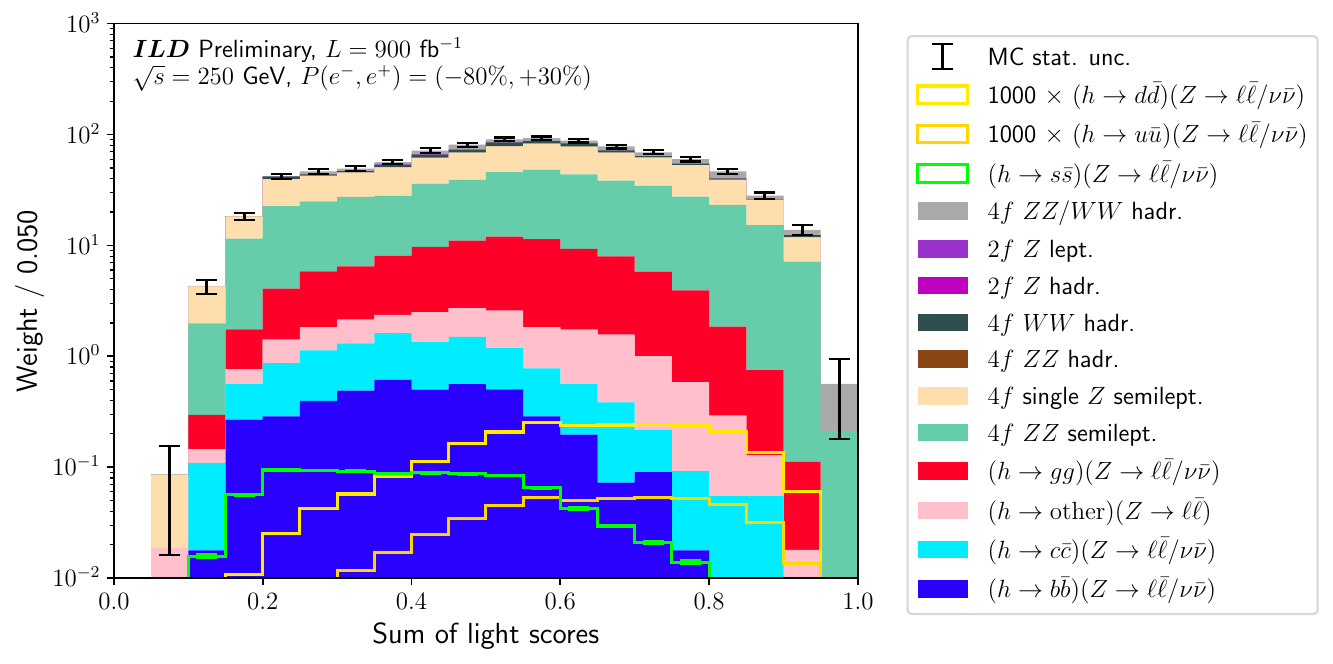}
        \caption{\Zll channel}
    \end{subfigure} \\
    \caption{Fit discriminants for each channel of the SM $h \rightarrow d\bar{d}$ and $h \rightarrow u\bar{u}$ analyses, Eq.~\ref{eqn:discriminant_light}. Each histogram is produced at the level of the last selection of their respective channel in Table~\ref{tab:selections}. The error bars represent the MC statistical uncertainties. The sum-of-weights per process is normalised to the SM cross section. N.B. the $h \rightarrow s\bar{s}$, $h \rightarrow d\bar{d}$, and $h \rightarrow u\bar{u}$ signals are unstacked, with the latter two scaled by a factor of 1,000.}
    \label{fig:discriminant_light}
\end{figure}

Without modifying the signal region selections (i.e., Table~\ref{tab:selections}) for the \Hss analysis presented in Section~\ref{sec:analysis}, we have estimated the 95\% upper confidence bounds on the Higgs-down quark Yukawa coupling, $\kappa_d$, and the Higgs-up quark Yukawa coupling, $\kappa_u$. The signal strength is modified as:

\begin{equation}
    \mu(\kappa_d) = \frac{\kappa_d^2}{\kappa_d^2 \times \BR[h \rightarrow d\bar{d}]_\textrm{SM} + (1 - \BR[h \rightarrow d\bar{d}]_\textrm{SM})} \,.
    \label{eqn:mu_d}
\end{equation}

\noindent for the $\kappa_d$ measurement, and as:
 
\begin{equation}
    \mu(\kappa_u) = \frac{\kappa_u^2}{\kappa_u^2 \times \BR[h \rightarrow u\bar{u}]_\textrm{SM} + (1 - \BR[h \rightarrow u\bar{u}]_\textrm{SM})} \,.
    \label{eqn:mu_u}
\end{equation}

\noindent for the $\kappa_u$ measurement. The signal region discriminant is modified to be the sum of the light-jet scores:

\begin{equation}
    \mathcal{D}(\vec{x}_{j_0}, \vec{x}_{j_1}) = \frac{1}{2} \times \left( [\vec{F}(\vec{x}_{j_0})]_\textrm{light} + [\vec{F}(\vec{x}_{j_1})]_\textrm{light} \right) \,,
    \label{eqn:discriminant_light}
\end{equation}

\noindent where the subscript ``light'' indicates the light-jet output node. The discriminant is shown for the \Zinv and \Zll channels in Fig.~\ref{fig:discriminant_light}.

The optimal cut on the discriminant is found to be $>$0.45 for both the \Zinv and \Zll channels. A two-bin measurement is constructed for each of the coupling strength modifiers, and the 95\% upper confidence bounds on $\kappa_d$ and $\kappa_u$ are found to be 146 and 313, respectively.

\begin{figure}[htbp]
    \centering
    \begin{subfigure}{\textwidth}
        \centering
        \includegraphics[width=0.6\textwidth]{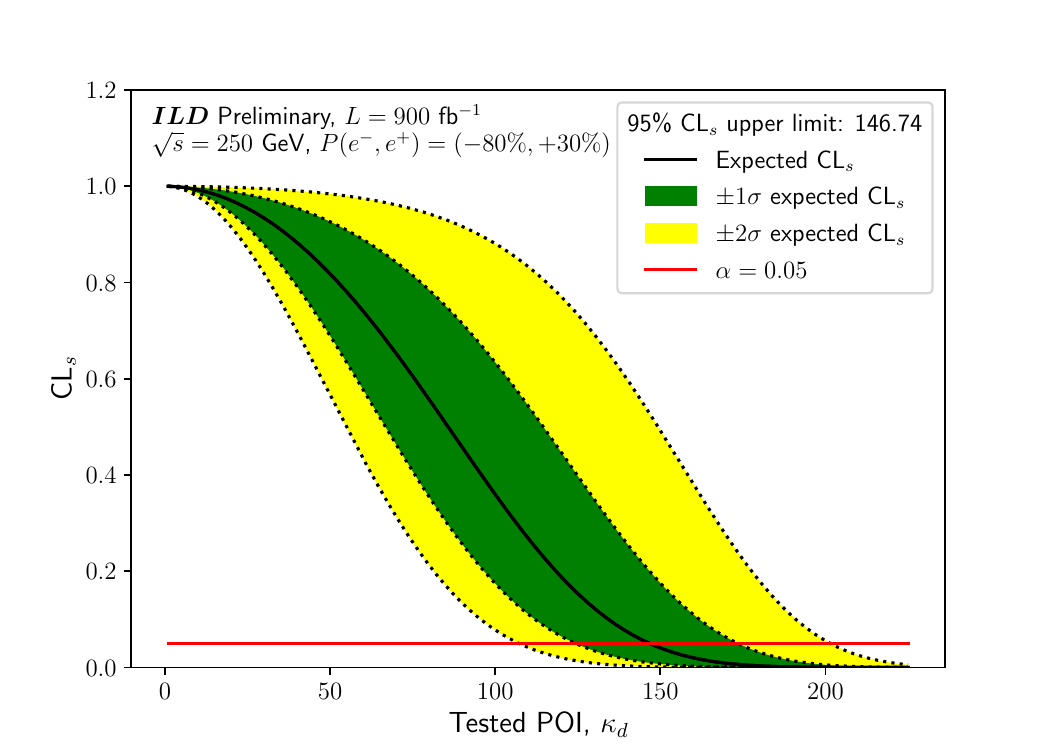}
        \caption{Limits on $\kappa_d$}
    \end{subfigure} \\
    \vspace{0.5em}
    \begin{subfigure}{\textwidth}
        \centering
        \includegraphics[width=0.6\textwidth]{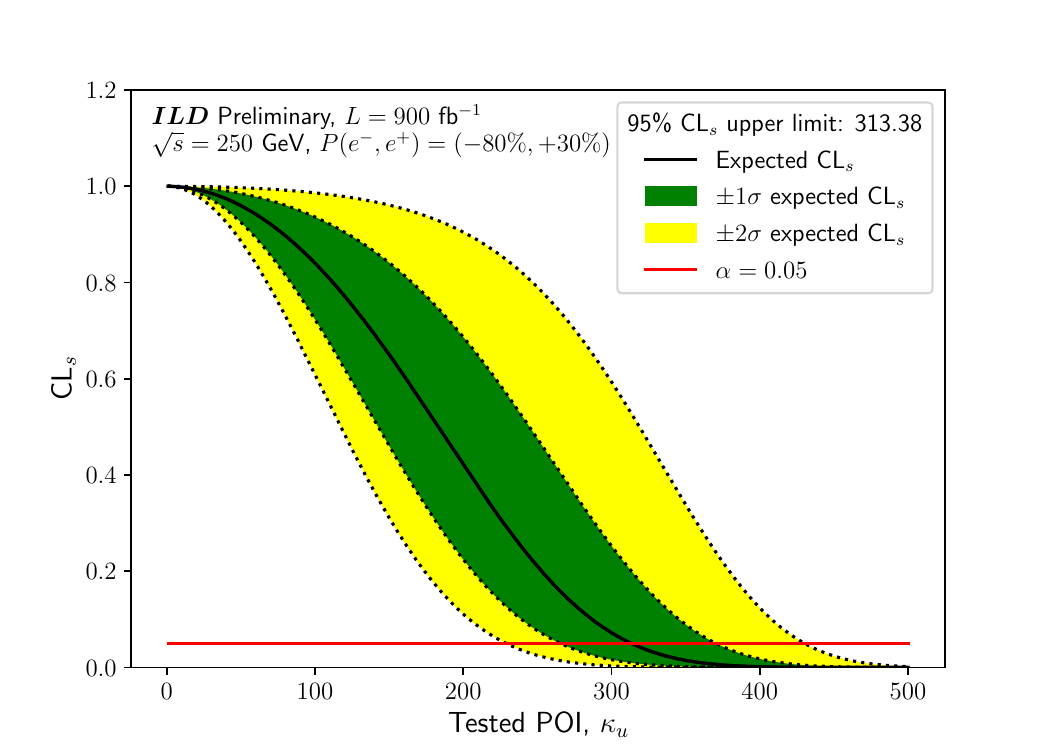}
        \caption{Limits on $\kappa_u$}
    \end{subfigure} \\
    \caption{\CLs upper limit plots for the measurement of the Higgs-down quark coupling strength modifier, $\kappa_d$, and the Higgs-up quark coupling strength modifier, $\kappa_u$. The results are obtained from fitting the signal discriminants in Fig.~\ref{fig:discriminant_light} in the regions where each discriminant is above 0.45. Only the combined fit of the \Zinv and \Zll channels is shown for each measurement. The crossing of the black and red lines indicates the 95\% confidence level.}
    \label{fig:limits_light}
\end{figure}

\begin{figure}[ht]
    \centering
    \includegraphics[width=0.8\textwidth]{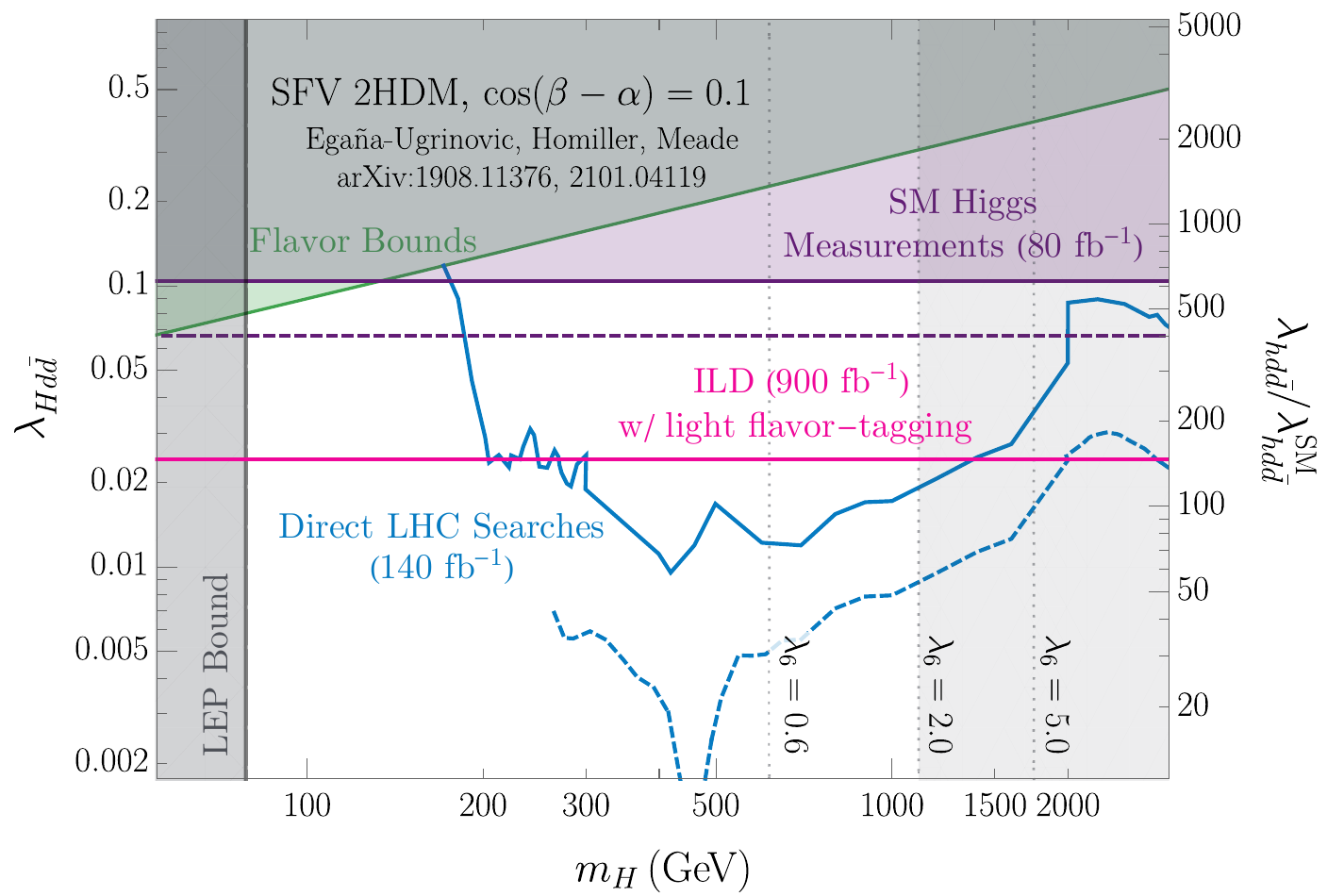}
    \caption{95\% CL bounds on the CP-even Higgs-down Yukawa coupling $\lambda_{Hd\bar{d}}$ as well as on \unit[125]{GeV} SM Higgs-down Yukawa coupling $\lambda_{hd\bar{d}}/\lambda_{hd\bar{d}}^\textrm{SM}$ (i.e., $\kappa_d$) for the SFV 2HDM described in Refs.~\cite{Egana-Ugrinovic:2019, Egana-Ugrinovic:2021}. The pink line shows the bounds obtained from the $h \rightarrow d\bar{d}$ analysis presented in this appendix.
    See the caption of Fig.~\ref{fig:exclusion_plot_2HDM} for further details.}
    \label{fig:exclusion_plot_2HDM_down}
\end{figure}

The relevance of the projected bounds on the light quark Yukawa couplings for a specific BSM model, an up-type SFV 2HDM (discussed in Section~\ref{sec:analysis}) with a CP-even Higgs $H$ with preferential couplings to the down quarks, is presented in Fig.~\ref{fig:exclusion_plot_2HDM_down}. The ILD bounds, based only on \unit[900]{\ifb} of the data foreseen at the ILC, compare favourably with current and future LHC limits and would provide the strongest limits for a second Higgs doublet with masses between approximately 80 and \unit[200]{GeV} within this model. 

\FloatBarrier

\section{Additional discussion on PID reach by various PID techniques}
\label{app:PID_reach}

\begin{figure}[htbp]
    \centering
    \includegraphics[width=0.8\textwidth]{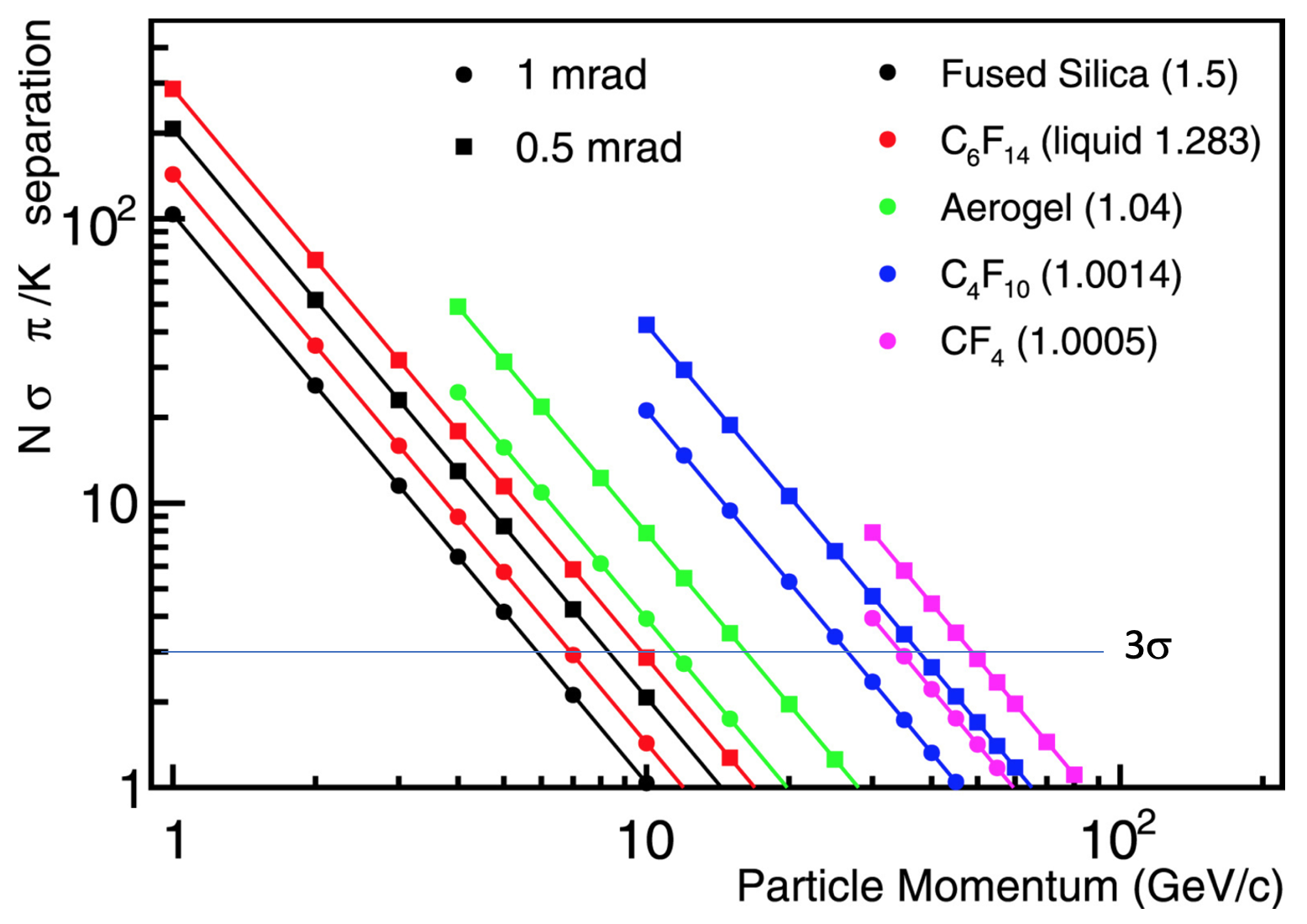}
    \caption{Expected $\pi$/$K$ separation reach in terms of number of sigma for various radiator choices and for two Cherenkov angle resolutions, $\sigma_\textrm{tot} = 0.5$ and \unit[1]{mrad}~\cite{Papanestis:2020, PapanestisConf}. In practice, the resolution tends to be worse when all contributions are included.}
    \label{fig:kaon_pion_separation}
\end{figure}

To reach a good $\pi$/$K$ separation at \unit[30--40]{GeV}, one has two choices: either a gaseous RICH detector or a very large TPC using $dE/dx$. To demonstrate this, we use Fig.~\ref{fig:kaon_pion_separation}, which shows the $\pi$/$K$ separation versus particle momentum for different radiators, solid, liquid, and gaseous, and two different values of total Cherenkov angle resolution, $\sigma_\textrm{tot} = 0.5$ and \unit[1]{mrad}~\cite{Papanestis:2020, PapanestisConf}. In practice, the resolution tends to be worse when all contributions are included.

\begin{figure}[htbp]
    \centering
    \includegraphics[width=\textwidth]{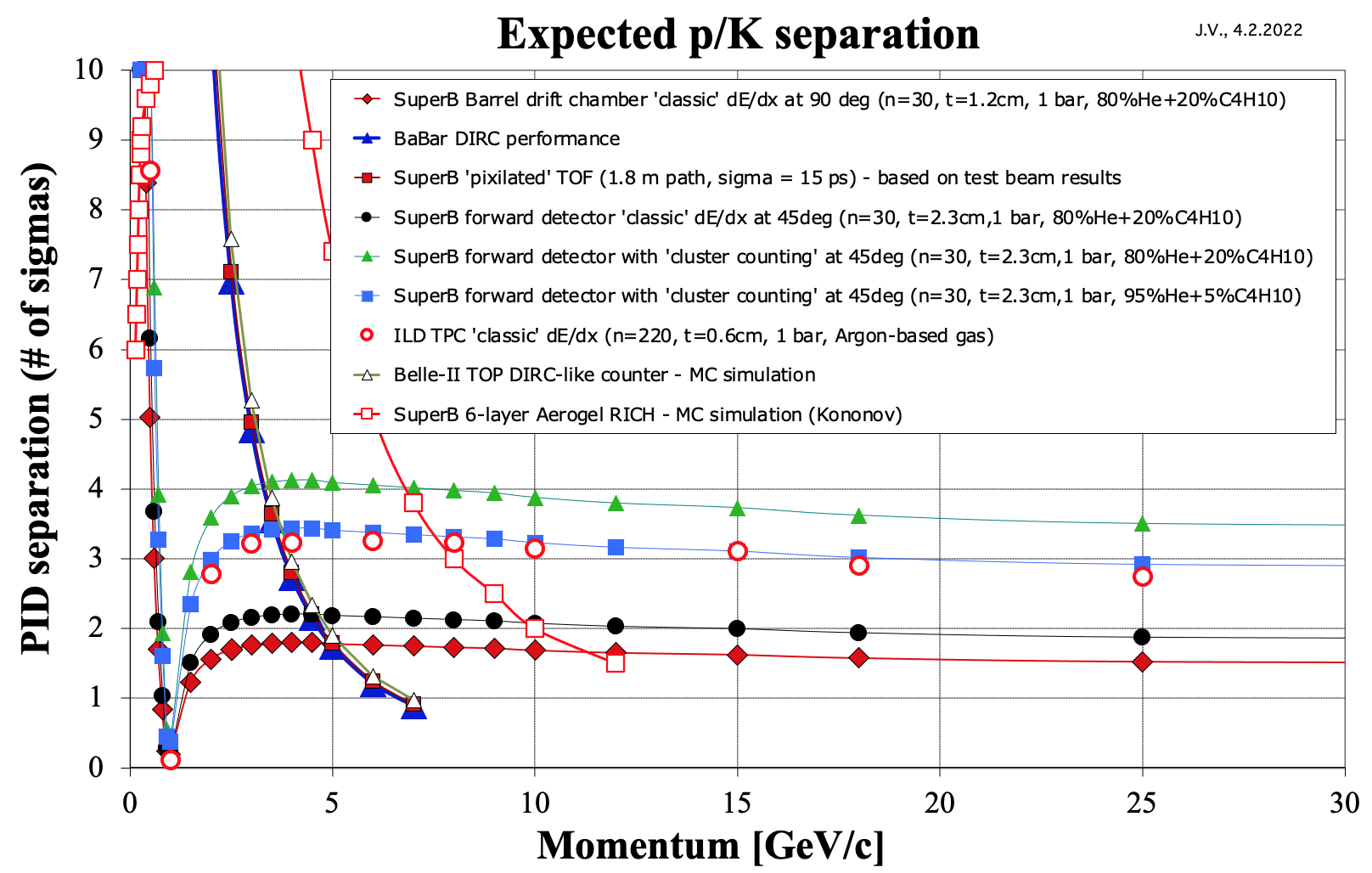}
    \caption{Expected $\pi$/$K$ separation~\cite{Vavra} reach in terms of number of sigma for TOF, $dE/dx$, BaBar DIRC, Belle-II TOP counter, aerogel detectors within SuperB~\cite{SuperB:2013} and Belle-II~\cite{BelleII:2010}, and ILD TPC $dE/dx$ MC simulation~\cite{ILDConceptGroup:2020}.}
    \label{fig:lots_of_kaon_pion_separation}
\end{figure}

Fig.~\ref{fig:lots_of_kaon_pion_separation} shows a PID performance~\cite{Vavra} for a TOF counter with \unit[1.8]{m} flight path, the SuperB drift chamber $dE/dx$, the BaBar DIRC, the Belle-II time-of-propagation (TOP) counter, aerogel detectors within SuperB~\cite{SuperB:2013} and Belle-II~\cite{BelleII:2010}, and ILD TPC $dE/dx$ MC simulation~\cite{ILDConceptGroup:2020}. We see that the cluster counting method improves the PID when compared to classical $dE/dx$ in the SuperB drift chamber; however, it has to be yet demonstrated that it is usable in the ILD environment (so far this technique was demonstrated only in test beams and is currently being investigated~\cite{Bedeschi:2022rnj} as an option for the FCC-ee detectors~\cite{Bernardi:2022hny}).

\FloatBarrier

\section{Additional discussion on SiPM noise}
\label{app:SiPM_noise}

\begin{figure}[htbp]
    \centering
    \includegraphics[width=\textwidth]{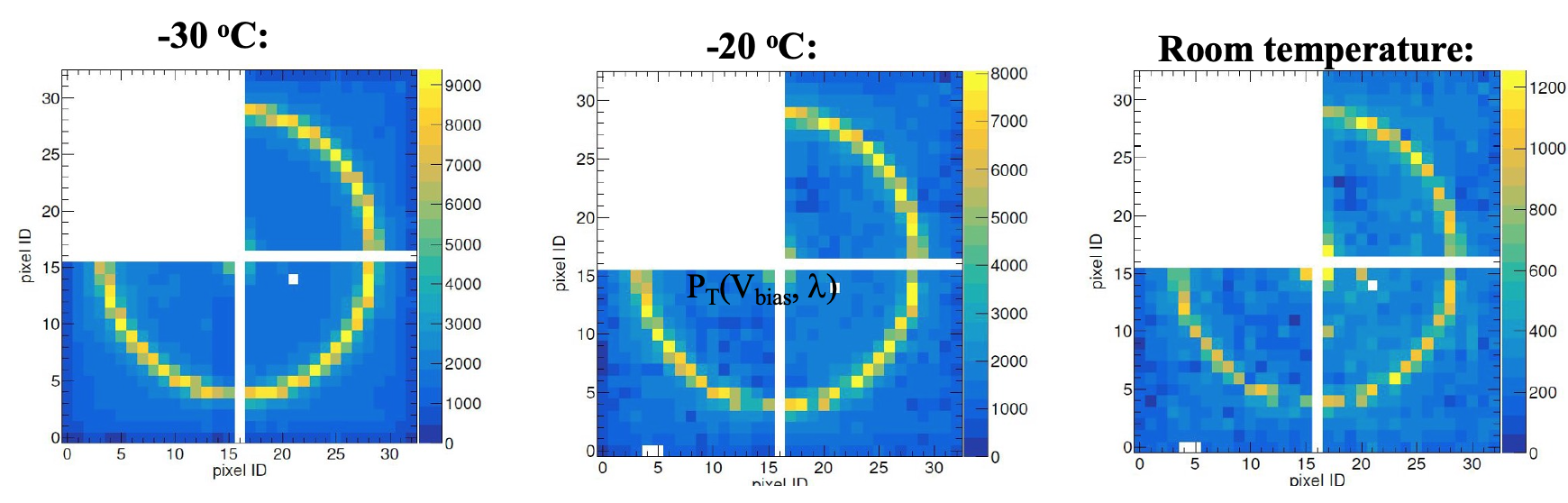}
    \caption{SiPM thermal random noise in images of Cherenkov rings as a function of temperature. These are results from EIC detector R\&D~\cite{HeAndSchwiening}. We are proposing to run the SiPMs at \unit[1--2]{\oC} and to eliminate the thermal noise by a timing cut.}
    \label{fig:SiPM_noise}
\end{figure}

The main advantage of SiPMs is that they can certainly operate at \unit[5]{T} and even at \unit[7]{T}~\cite{Espana:2017}. However, compared to an ideal photon-detector, the SiPM performance is affected by a random dark noise~\cite{Klanner:2018}. It was an open question until a few years ago if they are suitable for the RICH imaging application. However, several experiments proved that the noise can be managed by lowering the SiPM temperature. Fig.~\ref{fig:SiPM_noise} shows an example of aerogel electron-ion collider (EIC) RICH detector noise being controlled by temperature~\cite{HeAndSchwiening}. This noise gets worse if SiPMs are exposed to a total integrated neutron flux~\cite{Korpar:2020}. However, neutron backgrounds are predicted to be very low at SiD/ILD. Therefore, we believe that the SiPM thermal noise can be managed by a combination of running them at a relatively cool temperature of \unit[1--2]{\oC} and by a simple timing cut on a time difference between the SiPM and the beam crossing signals. The expected SiPM single photoelectron timing resolution of \unit[100--200]{ps} depends on the SiPM overvoltage and the electronics contribution~\cite{Korpar:2020} -- these parameters should be chosen carefully with a consideration to the noise and a large PDE.

\FloatBarrier

\section{Physics performance of the SLD CRID}
\label{app:SLD_CRID}

\begin{figure}[htbp]
    \centering
    \begin{subfigure}{0.49\textwidth}
        \centering
        \includegraphics[width=1.\textwidth]{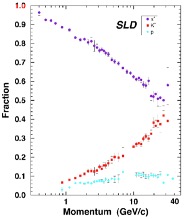}
        \caption{}
    \end{subfigure}
    \hfill
    \begin{subfigure}{0.49\textwidth}
        \centering
        \includegraphics[width=1.\textwidth]{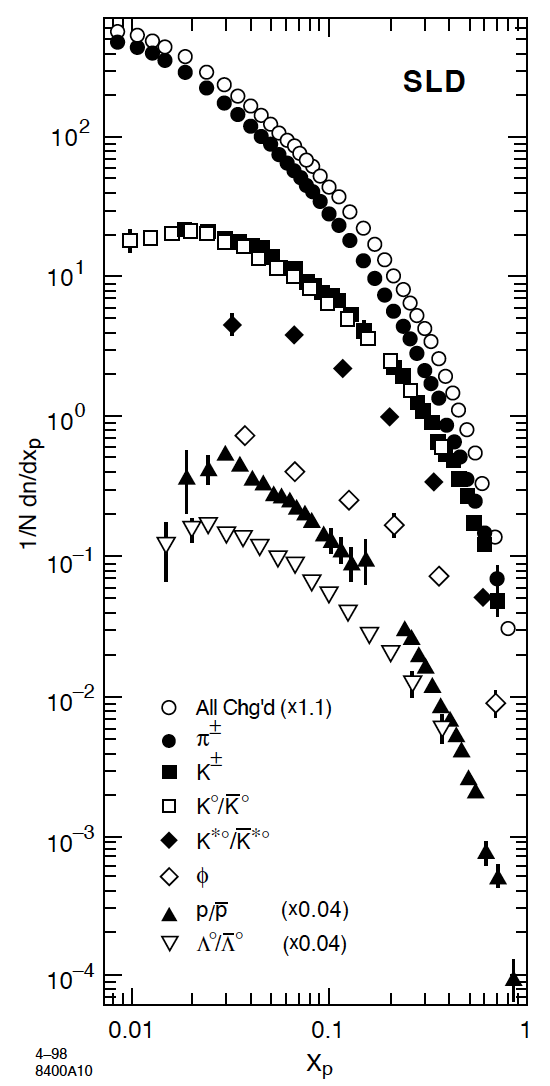}
        \caption{}
    \end{subfigure} \\
    \caption{(a) $\pi$/$K$/$p$ fractions determined by the SLD CRID~\cite{Muller}. (b) Differential cross sections as a function of hadronic momentum fraction $x_p$ per hadronic $Z^0$ decay, by all SLD detectors~\cite{SLD:1998}.}
    \label{fig:SLD_CRID_piKp}
\end{figure}

Fig.~\ref{fig:SLD_CRID_piKp} demonstrates the physics achieved with a \unit[4.3]{mrad} Cherenkov angle resolution at the SLD CRID~\cite{Muller, SLD:1998}.

\end{document}